\documentclass[superscriptaddress,showpacs,floatfix]{report}

\usepackage[numbers,sort&compress]{natbib}
\usepackage{amsmath}
\usepackage{amsthm}

\usepackage{authblk}
\usepackage{hyperref}
\usepackage{color}
\usepackage{graphicx}	
\usepackage[utf8]{inputenc} 
\usepackage[nottoc,notlot,notlof]{tocbibind}
\usepackage[toc,page]{appendix}

\usepackage{makeidx}
\graphicspath{%
  {./},%
}

\usepackage{amstext}
\usepackage{amssymb}
\usepackage{amsmath}
\usepackage{array}
\usepackage{enumitem} 
\usepackage{bm}
\usepackage{upgreek}

\usepackage{xspace}	

\newcommand{\mt}{\left(\nabla^\perp_\lambda u^\lambda\right)}

\makeindex

\begin{document}

\title{Relativistic Fluid Dynamics In and Out of Equilibrium\\\quad\\
  \large Ten Years of Progress in Theory and Numerical Simulations of Nuclear Collisions}

\author[1,2]{Paul Romatschke}
\author[1,3]{Ulrike Romatschke} 
\affil[1]{Department of Physics, University of Colorado, Boulder, Colorado 80309, USA}
\affil[2]{Center for Theory of Quantum Matter, University of Colorado, Boulder, Colorado 80309, USA}
\affil[3]{Earth Observing Laboratory, National Center for Atmospheric Research, Boulder, Colorado}
\date{Preprint version December 2017\\ updated May 2019}

\maketitle

\begin{abstract}
  Ten years ago, relativistic viscous fluid dynamics was formulated from first principles in an effective field theory framework, based entirely on the knowledge of symmetries and long-lived degrees of freedom. In the same year, numerical simulations for the matter created in relativistic heavy-ion collision experiments became first available, providing constraints on the shear viscosity in QCD.   The field has come a long way since then. We present the current status of the theory of non-equilibrium fluid dynamics in 2017, including the divergence of the fluid dynamic gradient expansion, resurgence, non-equilibrium attractor solutions, the inclusion of thermal fluctuations as well as their relation to microscopic theories. Furthermore, we review the theory basis for numerical fluid dynamics simulations of relativistic nuclear collisions, and comparison of modern simulations to experimental data for nucleus-nucleus, nucleus-proton and proton-proton collisions.
  \end{abstract}

\chapter*{Preface}

Strictly speaking, the subtitle of this work is somewhat misleading. Considerable progress on relativistic viscous fluid dynamics had been made earlier than 2007 both in the context of theoretical formulations as well as in numerical simulations. However, in particular for simulations of high energy nuclear collisions, typically an unrealistically high degree of symmetry had been assumed, making the resulting dynamics 0+1d or 1+1 dimensional. Only ten years ago, simulations in 2+1d became available, which is the minimum required to simulate the so-called elliptic flow observed in experiments. As one of the groups that first achieved 2+1d relativistic viscous fluid dynamics simulations ten years ago, we took the opportunity to celebrate this anniversary by compiling the present review of the current status of the field. Given the ongoing vibrant research activity on relativistic viscous fluid dynamics as well as continued experimental developments, we fully expect this review to be outdated in a few years. This is of course good news, and we hope it will require a new review subtitled “Twenty Years of Progress” when the time has come.

Happy anniversary, relativistic viscous fluid dynamics!

\tableofcontents

\chapter{Introduction}

Numerical simulations of relativistic non-equilibrium fluid dynamics\footnote{As is common in the literature, we will use the terms ``fluid dynamics'' and ``hydrodynamics'' synonymously.} have had enormous success in describing, explaining and predicting experimental data from relativistic nuclear collisions. At the same time, nuclear collision experiments have had a tremendous importance in pushing the development of relativistic fluid dynamics out of equilibrium, both on a formal level such as the first-principles derivation of the fluid dynamics equations of motion, as well as on the practical level through the development of algorithms for obtaining numerical solutions. This interplay between theory and experiment has led to the creation of a vibrant subfield of relativistic non-equilibrium (or viscous) fluid dynamics which unites research from traditionally separate disciplines such as string theory, classical gravity, computational physics, nuclear physics and high energy physics. While the early “gold-rush” years  aimed at constraining the shear viscosity of QCD may be coming to an end, new, previously unexpected avenues have opened up in the past 10 years, such as the  duality between fluids and gravity, the applicability of fluid dynamics to small systems below the femtoscale, the role of thermal fluctuations in relativistic systems, fluid dynamics in the presence of anomalies, anisotropic hydrodynamics, an action formulation for dissipative fluid dynamics, relativistic magneto-hydrodynamics and the role of non-hydrodynamic modes. It is probably fair to say that ten years ago, only very few people expected such rich and novel physics to emerge from the old discipline of hydrodynamics! At the time of writing, the research in relativistic viscous fluid dynamics is alive and well, with vibrant new ideas continuing to be proposed and new experimental data from the Relativistic Heavy-Ion Collider (RHIC) as well as the Large Hadron Collider (LHC) continuing to stream in. This wealth of experimental data is key to confirm or rule out theory predictions, and, sometimes, to challenge the relativistic hydrodynamics community, as has happened for instance through the discovery of flow-like signals in proton-proton collisions at the LHC. The borders between the traditionally separate high-energy physics and high-energy nuclear physics communities, never solid to begin with, have now started to disintegrate completely, with all of the present LHC experiments (ALICE, ATLAS, CMS and LHCb) having working groups directly or indirectly aimed at studying the properties of relativistic viscous fluids. Recently, gravitational wave observations from LIGO have added to the treasure trove of 
data by providing measurements of black-hole non-hydrodynamic modes as well as neutron star mergers, which likely will play a key role in calibrating future relativistic viscous fluid dynamics simulations of compact stars.

The influx of new experimental capabilities and manpower is a welcome addition to the field, which continues to grow and strengthen, with no obvious limit in sight. The future of relativistic fluid dynamics looks bright, indeed!

\section{Time-line of Major Events}

The current formulation of relativistic viscous fluid dynamics did not appear out of nowhere. The following time-line summarizes some of the major events (heavily biased by personal opinion!) that played a vital role in the development of the field. 

\begin{itemize}
\item
  pre-1950s: Work on relativistic equations of motion for viscous fluids by Maxwell  and Cattaneo \cite{Maxwell:1867,Cattaneo:1948}
\item
  1960s-1970s: Work on relativistic equations of motion for viscous fluids by M\"uller, Israel and Stewart \cite{Muller:1967zza,Israel:1976tn,Israel:1979wp}
\item
  1982: Analytic fluid modeling of heavy-ion collision by Bjorken \cite{Bjorken:1982qr}
\item
  1980s/1990s: First theoretical calculations for transport coefficients in gauge theories \cite{Hosoya:1983xm,Baym:1990uj,Heiselberg:1994vy}
\item
  1990s/early 2000s: Relativistic ideal fluid modeling of heavy-ion collisions by multiple groups \cite{Ollitrault:1992bk,Schnedermann:1993ws,Kolb:2000fha,Teaney:2000cw,Huovinen:2001cy,Hirano:2002ds,Nonaka:2006yn}, including predictions for observables, in particular for the magnitude of the so-called elliptic flow
\item
  2000: Calculation of shear viscosity in gauge theories to leading order in weak coupling by Arnold, Moore and Yaffe \cite{Arnold:2000dr}
\item
  2001: Following a break-through discovery in string theory \cite{Maldacena:1997re}, calculation of the shear viscosity for a gauge theory to leading order in strong coupling by Policastro, Son and Starinets \cite{Policastro:2001yc}
\item
  2001: The Relativistic Heavy Ion Collider (RHIC) at Brookhaven National Laboratory (BNL) starts operation
\item
  2003: Analytic calculation indicating a strong sensitivity of elliptic flow to changes in shear viscosity by Teaney \cite{Teaney:2003kp}
  \item
  2001-2003: Work on relativistic equations of motion for viscous fluids by Muronga \cite{Muronga:2001zk,Muronga:2003ta}
\item
  2004: Magnitude of elliptic flow measured by RHIC experiments, finding
overall agreement with relativistic ideal fluid dynamics simulations  \cite{Adcox:2004mh,Arsene:2004fa,Back:2004je,Adams:2005dq}
\item
2005:  BNL press release: 'RHIC Scientists Serve Up 'Perfect' Liquid' \cite{perfect1}: the  matter created in relativistic ion collisions behaves like a liquid with a very small viscosity approaching the string theory bound. No constraints on viscosity value 
\item
  2007: Theory of relativistic viscous fluid dynamics set up as an effective field theory of long-lived, long-wave excitations by two groups \cite{Baier:2007ix,Bhattacharyya:2008jc}
\item
  2007: 2+1 relativistic viscous fluid modeling of elliptic flow in heavy-ion collisions by several groups \cite{Romatschke:2007mq,Chaudhuri:2007qp,Song:2007fn,Dusling:2007gi}. First constraints on QCD shear viscosity value, placing it much closer to the strong coupling results than the weak coupling results
  \item
  2010: The Large Hadron Collider (LHC) at CERN starts operation
\item
  2010: Understanding that nuclear geometry fluctuations drive triangular flow in heavy-ion collisions by Alver and Roland \cite{Alver:2010gr}
  \item
  2010: Analytic fluid modeling of heavy-ion collisions with two-dimensional flow by Gubser \cite{Gubser:2010ze}
  \item
  2010: Formulation of relativistic anisotropic hydrodynamics by two groups \cite{Florkowski:2010cf,Martinez:2010sc}
\item
  2010: Fully 3+1d relativistic viscous hydrodynamic simulations of heavy-ion collisions by Schenke, Jeon and Gale \cite{Schenke:2010rr}
  \item
  2010: Elliptic flow measured in proton-proton collisions at the LHC \cite{Khachatryan:2010gv}
   \item
  2012: Elliptic flow measured in proton-lead collisions at the LHC \cite{CMS:2012qk,Abelev:2012ola,Aad:2012gla,Chatrchyan:2013nka}
\item
  2015: Discovery of a off-equilibrium 'hydrodynamic' attractor by Heller and Spali\'nksi \cite{Heller:2015dha}.
  \item
  2015: Elliptic flow measured in proton-gold, deuteron-gold and $^3{\rm He}$-gold collisions at RHIC, confirming relativistic viscous hydrodynamics predictions \cite{Adare:2017wlc}
\item
  2016: Observation of non-hydrodynamic mode ring-down of a black hole through gravitational wave data by LIGO \cite{TheLIGOScientific:2016wfe,TheLIGOScientific:2016src}
  \end{itemize}

\section{Notation and Conventions}

Throughout this work, natural units will be used in which Planck's constant, the speed of light and Boltzmann's constant will be set to unity, $\hbar=c=k_B=1$.

Much of the theoretical groundwork will be performed in $d$ space-time dimension ($d-1$ spatial dimensions and one time-like dimension). The convention used for the metric tensor is that of the mostly plus convention, and Greek letters  ($\mu,\nu,\lambda,\ldots$) will be used to denote indices of space-time vectors, with $\mu=0$ denoting time-like directions and the other entries denoting space-like directions. In general, curved space-time will be assumed, but the symbol $g_{\mu\nu}$ may refer to the metric tensor in both curved and flat space-times. The symbol $\nabla_\mu$ will indicate a geometric covariant derivative, while a plain coordinate derivative will be denoted as $\partial_\mu$. Gauge-covariant derivatives will be denoted as $D_\mu$.

For space vectors, Latin indices from the middle of the alphabet ($i,j,k$) will be used, and space-time indices may thus be decomposed as $\mu=(0,i)$. Space vectors also will be denoted by bold face letters, e.g. ${\bf v}$ or arrows, e.g. $\vec{v}$ whereas space gradients will be denoted by the symbol $\partial$ or $\vec{\partial}$.

\section{An Analogy To Fluid Dynamics: Particle Diffusion}
\label{sec:analogy}
\index{Diffusion}

As an introduction to many of the concepts discussed in the following, let us consider a simple analogy to fluid dynamics: non-relativistic diffusion. Specifically, consider a system with a conserved charge (such as electric charge). Local charge conservation can be written in terms of the charge density $n$ and the associated current density $\vec j$ as
\begin{equation}
  \label{eq:chargecons}
  \partial_t n+\vec \partial\cdot {\vec j}=0\,.
\end{equation}

Because charge is the only conserved quantity, the current density ${\vec j}$ is not a fundamental object. Therefore, $\vec{j}$ must in some way depend on the local charge density $n$.

In global equilibrium, one can expect the current density to vanish and the charge density to be constant. Deviations from global equilibrium will involve gradients of $n$ and non-vanishing currents. Hence it is natural to consider an expansion of $\vec{j}$ in terms of gradients of $n$, known as the gradient expansion\footnote{Note that no mention of a microscopic description giving rise to $\vec{j}$ is made, such as ``currents arise from moving electrons''. For this reason, the gradient expansion is universal in the sense that it contains all terms allowed by symmetries alone, which corresponds to the framework of effective field theory. On the other hand, effective field theory cannot be used to derive the quantitative value of the coefficients arising in the gradient expansion (e.g. the dependence of the diffusion constant on the electron charge).}.
The lowest order in this gradient expansion is zeroth order, where $\vec{j}_{(0)}=0$, because it is not possible to write down a vector in terms of the scalar $n$. Zeroth order diffusion therefore corresponds to $\partial_t n=0$, or static charge density distribution, which in contrast to its fluid dynamics equivalent (the Euler equation) is not very useful in practice.

The next order in the gradient expansion of ${\vec j}$ is first order in gradients. There are two possible structures to first order gradients of $n$ that one can write down: $\partial_t n$ and $\vec{\partial} n$, of which only the latter is a vector. Hence from the effective field theory expansion, we can expect $\vec{j}\propto \vec{\partial} n$.

However, there is one other structure to first order in gradients contributing to $\vec{j}$ that we could consider if we slightly enlarge our original setup. Because moving electric charges are associated with electromagnetic fields, we should consider gradients of the gauge potentials $\Phi,\vec{A}$ as building blocks in addition to gradients of $n$. The gauge potentials are typically referred to as ``sources'', and may be neglected if desired (for instance if the charge under consideration is not electric and/or the associated gauge fields are very weak). However, in full generality gradients of sources are to be included in the gradient expansion, which to first order in gradients suggests $\vec{j}\propto \vec{E}$ (Ohm's law) where $\vec{E}$ is the ``electric'' field associated with the gauge potentials\footnote{In principle, also ${\vec j}\propto \vec{B}$ is possible, where 
${\vec B}$ is the ``magnetic'' field. However, most systems in nature respect a symmetry known as parity (space direction flips), which forbids ${\vec j}\propto \vec{B}$. We taciturnly assume symmetry under parity in the following, even though the study of parity-violating fluids is a vibrant research subfield in itself.}.
Therefore, to first order in gradients one has
\begin{equation}
  \label{eq:j1}
  \vec{j}_{(1)}=-D \vec{\partial}n+\sigma \vec{E}\,,
\end{equation}
where $D,\sigma$ are two proportionality coefficients (``transport coefficients'') and the sign choices are convention. $D,\sigma$ are better known as diffusion constant and conductivity, respectively. The effective field theory framework cannot be used to calculate values for $D,\sigma$ (for this, a particular microscopic theory, such as kinetic theory, must be selected). However, effective field theory \textit{can} be used to obtain relations among transport coefficients, such as the Einstein relations
\begin{equation}
  \index{Einstein relations}
\label{eq:einsteinrelations}
 \sigma=D \chi\,,
\end{equation}
where $\chi$ is the static charge susceptibility\footnote{As a note to expert readers, a quick way to derive the Einstein relations is to calculate two-point retarded correlation functions $\langle \vec{j} \vec{j}\rangle$, once in the canonical approach (neglecting the sources $\vec{A}=0$ from the outset), and once in the variational approach where $\langle \vec{j} \vec{j}\rangle\propto \frac{\delta \vec{j}}{\delta \vec{A}}$. The Einstein relations then follow from the fact that both approaches must agree with each other.}.
Plugging (\ref{eq:j1}) in the equation for charge conservation (\ref{eq:chargecons}), neglecting the source $\vec{E}\rightarrow 0$, and assuming $D$ to be constant leads to
\begin{equation}
  \label{eq:diff}
  \partial_t n=D \partial^2 n\,,
\end{equation}
which is the familiar diffusion equation. Based on the derivation above, the diffusion equation (\ref{eq:diff}) can be expected to provide a good approximation to the actual evolution of the charge density as long as higher-order gradient corrections are small.

Considering higher order corrections in the gradient expansion, one encounters terms such as $\vec j\propto \vec{\partial} \partial^2 n$,  $j\propto \vec{\partial} \left(\vec{\partial} n\cdot \vec{\partial n}\right)$ $j\propto \vec{\partial} \partial^2\partial^2 n$, etc. Since the number of possible combinations of gradients of $n$ increases factorially, it is plausible that the gradient series diverges whenever $\vec{\partial}n\neq 0$. For a divergent series, higher-order corrections are not small for any $\vec{\partial}n\neq 0$, hence 
requiring small higher-order corrections  would imply that the diffusion equation (\ref{eq:diff}) is only applicable for static situations where $\partial_t n=0$.

Given the phenomenal success of using the diffusion equation in a large number of non-static situations, clearly the criterion of requiring small higher-order gradient corrections must be too strict. It should be replaced by a different criterion that does justice to the success of the diffusion equation, as well as correctly predicting its breakdown. The present work reviews the progress made towards formulating such a criterion for the case of relativistic fluid dynamics.

\chapter{Modern Theory of Fluid Dynamics}
\label{sec:modern}

What is fluid dynamics?

The answer to this question has been evolving over time. In this work, we will present our current understanding of fluid dynamics as a long-wavelength effective theory, without much reference to previous derivations or paradigms, which -- while probably not outdated -- seem to us less powerful than the modern version.

To build an effective theory, one has to have two main ingredients: knowledge of the effective degrees of freedom, and knowledge of the symmetries of the system under consideration. In the case of relativistic fluid dynamics, we aim for an effective description of the dynamics of conserved quantities of an underlying (quantum) system, such as its energy, momentum and charge. For a relativistic system, the required symmetry is that of the diffeomorphism group, consisting of general coordinate reparametrizations.

The modern effective theory of fluid dynamics is built up in stages: first fluid dynamics in the absence of fluctuations (``classical'' fluid dynamics) is derived, starting with fluid dynamics in equilibrium (ideal fluid dynamics). Then, small departures from equilibrium are introduced, giving rise to dissipative (or viscous) fluid dynamics. Finally, large deviations from equilibrium  are discussed using recent theory developments such as the resummation of the gradient expansion, resurgence and non-hydrodynamic modes. This treatment gives rise to the new theory of off-equilibrium fluid dynamics, which -- while still being formulated --- has the potential of becoming a powerful tool to study strongly coupled non-equilibrium phenomena. We close by discussing the emerging theory of relativistic fluid dynamics in the presence of fluctuations, including an effective action formulation of dissipative fluid dynamics.

\section{Classical Fluid Dynamics in Equilibrium}
\label{sec:equi}

The term ``classical fluid dynamics'' describes the dynamics of fluids in the absence of fluctuations such as thermal fluctuations. It offers the standard description for the expectation value (or one-point function) of bulk observables such as the energy density of a fluid. In the classical treatment, dissipation through viscous terms can be naturally derived, giving for instance rise to the damping of sound waves.

Let us consider a theory without conserved charges to simplify the discussion. Then the globally conserved quantities will be a Lorentz scalar (the energy) and a Lorentz vector (the momentum). For a local description, we are thus led to the presence of a Lorentz scalar and a Lorentz vector as the effective hydrodynamic fields. Let us denote these as $a$ and $b^\mu$, respectively.

Energy and momentum in a relativistic theory are encoded in the energy-momentum tensor $T^{\mu\nu}$, which is the response of a system to some perturbation of the metric tensor $g_{\mu\nu}$. Let us now write down an effective hydrodynamic description of the expectation value of the energy-momentum tensor of some underlying quantum field theory in terms of the effective hydrodynamic fields as well as the presence of a source (the metric tensor). The energy-momentum tensor is a symmetric rank two tensor which has to transform properly under Poincar\'e transformations, and it has to be built out of the effective hydrodynamic fields $a,b^\mu$ as well as the symmetric rank two tensor $g_{\mu\nu}$. In principle, one could associate $b^\mu$ with the local momentum. However, it is customary to instead trade the local momentum for a local velocity, since the velocity is a dimensionless quantity (see appendix \ref{chap:vel1} for a brief discussion of relativistic velocities). In the following, a fluid description of matter with time-like local momentum will be considered\footnote{Fluid descriptions of light-like particles have been discussed in Ref.~\cite{Salzer2012}.}, so that as a consequence also the local velocity vector should be time-like, and thus can be normalized as $b^\mu b_\mu=-1$.

To lowest (zeroth) order in effective field theory, only algebraic contributions in the fields and the source are considered. Considering all possible combinations of the available building blocks $a,b^\mu,g^{\mu\nu}$ leads to the form of the effective hydrodynamic energy-momentum tensor $T_{(0)}^{\mu\nu}=c_1\, a\, b^\mu b^\nu+c_2\, a\, g^{\mu\nu}$, where $c_1,c_2$ are pure numbers and the mass dimension of the scalar $a$ has to match that of the energy-momentum tensor (space-time dimension $d$).
In this form, the trace ${\rm Tr}\, T_{(0)}^{\mu\nu}=T^\mu_{(0) \mu}=(-c_1+c_2 d)a$ of the effective hydrodynamic energy-momentum tensor is a function of only the scalar $a$ and pure numbers. As will be described below, this would severely restrict the applicability of the effective hydrodynamic theory to systems with a high degree of symmetry, for instance conformal systems for which ${\rm Tr}\, T^{\mu\nu}=0$. It is possible to lift this restriction by allowing $c_2$ to also be a function of the scalar $a$, or equivalently allowing the presence of a second scalar function $f_2(a)$, such that
\begin{equation}
  \label{eq:hydro0}
  T_{(0)}^{\mu\nu}=a \left(c_1 b^\mu b^\nu+c_2 g^{\mu\nu}\right)+ f_2(a) \left(c_3 b^\mu b^\nu+c_4  g^{\mu \nu}\right)\,,  
\end{equation}
where $c_1,\ldots, c_4$ are numbers. Note that the time-like vector $b_\mu$ is an eigenvector of $T^{\mu\nu}_{(0)}$.
%
%
It should be stressed that $f_2$ is not a new scalar degree of freedom, but a function of the scalar $a$. The functional relation between $f_2(a)$ and $a$ (which will become the Equation of State) has to be provided by explicit computation in the underlying quantum system.

Let us now consider the underlying quantum system to be embedded in $d$ dimensional Minkowski space-time. If gradients of the hydrodynamic fields and metric perturbations can be neglected, this underlying quantum system must give rise to the effective zeroth order hydrodynamic description (\ref{eq:hydro0}).  The spatial and temporal gradients in the underlying system are small if the system is quasi-stationary, for instance close to a non-thermal fixed point \cite{Berges:2008wm}. While the present framework would likely be applicable in such situations, a more traditional approach is to assume this quasi-stationary state to be the equilibrium state\footnote{Note that here and in the following ``equilibrium state'' is meant to denote a local equilibrium state, not a global equilibrium state. Hydrodynamics of systems in global equilibrium becomes hydrostatics, which we will not consider in this work.}  of the underlying quantum system. We will pursue this traditional approach in the following.

Let us assume that the expectation value of the energy-momentum tensor $\langle T^{\mu\nu}\rangle$ can be calculated in an equilibrium quantum system in Minkowski space-time. In this case, $\langle T^{\mu\nu}\rangle$ possesses a time-like eigenvector $u^\mu$, normalized to $u^\mu u_\mu=-1$. A suitable Lorentz boost will then allow to bring the equilibrium energy-momentum tensor into the local rest frame (LRF), \index{Local rest frame (LRF)} where\footnote{Here and in the following, ${\rm diag}(x_0,x_1,\ldots)$ indicates a symmetric matrix with non-vanishing entries $x_0,x_1$ only along the diagonal.}
$$\langle T^{\mu\nu}\rangle_{\rm LRF}={\rm diag}\left(\epsilon,P,P,\ldots,P\right)\,.$$
Here $\epsilon$ and $P(\epsilon)$ are the local equilibrium energy density and pressure (see the demonstration in section \ref{sec:fieldtheory}). Also, the zeroth order hydrodynamic effective result (\ref{eq:hydro0}) possesses a time-like eigenvector $b^\mu$ which may be identified with the time-like velocity $u^\mu$ of the equilibrium energy-momentum tensor. Choosing coordinates where $g_{00}=-1$, a Lorentz boost to the LRF implies $b^\mu=\left(1,{\bf 0}\right)$
and as a consequence the form of (\ref{eq:hydro0}) in the LRF is given by
$$
T^{\mu\nu}_{(0),\rm LRF}={\rm diag}\left(a (c_1-c_2)+f_2 (c_3-c_4),a c_2+f_2 c_4,a c_2+f_2 c_4,\ldots,a c_2+f_2 c_4\right)\,.
$$
Requiring this result to match $\langle T^{\mu\nu}\rangle_{\rm LRF}={\rm diag}(\epsilon,P,P,P,\ldots)$ from above leads to $a c_2+f_2 c_4=P$, $a c_1+f_2 c_3=\epsilon+P$ such that
\begin{equation}
  \label{eq:hydro0f}
  T_{(0)}^{\mu\nu}=(\epsilon+P) u^\mu u^\nu+ P g^{\mu\nu}\,.
\end{equation}
Thus, matching the form of the hydrodynamic effective energy-momentum tensor in the limit of vanishing gradients to the expectation value of an equilibrium quantum system has fixed all previously undetermined constants $c_1,\ldots c_4$. The final zeroth order fluid dynamics result (\ref{eq:hydro0f}) involves the equilibrium energy density (a Lorentz scalar), a time-like Lorentz vector $u^\mu$ (which can be interpreted as the fluid velocity), the metric tensor $g^{\mu\nu}$ (the source) as well as another Lorentz scalar $P$ which was found to correspond to the equilibrium pressure. The equilibrium relation $P(\epsilon)$ which will be referred to as the ``Equation of State'' (EoS)\index{Equation of State (EoS)}, has to be provided by an explicit calculation in the underlying equilibrium quantum system.

The zeroth order hydrodynamic effective theory result (\ref{eq:hydro0}) constitutes by far the most widely used form of relativistic fluid dynamics. It provides the backbone of modern cosmology, supernova simulations, and many other astrophysical applications.

To appreciate the power of (\ref{eq:hydro0}), recall that in the absence of sources, the energy-momentum tensor is covariantly conserved,
\begin{equation}
  \label{eq:euler}
  \nabla_\mu T^{\mu\nu}_{(0)}=0\,,
\end{equation}
where $\nabla_\mu$ denotes the geometric covariant derivative. Allowing the hydrodynamic fields $\epsilon,P,u^\mu$ as well as the metric $g_{\mu\nu}$ to be arbitrary functions of space-time, the equations of motion (\ref{eq:euler}) then describe the bulk motion of the underlying quantum system. If the energy-momentum tensor is of the zeroth order hydrodynamic form (\ref{eq:hydro0f}), then the energy-momentum conservation equation (\ref{eq:euler}) is referred to as the relativistic Euler equation or ideal fluid dynamics equations.
\index{Euler equation}
\index{Fluid Dynamics! Ideal|see {Euler equation}}

\subsection{The Relativistic Euler Equation}
\label{chap:euler}

It is convenient to introduce a projection operator $\Delta^{\mu\nu}$ that is providing projections to the space-like part of a tensor through
\begin{equation}
  \Delta^{\mu\nu}=g^{\mu\nu}+u^\mu u^\nu\,.
\end{equation}
For arbitrary metric $g_{\mu\nu}$, the projector $\Delta^{\mu\nu}$ is orthogonal to the time-like velocity, $\Delta_{\mu\nu}u^\nu=\Delta_{\mu\nu}u^\mu=0$, and furthermore satisfies the relation $\Delta_{\mu\nu}\Delta^{\nu}_\rho=\Delta_{\mu\rho}$. With the use of $\Delta$, Eq.~(\ref{eq:hydro0f}) may be re-written as $T^{\mu\nu}_{(0)}=\epsilon u^\mu u^\nu+P \Delta^{\mu\nu}$.

The form of $\Delta$ is most easily visualized by an example in Minkowski space-time where $g_{\mu\nu}={\rm diag}\left(-1,1,1,\ldots\right)$. In this case, a Lorentz transformation to the LRF will lead to $u^\mu_{LRF}=(1,{\bf 0})$ and thus $\Delta^{\mu\nu}_{LRF}={\rm diag}(0,1,1,\ldots,1)$, making the space-like projection property of $\Delta$ explicit. Note that similarly, $u^\mu$ can be recognized to constitute a time-like projection operator.

As an application of these projectors, let us consider separately the co-moving time-like and space-like components of (\ref{eq:euler}), by projecting (\ref{eq:euler}) onto $u^\nu$ and $\Delta^{\nu\rho}$. For the time-like projection, one finds
\begin{eqnarray}
  u_\nu \nabla_\mu T^{\mu\nu}_{(0)}&=&-u^\mu \nabla_\mu \epsilon-\epsilon \nabla_\mu u^\mu+\epsilon u^\mu u_\nu \nabla_\mu u^\nu+P u_\nu \nabla_\mu \Delta^{\mu\nu}\,,\nonumber\\
  &=&-(\epsilon+P)\nabla_\mu u^\mu-u^\mu \nabla_\mu \epsilon=0\,,
  \label{eq:eulera}
\end{eqnarray}
where the identity $u_\nu \nabla_\mu u^\nu=\frac{1}{2}\nabla_\mu u_\nu u^\nu=- \frac{1}{2}\nabla_\mu 1=0$ has been used. For the space-like projection, one finds
\begin{eqnarray}
  \Delta_\nu^\rho \nabla_\mu T^{\mu\nu}_{(0)}&=& \epsilon \Delta_\nu^\rho u^\mu \nabla_\mu u^\nu+\Delta^{\mu\rho}\nabla_\mu P+P \Delta_\nu^\rho u^\mu \nabla_\mu u^\nu\,,\nonumber\\
  &=&(\epsilon+P) u^\mu \nabla_\mu u^\rho+\Delta^{\mu \rho}\nabla_\mu P=0\,.
  \label{eq:eulerb}
  \end{eqnarray}
Introducing the short-hand notations
\begin{equation}
  \label{eq:derivs}
  D \equiv u^\mu \nabla_\mu\,,\quad \nabla_\perp^\rho=\Delta^{\mu\rho}\nabla_\mu\,,
\end{equation}
for the co-moving time-like and space-like derivatives, Eqns.~(\ref{eq:eulera},\ref{eq:eulerb}) become
\begin{equation}
  \label{eq:eulerf}
  D \epsilon+(\epsilon+P)\nabla^\perp_\mu u^\mu=0\,,\quad
  (\epsilon+P)D u^\rho+\nabla_\perp^\rho P=0\,,
\end{equation}
where a little calculation shows that $\nabla_\mu u^\mu=\nabla_\mu^\perp u^\mu$. Using the equation of state, it is possible to further rewrite $\nabla_\perp^\rho P=\frac{\partial P(\epsilon)}{\partial \epsilon} \nabla_\perp^\rho \epsilon$ such that Eqns.~(\ref{eq:eulerf}) become
\begin{equation}
  \label{eq:eulerf2}
  D \epsilon+(\epsilon+P)\nabla^\perp_\mu u^\mu=0\,,\quad
  (\epsilon+P)D u^\rho+c_s^2\nabla_\perp^\rho \epsilon=0\,,
\end{equation}
%
%
where
\begin{equation}
  c_s(\epsilon)\equiv \sqrt{\frac{\partial P(\epsilon)}{\partial \epsilon}}\,,
\end{equation}
will later be recognized to correspond to the speed of sound. The speed of sound $c_s$ is the first example of a (``zeroth order'') transport coefficient.
\index{Speed of sound}
\index{Transport coefficients}

\subsubsection{Application: Non-Relativistic Euler Equation}

In the case of small (space) velocities $|{\vec v}|\ll 1$ and Minkowski space-time, the co-moving derivatives (\ref{eq:derivs}) become
\begin{equation}
  \label{eq:nrlimit}
  D\rightarrow \partial_t+\vec{v}\cdot \vec{\partial} +{\cal O}(|{\vec v}|^2)\,,
  \quad
  \nabla_\perp^i\rightarrow \partial^i+{\cal O}(|{\vec v}|)\,
\end{equation}
cf. Eq.~(\ref{eq:uexp}). In addition, recall that the energy-density $\epsilon$ includes the rest mass energy which typically is much larger than the pressure for non-relativistic systems, hence $\epsilon\gg P$. These approximations are referred to as non-relativistic limit, and for Eq.~(\ref{eq:eulerf}) lead to
\begin{equation}
\label{eq:eaul14}
  \partial_t \epsilon+ {\vec v}\cdot \vec \partial \epsilon+\epsilon \vec\partial \cdot {\vec v}=0\,,\quad
  \epsilon \partial_t {\vec v}+\epsilon \left({\vec v}\cdot \vec\partial\right) {\vec v}+\vec\partial P=0\,,
\end{equation}
which can be recognized as the non-relativistic continuity equation and non-relativistic (compressible) Euler equation \cite{Euler}, respectively\footnote{Note that the ``usual'' derivation of non-relativistic fluid dynamics assumes the presence of an additional conserved quantity: particle number. Particle number conservation  then leads to the continuity equation, which is of exactly the same form as Eq.~(\ref{eq:eaul14}).}.

\subsubsection{Application: Cosmology}
\index{Cosmology}
\index{FLRW|see {Cosmology}}

For a homogeneous and isotropic universe in 3+1 space-time dimensions, the metric will be of the Friedmann-Lemaitre-Robertson-Walker (FLRW) form $g_{\mu\nu}={\rm diag}\left(-1,a^2(t),a^2(t),a^2(t)\right)$, where $a(t)$ is the so-called scale-factor of the universe. Also, all fluid dynamic degrees of freedom can only depend on time, e.g. $\epsilon(t)$. Because the space-time is homogeneous, the local (space) velocity of the fluid must vanish, thus $u^\mu=(1,{\bf 0})$. The relativistic ideal fluid dynamics equations (\ref{eq:eulerf}) thus become
\begin{equation}
  \label{eq:cosmofluid}
  \partial_t \epsilon+(\epsilon+P)\frac{3 a^\prime(t)}{a(t)}=0\,,
\end{equation}
where the explicit form of the Christoffel symbol $\Gamma^\mu_{\mu t}=\frac{3 a^\prime(t)}{a(t)}$ for the FLRW metric has been used (see appendix \ref{chap:aGR} for a brief review of common objects in general relativity). It is customary to trade $a(t)$ with the Hubble ``constant'' $H(t)\equiv \frac{a^\prime(t)}{a(t)}$ such that the ideal fluid dynamics equations become
\begin{equation}
  \partial_t \epsilon=-3 H(\epsilon+P)\,.
\end{equation}
The fluid dynamics equations need to be supplemented by the Einstein equations (\ref{eq:einstein}), which in the case of vanishing cosmological constant become
\begin{equation}
  \label{eq:flrwee}
  R^t_t-\frac{1}{2}g^t_t R=-3 H^2=8 \pi G T^{t}_t=-8 \pi G \epsilon\,,
\end{equation}
or $H^2=\frac{8 \pi G}{3}\epsilon$. Plugging this into the fluid equation (\ref{eq:cosmofluid}) one finds
\begin{equation}
  \label{eq:cosmo2}
  \partial_t \epsilon=-3 (\epsilon+P) \sqrt{\frac{8 \pi G}{3} \epsilon}\,.
  \end{equation}
To solve this equation, one needs to supply an equation of state $P=P(\epsilon)$. If we consider the matter in the universe to be composed of dust, then its pressure will be negligible compared to its rest mass, and the equation of state is simply $P=0$. Equation (\ref{eq:cosmo2}) may then be integrated directly to give $\epsilon(t)=\frac{1}{6 \pi G t^2}$, or
\begin{equation}
  H=\frac{2}{3 t}\,,\quad a(t)\propto t^{2/3}\,.
  \end{equation}

\subsubsection{Application: Bjorken Flow}
\label{sec:idealbjork}
\index{Bjorken flow}
\index{Analytic Flows! {Bjorken}|see {Bjorken flow}}

Maybe somewhat surprisingly, a simple model for the bulk evolution of matter created in relativistic nucleus-nucleus collisions is in many ways similar to standard cosmology. Taking the collision axis to be the longitudinal z-axis, one assumes the nuclei to be homogeneous and of infinite transverse extent, thereby removing all dependence on the coordinates $x,y$.

Following Bjorken \cite{Bjorken:1982qr}, one additionally assumes that the matter produced in the collision is invariant with respect to boosts along the z-axis, e.g. $v^z=\frac{z}{t}$ for $|z|\leq t$ (this property is sometimes referred to as ``Bjorken flow''). In this case, it is convenient to introduce Milne coordinates proper time $\tau=\sqrt{t^2-z^2}$ and space-time rapidity $\xi={\rm arctanh}(z/t)$ for which $u^\xi=u^x=u^y=0$ and $u^\tau=1$, and all hydrodynamic fields are functions of proper time exclusively.

It is maybe useful to point out that the description is aimed at matter evolving in $d=4$ Minkowski space-time dimensions. Nevertheless, the coordinate transformation to Milne coordinates is non-linear. The resulting Milne metric is given by $g_{\mu\nu}={\rm diag}(-1,1,1,\tau^2)$ (see appendix \ref{sec:milnecoo}) and describes a one-dimensional expanding space-time in which the fluid is at rest (recall $u^\mu=(1,{\bf 0})$). So rather than describing a fluid expanding longitudinally in static Minkowski space-time, we have chosen an equivalent formulation of a static fluid in longitudinally expanding space-time. A quick check reveals that the Ricci tensor for the Milne metric indeed vanishes, so the space-time is expanding, but flat.
The only non-vanishing Christoffel symbols are $\Gamma_{\xi \tau}^\xi=\frac{1}{\tau}$ and $\Gamma^\tau_{\xi \xi}=\tau$. The ideal fluid dynamics equations (\ref{eq:eulerf}) become
\begin{equation}
  \label{eq:bjork0}
  \partial_\tau \epsilon+\frac{\epsilon+P}{\tau}=0\,.
\end{equation}
As before, an equation of state needs to be supplied. For a relativistic gas in three space dimensions one has $P=c_s^2\epsilon$ with $c_s^2=\frac{1}{3}$ the relativistic speed of sound squared. Generalizing this to the case of constant speed of sound squared $c_s^2={\rm const.}$, solving (\ref{eq:bjork0}) leads to 
\begin{equation}
  \label{eq:ebjor0}
  \frac{\epsilon(\tau)}{\epsilon(\tau_0)}=\left(\frac{\tau_0}{\tau}\right)^{1+c_s^2}\,,
\end{equation}
where $\tau_0$ is an integration constant.

\subsubsection{Application: Gubser Flow}
\index{Gubser flow}
\index{Analytic Flows! {Gubser}|see {Gubser flow}}

A generalization of Bjorken flow that includes transverse dynamics was introduced by Gubser \cite{Gubser:2010ze}. Similar to Bjorken flow, one introduces new coordinates in which the fluid will be taken to be at rest. Starting with a coordinate transformation similar to Bjorken flow using $\tau=\sqrt{t^2-z^2}, \xi={\rm arctanh}(z/t)$, the transverse coordinates $x,y$ may additionally be written in polar coordinate form $x=r \cos\phi$, $y=r \sin\phi$ with $r=\sqrt{x^2+y^2}$ and $\phi={\rm arctan}(y/x)$. So far, this is just Bjorken flow in polar coordinates. The key is then to consider new coordinates $\rho,\theta$ which are related to $\tau,r$ as
\begin{equation}
  \label{eq:gucotra}
\sinh\rho = - \frac{1-q^2 \tau^2+q^2 r^2}{2 q \tau}\,,\quad
\tan \theta = \frac{2 q r}{1+q^2\tau^2-q^2 r^2}\,,
\end{equation}
where $q^{-1}$ is a characteristic length scale which is taken to be constant.
The coordinate transformation from $t,x,y,z$ to $\rho,\theta,\phi,\xi$ then leads to a metric of the form
$$
  g_{\mu\nu}={\rm diag}\left(g_{\rho\rho},g_{\theta\theta},g_{\phi\phi},g_{\xi\xi}\right)={\rm diag}\left(-\tau^2,\tau^2 \cosh^2\rho,\tau^2 \cosh^2\rho \sin^2\theta,\tau^2\right)\,.
$$
Unlike the case of Bjorken flow, one now performs a Weyl rescaling of the metric (cf. Eq.~(\ref{eq:Weyltrafo})),
$$
g_{\mu\nu}\rightarrow \tilde g_{\mu\nu}=\frac{g_{\mu\nu}}{\tau^2}={\rm diag}\left(-1,\cosh^2\rho,\cosh^2\rho \sin^2\theta,1\right)\,,
$$
which -- upon calculating the Ricci scalar for the metric $\tilde g_{\mu\nu}$ -- may be recognized to be a metric of $dS_3\times \mathbb{R}^1$, where $dS_3$ denotes three-dimensional de-Sitter space, rather than d=4 Minkowski space ($\mathbb{R}^{3,1}$). Using the analogy to Bjorken flow and demanding that $u^\mu=(1,{\bf 0})$ in coordinates $\rho,\theta,\phi,\xi$ on $dS_3\times \mathbb{R}^1$, the ideal fluid dynamics equations (\ref{eq:eulerf}) become
\begin{equation}
  \label{eq:gubser1}
\partial_\rho \tilde \epsilon+2 (\tilde \epsilon+\tilde P)\tanh\rho =0\,,
\end{equation}
where the Christoffel symbols corresponding to $\tilde g_{\mu\nu}$ have been evaluated as $\Gamma^\rho_{\theta\theta}=\cosh \rho \sinh\rho$, $\Gamma^\rho_{\phi\phi}=\cosh\rho \sinh\rho \sin^2\theta$, $\Gamma^\theta_{\rho\theta}=\tanh\rho$, $\Gamma^\theta_{\phi\phi}=-\cos\theta \sin\theta$, $\Gamma^\phi_{\rho \phi}=\tanh\rho$, $\Gamma^\phi_{\theta\phi}=\cot{\theta}$. Using again an equation of state with $\tilde P=c_s^2 \tilde \epsilon$ and constant speed of sound squared $c_s^2$ (of which $c_s^2=\frac{1}{3}$ is a special case) one finds
\begin{equation}
  \label{eq:gubser2}
  \frac{\tilde\epsilon(\rho)}{\tilde \epsilon(\rho_0)}=\cosh^{-2 (1+c_s^2)}\rho\,,
\end{equation}
where $\rho_0$ is an integration constant, to be a solution of (\ref{eq:gubser1}) in $dS_3\times \mathbb{R}^1$, cf. Ref.~\cite{Gubser:2010ui}. In order to convert this solution for $\tilde \epsilon(\rho)$ to a solution in Minkowski space, one needs to undo the coordinate transformation to $\rho,\theta$ coordinates and Weyl rescaling of the metric. For scalars, such as the energy density, this is easy to do. Eq.~(\ref{eq:gucotra}) provides the explicit expression for $\sinh\rho$ and Weyl rescalings, implying that length and time were contracted by a factor $\tau^{-1}$ when transforming from Minkowski to $dS_3\times \mathbb{R}^1$, hence there should be a factor of $\tau^3$ for the transformation of volume when going from $dS_3\times \mathbb{R}^1$ to $\mathbb{R}^{3,1}$. Accordingly, energy should scale as $\tau^{-1}$ when transforming from $dS_3\times \mathbb{R}^1$ to $\mathbb{R}^{3,1}$, so that energy density in Minkowski space should be related to (\ref{eq:gubser2}) as \cite{Gubser:2010ui}
\begin{equation}
  \epsilon(\tau,r,\xi)=\frac{\tilde \epsilon(\rho)}{\tau^4}=\frac{\tilde \epsilon(\rho_0)}{\tau^4 \left(\frac{(2 q \tau)^2+(1-q^2 \tau^2+q^2 r^2)^2}{(2 q \tau)^2} \right)^{1+c_s^2}}\,.
\end{equation}
For vectors such as $u^\mu$, the explicit coordinate transformations $u_\mu=\tau \frac{\partial \rho}{\partial x^\mu}$ need to be considered and one finds \cite{Gubser:2010ui}
\begin{equation}
 v^r\equiv \frac{u^{r}}{u^\tau}=\frac{-\tau \frac{\partial \rho}{\partial \tau}}{\tau \frac{\partial \rho}{\partial r}}=\frac{2 q^2 r \tau}{1+q^2 r^2+q^2\tau^2}\,,
\end{equation}
which implies a non-vanishing (azimuthally symmetric) transverse expansion of the system.

\subsubsection{Application: Neutron Star Masses and Radii}
\index{Neutron Stars}

Considering a non-spinning neutron star in equilibrium, all gradients vanish, corresponding to a situation of relativistic hydrostatic equilibrium. The problem is spherically symmetric, so it is convenient to choose spherical coordinates $x^{\mu}=(t,r,\theta,\phi)$ for which the metric tensor may be parametrized as $g_{\mu\nu}={\rm diag}\left(-e^{\nu(r)},\frac{1}{1-r_s(r)/r},r^2,r^2 \sin^2\theta\right)$. Since no dynamics is at play, all spatial velocities vanish, and hence only $u^t\neq 0$. From $u^\mu u_\mu=g_{tt} (u^t)^2=-1$ one finds $u^\mu=(e^{-\nu(r)/2},{\bf 0})$. The star's structure is determined by solving Einstein's equations (\ref{eq:einstein}) with $T^{\mu\nu}$ given by the ideal fluid energy-momentum tensor (\ref{eq:hydro0f}).  One finds  $\partial_r \nu(r)=\frac{8 \pi G r^3 P(r)+r_s(r)}{r^2(1-r_s/r)}$ as well as
\begin{equation}
  \partial_r r_s(r)=8 \pi G r^2 \epsilon(r)\,,\quad
  \partial_r P(r)=-\frac{(\epsilon+P)(8 \pi G r^3 P+r_s)}{2 r ^2(1-r_s/r)}\,,
\end{equation}
which are known as the Tolman-Oppenheimer-Volkoff (TOV) structure equations \cite{Tolman:1939jz,Oppenheimer:1939ne}. \index{Tolman-Oppenheimer-Volkoff (TOV)} An equation of state is needed in order to obtain a solution to these equations. Once $P(\epsilon)$ is given, one typically picks a central energy density for the star $\epsilon(r=0)=\epsilon_0$ and then solves the TOV numerically by integrating out from the star's center to the point where the pressure vanishes $P(r=R)=0$. The star's radius $R$ is defined by the point at which the pressure vanishes. With the star's energy density profile $\epsilon(r)$ thus determined, the star's mass $M(R)$ can be calculated from $\partial_r M(r)=4 \pi r^2 \epsilon(r)$. (Note that the parameter $r_s(r)$ in the metric ansatz is related to the mass as $r_s=2 G M(r)$).

\section{Fluids Near Equilibrium}
\label{sec:navi}

As outlined in the previous chapters, truncating the effective hydrodynamic field theory expansion at the lowest (zeroth) order gives rise to the theory of ideal fluid dynamics, a tremendously successful theory with wide applications in many disciplines.

Despite its great success, ideal fluid dynamics possesses a key weakness: it does not give any indication of its own regime of validity. Thus, while having the advantage of being simple and predictive, ideal fluid dynamics leads to results whose accuracy is completely unknown (with the possible exception of hydrostatics where gradients are absent by construction).

To obtain an estimate of the regime of validity of ideal fluid dynamics, one needs to be able to quantify the size of the corrections to ideal fluid dynamics. This can be achieved naturally in the effective hydrodynamics framework by including higher order corrections to the zeroth order hydrodynamic energy-momentum tensor (\ref{eq:hydro0f}).

These corrections may be constructed in a systematic fashion by considering gradients of the fundamental hydrodynamic fields ($\epsilon,u^\mu$) and the source $g^{\mu\nu}$ order-by-order, subject to the symmetries of the underlying system, leading to
\begin{equation}
  \label{eq:tmunuseries}
  \langle T^{\mu\nu}\rangle =T^{\mu\nu}_{(0)}+T^{\mu\nu}_{(1)}+T^{\mu\nu}_{(2)}+\ldots\,,
\end{equation}
where the subscripts $(0),(1),(2),\ldots$ indicate the number of gradient terms in the respective part of $T^{\mu\nu}$. The resulting theory is generally referred to as 'viscous' or 'dissipative' fluid dynamics.

\subsection{First Order Hydrodynamics}

Let us again consider the case of an uncharged fluid for simplicity, and construct the correction $T^{\mu\nu}_{(1)}$ containing only first order gradients.
Since $\epsilon$ and $P(\epsilon)$ are related by the equation of state,
%
%
one can choose any one function of these for constructing gradient corrections. Here the choice will be made\footnote{Note that in $d$ space-time dimensions $\epsilon$ has mass dimension $d$, so $\ln \epsilon$ should be understood to mean $\ln (\epsilon/\epsilon_0)$ with $\epsilon_0$ a constant with mass dimension $d$ to make the argument of the logarithm dimensionless. Since $\epsilon_0$ will not contribute when taking gradients of $\ln (\epsilon/\epsilon_0)$, it is dropped in the following.} to consider gradients of $\ln \epsilon$.

Thus, the building blocks for $T^{\mu\nu}_{(1)}$ are $\nabla_\mu \ln \epsilon$ and $\nabla_\mu u_\nu$, both of which have mass dimension one\footnote{This was the reason for using $\ln \epsilon$ instead of $\epsilon$ in the expansion.}$^,$\footnote{Note that in principle also explicit first order gradients of $g_{\mu\nu}$ could be allowed, but these vanish identically, $\nabla_\mu g_{\nu\lambda}=0$, cf. Eq.~(\ref{eq:vanishingmetric}).}. However, some of these gradients are linearly related through the ideal fluid dynamics equations of motion (\ref{eq:eulerf2}). By inspecting the equations of motion, one is led to the conclusion that a linearly independent set of first order gradients is given by the co-moving spatial gradients $\nabla_\mu^\perp \ln \epsilon,\nabla_\mu^\perp u_\nu$.

For later convenience, gradients are sorted into three classes: scalars, vectors, and rank-two tensors. Furthermore, since scalars multiplied by $u^\mu$ would trivially give corresponding vectors, let us restrict independent vectors to mean vectors that are orthogonal to $u^\mu$, and impose a similar restriction (orthogonality with respect to $u^\mu$) on the tensors. To first order in gradients, there is one in each class (scalar, vector, tensor):
\begin{equation}
  \nabla_\mu^\perp u^\mu\,,\quad \nabla_\mu^\perp \ln \epsilon\,,\quad \nabla_\mu^\perp u_\nu\,.
  \end{equation}
 These independent first-order gradient building blocks must now be combined with the zeroth-order fields $\epsilon,u^\mu$ and sources $g^{\mu\nu}$ to form the first-order correction $T^{\mu\nu}_{(1)}$ to the ideal fluid dynamic energy momentum tensor $T^{\mu\nu}_{(0)}$.

 There is one more important point to consider. Namely, for ideal fluid dynamics, the zeroth-order fields $\epsilon,u^\mu$ were defined as the time-like eigenvalue and time-like eigenvector of the expectation value of the energy-momentum tensor $\langle T^{\mu\nu}\rangle$ for an underlying quantum system \textit{in equilibrium}. In the following, the same operational procedure is used for a (slightly) off-equilibrium system: the fundamental hydrodynamic fields $\epsilon,u^\mu$ are defined through the equation
 \begin{equation}
   \label{eq:umudef}
   u_\mu \langle T^{\mu\nu}\rangle = - \epsilon u^\nu\,,
 \end{equation}
 where $\langle T^{\mu\nu}\rangle$ is now the expectation value for the underlying quantum system slightly \textit{out of equilibrium}. While this trivially implies that $\epsilon$ still is the local energy density, this energy density is an out-of-equilibrium quantity. This immediately begs the question: does the relation between the out-of-equilibrium pressure $P(\epsilon)$ still match the equilibrium equation of state, or does it contain out-of-equilibrium corrections?
\index{Equation of State (EoS)!Non-Equilibrium}
 
 Before attempting to answer this question, let us first finish the construction of the first-order effective hydrodynamic field theory. The condition (\ref{eq:umudef}) together with the form of the ideal fluid dynamic tensor (\ref{eq:hydro0f}) implies that $u_\mu T^{\mu\nu}_{(1)}=0$. Since $T^{\mu\nu}_{(1)}$ is a second-rank symmetric tensor, the only possible combinations of building blocks are
 $   \Delta_{\mu\nu} \nabla_\lambda^\perp u^\lambda$ and $\nabla_{(\mu}^\perp u_{\nu)}=\frac{1}{2}\left(\nabla_{\mu}^\perp u_{\nu}+\nabla_{\nu}^\perp u_{\mu}\right)$. For later convenience, it turns out to be advantageous to consider two linear combinations of these two terms instead, namely
 \begin{equation}
\label{eq:tterms}
   \Delta^{\mu\nu} \nabla_\lambda^\perp u^\lambda\,,\quad
   \sigma^{\mu\nu}=2 \nabla^{<\mu}u^{\nu>}=2 \nabla^{(\mu}_\perp u^{\nu)}-\frac{2}{d-1}\Delta^{\mu\nu} \nabla_\lambda^\perp u^\lambda\,,
 \end{equation}
 where $\sigma_{\mu\nu}$ has the property of being traceless, $g^{\mu\nu} \sigma_{\mu\nu}=0$. In terms of these building blocks one then has
 \begin{equation}
   \label{eq:hydro1f}
   T^{\mu\nu}_{(1)}=-\eta \sigma^{\mu\nu}-\zeta \Delta^{\mu\nu} \nabla_\lambda^\perp u^\lambda\,,
   \end{equation}
 with two scalar functions $\eta(\epsilon),\zeta(\epsilon)$ that have to have mass dimension $d-1$. Later these functions will be recognized as shear viscosity coefficient $\eta$ and bulk viscosity coefficient $\zeta$, respectively. These viscosities are known as ``first order'' transport coefficients.
 \index{Shear viscosity}
 \index{Bulk viscosity}
 \index{Transport coefficients}
 
 It is customary to split the effective hydrodynamic corrections into a traceless and a trace part,
 \begin{eqnarray}
   &\langle T^{\mu\nu}\rangle =T_{(0)}^{\mu\nu}+\pi^{\mu\nu}+\Delta^{\mu\nu}\Pi&\,,\nonumber\\
   &\pi^{\mu\nu}=T_{(1)}^{<\mu\nu>}+T_{(2)}^{<\mu\nu>}+\ldots\,,\quad
   \Pi=\frac{1}{d-1}\left(T_{(1)\, \mu}^\mu+T_{(2)\, \mu}^\mu\right)+\ldots\,,&\,
 \end{eqnarray}
 where $\pi^{\mu\nu}$, $\Pi$ are referred to as the shear stress and bulk stress, respectively\footnote{
Note the definition of the projection operator $\langle \rangle$ which for a rank two tensor $A^{\mu\nu}$ is given as
 \begin{equation}
 A^{<\mu \nu>}\equiv \frac{1}{2}\Delta^{\mu \lambda}\Delta^{\nu\rho}\left(A_{\lambda \rho}+A_{\rho \lambda}\right)-\frac{1}{d-1}\Delta^{\mu\nu}\Delta^{\lambda \rho} A_{\lambda \rho}\,,
 \end{equation}
 cf. Eq.~(\ref{eq:tterms}). This projection operator should not be confused with the expectation value of a quantity $A$, also denoted as $\langle A \rangle$.}. To first order in gradients, Eq.~(\ref{eq:hydro1f}) implies the relations
 \begin{equation}
   \label{eq:NSconst}
   \pi^{\mu\nu}=-\eta \sigma^{\mu\nu}\,,\quad \Pi=-\zeta \nabla_\lambda^\perp u^\lambda\,,
   \end{equation}
 which are sometimes referred to as (first-order) constitutive equations.
\index{Fluid Dynamics! First-order|see {Navier-Stokes equations}}
 
 Unlike the case of ideal fluids, the constitutive equations (\ref{eq:NSconst}) in general imply that the energy-momentum tensor is no longer isotropic in the local rest-frame. In particular, the diagonal entries of the space-like part of $T^{\mu\nu}$, which for ideal fluids had the interpretation of the local pressure $P$, receive non-equilibrium corrections. One thus has to deal with an effective pressure tensor of the form $P\delta^i_j+\pi^{i}_{j, LRF}+\delta^i_j\Pi$, such that the effective pressures in the direction $i$ become
 \begin{equation}
   \label{eq:effpress}
   P_{\rm eff}^{(i)}=P+\pi^{\underline{i}}_{\underline{i},{LRF}}+\Pi\,,
 \end{equation}
 where $\pi^{\mu\nu}_{LRF}$ is the shear tensor transformed to the local rest frame and $\underline{i}$ is taken to mean that there is no summation over $i$.

 \subsection{Equation of State Near Equilibrium}
 \label{sec:neqeos}
\index{Equation of State (EoS)!Non-Equilibrium}
 
 In equilibrium, the equation of state may for instance be obtained by calculating the trace anomaly of the expectation value for the underlying quantum field theory $\langle T^{\mu}_{\mu}\rangle=T^\mu_{\mu,(0)}=-\epsilon+(d-1)P(\epsilon)$ as a function of $\epsilon$ (see e.g. Eq.~(\ref{eq:latticeeos1}) for lattice QCD).

 Out of equilibrium, the same procedure may be employed.  Eq.~(\ref{eq:umudef} defines the out-of-equilibrium energy density $\epsilon$ as the time-like eigenvalue of the local energy momentum tensor. Besides the time-like eigenvector, there typically will be $d-1$ space-like eigenvalues of $\langle T^{\mu\nu}\rangle $, which can be identified with the $d-1$ effective pressures (\ref{eq:effpress}). The trace of $\langle T^{\mu\nu}\rangle $ corresponds to the sum of all eigenvalues and hence near equilibrium may be written as
 \begin{equation}
   \langle T^{\mu}_{\mu}\rangle\simeq T^\mu_{\mu,(0)}+T^\mu_{\mu,(1)}=-\epsilon+(d-1)(P(\epsilon)+\Pi)\,, 
 \end{equation}
 because the conditions ${\rm Tr}\ \pi^{\mu\nu}=0$ and $u_\mu \pi^{\mu\nu}=0$ imply $\sum_{\underline{i}=1}^{d-1} \pi^{\underline{i}}_{\underline{i},{LRF}}=0$. In equilibrium, the bulk stress $\Pi$ vanishes, and hence $P(\epsilon)$ is just the equilibrium equation of state. For out of equilibrium situations where $\Pi\neq 0$, all non-equilibrium corrections to the trace anomaly can be absorbed into $\Pi$. Hence, without loss of generality, one may define $P(\epsilon)$ to obey the equilibrium equation of state relation even if the system is not in equilibrium.  This is the definition of $P(\epsilon)$ that will be used in the following. The system will experience a non-equilibrium pressure
 \begin{equation}
   \label{eq:peff}
   P_{\rm eff}\equiv \frac{1}{d-1}\sum_{\underline{i}=1}^{d-1} P_{\rm eff}^{\underline{i}}=P+\Pi\,,
 \end{equation}
 which depends on both the energy density $\epsilon$ as well as on the strength of the gradients via $\Pi$. Near equilibrium, the bulk stress $\Pi$ constitutes only a small correction to the equilibrium pressure.

\subsection{Second Order Hydrodynamics}

 The above program of identifying all building blocks containing a specific number of gradients for a symmetric rank two tensor can be systematically continued, cf. Refs.~\cite{Baier:2007ix,Romatschke:2009kr}. For instance, to obtain $T^{\mu\nu}_{(2)}$ one writes down all independent structures containing exactly two derivatives:
 $\nabla_\mu^\perp \nabla_\nu^\perp \ln \epsilon$, $\nabla_\mu^\perp \ln \epsilon \nabla_\nu^\perp \ln \epsilon$, $\nabla_\mu^\perp \nabla_\nu^\perp u_\lambda$, $\nabla_\mu^\perp u_\rho \nabla_\nu^\perp u_\lambda$, $\nabla_\mu^\perp u_\rho \nabla_\nu^\perp \ln \epsilon$ plus (which is new at second order) the Riemann tensor $R^{\lambda}_{\ \mu \nu \rho}$ (see appendix \ref{chap:aGR}). Next, classify all scalars and vectors orthogonal to $u^\mu$, and rank two tensors orthogonal to $u^\mu$. For an uncharged fluid, there are seven independent scalars: $\nabla_\mu^\perp \nabla^\mu_\perp \ln \epsilon$, $\nabla_\mu^\perp \ln \epsilon \nabla^\mu_\perp \ln \epsilon$, $\sigma_{\mu\nu} \sigma^{\mu\nu}$, $\Omega_{\mu\nu}\Omega^{\mu\nu}$, $\left(\nabla_\mu^\perp u^\mu\right)^2$, $u^\mu u^\nu R_{\mu\nu}$, and $R$, where
 \begin{equation}
 \label{eq:vortdef}
   \Omega_{\mu\nu}=\nabla^\perp_{[\mu} u_{\nu]}=\frac{1}{2}\left(\nabla_{\mu}^\perp u_{\nu}-\nabla_{\nu}^\perp u_{\mu}\right)
 \end{equation}
 is the fluid vorticity and $R_{\mu\nu},R$ are the Ricci tensor and scalar, respectively. There are six independent vectors orthogonal to $u^\mu$: $\nabla_\lambda^\perp \sigma^{\lambda \mu}$, $\nabla_\lambda^\perp \Omega^{\lambda \mu}$, $\sigma^{\lambda \mu} \nabla_\lambda^\perp \ln \epsilon$, $(\nabla_\lambda^\perp u^\lambda) \nabla^\mu_\perp \ln \epsilon$, $\Delta^{\lambda \mu} u^\nu R_{\lambda \nu}$\footnote{Note that $\Delta^{\lambda \mu} u^\nu R_{\lambda \nu}$ contains, and hence is not independent from, $\nabla^\mu_\perp \left(\nabla_\lambda^\perp u^\lambda\right)$.}. Multiplying by a metric factor, the scalars can be used to generate symmetric rank two tensors with non-vanishing trace. What remains to be done is to list all independent symmetric traceless rank-two tensors that are orthogonal to $u^\mu$, of which there are eight: 
 $\nabla_\perp^{<\mu}\nabla_\perp^{\nu>} \ln \epsilon$, $\nabla_\perp^{<\mu}\ln \epsilon \nabla_\perp^{\nu>} \ln \epsilon$, $\sigma^{\mu\nu}\left(\nabla^\perp_\lambda u^\lambda\right)$, $\sigma^{<\mu}_{\ \lambda} \sigma^{\nu>\lambda}$, $\sigma^{<\mu}_{\ \lambda} \Omega^{\nu>\lambda}$, $\Omega^{<\mu}_{\ \lambda} \Omega^{\nu>\lambda}$, $u_\lambda R^{\lambda <\mu \nu> \rho} u_{\rho}$, $R^{<\mu \nu>}$.  It is possible to write down an expression for $T^{\mu\nu}_{(2)}$ by including all the above structures multiplied by suitable ``second-order'' transport coefficients.

The resulting complete expressions for the shear and bulk stress, including first and second order gradients, become
 \begin{eqnarray}
   \label{eq:constf}
   \pi^{\mu \nu}&=&-\eta \sigma^{\mu \nu}+\eta\tau_\pi \left[^{<} D \sigma^{\mu \nu >}+\frac{\nabla_\lambda^\perp u^\lambda}{d-1} \sigma^{\mu \nu}\right]
+\kappa\left[R^{<\mu \nu>}-2 u_\lambda u_\rho R^{\lambda <\mu \nu>\rho}\right]
\nonumber\\
&&+\lambda_1 \sigma^{<\mu}_{\quad \lambda}\sigma^{\nu> \lambda}+\lambda_2 \sigma^{<\mu}_{\quad \lambda} \Omega^{\nu> \lambda}
+\lambda_3 \Omega^{<\mu}_{\quad \lambda}\Omega^{\nu> \lambda}\nonumber\\
&&+\kappa^* 2 u_\lambda u_\rho R^{\lambda <\mu \nu> \rho} + \eta \tau_\pi^* \frac{\nabla_\lambda^\perp u^\lambda}{d-1} \sigma^{\mu \nu}
+\bar{\lambda}_4 \nabla_\perp^{<\mu} \ln \epsilon \nabla_\perp^{\nu>}\ln \epsilon\,,\nonumber\\
\Pi&=&-\zeta\left(\nabla_\lambda^\perp u^\lambda\right)+\zeta \tau_\Pi D\mt
+\xi_1 \sigma^{\mu \nu} \sigma_{\mu \nu}+\xi_2 \left(\nabla_\lambda^\perp u^\lambda\right)^2
\nonumber\\
&&+\xi_3 \Omega^{\mu \nu} \Omega_{\mu \nu}
+\bar{\xi_4} \nabla^\perp_\mu \ln \epsilon \nabla_\perp^\mu \ln \epsilon+\xi_5 R
+\xi_6 u^\lambda u^\rho R_{\lambda \rho}\,.
   \end{eqnarray}
The careful reader will have noticed that (\ref{eq:constf}) contains expressions such as $\left< D \sigma^{\mu \nu}\right>+\frac{\nabla_\lambda^\perp u^\lambda}{d-1} \sigma^{\mu \nu}$ and $D\mt$ that were not part of the original building blocks identified above. It turns out that these combinations are related to particular combinations of seven independent tensors and seven independent scalars, respectively (see Refs.~\cite{Baier:2007ix,Romatschke:2009kr} for detail). These relations use the lower-order equations of motion such as (\ref{eq:eulerf2}) to rewrite time derivatives, and they are approximately true up to second-order in gradients. While convenient, there are some recent hints in the literature  that in particular situations these higher-order gradient terms may become important \cite{Attems:2018gou}.
However, as will be discussed in the following sections, it is necessary to perform a replacement of at least part of the second-order gradient terms by time-derivatives as in (\ref{eq:constf}) in order to obtain a set of equations of motions that are causal and stable.

 \index{BRSSS}
 \index{Fluid Dynamics! Second-order|see {BRSSS}}
 
 The coefficients $\tau_\pi,\kappa,\lambda_1,\lambda_2,\lambda_3,\kappa^*,\tau_\pi^*,\bar{\lambda}_4,\tau_\Pi,\xi_1,\xi_2,\xi_3,\bar{\xi}_4,\xi_5,\xi_6$ are so-called ``second-order'' transport coefficients\footnote{Note that the the transport coefficient $\bar{\lambda}_4,\bar{\xi}_4$ in Eq.~(\ref{eq:constf}) differ from $\lambda_4,\xi_4$ in the original reference \cite{Romatschke:2009kr} because there a different expansion basis was used.}. They can be thought of as 'non-Newtonian' (or curvature) corrections to the Navier-Stokes constitutive relations (\ref{eq:NSconst}).
 The second-order constitutive relations for an uncharged fluid slightly out of equilibrium (\ref{eq:constf}) contain several terms that vanish in flat space-times, e.g. the terms multiplied by the coefficients $\kappa,\kappa^*,\xi_5,\xi_6$. Nevertheless, these terms will turn out to contribute to the hydrodynamic correlation functions in flat space-time, as will be shown below. 

 \subsubsection{Third Order Hydrodynamics}
\index{Fluid Dynamics! Third-order}
 
 Third order correction terms were first considered in Ref.~\cite{El:2009vj} followed by the construction of third order gradient terms for a particular underlying microscopic theory, namely weakly coupled kinetic theory \cite{Jaiswal:2013vta,Chattopadhyay:2014lya}. All possible structures to third order in gradients were derived in Ref.~\cite{Grozdanov:2015kqa}. It was found that there are 68 independent structures contributing to $T^{\mu\nu}_{(3)}$, and we refer the interested reader to Ref.~\cite{Grozdanov:2015kqa} for details.

 \subsubsection{The Relativistic Navier-Stokes Equations}
\index{Navier-Stokes equations}
 
 If one truncates the hydrodynamic gradient expansion at first order, then the covariant conservation of the energy-momentum tensor implies
 \begin{equation}
   \label{eq:ns1}
   \nabla_\mu \left[T^{\mu\nu}_{(0)}+T^{\mu\nu}_{(1)}\right]=0\,,
 \end{equation}
 with zeroth and first order terms $T^{\mu\nu}_{(0)},T^{\mu\nu}_{(1)}$ given by Eqns.~(\ref{eq:hydro0f}), (\ref{eq:hydro1f}), respectively. Using again projection operators $u^\mu$ and $\Delta^{\mu\alpha}$ on the equations of motion (\ref{eq:ns1}) (cf. chapter \ref{chap:euler}), one finds\footnote{Note that $u_\mu \nabla_\nu \sigma^{\mu\nu}=\nabla_\nu (0)-\sigma^{\mu\nu}\nabla_\nu u_\mu=-\frac{1}{2}\sigma^{\mu\nu}\sigma_{\mu\nu}$ because $u_\mu \sigma^{\mu\nu}=0$, cf. Eq.(\ref{eq:tterms}).}
 \begin{eqnarray}
   \label{eq:NSf}
   D\epsilon+   (\epsilon+P)\mt&=&\frac{\eta}{2} \sigma^{\mu\nu}\sigma_{\mu\nu}+\zeta \left(\nabla_\lambda^\perp u^\lambda\right)^2\,,\nonumber\\
   (\epsilon+P)Du^\alpha+c_s^2 \nabla_\perp^\alpha \epsilon&=&\Delta^\alpha_\nu \nabla_\mu \left(\eta \sigma^{\mu\nu}+\zeta \Delta^{\mu\nu}\mt\right)\,.
   \end{eqnarray}
 These equations are known as the relativistic viscous fluid dynamics equations, or the relativistic Navier-Stokes equations. Unlike their non-relativistic counter-part, they are only useful for a small class of problems which can be treated analytically. Their practical numerical application is hampered by the fact that they violate causality, which in turn causes instabilities, as shall be discussed below in chapter \ref{sec:NSinst} \cite{Hiscock:1985zz}.

 \subsubsection{Application: Non-Relativistic Navier-Stokes Equations}

 In the case of small (space) velocities $|{\bf v}|\ll 1$ and for Minkowski space-time, the approximations (\ref{eq:nrlimit})  applied to Eqns.~(\ref{eq:NSf}) together with assuming $\eta,\zeta$ to be constant lead to \cite{LL}, \S15
 \begin{eqnarray}
   \partial_t \epsilon+ {\vec v}\cdot \vec\partial \epsilon+\epsilon \vec\partial \cdot {\vec v}&=&0\,,\nonumber\\
   \epsilon \partial_t {\vec v}+\epsilon \left({\vec v}\cdot \vec\partial\right) {\vec v}+\vec\partial P&=&\eta\, \partial^2 {\vec v}+\left(\zeta+\frac{d-3}{d-1}\eta\right)\vec\partial \left(\vec\partial\cdot {\vec  v}\right)\,,
   \end{eqnarray}
 which can be recognized as the continuity and non-relativistic Navier-Stokes equations in $d$ space-time dimensions \cite{Navier:1822,Stokes:1845}.

 We close by remarking that by analogy, the non-relativistic limit of the equations of motion for the second- and third- order theory, respectively, correspond to a complete form of the Burnett and super-Burnett equations \cite{Burnett}.

 \subsubsection{Conformal Second Order Hydrodynamics}
 \label{sec:confsym}
\index{Conformal fluid dynamics}
 
 There exists a class of theories for which the trace of the energy-momentum tensor expectation value vanishes in Minkowski space. These theories will be referred to as ``conformal'' theories. In curved space-time, the trace of the energy-momentum tensor is anomalous,
 \begin{equation}
   \label{eq:weyl}
   g_{\mu\nu}\langle T^{\mu\nu}\rangle=W_d[g_{\mu\nu}]\,,
 \end{equation}
 where $W_d$ is the called Weyl anomaly in $d$ space-time dimensions \cite{Duff:1993wm}. The Weyl anomaly vanishes in odd space-time dimensions and contains $d$ derivatives in even space-time dimensions\footnote{For d=4, one has $$W_d=-\frac{a}{16 \pi^2}\left(R_{\mu\nu \lambda \rho} R^{\mu\nu\lambda\rho}-4 R_{\mu\nu}R^{\mu\nu}+R^2\right)+\frac{c}{16 \pi^2}\left(R_{\mu\nu\lambda\rho}R^{\mu\nu\lambda \rho}-2 R_{\mu\nu}R^{\mu\nu}+\frac{1}{3}R^2\right)\,,$$ where $a=c=\frac{N_c^2-1}{4}$ for an SU($N_c$) gauge theory \cite{Aharony:1999ti}.}. Thus, when attempting to construct an effective theory of fluid dynamics to first or second order in gradients in $d>2$ space-time dimensions, the Weyl anomaly may effectively be neglected.

 The condition $g_{\mu\nu}\langle T^{\mu\nu}\rangle=0$ implies an additional symmetry for the theory, namely symmetry under Weyl transformations of the metric:
 \begin{equation}
   \label{eq:Weyltrafo}
   g_{\mu\nu}\rightarrow g_{\mu\nu}e^{-2 w(x)}\,,
 \end{equation}
 where $w(x)$ is a space-time dependent scale factor. Recalling that the energy-momentum tensor is defined as the metric derivative of an effective action \cite{Baier:2007ix} one expects (cf. Eq.~(\ref{eq:traforules})
 \begin{equation}
   \label{eq:effS}
   \frac{\delta S_{\rm eff}}{\delta g_{\mu\nu}}\propto \sqrt{-g}\  \langle T^{\mu\nu}\rangle\,,
 \end{equation}
 where $g\equiv {\rm det}g_{\mu\nu}$.  Since the action has to be invariant under Weyl transformations, this implies the condition 
 \begin{equation}
   \label{eq:weylsymm}
   \langle T^{\mu\nu}\rangle \rightarrow \langle T^{\mu\nu} \rangle e^{(d+2)w(x)}\,,
 \end{equation}
 for the energy-momentum tensor. The factor $(d+2)$ in the exponent of the Weyl transformed quantity is referred to as conformal weight. Since $u^\mu u_\mu=g_{\mu\nu}u^\mu u^\nu=-1$, the conformal weight of $u^\mu$ is $1$, because the conformal weight of the metric tensor is $-2$. Eq.~(\ref{eq:hydro0f}) then implies a conformal weight of $d$ for $\epsilon,P$.

 Conformal symmetry has immediate implications for the transport coefficients of the theory. For instance, if the system gradients are small enough such that the Weyl anomaly (\ref{eq:weyl}) can be neglected, the condition $g_{\mu\nu} T^{\mu\nu}_{(0)}=0$ implies the relations
 \begin{equation}
   \label{eq:confeos}
   P(\epsilon)=\frac{\epsilon}{d-1}\,,\quad c_s=\sqrt{\frac{1}{d-1}}
 \end{equation}
for the pressure \index{Equation of State (EoS)!Conformal} and energy density out of equilibrium. This finding is fully consistent with the discussion in section \ref{sec:neqeos}, because for conformal systems the tracelessness condition $g_{\mu\nu}T^{\mu\nu}_{(1)}=0$ implies
 \begin{equation}
   \zeta=0\,,
 \end{equation}
or a vanishing bulk viscosity coefficient.

 To first order in gradients, the symmetry (\ref{eq:weylsymm}) under Weyl transformations is non-trivial since gradients act on scale factor $w(x)$. For instance, $\nabla_\mu u^\nu=\partial_\mu u^\nu+\Gamma^\nu_{\mu\lambda} u^\lambda$ picks up derivatives of $w(x)$ from the first term as well as the definition of the Christoffel symbol (\ref{eq:christoffel}),
 \begin{equation}
   \Gamma^\nu_{\mu\lambda}\rightarrow \Gamma^\nu_{\mu \lambda}-\left(g^\nu_\lambda \partial_\mu w+g^\nu_\mu \partial_\lambda w- g_{\mu \lambda}\partial^\nu w\right)\,.
   \end{equation}
 Thus, one finds that the second term in Eq.~(\ref{eq:hydro1f}) does not transform properly under Weyl transformations, $\nabla_\lambda^\perp u^\lambda\rightarrow e^w\left(\nabla_\lambda^\perp u^\lambda-(d-1)D w\right)$. This implies that the second term in Eq.~(\ref{eq:hydro1f}) is not allowed in a conformal theory, which is consistent with the finding $\zeta=0$ from above. Interestingly, while $\nabla_\lambda^\perp u^\lambda$ does not transform properly under Weyl transformations, it is straightforward to show that $\sigma^{\mu\nu}$ does, obeying $\sigma^{\mu\nu}\rightarrow e^{3 w} \sigma^{\mu\nu}$. Therefore, non-vanishing shear viscosity is allowed for conformal systems.
 
While only a useful check for first-order hydrodynamics, conformal symmetry becomes a powerful organizational tool when considering second-order or higher-order hydrodynamic field theory. Requiring that building blocks of $T^{\mu\nu}_{(2)}$ transform properly under Weyl transformations, one finds that only three combinations out of the seven independent scalars to second order in gradients are Weyl invariant, two out of six vectors and five out of eight tensors. This implies that many of the second-order transport coefficients in Eq.~(\ref{eq:constf}), namely $\zeta,\xi_1,\xi_2,\xi_3,\bar{\xi}_4,\xi_5,\xi_6,\kappa^*,\tau_\pi^*$ and $\bar{\lambda}_4$ all vanish for conformal systems. The resulting form of the constitutive relations for conformal second order hydrodynamics is correspondingly simplified, and one finds
\begin{eqnarray}
  \label{eq:BRSSSpi}
  \pi^{\mu \nu}&=&-\eta \sigma^{\mu \nu}+\eta\tau_\pi \left[^{<} D \sigma^{\mu \nu>}+\frac{\nabla\cdot u}{d-1} \sigma^{\mu \nu}\right]
+\kappa\left[R^{<\mu \nu>}-2 u_\lambda u_\rho R^{\lambda <\mu \nu>\rho}\right]
\nonumber\\
&&+\lambda_1 \sigma^{<\mu}_{\quad \lambda}\sigma^{\nu> \lambda}+\lambda_2 \sigma^{<\mu}_{\quad \lambda} \Omega^{\nu> \lambda}
+\lambda_3 \Omega^{<\mu}_{\quad \lambda}\Omega^{\nu> \lambda}\,,
\end{eqnarray}
together with $\Pi=0$.

\subsubsection{Application: BRSSS/BHMR}
\index{BRSSS}

Conformal second-order hydrodynamics is defined by the equations of motion
\begin{equation}
  \label{eq:conf2d}
  \nabla_\mu \left[T^{\mu\nu}_{(0)}+\pi^{\mu\nu}\right]=0\,.
\end{equation}
where $T^{\mu\nu}_{(0)},\pi^{\mu\nu}$ are defined in Eqns.~(\ref{eq:hydro0f},\ref{eq:BRSSSpi}). Projecting onto time-like and space-like components, these equations take the form
\begin{eqnarray}
   \label{eq:BRSSSf}
   D\epsilon+   (\epsilon+P)\mt&=&-\frac{1}{2} \pi^{\mu\nu}\sigma_{\mu\nu}\,,\nonumber\\
   (\epsilon+P)Du^\alpha+c_s^2 \nabla_\perp^\alpha \epsilon&=&-\Delta^\alpha_\nu \nabla_\mu \pi^{\mu\nu}\,.
  \end{eqnarray}
Since $\pi^{\mu\nu}$ includes second-order gradients, Eqns.~(\ref{eq:conf2d}) are generalizations of the (conformal) Navier-Stokes equations (\ref{eq:NSf}), and are sometimes referred to as BRSSS or BHMR equations \cite{Baier:2007ix,Bhattacharyya:2008jc}. The BRSSS equations can be thought of as relativistic generalizations of the Burnett equations in non-relativistic fluid dynamics \cite{Burnett}. Similar to the Navier-Stokes equations, the BRSSS equations violate causality, limiting their practical application to analytic problems (see the discussion in section \ref{sec:NSinst}). Unlike the Navier-Stokes equations, a resummed version of the BRSSS equations does not suffer from causality violations or instabilities and has wide practical applications in numerical relativistic viscous fluid dynamics simulations (see section \ref{sec:rBRSSS}).

The BRSSS/BHMR equations trivially reduce to the conformal Navier-Stokes equations in the limit of $\tau_\pi\rightarrow 0,\lambda_1\rightarrow 0,\lambda_2\rightarrow 0,\lambda_3\rightarrow 0,\kappa\rightarrow 0$.

\subsubsection{Application: Viscous Cosmology}
\index{Cosmology!Viscous}

 While most current approaches to cosmology ignore viscous effects in the description, let us estimate the effect of a non-vanishing and constant bulk viscosity coefficient on the evolution of the universe\footnote{A constant bulk viscosity coefficient is not very realistic because $\zeta$ will generally be dependent on the energy-density \cite{Lu:2011df,Arnold:2006fz}, but this case is easy to solve analytically.}. This leads to the discussion of viscous cosmology (cf. Ref.~\cite{Brevik:2017msy} for a recent review). To simplify the discussion, let us furthermore consider only first-order hydrodynamics (Navier-Stokes). For the FLRW universe (see the discussion in section \ref{sec:idealbjork}), the relativistic Navier-Stokes equations (\ref{eq:NSf}) imply
 \begin{equation}
   \partial_t \epsilon+3(\epsilon+P)H=9 \zeta H^2\,,
 \end{equation}
 where $H(t)$ is the Hubble ``constant'' which again is related to the energy density via the Einstein equations (\ref{eq:flrwee}) as $H^2=\frac{8 \pi G}{3}\epsilon$.

 For dust, $P=0$ and the viscous cosmology equations of motion can be written in the form
 \begin{equation}
   \label{eq:simvisc}
   \partial_t \epsilon^{-1/2}=\sqrt{6 \pi G}-12\pi G \zeta \epsilon^{-1/2}\,.
 \end{equation}
 Clearly, viscous corrections are small only when the energy density is high $\epsilon\gg 1$. This implies that for viscous cosmology, the small gradient regime corresponds to the high-density (early-time) period. Eq.~(\ref{eq:simvisc}) may be solved exactly to yield
 \begin{equation}
   \epsilon(t)=\frac{24 \pi G \zeta^2}{\left(1-e^{-12 \pi G \zeta t}\right)^2}\simeq \frac{1}{6 \pi G t^2}+\frac{2 \zeta}{t}+\ldots\,,
 \end{equation}
 where the rhs is an expansion for early times. Clearly the energy density is dropping considerably more slowly in viscous cosmology than in standard cosmology. The scale factor $a(t)$ may be calculated from the Hubble constant $H=\frac{a^\prime(t)}{a(t)}$ as
 \begin{equation}
   \label{eq:viscscal}
   a(t)\propto\left(e^{12 G \pi \zeta t}-1\right)^{2/3} \propto t^{2/3}+4 \pi G \zeta t^{5/3}\,.
   \end{equation}
 Taken at face value, (\ref{eq:viscscal}) implies that the expansion of the universe accelerates exponentially for $t\gtrsim \frac{1}{12\pi G \zeta}$. This is known as viscous inflation. However, this interpretation is potentially misleading because the regime of applicability was identified to be that of early times. Indeed, the effective pressure (\ref{eq:effpress}) for this system is negative, and one can expect the phenomenon of cavitation to occur \cite{Rajagopal:2009yw,Bhatt:2010hu,Klimek:2011by,Habich:2014tpa,Sanches:2015vra}, leading to a break-down of fluid dynamics.  One thus needs at the very least an off-equilibrium formulation of fluid dynamics to properly study the properties of viscous inflation, cf.~\cite{Zimdahl:1996ka}.

Blindly assuming that Eq.~(\ref{eq:viscscal}) would somehow be applicable for such an off-equilibrium formulation of viscous cosmology, it is nevertheless remarkable that 
bulk viscous effects generically would lead to an acceleration of the universe and thus may offer a viable alternative to dark energy \cite{Gagnon:2011id}.

\subsubsection{Application: Modified Gravity Effects from Matter}
\index{Modified Gravity Effects}

An amusing consequence of the fact that the energy-momentum tensor $T^{\mu\nu}$ itself contains curvature terms at second-order in gradients is that standard Einstein's equations behave as if gravity has been modified. Specifically, consider Einstein's equations in matter (\ref{eq:einstein}) and consider the case of local thermodynamic equilibrium $\epsilon={\rm const}$, $u^\mu=\left(1,{\bf 0}\right)$. In this case, second-order hydrodynamics (\ref{eq:constf}) leads to
\begin{equation}
\label{eq:mimi}
R^{\mu\nu}-\frac{1}{2} R g^{\mu\nu}+\Lambda g^{\mu\nu}=8 \pi G \left[\epsilon u^\mu u^\nu +\left(P+\xi_5 R + \xi_6 u^\lambda u^\rho R_{\lambda \rho}\right)\Delta^{\mu\nu} +\pi^{\mu\nu}\right]\,,
\end{equation}
where $\pi^{\mu\nu}=\kappa\left[R^{<\mu \nu>}-2 u_\lambda u_\rho R^{\lambda <\mu \nu>\rho}\right]+\kappa^* 2 u_\lambda u_\rho R^{\lambda <\mu \nu> \rho}$. By dimensional reasons, the second-order transport coefficients $\kappa,\kappa^*,\xi_5,\xi_6$ are proportional to $\sqrt{\epsilon}$ and at least some of them do not vanish even for non-interacting particles, cf. Tab.~ \ref{tab:one2}. It is possible to formally write 
(\ref{eq:mimi}) as ``modified gravity'' with a perfect fluid energy-momentum tensor by subtracting the second-order matter terms from the Einstein tensor,
\begin{eqnarray}
\label{eq:mimi2}
8 \pi G \left[\left(\epsilon+P\right) u^\mu u^\nu+P g^{\mu\nu}\right] &=&
R^{\mu\nu}-\frac{1}{2} R g^{\mu\nu} +\Lambda g^{\mu\nu}\nonumber\\
&&- 8 \pi G \left(\xi_5 R+\xi_6 u^\lambda u^\rho R_{\lambda \rho}\right) \Delta^{\mu\nu}\nonumber\\
 &&- 8 \pi G \left[\kappa R^{<\mu \nu>}-2 (\kappa+\kappa^*) u_\lambda u_\rho R^{\lambda <\mu \nu>\rho}\right]\,.\nonumber
\end{eqnarray}
The above equation formally behaves as a modification from standard gravity because the coefficients multiplying the curvature terms depend on the local energy density, and additional structures not present in standard gravity appear. However, all of these effects are strongly suppressed as $\sqrt{\epsilon} G$, such that they would become important only if $\sqrt{\epsilon}$ is of order of the Plank mass squared. Nevertheless, since the above second-order curvature contributions to $T^{\mu\nu}$ are known to exist, it could be interesting to study their consequence for instance in theories of the early universe.

 \subsubsection{Application: Viscous Bjorken Flow}
 \label{sec:viscbjork}
 \index{Bjorken flow!Viscous}

 It is interesting to consider corrections to the ideal fluid result for Bjorken flow, cf. Eq.~(\ref{eq:ebjor0}). To keep the discussion transparent, we limit ourselves to the case of a conformal system in flat $d=4$ space-time where $c_s^2=\frac{1}{3},\zeta=0$ and many of the second-order transport coefficients are vanishing. Using again Milne coordinates as in the discussion in section \ref{sec:idealbjork}, one has $\nabla_\mu u^\nu=\Gamma_{\mu\tau}^\nu$, which one can readily evaluate using $\Gamma_{\xi \tau}^\xi=\frac{1}{\tau}$ and $\Gamma^\tau_{\xi \xi}=\tau$. This leads to
 \begin{equation}
   \sigma_{\mu\nu}={\rm diag}\left(\sigma_{\tau\tau},\sigma_{xx},\sigma_{yy},\sigma_{\xi\xi}\right)={\rm diag}\left(0,-\frac{2}{3\tau},-\frac{2}{3\tau},\frac{4\tau}{3}\right)\,.
   \end{equation}
 Further manipulations for the quantities appearing in (\ref{eq:BRSSSpi}) lead to $D \sigma^{\mu\nu}=\partial_\tau \sigma^{\mu\nu}+2 \Gamma^\mu_{\tau \xi}\sigma^{\xi \nu}={\rm diag}\left(0,\frac{2}{3\tau^2},\frac{2}{3\tau^2},-\frac{4}{3\tau^4}\right)$, $\sigma^{<\mu}_{\quad \lambda}\sigma^{\nu> \lambda}={\rm diag}\left(0,-\frac{4}{9 \tau^2},-\frac{4}{9 \tau^2},\frac{8}{9 \tau^4}\right)$ and $\Omega^{\mu\nu}=0$. Since the space-time is flat, $R=0=R^{\mu\nu}$ and hence the viscous fluid dynamics equations generalizing (\ref{eq:bjork0}) are given by \cite{Baier:2007ix}
 \begin{equation}
   \label{eq:bjork2}
   \partial_\tau \epsilon+\frac{\epsilon+p}{\tau}=\frac{4\eta}{3\tau^2}+\frac{8 \eta \tau_\pi}{9 \tau^3}-\frac{8\lambda_1}{9 \tau^3}\,,
 \end{equation}
 where it is recalled that for a conformal fluid $P=\frac{\epsilon}{3}$, cf. Eq.~(\ref{eq:confeos}). For a conformal system, all quantities scale with their mass dimension, so dimensionless quantities have to be independent of $\tau$. Dividing (\ref{eq:bjork2}) by $\epsilon$, it is customary to introduce the dimensionless combinations $C_\eta,C_\pi,C_\lambda$ (which are constants for conformal systems) through
 \begin{equation}
   \label{eq:defs}
   \frac{\eta}{\epsilon}=\frac{4}{3} C_\eta T^{-1}\,,\quad
   \frac{\eta\tau_\pi}{\epsilon}=\frac{4}{3}C_\eta C_\pi T^{-2}\,,\quad
   \frac{\lambda_1}{\epsilon}=\frac{4}{3}C_\eta C_\pi C_\lambda T^{-2}\,.
 \end{equation}
 (The reason for these particular combinations will become apparent once results for $\eta,\tau_\pi,\lambda_1$ are discussed in sections \ref{sec:kin},\ref{sec:ads}).
 Since for $d=4$ the energy density has mass dimension four, a new quantity
 $T(\tau)\equiv\left(\frac{\epsilon(\tau)}{\rm const}\right)^{1/4}$ was introduced which has mass dimension one. It is customary, but not necessary, to interpret $T$ as the non-equilibrium ``pseudo-temperature'' of the conformal system. Using (\ref{eq:defs}), Eq.~(\ref{eq:bjork2}) becomes
 \begin{equation}
   \label{eq:bjork3}
   \partial_\tau \ln\left(\frac{\epsilon}{\rm const}\right)=-\frac{4}{3\tau}+\frac{16 C_\eta}{9\tau^2} \left(\frac{\epsilon}{\rm const}\right)^{-\frac{1}{4}}+\frac{32}{27 \tau^3}\left(\frac{\epsilon}{\rm const}\right)^{-\frac{1}{2}}
   C_\eta C_\pi\left(1-C_\lambda\right)\,.
 \end{equation}
 From the ideal Bjorken flow solution Eq.~(\ref{eq:ebjor0}), it is obvious that the viscous terms involving $C_\eta,C_\pi,C_\lambda$ in (\ref{eq:bjork3}) are small corrections if $\frac{\tau}{\tau_0}\gg 1$. This implies that for Bjorken flow, the regime of small gradients corresponds to the late-time limit. This suggests one can expand the solution for $T(\tau)$ as a series of inverse powers of $\tau$. One finds \cite{Baier:2007ix}
 \begin{equation}
   \label{eq:bjorsol1}
   T(\tau)\simeq \tau_0^{-1}\left(\frac{\tau_0}{\tau}\right)^{1/3}\left(1- \frac{2 C_\eta}{3 } \left(\frac{\tau_0}{\tau}\right)^{2/3}-\frac{2 C_\eta C_\pi\left(1-C_\lambda\right)}{9 }\left(\frac{\tau_0}{\tau}\right)^{4/3}\right)\,,
 \end{equation}
 where $\tau_0$ is an integration constant.

 It is also interesting to consider the effective pressure anisotropy $\frac{P_L}{P_T}\equiv P_{\rm eff}^{(\xi)}/P^{(x)}_{\rm eff}$, cf. Eq.~(\ref{eq:effpress}). To this end, it is useful to realize that the high degree of symmetry for the Bjorken flow in a conformal system implies that the effective pressures (and derived quantities such as the anisotropy) can be recast in terms of the energy density and its derivatives. Specifically, $\nabla_\mu T^{\mu\nu}=0$ and $T^\mu_\mu=0$ for conformal Bjorken flow imply \cite{Janik:2005zt}
 \begin{equation}
   \tau \partial_\tau T_{\tau\tau}+T_{\tau\tau}+\frac{T_{\xi \xi}}{\tau^2}=0\,,\quad
   -T_{\tau\tau}+2 T_{xx}+\frac{T_{\xi \xi}}{\tau^2}=0\,,
 \end{equation}
 which using $T_{\tau\tau}=\epsilon(\tau)$ may be recast as $P_{\rm eff}^{(\xi)}=T_\xi^\xi=-\epsilon(\tau)-\tau \partial_\tau \epsilon(\tau)$ and $P_{\rm eff}^{(x)}=T_x^x=\epsilon+\frac{\tau}{2}\partial_\tau \epsilon$. Thus the pressure anisotropy $P_{\rm eff}^{(\xi)}/P^{(x)}_{\rm eff}$ for conformal Bjorken flow takes the form
 \begin{equation}
   \label{eq:bjorkani}
   \frac{P_L}{P_T}=\frac{-1-\tau \partial_\tau \ln \epsilon}{1+\frac{\tau}{2}\partial_\tau \ln \epsilon}\simeq 1-\frac{8 C_\eta}{\tau T}+\frac{16 C_\eta \left(4 C_\eta-C_\pi\left(1-C_\lambda\right)\right)}{3 \tau^2 T^2}+{\cal O}\left(\left(\tau T\right)^{-3}\right)\,,
 \end{equation}
 where the rhs has been evaluated using (\ref{eq:bjork3}),
 \begin{equation}
   \label{eq:dtbjork}
   \tau \partial_\tau \ln \epsilon = -\frac{4}{3}+\frac{16 C_\eta}{9 \tau T}+\frac{32 C_\eta C_\pi\left(1-C_\lambda\right)}{27 \tau^2 T^2}\,.
 \end{equation}
 For the pressure anisotropy (\ref{eq:bjorkani}), the hydrodynamic expansion becomes an expansion in the dimensionless combination $\frac{1}{\tau T}\propto \tau^{-2/3}$. For $\frac{1}{\tau T}\ll 1$, $P_L=P_T$, and the system has a locally isotropic pressure tensor. However, viscous corrections lead to effectively anisotropic longitudinal and transverse pressures, and the corrections in (\ref{eq:bjorkani}), (\ref{eq:dtbjork}) become large for $\tau T\propto C_\eta$. The hydrodynamic gradient expansion can no longer be trusted for times earlier than this.

\subsubsection{Application: Flows with Vorticity}
\index{Vorticity}
\index{Analytic Flows! {BIR}|see {Vorticity}}

\begin{figure*}[t]
  \centering
  \includegraphics[width=.7\linewidth]{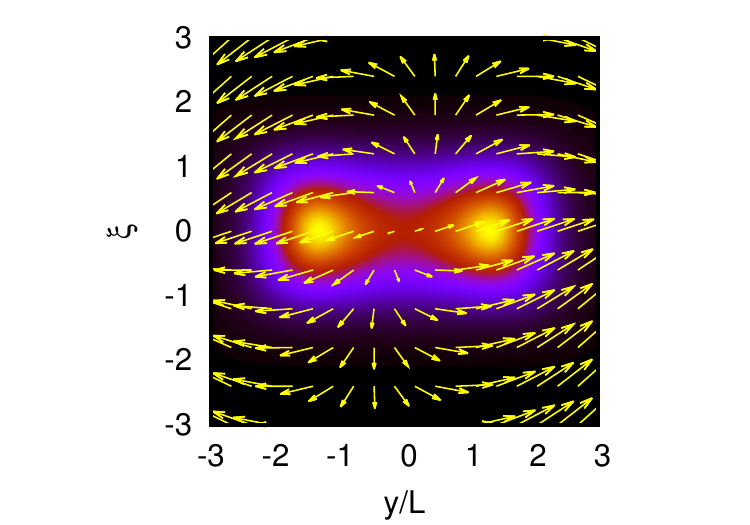}
  \caption{\label{fig:BIR} Example for the analytic fluid flow solution (\ref{eq:oursol}) with $\omega_1=0.5$, $\omega_2=0.05$ and $\tau=1.5 L$. Shown are flow velocities (vectors) and iso-contours for the energy density in the $y-\xi$ plane for $x=0$. Figure adapted from Ref.~\cite{Bantilan:2018vjv}.}
\end{figure*}

A further generalization of Bjorken and Gubser flows that includes full three-dimensional dynamics and non-vanishing vorticity has been considered in Ref.~\cite{Bantilan:2018vjv}. Similar to Gubser flow, this flow arises as a coordinate-transformation from a stationary solution. While the interested reader is referred to the original reference for details about the derivation, we note that the resulting energy density and fluid velocities $u^\mu=\gamma \left(1,v^x,v^y,v^\xi/\tau\right)$ are given by
\begin{eqnarray}
  \label{eq:oursol}
   \epsilon &=&16 L^8 T_0^4 \left[(L^4+2 L^2 {\bf x}_\perp^2+(\tau^2-{\bf x}_\perp^2)^2)(1-\omega_2^2)\right.\nonumber\\
   &&\left.+2 L^2 (\tau^2-2 y^2)(\omega_1^2-\omega_2^2)+2 L^2 \tau^2(1-\omega_1^2)\cosh 2\xi\right]^{-2}\,,\nonumber\\
\gamma&=&\frac{\left[(L^2+\tau^2+{\bf x}_\perp^2)\cosh \xi+2 (\tau \omega_2 x-L \omega_1 y\sinh\xi)\right]}{\left(16 L^8 T_0^4/\epsilon\right)^{1/4}} \nonumber\\
v^x&=&\frac{2 \tau x \cosh\xi+\omega_2(L^2+\tau^2+x^2-y^2)}{(L^2+\tau^2+{\bf x}_\perp^2)\cosh \xi+2 (\tau \omega_2 x-L \omega_1 y \sinh\xi)}\,,\nonumber\\
  v^y&=&\frac{2 \tau y \cosh\xi+2 \omega_2 x y-2 L \tau \omega_1 \sinh\xi}{(L^2+\tau^2+{\bf x}_\perp^2)\cosh \xi+2 (\tau \omega_2 x-L \omega_1 y \sinh\xi)}\,,\nonumber\\
v^\xi&=&-\frac{(L^2-\tau^2+{\bf x}_\perp^2)\sinh\xi-2 L \omega_1 y \cosh\xi}{(L^2+\tau^2+{\bf x}_\perp^2)\cosh \xi+2 (\tau \omega_2 x-L \omega_1 y \sinh\xi)}\,,
  \end{eqnarray}
  in Milne coordinates $\tau=\sqrt{t^2-z^2},\xi={\rm arctanh}(z/t)$ and ${\bf x}_\perp^2=x^2+y^2$. In the solution (\ref{eq:oursol}), $T_0$ denotes the overall energy scale, $L$ is a length scale that corresponds to a choice of units and $|\omega_{1,2}|<1$ are two angular rotation frequencies.

The flow solution (\ref{eq:oursol}) has several curious properties. For instance, it is conformal, and one may verify explicitly that (\ref{eq:oursol}) is a solution to \textit{both} the relativistic conformal Navier-Stokes equations (\ref{eq:NSconst}) and the relativistic conformal Euler equation (\ref{eq:eulera}). This is possible because the shear-stress $\sigma_{\mu\nu}$ defined in Eq.~(\ref{eq:tterms}) exactly vanishes for the solution (\ref{eq:oursol}) and hence the Navier-Stokes equations revert to the Euler equation for arbitrary values of the shear-viscosity coefficient $\eta$. Furthermore, the solution 
(\ref{eq:oursol}) describes a fully dynamical three-dimensional fluid evolution profile in terms of two parameters $\omega_{1,2}$, where $\omega_1$ controls rotations in the $x-\xi$ plane and $\omega_2$ controls asymmetries in the $x-y$ plane. Finally, it may be verified explicitly that (\ref{eq:oursol}) has non-vanishing vorticity $\Omega_{\mu\nu}$ defined in (\ref{eq:vortdef}) for $\omega_1\neq 0$.

 \section{Out of Equilibrium Fluid Dynamics}
\label{sec:offeq}

In sections \ref{sec:equi} and \ref{sec:navi}, the theory of fluid dynamics was constructed through a systematically improvable gradient expansion of the fundamental hydrodynamic degrees of freedom. This effective theory is applicable whenever the correction term $T^{\mu\nu}_{(1)}$ is small compared to the leading order, ideal fluid part $T^{\mu\nu}_{(0)}$. Since $T^{\mu\nu}_{(1)}$  contains first order gradients, and $T^{\mu\nu}_{(0)}$ does not, this implies that in order for hydrodynamics to be valid, gradients are required to be small.

It is possible to recast the requirement of small gradients in different form, which may be more familiar to some readers. For instance, using Eqns.~(\ref{eq:hydro0f},\ref{eq:hydro1f}) the size of the gradient $\sigma^{\mu\nu}$ in $T^{\mu\nu}_{(1)}$ may be estimated as the inverse system size $L^{-1}$, and the relative size of $\eta$ to $T^{\mu\nu}_{(0)}$ may be estimated as
\begin{equation}
  \label{eq:lambdamfp}
  \frac{\eta}{\epsilon+P}\equiv \lambda_{\rm mfp}\,,
\end{equation}
which \index{Mean free path} has the interpretation of a mean free path. Thus the requirement of small gradients is synonymous with requiring the mean free path in the fluid to be small compared to the system size $\lambda_{\rm mfp}\ll L$ \cite{Belenkij:1956cd}. A trivial rearrangement of this statement in terms of the Knudsen number ${\rm Kn}\equiv \frac{\lambda_{\rm mfp}}{L}$ leads to
\begin{equation}
  \label{eq:landaucrit}
  {\rm Kn}\ll 1\,.
\end{equation}
\index{Knudsen number}

These requirements for the validity of fluid dynamics have been used for decades if not centuries, and are the de-facto standard. This is the reason why most current treatise on fluid dynamics theory limit the applicability of fluid dynamics to the description of near-equilibrium systems.

However, there is mounting evidence that fluid dynamics can successfully be applied to genuine out-of-equilibrium situations \cite{Chesler:2009cy,Heller:2011ju,Wu:2011yd,vanderSchee:2012qj,Casalderrey-Solana:2013aba,Kurkela:2015qoa,Keegan:2015avk,Attems:2017ezz}. The present section deals with setting up such a theory of genuine ``out of equilibrium fluid dynamics''.

To get started, let us consider a system in a state where the gradients of the hydrodynamic fields are \textit{not} small. Then the first order correction $T^{\mu\nu}_{(1)}$ would generally be of the same size as $T^{\mu\nu}_{(0)}$, $T^{\mu\nu}_{(2)}$ of the same size as $T^{\mu\nu}_{(1)}$, and so on. The standard fluid dynamics requirement ${\rm Kn}\ll 1$ would not apply in this case, implying that one is not dealing with a near-equilibrium situation. At least in principle, one could nevertheless try to obtain a valid description by including higher and higher order gradient terms, even though in practice such an undertaking would be exceedingly difficult. If the gradient expansion was convergent, then at least in principle one could succeed in writing down such a high-order hydrodynamic theory which would accurately capture true off-equilibrium dynamics\footnote{The situation is akin to performing a Taylor expansion of the function $e^{x}$ around $x=0$ to obtain accurate results for $x\simeq 1$. The gradient expansion $e^x=1+x+\frac{x^2}{2}+\frac{x^3}{6}+\ldots$ is convergent, but for $x\simeq 1$ the first few terms in the series are of the same size as the leading term. Only when including high enough orders does the convergence of the series become apparent.}.

Unfortunately, the hydrodynamic gradient expansion does not lead to a convergent series. For particular situations with a high degree of symmetry, it has been possible to push the hydrodynamic gradient expansion to very high orders \cite{Heller:2013fn,Heller:2015dha,Buchel:2016cbj,Denicol:2016bjh,Heller:2016rtz,Casalderrey-Solana:2017zyh}. In all of these examples, it was found that the relative size of the correction term with $n$ gradients $\left| T^{\mu\nu}_{(n)}\right|$ compared to that with $n-1$ gradients is $\left| T^{\mu\nu}_{(n)}\right|/\left| T^{\mu\nu}_{(n-1)}\right|\simeq n$, implying factorial growth of the hydrodynamic coefficients\footnote{In non-relativistic systems, it has been known for quite some time that the hydrodynamic gradient expansion diverges, cf. Ref.~\cite{PhysRevLett.56.1571}.}.
\index{Fluid Dynamics! Gradient series}

It is possible to understand the origin of this factorial growth as follows: At a given order $n$, the structures contributing to the traceless part of $T^{\mu\nu}_{(n)}$ involve combinations of $n$ gradient terms $\nabla^\perp_{\mu_1},\nabla^\perp_{\mu_2},\nabla^\perp_{\mu_3},\ldots,\nabla^\perp_{\mu_{n}}$ with all but two of the indices $\mu_1,\mu_2,\ldots, \mu_{n}$ contracted. This amounts to $(n-2)!$ possible index contractions. There may also be additional structures (e.g. gradients acting on $u^\mu$), but in any case one can expect the number of terms in $T^{\mu\nu}_{(n)}$ to grow factorially with $n$ for large $n$. Thus, unless the value of most of the transport coefficients multiplying these structures are exponentially suppressed for large $n$, one can expect factorial growth of $\left| T^{\mu\nu}_{(n)}\right|$, and hence a divergent gradient series.

\subsection{Borel-Resummed Fluid Dynamics}
\index{Borel resummation}

The fact that the hydrodynamic gradient expansion does not form a convergent series is inconvenient, but does not invalidate the search for an off-equilibrium formulation of hydrodynamics. This is because the divergent series turns out to be Borel summable (in a generalized sense) in some of the examples where the hydrodynamic series coefficients are known to high orders \cite{Heller:2013fn,Heller:2015dha,Aniceto:2015mto,Buchel:2016cbj,Florkowski:2016zsi}. Before delving into the Borel summation of hydrodynamics let us consider a simple warm-up example on how Borel summation itself works.

\subsubsection{Digression: Borel Summation Example}

\begin{figure*}[t]
  \centering
  \includegraphics[width=.7\linewidth]{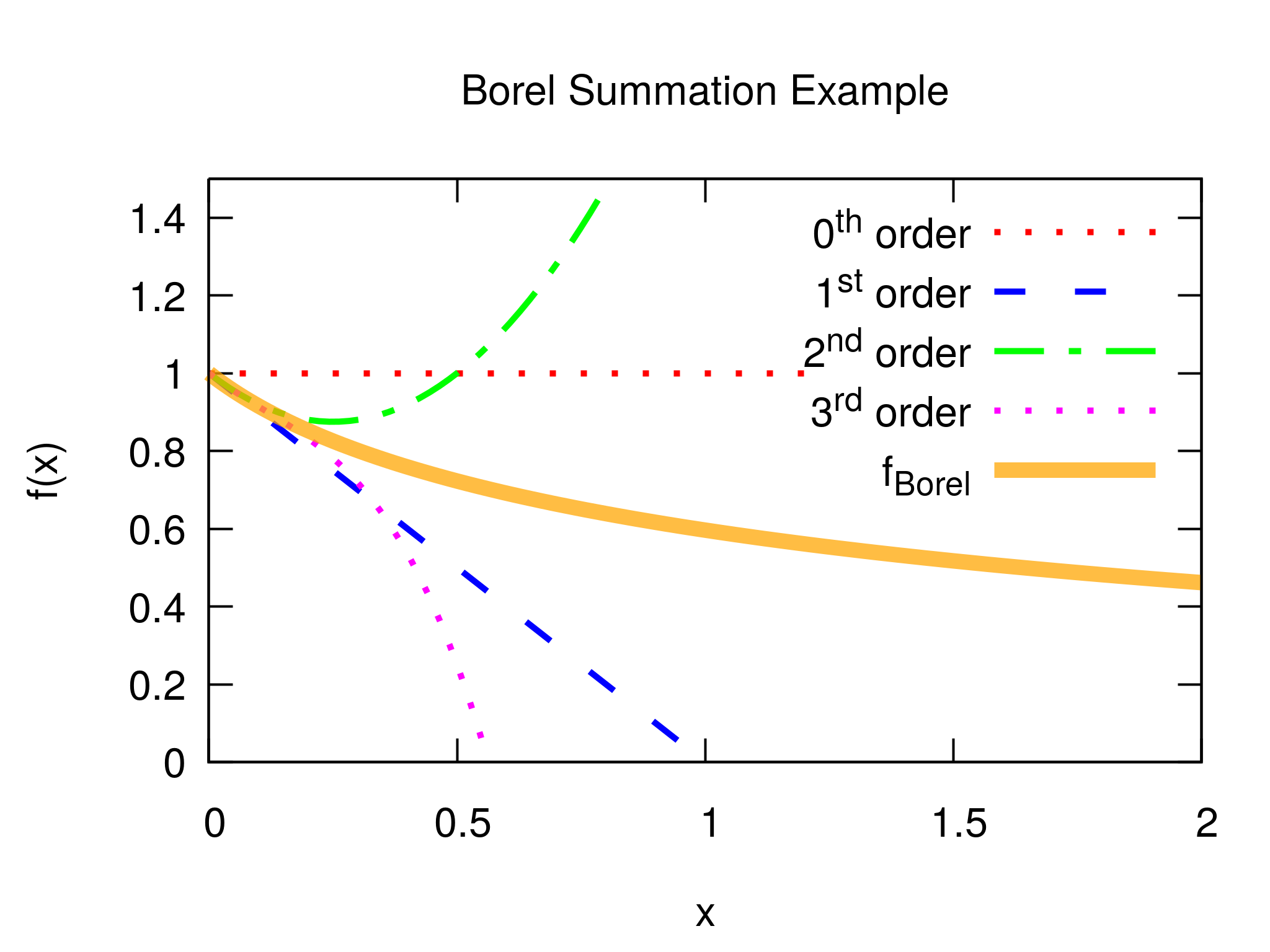}
  \caption{\label{fig:zero} Simple example of Borel summation. Shown are truncation of the divergent series (\ref{eq:expbor1}) at leading, first, second and third order as well as the Borel-resummed function (\ref{eq:invborel}).}
\end{figure*}

Consider the case of a function which is known in terms of a power series
\begin{equation}
\label{eq:expbor1}
f(x)=\sum_{n=0}^\infty \alpha_n x^n\,, \quad \alpha_n=(-1)^n n!\,,
\end{equation}
which clearly diverges for any $x\neq 0$, see Fig. ~\ref{fig:zero}.
One may now consider the Borel transform $B[f](x)$ of this series, defined as
\begin{equation}
  \label{eq:boreltrafo}
  B[f](x)=\sum_{n=0}^\infty \frac{\alpha_n}{n!} x^n=\frac{1}{1+x}\,,
\end{equation}
which has a finite radius of convergence $|x|<1$ and a pole located at $x=x_0=-1$. A Borel-resummed function of the original series $f(x)$ may be obtained from $B[f]$ as 
\begin{equation}
\label{eq:invborel}
  f_{\rm Borel}(x)=\int_0^\infty dz e^{-z} B[f]( x z)=\frac{e^{1/x}}{x} \Gamma\left(0,\frac{1}{x}\right)\,,
\end{equation}
where $\Gamma(0,x)$ is the incomplete Gamma-function. It is easy to verify that the Taylor-series expansion of $f_{\rm Borel}(x)$ around $x=0$ coincides with the original series (\ref{eq:expbor1}). Unlike any finite-order truncation of (\ref{eq:expbor1}), $f_{\rm Borel}(x)$ is well-defined for all $x\geq 0$, thus providing a closed-form ``correct'' resummation of the original divergent series (\ref{eq:expbor1}), cf. Fig.~\ref{fig:zero}.

For many cases, the coefficients $\alpha_n$ are only known up to a finite order $n=N$. In this case, the Borel transform $B[f](x)=\sum_{n=0}^N\frac{\alpha_n}{n!}x^n$ can still be defined, but usually can not be expressed in closed form. However, it has been found that analytic approximations of $B[f](x)$, for instance Pad\'e approximations, are acceptable to obtain an approximation for $f_{\rm Borel}$ as long as $N\gg 1$.

\subsubsection{Resumming the Divergent Hydrodynamic Series}
\label{sec:borel1}

Let us now return to the question of the Borel summation of hydrodynamics, giving rise to the program of resurgence \index{Resurgence}. Consider a system for which the coefficients $\alpha_n$ of the hydrodynamic gradient series have been calculated for high orders $N\gg 1$. Let $x$ be the dimensionless strength of a gradient. Then the hydrodynamic gradient series for a quantity $h$ takes the form
\begin{equation}
  \label{eq:divser}
  h(x)=\sum_{n=0}^N \alpha_n x^n\,.
 \end{equation}
(To be even more concrete, it may be useful to think of $x$ as $x\sim \tau^{-2/3}$ and $h=T(\tau)$ for the case of a system undergoing Bjorken-like expansion, cf. Eq.~(\ref{eq:bjorsol1})).
The resulting series will generally be divergent, with coefficients $\alpha_n\propto n!$ for $n\gg 1$. The Borel transform $B[h](x)$ of the series (\ref{eq:divser}), defined as in (\ref{eq:boreltrafo}), will typically have a finite radius of convergence. An analytic continuation of $B[h](x)$ may then be obtained from a symmetric Pad\'e approximation to $B[h](x)$,
\begin{equation}
\label{eq:borel234}
  \tilde B[h](x)=\frac{\sum_{n=0}^{N/2} u_n x^n}{1+\sum_{n=1}^{N/2} d_n x^n}\,,
\end{equation}
with coefficients $u_n,d_n$ determined such that re-expanding $\tilde B[h](x)$ to order $N$ around $x=0$ matches Eq.~(\ref{eq:boreltrafo}). The Pad\'e-Borel transform $\tilde B[h](x)$ will in general possess poles in the complex $x$ plane, determined by the zeros of the denominator of Eq.~(\ref{eq:borel234}), and we denote the pole closest to the origin as $x=x_0$. Note that these poles may cluster into branch-cuts, and in particular it can happen that poles occur on the positive real axis, which prohibits the use of standard Borel summation. However, it is possible to define a generalized Borel-resummed function $S[h](x)$ as
\begin{equation}
  h_{\rm Borel}(x)=S[h](x)=\int_{\cal C} dz e^{-z} \tilde B[h]( x z)\,,
\end{equation}
where the contour ${\cal C}$ starts at $z=0$ and ends at $z=\infty$. Different choices for the contour ${\cal C}$ imply an ambiguity for $S[h](x)$ because depending on the choice of ${\cal C}$, the presence of singularities of $\tilde B[h](x z)$ will lead to different results. In particular, the pole closest to the origin will lead to an ambiguity of the form $\delta S_h(x)\propto e^{-x_0/x}$ for the Borel-resummed function $S[h](x)$. It is possible to resolve the ambiguity by promoting the original hydrodynamic gradient series (\ref{eq:divser}) to a transseries,
\begin{equation}
  \label{eq:transser}
  t(x)=\sum_{m=0}^\infty c^m \Omega^m(x) t^{(m)}(x) =\sum_{m=0}^\infty c^m \Omega^m(x) \sum_{n=0}^\infty \alpha_{n,m} x^n\,,
\end{equation}
with $\Omega(x)=x^\gamma e^{-x_0/x}$. For each value of $m$, the series over $n$ in Eq.~(\ref{eq:transser}) is expected to be divergent, and needs to be Borel-resummed to a function $S[t^{(m)}](x)$. 
The ambiguity $\delta S[t^{(0)}](x)$ arising from the $m=0$ term in Eq.~(\ref{eq:transser}) may then be canceled against the $\Omega^1(x)$ term from $m=1$ by demanding that the overall result $S[t](x)$ be real. (See e.g. Ref.~\cite{Heller:2015dha} where this program has been carried out in practice.)

To summarize, whenever the strength of a gradient $x$ is small compared to the scale $x_0$ one finds
\begin{equation}
  \label{eq:resummed}
  h_{\rm Borel}(x)=S[t](x)=S[t^{(0)}](x)+c x^\gamma e^{-x_0/x} S[t^{(1)}](x)+{\cal O}\left(e^{-2 x_0/x}\right)\,,
\end{equation}
because terms of order $e^{-x_0/x}$ are exponentially suppressed for $x\ll x_0$. The leading order contribution $S[t^{(0)}](x)$ in this expression corresponds to the Borel-resummed hydrodynamic gradient series. It typically is a non-linear function of the gradient $x$, and remains bounded even for large values of $x\rightarrow \infty$. The function $S[t^{(0)}](x)$ has been referred to as the ``hydrodynamic attractor'' or ``all-orders hydrodynamics'', respectively, cf. Ref.~\cite{Lublinsky:2007mm,Heller:2015dha,Bu:2014ena}. \index{Fluid Dynamics! Attractor} It corresponds to a (non-linear) function of the gradients which generalizes the concept of near-equilibrium hydrodynamics to far-from-equilibrium situations. Some of the properties of the hydrodynamic attractor will be discussed below.

Of equal importance is the second term in (\ref{eq:resummed}), which contains a term proportional to $e^{-x_0/x}$. Since this term has an essential singularity at $x=0$, it is clear that the presence of such a term will give rise to a divergent contribution to the hydrodynamic gradient expansion around $x=0$. Since terms of this form cannot be described using the hydrodynamic gradient expansion, they are referred to as non-hydrodynamic contributions. The relative importance of the non-hydrodynamic contribution in (\ref{eq:resummed}) is controlled by the location $x_0$ of the singularity of $\tilde B[h](x)$ closest to the origin. This singularity will be referred to as the dominant non-hydrodynamic singularity or dominant non-hydrodynamic mode in the following. \index{Non-hydrodynamic mode} For gradients that are weak compared to the scale of the dominant non-hydrodynamic singularity $x\ll x_0$, all non-hydrodynamic contributions in (\ref{eq:resummed}) are exponentially suppressed, and hence the system evolution is dominated by the hydrodynamic attractor $S[t^{(0)}](x)$. Note that while $x\ll x_0$, this regime can include large gradients $x\gg 1$ as long as $x_0\gg x\gg 1$. 

\subsubsection{Out of Equilibrium Fluid Dynamics}
\label{sec:outoff}
\index{Fluid Dynamics! Out of Equilibrium}

Let us now attempt to generalize the example of the preceding section to the full theory of fluid dynamics in a formal sense. Starting with the hydrodynamic gradient series for the expectation value of the energy-momentum tensor (\ref{eq:tmunuseries}), one expects this series to be divergent because the number of possible gradient terms contributing to $T^{\mu\nu}_{(n)}(\nabla_\perp)$ is of order $n!$ (see preceding sections)\footnote{Here, the meaning of $T^{\mu\nu}_{(n)}(\nabla_\perp)$ is that of the contribution to the energy-momentum tensor as a formal function of the gradient $\nabla_\perp$.}.
Because the number of terms in $\langle T^{\mu\nu}\rangle$ does not grow faster than factorial, the Borel transform $B[\langle T^{\mu\nu}\rangle]$ of this series should exist within a finite radius of convergence for $\nabla_\perp$. Then let us analytically continue $B[\langle T^{\mu\nu}\rangle]$ via a Pad\'e approximation $\tilde B[\langle T^{\mu\nu}\rangle]$. This Pad\'e approximation will generically have non-hydrodynamic singularities located in the complex $\nabla_\perp$ plane, with the singularity closest to the origin given by the dominant non-hydrodynamic mode. Promoting (\ref{eq:tmunuseries}) to a transseries will then allow resolving the ambiguity in the Borel-transform of $\tilde B[\langle T^{\mu\nu}\rangle]$ arising from the dominant non-hydrodynamic mode and lead to a result of the form
\begin{equation}
  \label{eq:formalhydro}
  \langle T^{\mu\nu}\rangle_{Borel}=T^{\mu\nu}_{\rm hydro}+T^{\mu\nu}_{\rm non-hydro}\,,
\end{equation}
where $T^{\mu\nu}_{\rm hydro}(\nabla_\perp)$ is the hydrodynamic attractor contribution and all (dominant plus sub-dominant) non-hydrodynamic contributions have been lumped into $T^{\mu\nu}_{\rm non-hydro}$.

Similar to the example given in section \ref{sec:borel1}, the hydrodynamic attractor contribution is expected to be well-defined for large gradients $\nabla_\perp$. However, the non-hydrodynamic contribution can only be neglected if gradients are small compared to the scale given by the dominant non-hydrodynamic mode. This leads to the following requirement for the applicability of (Borel-resummed) fluid dynamics \cite{Romatschke:2016hle}:

\textbf{Central Lemma of Out of Equilibrium Fluid Dynamics}
\textit{Given the existence of a local rest frame, Borel-resummed fluid dynamics offers a valid and quantitatively reliable description  of  the  energy-momentum tensor even  in  off-equilibrium situations  as  long as  the contribution  from  all  non-hydrodynamic modes can be neglected.}
\index{Fluid Dynamics! Central Lemma of}

Phrasing the applicability of hydrodynamics in this fashion offers a key advantage compared to the ``old'' definition of small mean free path, namely being able to describe genuine non-equilibrium situations. Note that notions of equilibrium such as temperature have not even appeared in the derivation above! 

However, the above formal derivation also has a number of disadvantages that need to be addressed in the following. For instance:
\begin{itemize}
\item
  What is the exact form of the Borel-resummed hydrodynamic attractor $T^{\mu\nu}_{\rm hydro}$ in Eq.~(\ref{eq:formalhydro}) and how is it related -- if at all -- to ideal, first and second-order hydrodynamics?\item
  What are non-hydrodynamic modes, and, given their importance in determining the regime of validity of hydrodynamics, how can the dominant non-hydrodynamic mode for a given system be determined?
\end{itemize}
Answers to these questions will be given in sections \ref{sec:cfatt}, \ref{sec:howgen}, \ref{sec:nonhydro}, \ref{sec:ktnonhydro} and \ref{sec:ggnonhydro}.

\subsection{Conformal Out of Equilibrium Fluid Dynamics}
\label{sec:confborel}

It seems that we have done something terrible. We have replaced the well-defined (but divergent) hydrodynamic gradient series by a formal 'hydrodynamic attractor' solution $T^{\mu\nu}_{\rm hydro}$ about which we know essentially nothing, except that it should exist for large gradients.

At second glance, however, things are not as bleak. We do know that for vanishing gradients, the form of $T^{\mu\nu}_{\rm hydro}$ should match that from ideal fluid dynamics (\ref{eq:hydro0f}). For infinitesimally small gradients, we have empirical proof that first order hydrodynamics (\ref{eq:hydro1f}) gives a good description of the system dynamics, so one can reasonably expect the form of $T^{\mu\nu}_{\rm hydro}$ to coincide with the Navier-Stokes expression for small gradients  or very close to equilibrium. The hydrodynamic transport coefficients will now be (non-analytic) functions of the gradients of the hydrodynamic variables, but as long as these gradients are small, these expressions should be close to their equilibrium values. This suggests that the form of the hydrodynamic attractor in Borel-resummed fluid dynamics could be given by
\index{Fluid Dynamics! Attractor}
\begin{equation}
  \label{eq:hattrac}
  T^{\mu\nu}_{\rm hydro}\simeq (\epsilon+P) u^\mu u^\nu+ P g^{\mu\nu}-\eta_B \sigma^{\mu\nu}-\zeta_B \Delta^{\mu\nu}\nabla^\perp_\lambda u^\lambda\,,
  \end{equation}
where $\eta_B,\zeta_B$ are now functions of both $\epsilon$ as well as the gradients of $\epsilon,u^\mu,g^{\mu\nu}$ from the Borel resummation. \index{Bulk viscosity! Borel-resummed}\index{Shear viscosity! Borel-resummed} Note that $\epsilon$ and $u^\mu$ do not depend on the gradients because they are defined as the time-like eigenvalue and eigenvector of $\langle T^{\mu\nu}\rangle$, cf. Eq.~(\ref{eq:umudef}). Similarly, $P(\epsilon)$ does not depend on gradients because the effective non-equilibrium pressure (\ref{eq:peff}) can again be separated into $P(\epsilon)$ (fulfilling the equilibrium equation of state relation) and the Borel-resummed bulk stress $\zeta_B \nabla^\perp_\lambda u^\lambda$.\index{Equation of State (EoS)!Non-Equilibrium}  Thus even out-of-equilibrium, the relation $P(\epsilon)$ is the same as in equilibrium because the non-equilibrium corrections have been entirely absorbed into the bulk stress. Therefore, one can think of $P(\epsilon)$ as a kind of pseudo-equilibrium quantity that corresponds to the pressure of the system with a given energy density $\epsilon$ if all non-equilibrium corrections were suddenly turned off.  Despite $P(\epsilon)$ fulfilling the equilibrium equation of state, it should be stressed that the system as a whole nevertheless evolves under a non-equilibrium equation of state with the effective pressure given by
\begin{equation}
  \label{eq:Borelpeff}
  P_{\rm eff}=P-\zeta_B \nabla^\perp_\lambda u^\lambda\,.
  \end{equation}
Unlike the near-equilibrium discussion in section \ref{sec:neqeos}, Eq.~(\ref{eq:Borelpeff}) can reasonably be expected to hold as long as a hydrodynamic attractor for the trace anomaly exists. This implies that if $\zeta_B \nabla^\perp_\lambda u^\lambda$ becomes sizable, $P_{\rm eff}$ can become considerably smaller than the (pseudo-) equilibrium pressure $P$. One can imagine that a reduction of $P_{\rm eff}$ below the effective pressure of some other phase that is thermodynamically disfavored in equilibrium would eventually lead to the formation of bubbles in the fluid (``cavitation''), cf. the discussion in Refs.~\cite{Rajagopal:2009yw,Habich:2014tpa,Klimek:2011by,Bhatt:2011kr,Sanches:2015vra,Fogaca:2016eat}.
\index{Cavitation}

If Eq.~(\ref{eq:hattrac}) correctly defines the Borel-resummed fluid energy-momentum tensor, then the resulting equations of motion are
\begin{eqnarray}
  D \epsilon+(\epsilon+P) \nabla_\lambda^\perp u^\lambda&=&\frac{\eta_B}{2}\sigma^{\mu\nu}\sigma_{\mu\nu}+\zeta_B \left(\nabla_\lambda^\perp u^\lambda\right)^2\,,\\
  (\epsilon+P) D u^\alpha+\nabla_\perp^\alpha P&=&\Delta^\alpha_\nu \nabla_\mu \left(\eta_B\sigma^{\mu\nu}+\zeta_B \Delta^{\mu\nu} \nabla_\lambda^\perp u^\lambda\right)\,.
\end{eqnarray}
Since $\epsilon, P(\epsilon)$ are related by the equilibrium equation of state 
(cf. the discussion in section \ref{sec:neqeos}), we may introduce the quantities of pseudo-temperature $T$ and pseudo-entropy density $s$ by employing the equilibrium thermodynamic relations (\ref{eq:basicthermo}). \index{Pseudo-temperature} While these quantities are not the ``actual'' temperature and entropy density because the system is out of equilibrium, for calculational purposes the distinction between the pseudo- and actual equilibrium quantities is unimportant.

Relativistic fluids are never incompressible, and therefore $\nabla_\lambda^\perp u^\lambda$ is generally non-vanishing. As a consequence, $D\epsilon$ may be divided by $\nabla_\lambda^\perp u^\lambda$, leading to
\begin{equation}
  \label{eq:noncfatt}
  \frac{D \epsilon}{(\epsilon+P)\nabla_\lambda^\perp u^\lambda}=\frac{D \ln s}{\nabla_\lambda^\perp u^\lambda}=-1+\frac{\eta_B}{2 s}\frac{\sigma^{\mu\nu}\sigma_{\mu\nu}}{T \nabla_\lambda u^\lambda}+\frac{\zeta_B}{s} \frac{\nabla_\lambda^\perp u^\lambda}{T}\,.\\
\end{equation}
Eq.~(\ref{eq:noncfatt}) suggests that the quantity
\begin{equation}
  \label{eq:attvar}
  A_1=\frac{D \ln s}{\nabla_\lambda^\perp u^\lambda}\,,
\end{equation}
tends to $-1$ regardless of initial conditions if gradients are small, which is indicative of a kind of attractor behavior. 
Furthermore, even for large gradients, Eq.~(\ref{eq:noncfatt}) suggests that $A_1$ is independent from initial conditions if expressed as a function of $\frac{\eta_B}{2 s}\frac{\sigma^{\mu\nu}\sigma_{\mu\nu}}{T \nabla_\lambda u^\lambda}+\frac{\zeta_B}{s} \frac{\nabla_\lambda^\perp u^\lambda}{T}$. Since $\eta_B,\zeta_B$ are generally not known, an acceptable proxy for this quantity at moderately strong gradients is the (dimensionless)  inverse gradient strength
\begin{equation}
  \label{eq:attvar2}
  \Gamma=\left[\frac{\eta}{2 s}\frac{\sigma^{\mu\nu}\sigma_{\mu\nu}}{T \nabla_\lambda u^\lambda}+\frac{\zeta}{s} \frac{\nabla_\lambda^\perp u^\lambda}{T}\right]^{-1}\,,
\end{equation}
formulated in terms of the un-resummed transport coefficients $\eta,\zeta$ instead of $\eta_B,\zeta_B$.

\subsection{Hydrodynamic Attractor for Conformal Fluids}
\label{sec:cfatt}

Let us now come back to the question raised in section \ref{sec:outoff} concerning the exact form of the hydrodynamic attractor. Arguably the simplest choice to try to answer this question is to study conformal systems in flat space-time where $\langle T^\mu_\mu\rangle=0$, and hence $\zeta_B=0$.  In order to constrain the Borel-resummed shear viscosity coefficient $\eta_B$, let us start with a warm-up problem where the hydrodynamic attractor solution can be calculated analytically. To this end, let us consider a 'mock' microscopic theory with parameters $\eta,\tau_\pi,\lambda_1$. Setting up Bjorken flow for this mock microscopic theory\footnote{Note that this mock microscopic theory is actually rBRSSS, which will be discussed in section \ref{sec:rBRSSS}.}, the evolution equations for the energy-density are given by \cite{Baier:2007ix}
\begin{equation}
  \label{eq:Bjorrbrsss}
  \tau \partial_\tau \ln \epsilon=-\frac{4}{3}+\frac{\Phi}{\epsilon}\,,\quad
  \tau_\pi \partial_\tau \Phi = \frac{4 \eta}{3 \tau}-\Phi-\frac{4}{3}\frac{\tau_\pi}{\tau} \Phi-\frac{\lambda_1}{2 \eta^2}\Phi^2\,.
  \end{equation}
The energy density, defined by the solution of Eq.~(\ref{eq:Bjorrbrsss}) possesses a late-time (hydrodynamic) solution given by Eq.~(\ref{eq:bjork3}), with parameters $C_\eta,C_\pi,C_\lambda$ calculated from $\eta,\tau_\pi,
\lambda_1$ as defined in (\ref{eq:defs}).

Unlike the hydrodynamic gradient solution (\ref{eq:bjork3}), the solution of Eq.~(\ref{eq:Bjorrbrsss}) is well defined also for early times $\tau$. In fact, Eq.~(\ref{eq:Bjorrbrsss}) can be written in terms of a single equation for $f(\tau T)\equiv 1+\frac{1}{4}\tau \partial_\tau \ln \epsilon$ as \cite{Heller:2015dha}
\begin{eqnarray}
  \label{eq:fullMIS}
  C_\pi w f(w) f'(w)+4 C_\pi\left(1+\frac{3 C_\lambda w}{8 C_\eta}\right)f^2(w)
  +\left(w-\frac{16 C_\pi}{3}\left(1+ \frac{3 C_\lambda w}{8C_\eta}\right)\right)f(w)
  &&\nonumber\\  
  -\frac{4 C_\eta}{9}+\frac{16 C_\pi}{9}\left(1+\frac{3 C_\lambda w }{8 C_\eta} \right)
  -\frac{2 w}{3}=0\,,\quad &&
\end{eqnarray}
where the shorthand notation $w=\tau T$ has been used. Note that for large $\tau T$, Eq.~(\ref{eq:fullMIS}) can be solved as
\begin{equation}
  \label{eq:bjorhyagain}
  A_1=\frac{3\partial \ln \epsilon}{4\partial \ln \tau}=3f(\tau T)-3\simeq
  -1+\frac{4 C_\eta}{3 \tau T}+\frac{8 C_\eta C_\pi\left(1-C_\lambda\right)}{9 \tau^2 T^2}+\ldots\,,
\end{equation}
which matches the hydrodynamic expansion given in Eq.~(\ref{eq:dtbjork}), cf. Ref.~\cite{Heller:2015dha}.

\begin{figure*}[t]
  \centering
  \includegraphics[width=.7\linewidth]{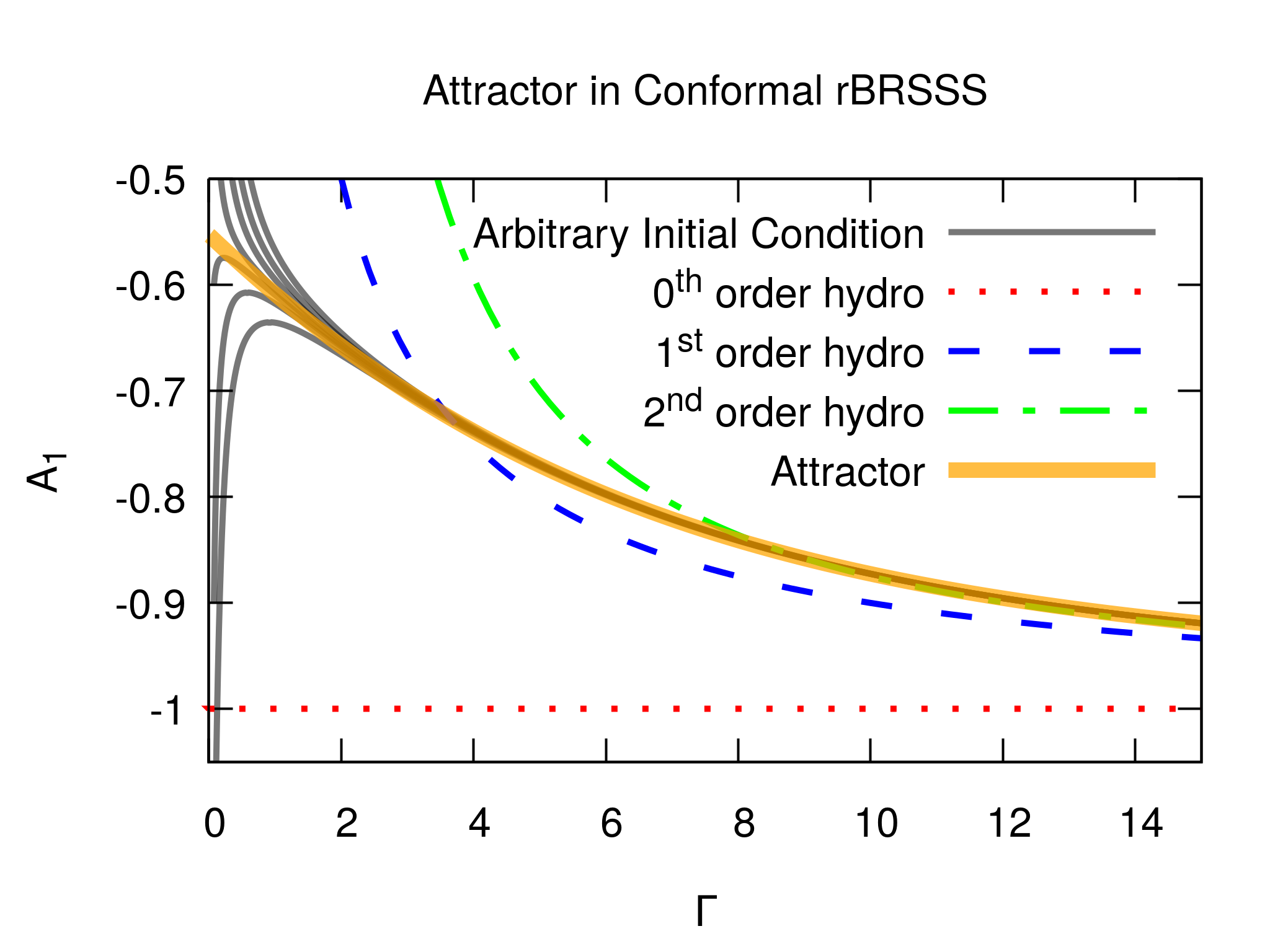}
  \caption{\label{fig:one} Hydrodynamic attractor solution in conformal rBRSSS theory undergoing Bjorken flow for $C_\eta=1$, $C_\pi=5 C_\eta$ and $C_\lambda=0$ (figure adapted from Ref.~\cite{Romatschke:2017vte}). See text for details.}
\end{figure*}

Choosing values for $C_\eta,C_\pi,C_\lambda$ and initial conditions $f(w=w_0)=f_0$, Eq.~(\ref{eq:Bjorrbrsss}) may be solved numerically, finding the results shown in Fig.~\ref{fig:one}. Also shown in Fig.~\ref{fig:one} are the late-time hydrodynamic series results to zeroth, first and second order in gradients (corresponding to Euler, Navier-Stokes and BRSSS equations, respectively). One observes from Fig.~\ref{fig:one} that the first-order and second-order solutions are close to each other, but are quite different from the zeroth order fluid dynamic result for the range of values $\tau T$ shown. The set of exact numerical solutions to Eq.~(\ref{eq:Bjorrbrsss}) seem to converge to the viscous hydrodynamic gradient series solution, and then follow the hydrodynamic solutions for late times as expected.

However, another feature apparent from Eq.~(\ref{eq:Bjorrbrsss}) is that the set of numerical solutions from Eq.~(\ref{eq:Bjorrbrsss}) seem to converge towards a unique curve well before converging to the viscous hydrodynamic series solutions. It is this feature that has given rise to the name 'hydrodynamic attractor' solution in Ref.~\cite{Heller:2015dha}. Remarkably, the hydrodynamic attractor solution to Eq.~(\ref{eq:Bjorrbrsss}) may be approximated through a 'slow-roll' type of approximation. Assuming $f'$ to be small in Eq.~(\ref{eq:Bjorrbrsss}) leads to a solution which to lowest order is given by
\begin{equation}
  f^{\rm att}(\tau T)\simeq\frac{2}{3}-\frac{\tau T}{8 C_\pi\left(1 +  \frac{3 C_\lambda \tau T}{8C_\eta} \right)}+\frac{\sqrt{64 C_\eta C_\pi\left(1+\frac{3 C_\lambda \tau T}{8 C_\eta}\right)+9 (\tau T)^2}}{24  C_\pi\left(1+\frac{3 C_\lambda \tau T}{8 C_\eta}\right)}\,.
\end{equation}
Moderate improvements of this solution can be obtained by iteration, but one should be aware that the slow-roll expansion is also typically divergent \cite{Denicol:2017lxn}. The solution $f^{\rm att}(\tau T)$ has the property that it reduces to the hydrodynamic series solution (\ref{eq:bjorhyagain}) when expanded in a power series of the argument. However, unlike the hydrodynamic series solution (\ref{eq:bjorhyagain}), $f^{\rm att}(\tau T)$ is well defined even for early times $\tau T\rightarrow 0$ when gradients become large.

\subsection{How General are Hydrodynamic Attractors?}
\label{sec:howgen}

The presence of a hydrodynamic attractor solution is not limited to spatially homogeneous systems. In Ref.~\cite{Romatschke:2017acs}, an attractor solution for rBRSSS theory has been found numerically for a system with non-trivial dynamics in the transverse coordinates ${\bf x}_\perp=(x,y)$ in addition to Bjorken flow. Also, attractor solutions are not limited to the case of rBRSSS theory. Rather, when studying microscopic dynamics originating from weakly-coupled kinetic theory (see section \ref{sec:kin}) or strongly coupled gauge theory via AdS/CFT (see section \ref{sec:ads}) one also finds such attractor solutions, see the discussions in Refs. \cite{Romatschke:2017vte,Spalinski:2017mel,Casalderrey-Solana:2017zyh}. Finally, attractor solutions for non-conformal systems have been found in kinetic theory in Ref.~\cite{Romatschke:2017acs,Florkowski:2017jnz} and several other systems \cite{Strickland:2017kux,Behtash:2017wqg,Blaizot:2017ucy}.

The apparent ubiquity of attractor solutions suggests that hydrodynamic attractors are a robust feature of a broad class of microscopic theories not restricted to specific symmetry assumptions. As a consequence, the attractor solutions give quantitative meaning to Borel-resummed hydrodynamics not only for near-equilibrium systems, but also for out of equilibrium situations. Put differently, the presence of a hydrodynamic attractor provides the foundation for out of equilibrium fluid dynamics.

\subsection{Towards Explaining The Unreasonable Success of Viscous Hydrodynamics}
\label{sec:unreasonable}
\index{Hydrodynamics |see {Fluid Dynamics}}
\index{Fluid Dynamics! Unreasonable success of}

As pointed out at the beginning of section \ref{sec:offeq}, the naive hydrodynamic gradient series is divergent. For this reason, one cannot expect any finite truncation of this gradient series to deliver accurate quantitative approximations to system dynamics except for cases extremely close to equilibrium, consistent with the textbook requirement (\ref{eq:landaucrit}). Yet, results based on low-order truncation of the gradient series are apparently very successful in moderately out of equilibrium situations \cite{Chesler:2009cy,Heller:2011ju,Wu:2011yd,vanderSchee:2012qj,Casalderrey-Solana:2013aba,Kurkela:2015qoa,Keegan:2015avk,Attems:2017ezz}, which sometimes is referred to as the ``unreasonable success of viscous hydrodynamics''.

In an attempt to elucidate this ``unreasonable success'', we would like to draw the reader's attention to another feature shown in Fig.~\ref{fig:one}. Specifically, while the attractor solution shown in Fig.~\ref{fig:one} is poorly approximated by ideal hydrodynamics or high orders of the divergent hydrodynamic gradient series (not shown in Fig.~\ref{fig:one}), it so happens that the first and second-order gradient series results are quantitatively close to the non-perturbative attractor for gradient strengths of $\Gamma\gtrsim 5$. A trivial re-writing of the gradient strength in terms of the mean free path (\ref{eq:lambdamfp}) leads to $\Gamma \simeq {\rm Kn}^{-1}$ in terms of the Knudsen number, such that 
\begin{equation}
  {\rm Kn}\lesssim 0.2\,,
\end{equation}
replaces the more restrictive textbook criterion (\ref{eq:landaucrit}) for low-order viscous hydrodynamics to give an accurate quantitative approximation of the non-perturbative system dynamics. The numerical values $\Gamma\simeq 5$ and ${\rm Kn}\simeq 0.2$ imply a system that is neither very far from, nor very close to equilibrium, but somewhere in between. Apparently, low-orders of the divergent gradient series (Navier-Stokes, BRSSS) provide a good quantitative approximation to the full non-perturbative attractor solution in moderately out of equilibrium situations. Similar observations can be made in all other cases where hydrodynamic attractor solutions are known, cf. Refs.~\cite{Heller:2015dha,Romatschke:2017acs,Spalinski:2017mel,Florkowski:2017jnz,Strickland:2017kux,Behtash:2017wqg}.

It is not known how generic this feature is, but it should be mentioned that an equivalent situation can be observed in the case of perturbative QCD at high temperature, where low-order terms of the gradient series provide good quantitative approximations to exact (lattice QCD) results \cite{Blaizot:2003iq}.

The observation that low-order truncations of a divergent series provide reasonably good approximations to the full non-perturbative result would indeed explain the ``unreasonable'' quantitative success of low-order viscous hydrodynamics (Navier-Stokes, BRSSS) in moderately out of equilibrium systems.

\subsection{Far From Equilibrium Fluid Dynamics}
\label{sec:faraway}
\index{Fluid Dynamics! Far from equilibrium}

Armed with the knowledge outlined in the previous sections, the following picture for out of equilibrium fluid dynamics (\ref{eq:formalhydro}) emerges: Borel resummation leads to the presence of a non-analytic hydrodynamic attractor solution $T^{\mu\nu}_{\rm hydro}$ which exists for large gradients and may be calculated directly by other methods. For moderately strong gradients, this attractor solution becomes indistinguishable from the results of a low-order series expansion around equilibrium. Therefore, even though the hydrodynamic gradient series diverges, low-order hydrodynamics (e.g. first or second-order) are quantitatively close to the full non-analytic attractor solution and therefore can be considered good approximations even in out of equilibrium situations. In the terminology of Ref.~\cite{CasalderreySolana:2011us}, the system has ``hydrodynamized''.

Judging from the results shown in Fig. ~\ref{fig:one}, the hydrodynamic attractor solution seems to exist not only moderately out of equilibrium, but extends even far away from equilibrium where gradients are large. In this regime, the hydrodynamic gradient series does not offer a quantitatively accurate description of the system dynamics. But the existence of the attractor solution does suggest the existence of a notion of fluid dynamics far from equilibrium.

\index{Non-hydrodynamic mode}
Arbitrary initial data for $\langle T^{\mu\nu}\rangle$ will typically involve a non-hydrodynamic component $T^{\mu\nu}_{\rm non-hydro}$ and thus not initially lie on the attractor. However, at least according to the example given in (\ref{eq:resummed}), one can expect the non-hydrodynamic mode component to decrease exponentially such that the system evolution will bring $\langle T^{\mu\nu}\rangle$ close to the attractor solution quickly (see Fig.~\ref{fig:one} for examples of this behavior)\footnote{Note that in the example of AdS/CFT studied in Refs.~\cite{Romatschke:2017vte,Spalinski:2017mel}, the dominant non-hydrodynamic singularity is complex, e.g. ${\rm Im}\,x_0\neq0$, implying oscillatory, but nevertheless exponentially fast approach of the attractor solution.}.

\begin{figure*}[t]
  \begin{center}
    \includegraphics[width=0.7\linewidth]{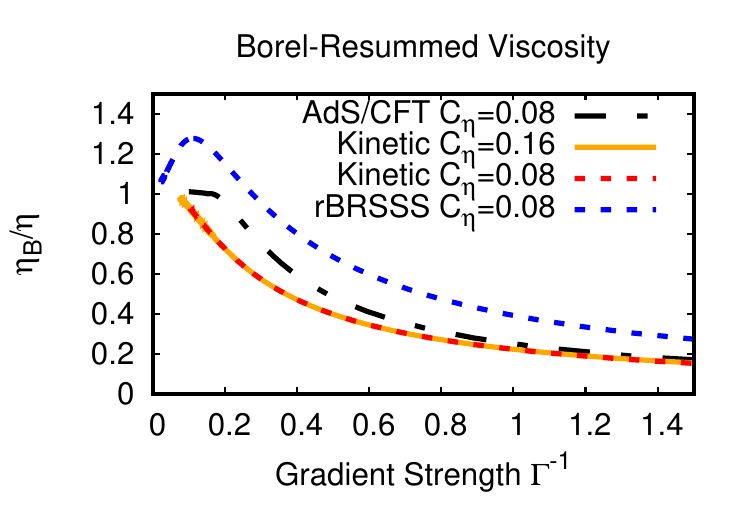}
    \end{center}
  \caption{\label{fig:two} Effective viscosity $\eta_B$ divided by near-equilibrium viscosity $\eta$ versus gradient strength $\Gamma^{-1}$ for attractor solutions in different theories (rBRSSS, kinetic theory and AdS/CFT). For small gradients, one recovers $\frac{\eta_B}{\eta}=1$, while $\frac{\eta_B}{\eta}\rightarrow 0$ for large gradient strength. Note that when gradient strength is expressed as $\Gamma^{-1}$, the kinetic theory curves for $\frac{\eta}{s}=0.08$ and $\frac{\eta}{s}=0.16$ collapse onto a single line. Figure adapted from Ref.~\cite{Romatschke:2017vte}.}
\end{figure*}

In the example case of the conformal rBRSSS attractor solution discussed in section \ref{sec:cfatt} (as well as other theories within this symmetry class, cf. Ref.~\cite{Romatschke:2017vte,Strickland:2017kux}), it is possible to calculate the Borel-resummed viscosity coefficient $\eta_B$. Requiring $A_1$ in Eq.~(\ref{eq:noncfatt}) to match the calculated non-analytic attractor solution $A_1(\Gamma)$ one finds $\eta_B(\Gamma)$ with $\Gamma$ given in Eq.~(\ref{eq:attvar2}). Results for $\eta_B$ for the attractors shown in Fig.~\ref{fig:one} are shown in Fig.~\ref{fig:two}. One finds that while $\eta_B\simeq \eta$ near equilibrium as expected, $\eta_B$ tends to zero for large gradient strength.

It is conceivable that the effective theory (\ref{eq:hattrac}) may be useful for the description of a variety of out-of-equilibrium systems.

\section{Hydrodynamic Collective Modes}
\label{sec:hydrocoll}

In the previous sections, the effective theory of fluid dynamics out of equilibrium has been derived from first principles based on symmetries, gradient series expansion and Borel resummation. Ideal fluid dynamics, Navier-Stokes theory, BRSSS and far-from-equilibrium fluid dynamics were introduced in this fashion.

In this section, the collective mode structure of these theories will be considered, leading to sound waves, shear modes, and the discovery of causality violations in Navier-Stokes and BRSSS.

\subsection{Relativistic Euler Equation}
\label{sec:NEuler}
\index{Euler equation! Collective modes}

To set the stage, let us first study relativistic ideal fluid dynamics, defined by the conservation of Eq.~(\ref{eq:hydro0f}), using the widely employed canonical approach. The canonical approach to calculating the collective modes of the system is to introduce sources for conserved quantities (e.g. energy-density, momentum) and then study the response of the system when these sources act as small perturbations. For the case at hand, let us consider small perturbations $\delta \epsilon, \delta u^\mu$ around a background of constant energy density $\epsilon_0$ and vanishing (space) velocity $u^\mu_0=\left(1,{\bf 0}\right)$ in Minkowski space-time:
\begin{equation}
  \label{eq:linearresp}
  \epsilon(t,{\bf x})=\epsilon_0+\delta \epsilon(t,{\bf x})\,,\quad
  u^\mu=u_0^{\mu}+\delta u^\mu(t,{\bf x})\,.
\end{equation}
In linear perturbation theory, the relativistic Euler equations (\ref{eq:eulerf2}) then become
\begin{equation}
  \label{eq:colleuler1}
  \partial_t \delta \epsilon+\left(\epsilon_0+P_0\right) \partial \cdot \delta {\bf u} =0\,,\quad
  \left(\epsilon_0+P_0\right) \partial_t \delta u^i+c_s^2\partial^i \delta \epsilon=0\,.
\end{equation}
Since perturbations are taken around constant backgrounds, it is useful to introduce
\begin{equation}
  Q(\omega,{\bf k})=\int_{-\infty}^\infty dt \int d^{d-1}x e^{i \omega t-i {\bf k}\cdot {\bf x}} Q(t,{\bf x})\,,
\end{equation}
as the Fourier transform of a quantity $Q(t,{\bf x})$. Introducing $\delta \epsilon(\omega,{\bf k}), \delta {\bf u}(\omega, {\bf k})$ in the above manner, Eq.~(\ref{eq:colleuler1}) is readily solved in Fourier-space as
\begin{equation}
  \label{eq:colleuler2}
  \delta \epsilon=\left(\epsilon_0+p_0\right) \frac{{\bf k}\cdot {\delta \bf u}}{\omega}\,,\quad \delta u^i=c_s^2 k^i \frac{{\bf k}\cdot \delta {\bf u}}{\omega^2}\,.
  \end{equation}
The first of these equations implies that density perturbations must involve longitudinal velocity perturbations, and the second equation requires transverse velocity perturbations to be absent. Furthermore, the longitudinal velocity perturbation ${\bf k}\cdot \delta {\bf u}$ must fulfill the dispersion relation
\begin{equation}
  \label{eq:euler-disp}
  \omega^2=k^2 c_s^2\,,
\end{equation}
which leads to the conclusion that the collective modes of the relativistic Euler equations are longitudinally propagating waves with group speed 
\begin{equation}
  \label{eq:euler-vel}
v_g\equiv \left|\frac{d\omega}{d k}\right| =c_s\,,
\end{equation}
equal to the velocity of sound. \index{Speed of sound}

In other words, the only collective modes of the Euler equation are sound waves.

\subsection{Relativistic Navier Stokes Equation}
\label{sec:NScoll}
\index{Navier-Stokes equations! Collective modes}

It is straightforward to repeat the above analysis for the case of the relativistic Navier-Stokes equations, cf.~\cite{Romatschke:2009im}. For further convenience, however, it is useful to consider an alternative to the canonical method to calculate collective modes, which is known as the variational approach. In the variational approach, one introduces sources that couple directly to the energy-momentum tensor, rather than just the conserved quantities \cite{Kovtun:2012rj}. Since the metric tensor is the source for the energy-momentum tensor, cf. Eq.~(\ref{eq:effS}), this implies that the retarded two-point correlator for the energy-momentum tensor in Minkowski space 
\begin{equation}
\label{eq:2ptcorr}
G^{\mu\nu,\gamma\delta}=-2\left.\frac{\delta T^{\mu\nu}\left(t,{\bf x},g\right)}{\delta g_{\gamma \delta}}\right|_{g={\rm Minkowski}}
  \end{equation}
may be calculated from the variation of $T^{\mu\nu}$ with respect to the metric $g^{\gamma\delta}$. The retarded correlator will determine the linear response of the energy-momentum tensor in Minkowski space-time originating from a source $S^{\gamma\delta}$ as
\begin{equation}
  \label{eq:linearrepgen}
  \delta T^{\mu\nu}(t,{\bf x})=\int d\omega d^3{\bf k} e^{-i \omega t + i {\bf k}\cdot {\bf x}} G^{\mu\nu,\gamma\delta}(\omega, {\bf k}) S_{\gamma\delta}\,.
\end{equation}
For sources $S_{\gamma \delta}$ that can be set up through the hydrodynamic variables $\epsilon,u^\mu$, the variational method contains the same information as the canonical approach. The converse is not true: some entries of the two-point correlator may not be accessed through hydrodynamic sources, which implies that the variational approach is more general than the canonical approach.

As an application, let us calculate the collective modes of the relativistic Navier-Stokes equation using the variational approach. To this end, consider Eqns.~(\ref{eq:hydro0f}), (\ref{eq:hydro1f}) in the presence of small fluctuations $\delta g_{\mu\nu}$ around $d=4$ dimensional Minkowski space-time, which will induce small fluctuations $\delta \epsilon,\delta u^\mu$ (proportional to $\delta g_{\mu\nu}$) around a constant background. To linear order in perturbations, one finds
\begin{equation}
  \label{eq:NSlinearresp}
  \delta T^{\mu\nu}=\delta \epsilon u_0^\mu u_0^\nu
 +\left(c_s^2 \delta \epsilon+\delta \Pi\right) \Delta_0^{\mu\nu}+2 (\epsilon_0+P_0) \delta u^{(\mu} u^{\nu)}_0
    + P_0 \delta g^{\mu\nu}+\delta \pi^{\mu\nu}\,,
\end{equation}
where
\begin{equation}
  \label{eq:NSlinhelp}
  \delta \Pi=-\zeta \delta \nabla \cdot u\,,\quad
  \delta \pi^{\mu\nu}=-\eta \delta \sigma^{\mu\nu}\,.
  \end{equation}
Note that unlike the background used in the canonical approach (\ref{eq:linearresp}), for metric perturbations the condition $u^\mu u_\mu=-1$ implies that $u^0=1+\frac{\delta g_{00}}{2}$. 

To obtain the retarded correlation function (\ref{eq:2ptcorr}), the terms $\delta \Pi,\delta \pi^{\mu\nu}$ must be expanded to first order in $\delta g_{\mu\nu}$ both explicitly (e.g through the Christoffel symbols appearing in $\nabla_\mu u^\nu=\partial_\mu u^\nu+\Gamma^\nu_{\mu\lambda} u^\lambda$) and implicitly (in terms of $\delta \epsilon,\delta u^\mu$). This somewhat tedious task can be simplified by choosing a specific dependence of the metric on coordinates, e.g. $\delta g^{\mu\nu}=\delta g^{\mu\nu}(t,x_3)$, which is equivalent to picking a direction of the wave-vector ${\bf k}=k {\bf e}_3$ in (\ref{eq:colleuler2}). With this choice one finds the following results for the correlators in momentum space \cite{Kovtun:2012rj}:
\begin{eqnarray}
  \label{eq:NS2pt}
  G^{00,00}(\omega,k)&=&-2\epsilon_0+\frac{k^2 (\epsilon_0+P_0)}{\omega^2-c_s^2 k^2+i \omega k^2 \gamma_s}\,,\nonumber\\
  G^{01,01}(\omega,k)&=&\epsilon_0+\frac{k^2 \eta }{i \omega-\gamma_\eta k^2}\,,\nonumber\\
  G^{12,12}(\omega,k)&=&P-i \eta \omega\,,
\end{eqnarray}
where the shear damping length $\gamma_\eta$ and sound attenuation lengths $\gamma_s$ are defined as\footnote{Note that the definition for the shear damping length $\gamma_\eta$ is identical to the classical mean free path $\lambda_{\rm mfp}$ defined in Eq.~(\ref{eq:lambdamfp}). In order to be consistent with the existing literature, both notations will be used.} \index{Sound attenuation length}
\begin{equation}
  \label{eq:sound attenuation}
  \gamma_\eta\equiv \frac{\eta}{\epsilon_0+P_0}\,,\quad \gamma_s=\frac{4\eta }{3 (\epsilon_0+P_0)}+\frac{\zeta}{(\epsilon_0+P_0)}\,.
\end{equation}
The zero-frequency terms $-2\epsilon_0,\epsilon_0$ and $P$ in Eqns.~(\ref{eq:NS2pt}), respectively, are so-called ``contact terms'', because they correspond to $\delta$-functions in the position-space representation of $G^{\mu\nu,\gamma\delta}$. Neglecting these contact terms, note that, being two-point functions of the energy-momentum tensor, the correlators $G^{\mu\nu,\gamma\delta}$ fulfill Ward identities resulting from $\partial_\mu T^{\mu\nu}=0$,
\begin{equation}
  \label{eq:ward}
  G^{00,03}(\omega,k)=\frac{\omega}{k}G^{00,00}(\omega,k)\,,\quad G^{03,03}(\omega,k)=\frac{\omega^2}{k^2}G^{00,00}(\omega,k)\,.
\end{equation}
Using the Ward identities, the correlator $G^{0i,0j}(\omega,k)$ for any orientation of ${\bf k}$ may compactly be written as 
\begin{equation}
  \label{eq:gijcorr}
  G^{0i,0j}(\omega,{\bf k})=\left(\epsilon_0+P_0\right)\left[\frac{k^i k^j}{{\bf k}^2}\frac{\omega^2}{\omega^2-c_s^2 k^2+i \omega k^2 \gamma_s}+\left(\delta^{ij}-\frac{k^i k^j}{{\bf k}^2}\right)\frac{k^2 \gamma_\eta}{i \omega-\gamma_\eta k^2}\right]\,,
  \end{equation}
where contact terms have been neglected.

A few remarks are in order. First note that the results in Eqns.(\ref{eq:NS2pt}) are only accurate to first order in $\omega,k$ because they were derived from the first-order hydrodynamic gradient series result for the energy-momentum tensor Eqns.~(\ref{eq:hydro0f}), (\ref{eq:hydro1f}). 
Second, since (\ref{eq:NS2pt}) are valid for small frequencies and momenta, they can be used to define the so-called Kubo-relations such as 
\begin{equation}
  \label{eq:Kubo}
  \lim_{\omega\rightarrow 0}\frac{\partial G^{12,12}(\omega,k=0)}{\partial \omega}=-i \eta\,.
\end{equation}
These Kubo relations can be used to extract the equilibrium value of transport coefficients (such as $\eta$) by measuring correlators (such as $G^{12,12}$) in some underlying microscopic theory.
\index{Kubo formula}

Third, the singularities of $G^{\mu\nu,\gamma\delta}(\omega,k)$ determine the linear response of $\delta T^{\mu\nu}$ through (\ref{eq:linearrepgen}). As a consequence, the singularities of $G^{\mu\nu,\gamma\delta}(\omega,k)$ determine the collective modes in the system. For instance, the singularities of $G^{00,00}(\omega,k)$ are simple poles with a dispersion relation of
\begin{equation}
  \label{eq:soundmode}
  \omega^2 = c_s^2 k^2-i \omega k^2 \gamma_s\,.
\end{equation}
For small values of $k$ (which is the regime of applicability of the calculation above), this implies $\omega\simeq \pm c_s k-\frac{i \gamma_s k^2}{2}$ or
\begin{equation}
  \label{eq:sound}
  \frac{\delta T^{00}}{d^3k}\propto  e^{\pm i c_s k t- \frac{\gamma_s}{2} k^2 t + i{\bf k}\cdot {\bf x}}\,,
\end{equation}
which can be recognized as corresponding to sound waves with a damping rate of $\frac{\gamma_s}{2}k^2$. This generalizes the result (\ref{eq:euler-disp}) from the Euler equation to the case of small, but non-vanishing wave-vector $k$. In addition to sound waves, Eqns.~(\ref{eq:NS2pt}) imply the presence of additional collective modes through the singularity structure of $G^{01,01}$ with dispersion relation
\begin{equation}
  \label{eq:dispNSshear}
  \omega =-i k^2 \gamma_\eta\,,
\end{equation}
corresponding to purely damped excitations $e^{-k^2 \gamma_\eta t+i {\bf k}\cdot {\bf x}}$ in $\delta T^{01}$. These modes are known as shear modes.

Finally, let us conclude by pointing out that $G^{00,00},G^{01,01}$ can be accessed by the canonical approach, while $G^{12,12}$ cannot.

\subsection{Acausality of the Navier Stokes Equation}
\label{sec:NSinst}
\index{Causality ! Violation}

The above treatment of collective modes in ideal fluid dynamics and in the Navier-Stokes equation is completely standard and can be found in many textbooks on (non-relativistic) fluid dynamics.

In the case of relativistic systems, there is, however, a slight complication that arises when inspecting the collective modes of the Navier-Stokes equations in more detail. Specifically, calculating the group velocity $ v_g\equiv \left|\frac{d\omega}{d k}\right|$ for the shear mode dispersion relation from (\ref{eq:dispNSshear}), one finds
\begin{equation}
  v_g=2 \gamma_\eta k\,,
\end{equation}
which grows without bound if $k\rightarrow \infty$. In particular, these modes have group velocities larger than the speed of light, indicating a violation of causality in the relativistic Navier-Stokes equations (see e.g. the appendix of Ref.~\cite{Romatschke:2009im} for a proof of the causality violation). At first inspection, this does not seem to constitute a principal problem of the theory given that the first-order hydrodynamic treatment only applies for small wavenumbers $k$. However, it turns out that modes that travel with speeds greater than the speed of light lead to the presence of instabilities which render the theory unusable in practice \cite{Hiscock:1983zz,Hiscock:1985zz}. In a nutshell, the problem can be understood through the fact that modes that travel faster than the speed of light in one Lorentz frame correspond to modes that travel backwards in time in a different Lorentz frame. The hydrodynamic equations of motion require the specification of initial conditions, but because of the presence of modes that travel backwards in time, it is not sufficient to specify initial conditions at some initial time-slice \cite{Kostadt:2000ty}. This prevents the solution of the relativistic Navier-Stokes equations as a numerical initial value problem in practice.

Note that for simple problems that can be solved analytically, the issue of causality violation can be evaded by restricting the solution to the small $k$ region only.

\section{Resummed Second Order Hydrodynamics}
\label{sec:rBRSSS}

\subsection{Maxwell-Cattaneo}

The violation of causality of the relativistic Navier-Stokes equation and the related problems of the diffusion equation can be tamed by considering a modified theory that leaves the low-wavenumber behavior unchanged, but changes the high-wavenumber behavior through the introduction of a regulator. Specifically, Maxwell \cite{Maxwell:1867} and later Cattaneo \cite{Cattaneo:1948} proposed that the algebraic relations (\ref{eq:NSconst}) should be replaced by a dynamic equation for the viscous stresses
\begin{equation}
  \label{eq:MaxwellCattaneo}
  \tau_\pi D \pi^{\mu\nu}+\pi^{\mu\nu}=-\eta \sigma^{\mu\nu}\,,\quad
  \tau_\Pi D \Pi+\Pi=-\zeta \nabla_\lambda^\perp u^\lambda\,.
\end{equation}
While the algebraic relations (\ref{eq:NSconst}) in the Navier-Stoke constitutive equations imply instantaneous propagation of gradients to the viscous stresses (violating causality), the Maxwell-Cattaneo modifications (\ref{eq:MaxwellCattaneo}) demand that the viscous stresses relax to their Navier-Stokes values on finite relaxation times $\tau_\pi,\tau_\Pi$. Thus, any non-zero value for the regulators $\tau_\pi,\tau_\Pi$ thus implies that the maximum propagation speed is bounded, but the Navier-Stokes equations are recovered in the limit $\tau_\pi\rightarrow 0,\tau_\Pi\rightarrow 0$.
The mathematics of this salient modification that Maxwell-Cattaneo theory brings to the collective mode structure of relativistic Navier-Stokes is easy to understand. Considering for example small perturbations $\delta \pi^{\mu\nu}$ in Eqns.~(\ref{eq:MaxwellCattaneo})  leads to 
\begin{equation}
  \delta \pi^{\mu\nu}=-\eta\frac{\delta \sigma^{\mu\nu}}{1-i \omega \tau_\pi}\,,
\end{equation}
in Fourier-space, which replaces the corresponding Navier-Stokes result (\ref{eq:NSlinhelp}). Thus, effectively, the collective mode structure of the Maxwell-Cattaneo theory can be gleaned by replacing $\eta\rightarrow \frac{\eta}{1-i \omega \tau_\pi}$ and $\zeta\rightarrow \frac{\zeta}{1-i \omega \tau_\Pi}$ in (\ref{eq:NS2pt}).

Let us first focus on the shear mode channel in the Maxwell-Cattaneo theory, for which the retarded correlator is given by
\begin{equation}
  G^{01,01}(\omega,k)=\epsilon_0+\frac{k^2 \eta}{i \omega(1-i \omega \tau_\pi)-\gamma_\eta k^2}\,.
  \end{equation}
This correlator has two single poles located at
\begin{equation}
  \label{eq:sheardisprels}
  \omega^{+}=\frac{-i+\sqrt{4 \gamma_\eta k^2 \tau_\pi-1}}{2 \tau_\pi}\,,\quad
  \omega^{-}=\frac{-i-\sqrt{4 \gamma_\eta k^2 \tau_\pi-1}}{2 \tau_\pi}\,.
\end{equation}
In the small $k$ limit, one of these poles (namely $\omega^{+}$) can be recognized as the shear mode that is also present in the Navier-Stokes equation, cf. Eq.~(\ref{eq:dispNSshear}). Unlike the Navier-Stokes result, however, the group velocity in the shear channel
\begin{equation}
  v_g^{\rm shear}(k)\equiv \left|\frac{d \omega^\pm}{dk}\right|=\frac{2 \gamma_\eta k}{|\sqrt{4 \gamma_\eta k^2 \tau_\pi-1}|}\,,
\end{equation}
is bounded for any $k$. The maximal signal propagation speed 
in the shear channel  is smaller than the speed of light if $\tau_\pi>\gamma_\eta$ \cite{Baier:2007ix,Romatschke:2009im}:
\begin{equation}
  \label{eq:maxshear}
   v_g^{\rm shear, max}\equiv\lim_{k\rightarrow \infty}v_g^{\rm shear}(k)=\sqrt{\frac{\gamma_\eta}{\tau_\pi}}\,.
  \end{equation}
Similar considerations lead to the maximal signal propagation speed in the sound channel $G^{00,00}$  \cite{Romatschke:2009im}:
\begin{equation}
  \label{eq:maxsound}
  v_g^{\rm max, sound}=\sqrt{c_s^2+\frac{4 \eta}{3 \tau_\pi (\epsilon_0+P_0)}+\frac{\zeta}{\tau_\Pi (\epsilon_0+P_0)}}\,.
\end{equation}
For a conformal theory in $d=4$ where $\zeta=0$ and $c_s^2=\frac{1}{3}$, this implies that Maxwell-Cattaneo theory is a causal theory if \index{Maximal Signal Speed}
\begin{equation}
  \tau_\pi\geq \frac{2 \eta}{\epsilon_0+P_0}\,,\quad {\rm or}\quad C_\pi\geq 2 C_\eta\,,
  \end{equation}
cf. Eq.~(\ref{eq:defs}). We will comment below on whether or not this requirement is fulfilled for actual microscopic theories.

For intermediate values of $k>\sqrt{\frac{1}{4 \gamma_\eta \tau_\pi}}$, Eq.~(\ref{eq:sheardisprels}) predicts that shear modes become propagating, similar to the phenomenon of second sound in heat conduction and shear waves in seismology. In fact, these transverse propagating modes have been observed in various systems, including water \cite{PhysRevLett.62.2616,PhysRevB.70.054203,PhysRevE.71.011501} (see Refs.~\cite{Baggioli:2018vfc,0034-4885-79-1-016502} for recent reviews).

\subsection{The Appearance of Non-Hydrodynamic Modes}
\label{sec:nonhydro}
\index{Non-hydrodynamic mode}

Another feature apparent from the Maxwell-Cattaneo theory is the appearance of new modes that were not present in the original Navier-Stokes equations, as can be seen by expanding $\omega^{-}$ in Eq.~(\ref{eq:sheardisprels}) for small wavenumbers $k$:
\begin{equation}
  \label{eq:nonh1}
  \omega^{-}(k)\simeq -\frac{i}{\tau_\pi}+i \gamma_\eta k^2+\ldots
  \end{equation}
This dispersion relation implies that this extra mode is over-damped with a damping rate that does not vanish as $k\rightarrow 0$, unlike the sound and shear modes discussed above. In fact, the behavior $\lim_{k\rightarrow 0}\omega^{-}\rightarrow {\rm const}$ seems inconsistent with the regime of applicability of the gradient expansion, which requires both $\omega,k$ to be small.

Because the $\omega^-$ mode in Eq.~(\ref{eq:nonh1}) is so unlike the familiar hydrodynamic shear and sound modes, it has been dubbed ``non-hydrodynamic mode'', which originally was meant as a catch-all for things that no one wanted to deal with. Indeed, a working definition for a non-hydrodynamic mode is that it is not contained in the Navier-Stokes equations. While it is terrible practice to define something in terms of what it is not, so far no accepted alternative definition (and name!) for non-hydrodynamic modes has arisen, leaving us little choice than to stick with this working definition.

While clearly not at the focus of hydrodynamic theory for centuries, it is easy to see that the presence of non-hydrodynamic modes should generally be expected in causal theories of hydrodynamics. In particular, the reason why the Maxwell-Cattaneo equations (\ref{eq:MaxwellCattaneo}) leads to finite signal propagation speed while the Navier-Stokes constitutive relations (\ref{eq:NSconst}) did not, is the fact that the equations of motion for Maxwell-Cattaneo theory are both first order in time and space derivatives. This makes the resulting equations of motion hyperbolic, whereas the Navier-Stokes equations are of  parabolic type. The dynamic (rather than algebraic) constitutive relations for Maxwell-Cattaneo theory imply a new mode in the linear regime, which is precisely the non-hydrodynamic mode discussed above. Turning the argument around, without an additional, non-hydrodynamic mode, the equations of motion of hydrodynamics will typically be of a parabolic type that implies acausal behavior.

The study of non-hydrodynamic modes is still in its infancy, but recent studies suggest that they are an ubiquitous
feature of microscopic theories and that they might offer unique insights into the nature of the relevant microscopic degrees of freedom for the theory under consideration \cite{Starinets:2002br,Heller:2013fn,Heller:2015dha,Brewer:2015ipa,Bantilan:2016qos,Romatschke:2015gic,Buchel:2016cbj,Grozdanov:2016vgg}. Examples for non-hydrodynamic modes present in kinetic theory and gauge/gravity duality are discussed in sections \ref{sec:ktnonhydro}, \ref{sec:ggnonhydro}, respectively.

\subsection{The Workhorse: Resummed BRSSS (rBRSSS)}
\label{sec:workhorse}
\index{BRSSS! resummed (rBRSSS)}

The Maxwell-Cattaneo equations (\ref{eq:MaxwellCattaneo}) lead to a causal theory of viscous hydrodynamics, but they are completely ad-hoc. In order to systematically construct a theory of causal viscous hydrodynamics from first principles, an important clue from Maxwell-Cattaneo theory is that it seems necessary to include second-order gradients in the description since otherwise the equations of motion would not become hyperbolic.

Thus let us have another look at the complete conformal second-order hydrodynamic theory (BRSSS) defined by the constitutive relations (\ref{eq:BRSSSpi}). These constitutive relations for $\pi^{\mu\nu}$ contain all possible allowed structures to second-order in gradients allowed by conformal symmetry. In writing these constitutive relations, some of the terms have been regrouped by using comoving time-derivatives $D\sigma^{\mu\nu}$. The low-order relations $\pi^{\mu \nu}=-\eta \sigma^{\mu \nu}$ may be used to rewrite Eqns.~(\ref{eq:constf}) to second-order gradient accuracy as \cite{Baier:2007ix}
 \begin{eqnarray}
   \label{eq:rBRSSS}
   \pi^{\mu \nu}&=&-\eta \sigma^{\mu \nu}-\tau_\pi \left[^{<} D \pi^{\mu \nu>}+\frac{d}{d-1} \pi^{\mu \nu} \nabla_\lambda^\perp u^\lambda\right]
+\kappa\left[R^{<\mu \nu>}-2 u_\lambda u_\rho R^{\lambda <\mu \nu>\rho}\right]
\nonumber\\
&&+\frac{\lambda_1}{\eta^2} \pi^{<\mu}_{\quad \lambda}\pi^{\nu> \lambda}-\frac{\lambda_2}{\eta} \pi^{<\mu}_{\quad \lambda} \Omega^{\nu> \lambda}
+\lambda_3 \Omega^{<\mu}_{\quad \lambda}\Omega^{\nu> \lambda}\,.
 \end{eqnarray}
 Re-expanding the rhs of Eq.~(\ref{eq:rBRSSS}) in a gradient series, one finds that this re-writing effectively resums an infinite number of terms into a closed form expression. This resummation is related to, but different from, the Borel-resummation discussed in section \ref{sec:borel1} in that the Borel resummation leads to a non-analytic function in spatial gradients (or wave-numbers $k$), while the resummation of BRSSS (sometimes referred to as ``rBRSSS'') leads to an analytic function expressed in time-like gradients (or frequency $\omega$).

 Eq.~(\ref{eq:rBRSSS}) has several attractive features which have made it (and its descendants) the workhorse of relativistic viscous hydrodynamics simulations. First, Eq.~(\ref{eq:rBRSSS}) is still a complete conformal theory to second order in gradients, which guarantees that all terms consistent with symmetries are present up to second-order accuracy. Thus, Eq.~(\ref{eq:rBRSSS}) describes the universal low-momentum behavior of any viable microscopic theory in the absence of fluctuations. Different theories will only differ in their values for the coefficients $\tau_\pi,\kappa,\lambda_1,\lambda_2,\lambda_3$. Second, rBRSSS leads to hyperbolic equations of motion for fluid dynamics, guaranteeing that signals propagate with speed less than
 $$
 v^{\rm max, sound}_{g}=\sqrt{c_s^2+\frac{2 (d-2) C_\eta}{(d-1)C_\pi}}\,.
 $$
 This implies that Eq.~(\ref{eq:rBRSSS}) may be solved numerically for arbitrary initial conditions in arbitrary space-time dimension $d>2$ without any inherent instabilities if
 \begin{equation}
\label{eq:stability}
   C_\pi> 2 C_\eta\,.
   \end{equation}

\index{BRSSS! Collective modes}
Numerical schemes to solve the rBRSSS equations of motion will be discussed in chapter \ref{sec:numsim}. Also, the collective modes for rBRSSS can be calculated in complete analogy to the treatment of the Navier-Stokes equation in section \ref{sec:NScoll}. Considering the case of $d=4$, one finds for instance in the variational approach
\begin{eqnarray}
  \label{eq:rBRSSS2pt}
  G^{00,00}(\omega,k)&=&-2\epsilon_0+\frac{k^2 \left[(\epsilon_0+P_0)(1-i\tau_\pi\omega)-\frac{2 \kappa k^2}{3}\right]}{(\omega^2-c_s^2 k^2)(1-i \tau_\pi \omega)+\frac{4 i \omega k^2 \gamma_\eta}{3}}\,,\nonumber\\
  G^{01,01}(\omega,k)&=&\epsilon_0+\frac{k^2 \left[\eta-\frac{i \omega \kappa}{2}\right]}{i \omega(1-i \tau_\pi \omega)-\gamma_\eta k^2}\,,\nonumber\\
  G^{12,12}(\omega,k)&=&P-\frac{i \eta \omega+\frac{\kappa}{2}(k^2+\omega^2)}{1-i \tau_\pi \omega}\,,
\end{eqnarray}
where $\gamma_\eta\equiv \frac{\eta}{\epsilon_0+P_0}$. Similar to the case of Maxwell-Cattaneo theory, one finds in addition to the sound and shear mode one non-hydrodynamic mode within each of the channels, including the ``scalar'' channel described by $G^{12,12}$, with a dispersion relation
\begin{equation}
  \label{eq:rbrsssnonhydro}
  \lim_{k\rightarrow 0}\omega=-\frac{i}{\tau_\pi}\,.
\end{equation}
(See Ref.~\cite{Heller:2014wfa} for a modified version of rBRSSS that exhibits pairs of non-hydrodynamic modes with non-vanishing real parts reminiscent of those found in holography, see section \ref{sec:ggnonhydro}.)

The retarded correlators (\ref{eq:rBRSSS2pt}) give rise to Kubo relations that can be used to determine the value of the transport coefficients $\tau_\pi,\kappa$ near equilibrium. In order to determine $\lambda_1,\lambda_2,\lambda_3$, it is necessary to go beyond linear-response and determine higher-point correlation functions. Table \ref{tab:one2} summarizes the current knowledge about the values for $\epsilon(P),\eta,\tau_\pi,\lambda_1,\lambda_2,\lambda_3$ in different approaches.
\begin{table}[t]
\begin{center}
  \begin{tabular}{c|cccc}
    \hline
        &    Gauge/Gravity & Kinetic (BGK) & pQCD & Lattice QCD\\
    \hline
    $\epsilon(P)$ & 3 P & Eq.~(\ref{eq:KTthermo}) & 3 P& Eq.~(\ref{eq:latticeeos1})\\
    $\eta$   & $\frac{\epsilon+P}{4 \pi T}$           &    $\frac{(\epsilon+P) \tau_R}{5}$           &   $\frac{3.85 (\epsilon+P)}{g^4 \ln\left(2.765 g^{-1}\right) T}$    & $0.17(2)\frac{\epsilon+P}{T}$\\
    $\tau_\pi$ & $\frac{2 -\ln 2}{2 \pi T}$           &     $\tau_R$          &   $\frac{5.9 \eta}{\epsilon+P}$    & \\
    $\lambda_1$ & $\frac{\eta}{2 \pi T}$           &   $\frac{5}{7}\eta \tau_R$            &   $\frac{5.2 \eta^2}{\epsilon+P}$    & \\
    $\lambda_2$ &  $2 \eta \tau_\pi-4 \lambda_1$          &    $-2 \eta \tau_R$           &    $-2 \eta \tau_\pi$   & \\
    $\lambda_3$ &  $0$          &    $0$           &   $\frac{30 (\epsilon+P)}{8 \pi^2 T^2}$    & \\
    $\kappa$ & $\frac{\epsilon+P}{4 \pi^2 T^2}$           &      $0$         &    $\frac{5 (\epsilon+P)}{8 \pi^2 T^2}$   & $0.36(15)T^2$ \\
    \hline
    Refs. &   \cite{Policastro:2001yc,Baier:2007ix,Bhattacharyya:2008jc}      &  \cite{Baier:2007ix,Romatschke:2011qp,Jaiswal:2014isa}  & \cite{Arnold:2003zc,York:2008rr,Romatschke:2009ng} & \cite{Borsanyi:2013bia,Bazavov:2014pvz,Nakamura:2004sy,Meyer:2007ic}\\
    &\cite{Haack:2008xx,Arnold:2011ja} && \cite{Moore:2012tc}&\cite{Meyer:2011gj,Philipsen:2013nea,Pasztor:2018yae}\\
    \hline
\end{tabular}
\end{center}
\caption{\label{tab:one2} Compilation of leading-order results for transport coefficients in various calculational approaches, see text for details.}
\end{table}
\index{Transport coefficients}
In this table, the column labeled ``Gauge/Gravity'' corresponds to ${\cal N}=4$ SYM in the limit of infinite 't Hooft coupling, cf. section \ref{sec:ads}. Finite coupling corrections to the gauge/gravity results have been calculated in Refs.~\cite{Buchel:2008sh,Buchel:2008bz,Buchel:2008kd,Saremi:2011nh,Grozdanov:2014kva,Grozdanov:2016fkt}. The column labeled ``Kinetic (BGK)'' refers to results from the Boltzmann equation in the relaxation time approximation, cf. section \ref{sec:kin}. \index{BGK} \index{Relaxation time approximation|see {BGK}} The column labeled ``pQCD'' refers to results obtained from perturbative QCD in the limit of small gauge coupling. Finite coupling corrections to some pQCD results have been calculated in Ref.~\cite{York:2008rr}. The column labeled ``Lattice QCD'' corresponds to results obtained from simulating $SU(3)$ Yang-Mills theory on the lattice, cf. section \ref{sec:lattice}. Note that lattice gauge theory results are not available for all second order transport coefficients and that existing results may additionally have considerable systematic errors, cf. Ref.~\cite{Ding:2015ona}. Equation numbers have been supplied in case the functional dependence would have been too lengthy to fit in Tab.~\ref{tab:one2}.

Inspecting the entries in Tab.~\ref{tab:one2}, one finds that the stability criterion (\ref{eq:stability}) is fulfilled for all known microscopic theories in $d=4$, with values for the ratio of $C_\pi$ to $C_\eta$ ranging from
\begin{equation}
  \label{eq:causalityrange}
C_\pi/C_\eta\in[2.61,5.9]\,,
\end{equation}
with the lower value originating from strong coupling results in gauge/gravity duality and the higher value originating from weak coupling pQCD results.

\subsection{Non-conformal Second Order Hydrodynamics}
\label{sec:nonconfwork}

In the previous subsection, a resummation of the conformal second-order hydrodynamic series was introduced
that leads to causal equations of motion while preserving the complete universal low-momentum behavior
inherent to hydrodynamics. One may ask if this program can be carried over to systems that do not have conformal symmetry
and/or have  additional conserved charges.

\begin{table}[t]
\begin{center}
  \begin{tabular}{c|ccc}
    \hline
        &Gauge/gravity & Kinetic(BGK) & pQCD \\
    \hline
    $\zeta$ & $2\eta\left(\frac{1}{3}-c^2_s\right)$ & Eq.~(\ref{eq:etazetaKT}) & $15 \eta \left(\frac{1}{3}-c^2_s\right)^2$ \\
    $\tau_\Pi$ & $\tau_\pi$ & $\tau_\pi$ &  \\
    $\kappa^*$ &  $-\frac{\kappa}{2 c^2_s}(1-3c^2_s)$  & &$\kappa-\frac{T}{2}\frac{d\kappa}{dT}$   \\
    $\xi_1$ & $\frac{\lambda_1}{3}(1-3c^2_s)$ & &  \\
    $\xi_2$ & $\frac{2\eta\tau_\pi c^2_s}{3}(1-3c^2_s)$ & &  \\
    $\xi_3$ & $\frac{\lambda_3}{3}(1-3c^2_s)$  & & Ref.~\cite{Moore:2012tc} \\
    $\xi_4$ & $0$  & &  Ref.~\cite{Moore:2012tc}  \\
    $\xi_5$ & $\frac{\kappa}{3}(1-3c^2_s)$  & & $-c_s^2 \kappa^*-\frac{\kappa}{6}(1-3c_s^2)$ \\
    $\xi_6$ & $\frac{\kappa}{3c^2_s}(1-3c^2_s)$ & & Ref.~\cite{Moore:2012tc} \\
    \hline
    Refs. &   \cite{Buchel:2007mf,Kanitscheider:2009as,Romatschke:2009kr}   & \cite{Romatschke:2011qp,Denicol:2012cn,Jaiswal:2014isa}  &  \cite{Arnold:2006fz,Banerjee:2012iz,Moore:2012tc}   \\
    &\cite{Bigazzi:2010ku,Finazzo:2014cna,Wu:2016erb} && \cite{Horsley:1985dz}\\
    \hline
\end{tabular}
\end{center}
\caption{\label{tab:one3} Compilation of non-conformal transport coefficient results in various calculational approaches, complementing the information in Tab.~\ref{tab:one2}. See text for details.}
\end{table}

The answer is affirmative. Rewriting the constitutive equations Eq.~(\ref{eq:constf}) in a manner similar to rBRSSS leads to\footnote{Note that the coefficients $\bar \tau_\pi^*,\bar \xi_2$ are different from the original coefficients $\tau^*,\xi_2$ in Ref.~\cite{Romatschke:2009kr} because a different expansion basis was used. }
 \begin{eqnarray}
   \label{eq:r2h}
   \pi^{\mu \nu}&=&-\eta \sigma^{\mu \nu}-\tau_\pi \left[^{<} D \pi^{\mu \nu>}+\frac{d}{d-1} \pi^{\mu \nu}\nabla_\lambda^\perp u^\lambda  \right]
+\kappa\left[R^{<\mu \nu>}-2 u_\lambda u_\rho R^{\lambda <\mu \nu>\rho}\right]
\nonumber\\
&&+\frac{\lambda_1}{\eta^2} \pi^{<\mu}_{\quad \lambda}\pi^{\nu> \lambda}-\frac{\lambda_2}{\eta} \pi^{<\mu}_{\quad \lambda} \Omega^{\nu> \lambda}
+\lambda_3 \Omega^{<\mu}_{\quad \lambda}\Omega^{\nu> \lambda}\nonumber\\
&&+\kappa^* 2 u_\lambda u_\rho R^{\lambda <\mu \nu> \rho} -\bar{\tau}_\pi^* \frac{\nabla_\lambda^\perp u^\lambda}{d-1} \pi^{\mu \nu}
+\bar{\lambda}_4 \nabla_\perp^{<\mu} \ln \epsilon \nabla_\perp^{\nu>}\ln \epsilon\,,\nonumber\\
\Pi&=&-\zeta\left(\nabla_\lambda^\perp u^\lambda\right)-\tau_\Pi D\Pi
+\frac{\xi_1}{\eta^2} \pi^{\mu \nu} \pi_{\mu \nu}+\frac{\bar{\xi_2}}{\zeta^2} \Pi^2
\nonumber\\
&&+\xi_3 \Omega^{\mu \nu} \Omega_{\mu \nu}
+\bar{\xi_4} \nabla^\perp_\mu \ln \epsilon \nabla_\perp^\mu \ln \epsilon+\xi_5 R
+\xi_6 u^\lambda u^\rho R_{\lambda \rho}\,.
 \end{eqnarray}
The equations (\ref{eq:r2h}) correspond to a complete theory up to second order gradients for uncharged fluids with causal signal propagation for sufficiently large $\tau_\pi,\tau_\Pi$. As was the case for rBRSSS, different microscopic theories would lead to different values of the transport coefficients in Eq.~(\ref{eq:r2h}). 

 \index{M\"uller-Israel-Stewart theory}
 Historically, parts of Eq.~(\ref{eq:r2h}) have been derived. For instance, Israel and Stewart's theory \cite{Israel:1979wp} (see also M\"uller's \cite{Muller:1967zza}) corresponds to Eq.~(\ref{eq:r2h}) with $\kappa=\kappa^*=\bar\tau^*_\pi=\xi_1=\bar \xi_2=\xi_3=\bar \xi_4=\xi_5=\xi_6=\lambda_1=\lambda_3=\bar\lambda_4=0$. Muronga's generalization of M\"uller-Israel-Stewart theory corresponds to Eq.~(\ref{eq:r2h}) with $\kappa=\kappa^*=\xi_1=\xi_3=\bar \xi_4=\xi_5=\xi_6=\lambda_1=\lambda_3=\bar\lambda_4=0$ but 
 $\bar\xi_2\neq 0,\bar \tau_\pi^*\neq 0$ \cite{Muronga:2001zk}. Denicol, Niemi, Molnar and Rischke (DNMR) derived a version of Eq.~(\ref{eq:r2h}) valid for kinetic theory in flat space (e.g. $R^{\mu\nu\lambda\delta}=0$), but chose to go beyond second-order accuracy by distinguishing between terms such as $\zeta \pi^{\mu\nu} \nabla^\perp_\lambda u^\lambda $ and $\eta \sigma^{\mu\nu}\Pi$ \cite{Denicol:2012cn}.
\index{DNMR}
 
 Tab.~\ref{tab:one3} summarizes the current knowledge about the non-conformal transport coefficients in Eq.~(\ref{eq:r2h}), complementing the information found in Tab.~\ref{tab:one2}. The column labeled ``Gauge/Gravity'' corresponds to results obtained from classical gravity calculations that are conjectured to be dual to particular non-conformal field theories at infinite 't Hooft coupling, cf. section \ref{sec:ads}. It should be pointed out that in the context of strongly coupled field theories, the relation
 \begin{equation}  \index{Transport coefficients}
   \kappa^*=\kappa - \eta \tau_\pi+\frac{\lambda_2}{2}
 \end{equation}
 has been shown to hold for a broad class of theories \cite{Kleinert:2016nav}.
The column labeled ``Kinetic (BGK)'' in Tab.~\ref{tab:one3} refers to results from the Boltzmann equation in the relaxation time approximation, cf. section \ref{sec:kin}. The column labeled ``pQCD'' refers to results obtained from perturbative QCD in the limit of small gauge coupling. In Tab.~\ref{tab:one3}, 
equation and reference numbers have been supplied in case the functional dependence would have been too lengthy to fit. 
Note that several results for transport coefficients in Tab.~\ref{tab:one3} are not available at the time of writing. 

The transport coefficients given in Tab.~\ref{tab:one3} may be used to evaluate the maximal propagation speeds (\ref{eq:maxshear}), (\ref{eq:maxsound}) for non-conformal theories. One finds that the maximal propagation speed decreases when decreasing $c_s^2<\frac{1}{3}$, suggesting that non-conformal fluid dynamics is causal also for this case.

\index{Causality ! Fluid Speed}

Finally, let us mention that, following the treatment in section \ref{sec:NScoll}, the correlation functions for non-conformal second order hydrodynamics in $d=4$ may be obtained as
\begin{eqnarray}
  \label{eq:non2hypt}
  G^{00,00}(\omega,k)&=&-2\epsilon_0+\frac{k^2\left[(\epsilon_0+P_0)-k^2\frac{2 (\kappa-2 \kappa^*)}{3(1-i \tau_\pi \omega)}+k^2\frac{2 \xi_5-\xi_6}{1-i \tau_\Pi \omega}\right]}{\omega^2-c_s^2 k^2+ i \omega k^2 \gamma_s(\omega)}\,,\nonumber\\
  G^{01,01}(\omega,k)&=&\epsilon_0+\frac{k^2\left(\eta -i \omega \left(\frac{\kappa}{2}-\kappa^*\right)\right)}{i \omega(1-i \tau_\pi \omega)-\gamma_\eta k^2}\nonumber\\
  G^{12,12}(\omega,k)&=&P-\frac{i \eta \omega+\frac{\kappa}{2}(k^2+\omega^2)-\kappa^* \omega^2}{1-i\tau_\pi \omega}\,,
\end{eqnarray}
where the sound attenuation length $\gamma_s$ in Eq.~(\ref{eq:sound attenuation}) is generalized to 
\begin{equation}
\label{gs2}
\gamma_s(\omega)=\frac{4 \eta}{3 (\epsilon_0+P_0) (1-i \tau_\pi \omega)}+\frac{ \zeta}{(\epsilon_0+P_0) (1-i \tau_\Pi \omega)}\,.
\end{equation}

\section{Fluctuating Hydrodynamics}
\label{sec:hydrofluc}

So far, the treatment of fluid dynamics has been limited to that of classical fluids, including the effect of viscous corrections. However, it is well-known from non-relativistic systems that the hydrodynamic correlation functions such as Eq.~ (\ref{eq:NS2pt}) do not match experiment even in the limit of small frequency and wave-number \cite{Pomeau197563}, because of the presence of so-called long-time tails. The reason for this mismatch is that while the effects of dissipation have been accounted for, the presence of fluctuations, which are required by the fluctuation-dissipation theorem, have not  been treated on equal footing.

Therefore, including the effects of fluctuations in the fluid dynamic treatment is desirable, giving rise to the theory of ``fluctuating hydrodynamics'' (as opposed to ``classical hydrodynamics'' without fluctuations).

\subsection{Bottom-up Approach of Fluctuating Hydrodynamics}

To illustrate the procedure, let us first consider supplementing \textit{linearized} hydrodynamics by a thermal noise term. To this end, start with the (first-order) classical hydrodynamic energy-momentum tensor linearized around a state with constant energy density $\epsilon_0$ and vanishing fluid velocity $u_0^\mu=(1,{\bf 0})$ in Minkowski space-time given in Eq.~(\ref{eq:NSlinearresp}). Even in the presence of fluctuations, $\epsilon,u^\mu$ are defined as the time-like eigenvalue and eigenvector of the full (fluctuating) energy-momentum tensor, thus to linear order in perturbations only the space-like part of $T^{ij}$ may fluctuate. This can be parametrized as
\begin{eqnarray}
  \label{eq:noisytmunu}
  T_{00,\xi}&=&\delta \epsilon\,,\nonumber\\
  T_{0i,\xi}&=&-(\epsilon_0+P_0)\delta u_i\,,\\
  T_{ij,\xi}&=&\delta_{ij}c_s^2 \delta \epsilon-\eta \left(\partial_i \delta u_j+\partial_j \delta u_i-\frac{2}{d-1}\delta_{ij}\partial_k \delta u_k\right)-\zeta \delta_{ij}\partial_k \delta u_k+\xi_{ij}\,,\nonumber
\end{eqnarray}
where $T_{\mu\nu,\xi}$ denotes the energy-momentum tensor in the presence of fluctuations $\xi_{ij}$. The noise term $\xi_{ij}$ itself is taken to be Gaussian \cite{LL9}
\begin{equation}
  \langle \xi_{ij}(x) \xi_{kl}(y)\rangle_\xi=2 T \left[\eta \left(\delta_{ik}\delta_{jl}+\delta_{il}\delta_{jk}-\frac{2}{d-1}\delta_{ij}\delta_{kl}\right)+ \zeta\, \delta_{ij}\delta_{kl}\right]\delta(x-y)\,,
\end{equation}
and the coefficients $2 T \eta,2 T \zeta$ have been chosen in hindsight of obtaining the correct expression for the hydrodynamic correlation functions (\ref{eq:NSlinearresp}). Because the noise is Gaussian, the noise average $\langle \rangle_\xi$ may be written as a functional integral,
\begin{equation}
  \label{eq:noiseav}
  \langle {\cal O}\rangle_\xi = \int{\cal D}\xi_{ij}\!\!\!\!\!\!\!\!\!\!\!\raisebox{-2.5ex}{ ${\scriptstyle i\leqslant j}$ } e^{-S_\xi} {\cal O}\,,
\end{equation}
where \cite{Kovtun:2014hpa}
$$
S_\xi=\int dx^d\,\xi_{ij}\left[\frac{1}{16 T \eta}\left(\delta_{ik}\delta_{jl}+\delta_{il}\delta_{jk}-\frac{2}{d-1}\delta_{ij}\delta_{kl}\right)+\frac{1}{4 T (d-1)^2 \zeta}\delta_{ij}\delta_{kl}\right]\xi_{kl}\,.
$$
One may for instance be interested in calculating the noise average of the energy-momentum tensor,
\begin{equation}
  \langle T_{ij,\xi}\rangle_\xi=\delta_{ij}c_s^2 \delta \epsilon-\eta \left(\partial_i \delta u_j+\partial_j \delta u_i-\frac{2}{d-1}\delta_{ij}\partial_k \delta u_k\right)-\zeta \delta_{ij}\partial_k \delta u_k\,,
\end{equation}
which is trivial because $\int {\cal D}\xi e^{-S_\xi}\xi_{ij}=0$. More interesting could be the noise average of
\begin{equation}
  \label{eq:noisy2}
  \langle T_{00,\xi}(x) T_{00,\xi}(y)\rangle_\xi =\langle \delta \epsilon(x) \delta \epsilon(y) \rangle_\xi = G_{00,00}^{\rm sym}(x-y)\,,
\end{equation}
where $G_{\mu\nu,\kappa \lambda}^{\rm sym}(x)$ is the \textit{symmetric} correlation function\footnote{Since $T_{00,\xi}(x)$ is not an operator, the order of $T_{00,\xi}(x) T_{00,\xi}(y)$ in the noise average is not important, and hence the average must be symmetric under the interchange $T_{00,\xi}(x)\leftrightarrow T_{00,\xi}(y)$.}. In equilibrium at temperature $T$, the symmetric correlator is related to the \textit{retarded} correlator introduced earlier in Eq.~(\ref{eq:2ptcorr}) by the Kubo-Martin-Schwinger (KMS) relation \index{Kubo-Martin-Schwinger (KMS)}
\begin{equation}
  \label{eq:KMS}
  G_{\mu\nu,\kappa \lambda}^{\rm sym}(\omega,k)=-\left(1+2 n_B(\omega)\right){\rm Im}\,G_{\mu\nu,\kappa \lambda}(\omega,k)\,,
\end{equation}
where $n_b(\omega)=(e^{\omega/T}-1)^{-1}$ is the bosonic thermal distribution, cf. Ref.~\cite{Kovtun:2011np}.

To evaluate (\ref{eq:noisy2}),  $\delta \epsilon(x)$ is to be taken as the energy-density perturbation that is a solution of the \textit{fluctuating} equations of motion $\nabla_\mu T_\xi^{\mu\nu}=0$ with the noise-dependent part of $T^{\mu\nu}_\xi$ given by Eq.~(\ref{eq:noisytmunu}). An explicit solution for $\delta \epsilon$ for the case at hand can easily be obtained following the procedure outlined in section \ref{sec:NEuler}, finding \cite{Kovtun:2014hpa}
\begin{equation}
  \delta \epsilon(\omega,k)=\frac{k_i k_j \xi_{ij}}{\omega^2-c_s^2 k^2 +i \gamma_s \omega k^2 }\,,
\end{equation}
where for arbitrary space-time dimension $d$, the sound attenuation length from Eq.~(\ref{eq:sound attenuation}) generalizes to $\gamma_s=\frac{\eta}{\epsilon_0+P_0}(2-\frac{2}{d-1})+\frac{\zeta}{(\epsilon_0+P_0)}$. \index{Sound attenuation length} Plugging this result into (\ref{eq:noisy2}) leads to 
\begin{eqnarray}
  \label{eq:noisy3}
 G_{00,00}^{\rm sym}(\omega, k)&=&\frac{2 T k_i k_j k_n k_l \left[\eta \left(\delta_{in}\delta_{jl}+\delta_{il}\delta_{jn}-\frac{2}{d-1}\delta_{ij}\delta_{nl}\right)+ \zeta\, \delta_{ij}\delta_{nl}\right]}{\left(\omega^2-c_s^2 k^2\right)^2 +\left(\gamma_s \omega k^2\right)^2 }\,,\nonumber\\
&=& - \frac{2 T}{\omega} {\rm Im}\frac{(\epsilon_0+P_0)k^2}{\omega^2-c_s^2 k^2 +i \gamma_s \omega k^2}\,.
\end{eqnarray}
Using (\ref{eq:KMS}), this result for $G_{00,00}^{\rm sym}$ matches the result for the retarded correlator $G_{00,00}$ from (\ref{eq:NS2pt}) in the hydrodynamic small frequency limit where
$$1+2 n_B(\omega)\simeq \frac{2 T}{\omega}+{\cal O}(\omega)\,.
$$

\subsection{Effective Action Formulation of Bottom-up Approach}
\index{Fluid Dynamics! Action formulation}

It is possible to recast the stochastic approach outlined above into field-theoretic language by employing an effective action formulation. To this end, start with the noise averaging definition (\ref{eq:noiseav}) and recall that every operator needs to be evaluated as a solution to the fluctuating equations of motion $\partial_\mu T^{\mu\nu}_\xi=0$. Introducing a collective field variable $\phi^\mu=\left(\delta \epsilon,(\epsilon_0+P_0)\delta u^i\right)=\left(T^{00},T^{0i}\right)$ for the $d$ hydrodynamic variables in $d$ space-time dimensions, this
 requirement can be enforced by a $\delta$ function:
\begin{equation}
\label{eq:tonoisea}
  \langle \ldots \rangle_\xi = \int{\cal D}\xi_{ij}\!\!\!\!\!\!\!\!\!\!\!\raisebox{-2.5ex}{ ${\scriptstyle i\leqslant j}$ } e^{-S_\xi} \int {\cal D}\phi^\mu \delta\left(\partial_\mu T^{\mu\nu}_\xi\right)\ldots=\int{\cal D}\xi_{ij}\!\!\!\!\!\!\!\!\!\!\!\raisebox{-2.5ex}{ ${\scriptstyle i\leqslant j}$ } {\cal D \phi^\mu} {\cal D\tilde \phi^\nu} e^{-S_\xi+i \int d^d x \tilde \phi_\nu \partial_\mu T^{\mu\nu}_\xi} \,,
\end{equation}
where an auxiliary field $\tilde \phi^\nu$ has been introduced to exponentiate the constraint of the fluctuating equations of motion.
Note that in this effective theory there are $d$ hydrodynamic degrees of freedom ($\delta \epsilon,\delta u^i$) and there are $d$ auxiliary scalar fields $\tilde \phi^\nu$.

The noise term on the rhs of Eq.~(\ref{eq:tonoisea}) may be readily integrated out, leading to a hydrodynamic partition function
\begin{equation}
  Z=\int {\cal D\phi^\nu} {\cal D\tilde \phi^\nu} e^{iS_{\rm eff}}\,,
\end{equation}
with the effective hydrodynamic action \cite{Kovtun:2014hpa}
\begin{equation}
\label{eq:seff1}
  S_{\rm eff}=\int d^dx \left[-\partial_\mu \tilde \phi_\nu T^{\mu\nu}
    +i T \partial_i \tilde \phi_j \left(\eta \left(\delta_{il}\delta_{jk}+\delta_{ik}\delta_{jl}-\frac{2}{d-1}\delta_{ij}\delta_{kl}\right)+\zeta \delta_{ij}\delta_{kl}\right) \partial_k\tilde \phi_l\right]\,,
\end{equation}
where $T^{\mu\nu}$ is the ``classical'' fluid energy-momentum tensor (that is, (\ref{eq:noisytmunu}) without noise term $\xi_{ij}=0$). Fluctuating hydrodynamics basically corresponds to a theory with double the number of variables of the corresponding classical hydrodynamic theory, namely $2 d$ in $d$ space-time dimensions.

It is interesting to consider the tree-level approximation to the theory defined by the effective action (\ref{eq:seff1}). This can be accomplished by considering only the contributions to $S_{\rm eff}^{(2)}$ that are quadratic in the fields $\phi,\tilde\phi$. In Fourier space, $S_{\rm eff}^{(2)}$ then becomes
\begin{equation}
  S_{\rm eff}^{(2)}=\frac{i}{2}\int d\omega d^{d-1}k
  \left(\begin{array}{c}
    \phi^{\mu,*}(\omega,k)\\
    \tilde \phi^{\mu,*} (\omega,k)
\end{array}
  \right)
  \left(\begin{array}{cc}
    M_{\phi\phi} & M_{\phi\tilde \phi}\\
    M_{\tilde \phi \phi} & M_{\tilde \phi \tilde\phi}
\end{array}
  \right)
  \left(\begin{array}{c}
    \phi^{\mu}(\omega,k)\\
    \tilde \phi^{\mu} (\omega,k)
\end{array}
  \right)\,,
\end{equation}
where 
\begin{eqnarray}
M_{\phi\tilde\phi}&=&\left(\begin{array}{cc}
    -\omega & -c_s^2 k^i\\
    k^i & \left(\omega-\frac{ i \eta k^2}{\epsilon_0+P_0}\right)\delta^{ij}-\frac{i k^i k^j}{\epsilon_0+P_0}\left(
    \frac{\eta (d-3)}{d-1}+\zeta \right)
\end{array}\right)\,,\nonumber\\
M_{\phi\phi}&=&\left(\begin{array}{cc}
    0 & 0\\
    0 & 2 T \left(\eta k^2 \delta^{ij}+k^i k^j \left(\frac{\eta (d-3)}{d-1}+\zeta\right)\right)
\end{array} \right)\,,\quad
\end{eqnarray}
$M_{\phi\phi}$ vanishes, and $M_{\tilde \phi \phi}=-M_{\phi\tilde\phi}^{T *}$. From this expression, 
it is straightforward to evaluate the two-point correlation functions of the theory as the inverse of the matrix $M$.
Specifically, the symmetric two-point correlator of $\delta \epsilon$ may be obtained from
\begin{equation}
  G^{00,00}_{\rm sym}(\omega,k)=\frac{\int {\cal D\phi^\nu} {\cal D\tilde \phi^\nu} e^{iS_{\rm eff}} |\phi^{0}(\omega,k)|^2}{\int {\cal D\phi^\nu} {\cal D\tilde \phi^\nu} e^{iS_{\rm eff}} }
  =\frac{2 T (\epsilon_0+P_0)\gamma_s k^4}{\left(\omega^2-c_s^2 k^2\right)^2+\gamma_s^2 \omega^2 k^4}\,,
\end{equation}
matching the earlier result (\ref{eq:noisy3}). Choosing again ${\bf k}=k{\bf e}_3$, it is also straightforward to calculate
\begin{equation}
  \label{eq:g0003sym}
  G^{00,03}_{\rm sym}(\omega,k)=\frac{\int {\cal D\phi^\nu} {\cal D\tilde \phi^\nu} e^{iS_{\rm eff}} \phi^{0,*}\phi^{3}}{\int {\cal D\phi^\nu} {\cal D\tilde \phi^\nu} e^{iS_{\rm eff}}}
  =\frac{2 T (\epsilon_0+P_0)\gamma_s \omega k^3}{\left(\omega^2-c_s^2 k^2\right)^2+\gamma_s^2 \omega^2 k^4}\,,
\end{equation}
which is related to $G^{00,00}_{\rm sym}(\omega,k)$ via the Ward identity (\ref{eq:ward}).
 Besides the symmetric two-point function, the effective action formalism also allows to calculate mixed correlation functions such $\langle \delta \epsilon(x)\tilde \phi^i(y)\rangle$, e.g.
 \begin{equation}
   \label{eq:g0003ret}
  \frac{\int {\cal D\phi^\nu} {\cal D\tilde \phi^\nu} e^{iS_{\rm eff}}\phi^{0,*} \tilde \phi^i}{\int {\cal D\phi^\nu} {\cal D\tilde \phi^\nu} e^{iS_{\rm eff}}}=\frac{-k^i}{\omega^2-c_s^2k^2+i \omega k^2 \gamma_s}=-\frac{1}{(\epsilon_0+P_0)\omega} G^{00,03}(\omega,k)\,,
  \end{equation}
 which almost looks like the retarded correlator $G^{00,03}(\omega,k)$ corresponding to Eq.~(\ref{eq:g0003sym}).  Indeed, when using Eq.~(\ref{eq:g0003ret}) as the \textit{definition} of the retarded correlator, 
Eq.~(\ref{eq:g0003sym}) and Eq.~(\ref{eq:g0003ret}) are related as
 \begin{equation}
   G^{00,03}_{\rm sym}(\omega,k)=-\frac{2 T}{\omega}{\rm Im}\, G^{00,03}(\omega,k)
 \end{equation}
 which is nothing \index{Kubo-Martin-Schwinger (KMS)} but the hydrodynamic KMS relation (\ref{eq:KMS}). Thus, the action formulation provides direct access to the retarded correlation functions (as well as to the advanced correlation functions), cf. Ref.~\cite{Kovtun:2012rj}.

 \subsection{Top-down Approach}

 In the preceding sections, the bottom-up approach to fluctuating hydrodynamics in the linear regime has been explored. This approach has its merits, notably its intuitive interpretation in terms of stochastic forces based on thermal fluctuations. However, there are also some drawbacks. For instance, when promoting the bottom-up approach to full (non-linear) hydrodynamics, it quickly becomes apparent that one has to deal with so-called multiplicative noise, cf. Refs.~\cite{Arnold:1999va,Kovtun:2014hpa}. Also, while the bottom-up approach is useful in order to calculate two-point correlation functions, it is not immediately obvious how it should be generalized for the calculation of n-point correlation functions.

For this reason, it is interesting to consider a different, top-down approach, that directly deals with the derivation of the hydrodynamic effective action. The guiding principle is similar to that of the derivation of the classical hydrodynamic equations of motion, namely symmetry and gradient expansion. The main difference between classical hydrodynamics and fluctuating hydrodynamics is that -- as is apparent in the bottom-up approach (\ref{eq:seff1}) -- the field content for the effective action of fluctuating hydrodynamics is double that of classical hydrodynamics.  Since hydrodynamics is effectively an out-of-equilibrium framework, it is tempting to interpret these fields as operator insertions on the two parts of a time-contour in the closed time path formalism \cite{Chou:1984es,Wang:1998wg}.

Following Refs.~\cite{Grozdanov:2013dba,Harder:2015nxa} (see also Refs.\cite{Crossley:2015evo,Haehl:2017zac} for a more mathematically-inclined approach), the symmetry for the hydrodynamic effective action can be identified as diffeomorphism invariance, since the energy-momentum tensor is the response of a theory to metric fluctuations. This diffeomorphism invariance is imposed on both types of fields, which will be taken as the a-type and r-type fields of Ref.~\cite{Wang:1998wg}, such that
\begin{equation}
  \label{eq:seff3}
  S_{\rm eff}[g_{\mu\nu}^r,g_{\mu\nu}^a]=S_{\rm eff}[g_{\mu\nu}^r,\chi_{\mu\nu}^a]\,,\quad \chi_{\mu\nu}^a=g_{\mu\nu}^a-\nabla_\mu \tilde \phi_\nu^a-\nabla_\nu \tilde \phi_\mu^a\,,
\end{equation}
where on the rhs invariance w.r.t. transformations of $g_{\mu\nu}^a$ under diffeomorphisms
\begin{equation}
  g_{\mu\nu}^a\rightarrow g_{\mu\nu}^a+g_{\mu\lambda}^a\partial_\nu \tilde \phi^\lambda_{a}+g_{\nu\lambda}^a\partial_\mu \tilde \phi^\lambda_{a}+\partial_\lambda g_{\mu\nu}^a \tilde \phi_a^\lambda\,,
\end{equation}
has been made explicit. The diffeomorphism field $\tilde \phi_a^\mu$ is sometimes referred to as a Goldstone-like mode. The action (\ref{eq:seff3}) may now be expanded in a power series in the fields $\tilde \phi^a$, leading to
\begin{equation}
  S_{\rm eff}=\int \frac{\delta S_{\rm eff}}{\delta g_{\mu\nu}^a} \chi_{\mu\nu}^a+\frac{\delta^2 S_{\rm eff}}{\delta g_{\mu\nu}^a\delta g_{\gamma\delta}^a}\chi_{\mu\nu}^a\chi_{\gamma\delta}^a+{\cal O}(\chi^3)\,.
\end{equation}
In this expression, it is easy to recognize
$$
\frac{\delta S_{\rm eff}}{\delta g_{\mu\nu}^a}=\frac{1}{2}\sqrt{-{\rm det}\,g_{\mu\nu}^r}T^{\mu\nu}_{r}
$$
as the retarded one-point function of the theory, cf. Eqns.~(\ref{eq:effS}), (\ref{eq:traforules}). In the absence of sources $g_{\mu\nu}^a\rightarrow 0$, $g_{\mu\nu}^r\rightarrow g_{\mu\nu}$ and for Minkowski space, the term linear in $\chi$ in the effective action becomes
$$
\int \frac{\delta S_{\rm eff}}{\delta g_{\mu\nu}^a} \chi_{\mu\nu}^a\rightarrow \int d^d x \left[-\partial_\mu \tilde \phi_\nu^a  T^{\mu\nu}\right]\,,
$$
which matches the expression (\ref{eq:seff1}) derived from the bottom-up approach. The form of the energy-momentum tensor itself (and that of the correlation functions $\frac{\delta S_{\rm eff}}{\delta^2 g}$) can be obtained by a gradient expansion. To leading order in gradients, and to linear order in a-type fields, $S_{\rm eff}$ must be given by
$$
S_{\rm eff}\propto \left(c_1 g^{\mu\nu}_r +c_2 u^\mu u^\nu\right)\chi_{\mu\nu}^a\,,
$$
which directly leads to the form of the energy-momentum tensor for ideal hydrodynamics (\ref{eq:hydro0f}). First order gradient corrections linear in a-fields then lead to viscous corrections to $T^{\mu\nu}$, whereas second-order terms in a-fields lead to two-point correlation functions \cite{Harder:2015nxa,Crossley:2015evo}. Of course, in the approach taken so far, these would be \textit{tree-level} two-point functions, whereas higher-order terms in a-fields would lead to effective vertices that can lead to a dressing of these two-point functions.

\subsection{Phenomenological Consequences}

In order to understand the consequences that fluctuating hydrodynamics will introduce with respect to classical hydrodynamics, let us return to the original stochastic formulation and study the conformal hydrodynamic one-loop correction to the correlation function $G^{12,12}(\omega,k)$ in the low-momentum limit (${\bf k}=k {\bf e}_3\rightarrow 0$). The relevant terms contributing to $T^{12}$ up to \textit{second order} in perturbation will then be given by \cite{Kovtun:2011np}
\begin{equation}
  T^{12}_\xi=(\epsilon_0+P_0)\delta u^1 \delta u^2+\xi^{12}
  \end{equation}
and hence the retarded correlator becomes (up to contact terms)
$$
G^{12,12}_{\rm sym}(x,y)=\langle\xi^{12}(x)\xi^{12})(y)\rangle_\xi+(\epsilon_0+P_0)^2\langle \delta u^1(x) \delta u^2(x)\delta u^1(y) \delta u^2(y)\rangle_\xi +{\cal O}(\partial^i)\,,
$$
where ${\cal O}(\partial^i)$ is meant to imply that contributions with non-vanishing $k$ have been neglected. Transforming to Fourier space leads to
\begin{eqnarray}
  G^{12,12}_{\rm sym}(\omega,k\rightarrow 0)=2 T \eta +\int \frac{d\omega^\prime}{2\pi}\frac{d^{d-1}k^\prime}{(2\pi)^{d-1}} &&\left[G_{\rm sym}^{01,01}(\omega^\prime,{\bf k}^\prime) G_{\rm sym}^{02,02}(\omega-\omega^\prime,-{\bf k}^\prime)\right.\nonumber\\
    &&\left.+G_{\rm sym}^{01,02}(\omega^\prime,{\bf k}^\prime) G_{\rm sym}^{02,01}(\omega-\omega^\prime,-{\bf k}^\prime)\right]\,.\nonumber
\end{eqnarray}
Using again the KMS relation (\ref{eq:KMS}), the symmetric and retarded two-point correlation functions are related as
$$
  G_{\rm sym}^{0i,0j}(\omega,{\bf k})=\int dt d^{d-1}x e^{i \omega t-i {\bf k}\cdot {\bf x}} \langle T^{0i}(x) T^{0j}(0)\rangle_\xi=-\frac{2 T}{\omega}{\rm Im}G^{0i,0j}(\omega,{\bf k})\,,
  $$
where the explicit form for $G^{0i,0j}(\omega,{\bf k})$ is given in Eq.~(\ref{eq:gijcorr}).
This leads to a retarded correlator of the form
\begin{eqnarray}
  G^{12,12}_{\rm sym}(\omega,k\rightarrow 0)=-i \eta \omega  +\int \frac{d\omega^\prime}{2\pi}\frac{d^{d-1}k^\prime}{(2\pi)^{d-1}} &&\left[G_{\rm sym}^{01,01}(\omega^\prime,{\bf k}^\prime) G^{02,02}(\omega-\omega^\prime,-{\bf k}^\prime)\right.\nonumber\\
    &&\left.+G_{\rm sym}^{01,02}(\omega^\prime,{\bf k}^\prime) G^{02,01}(\omega-\omega^\prime,-{\bf k}^\prime)\right.\,,\nonumber\\
    &&\left.+G^{01,01}(\omega^\prime,{\bf k}^\prime) G_{\rm sym}^{02,02}(\omega-\omega^\prime,-{\bf k}^\prime)\right.\,,\nonumber\\
    &&\left.+G^{01,02}(\omega^\prime,{\bf k}^\prime) G_{\rm sym}^{02,01}(\omega-\omega^\prime,-{\bf k}^\prime)\right]\,,\nonumber
\end{eqnarray}
which in the case of $d=4$ evaluates to \cite{Kovtun:2011np}
\begin{equation}
  \label{eq:goneloop}
  G^{12,12}(\omega,k\rightarrow 0)=-i \omega\left(\eta+\frac{17 T \Lambda_{UV}}{120 \pi^2 \gamma_\eta}\right)+(1+i) \omega^{3/2}\frac{\left(7+\left(\frac{3}{2}\right)^{3/2}\right) T}{240 \pi \gamma_\eta^{3/2}}\,,
\end{equation}
where $\Lambda_{UV}$ is the scale below which the hydrodynamic effective theory is valid. Recalling that $\gamma_\eta \equiv \frac{\eta}{\epsilon_0+P_0}$ equals the mean-free path (\ref{eq:lambdamfp}), (\ref{eq:sound attenuation}), the cut-off scale may be estimated as its inverse $\Lambda_{UV}\simeq \gamma_\eta^{-1}$.

The one-loop result (\ref{eq:goneloop}) has two features that are absent in classical hydrodynamics. First, the $\omega^{3/2}$ term implies a non-analytic behavior of the correlation function, corresponding to power-like (rather than exponential) decay of perturbations. Being slower than exponential, this implies that correlation functions have long-time tails.

\index{Shear viscosity! Lower bound (fluid dynamics)}
Second, if the effective viscosity of a system is calculated by a Kubo relation such as Eq.~(\ref{eq:Kubo}) then
Eq.~(\ref{eq:goneloop}) implies that this effective viscosity will differ from the ``classical'' viscosity $\eta$ by a correction term that is \textit{inversely proportional} to $\eta$. This corresponds to a renormalization of the viscosity. The particular form of Eq.~(\ref{eq:goneloop}) implies that there is a minimum effective viscosity for any fluid. The physics of this minimum viscosity is purely contained within fluid dynamics. It arises from the fact that for a very good classical fluid with $\eta\rightarrow 0$, sound waves are very long-lived (cf. Eq.~\ref{eq:sound}), allowing them to interact very efficiently. This sound-wave interaction (basically sound-sound scattering) leads to an effective viscosity correction in Eq.~(\ref{eq:goneloop}). For further reading on effective shear and bulk viscosities in heavy-ion collisions and cold atom systems consult Refs.~\cite{PeraltaRamos:2011es,Nahrgang:2011vn,Chafin:2012eq,Romatschke:2012sf,Murase:2013tma,Young:2014pka,Kovtun:2014nsa,Akamatsu:2016llw,Akamatsu:2017rdu,Martinez:2017jjf}.


\chapter{Microscopic Theory Background}

The effective field theory of fluid dynamics fixes the equations
of motion, but not the value of the transport coefficients.
These transport coefficients must be determined by calculations
in the underlying microscopic theory, e.g. by calculating
the correlation functions $G^{\mu\nu,\lambda\sigma}(\omega,k)$ and
comparing to the known fluid dynamic results, cf. Eqns.~(\ref{eq:NS2pt}).

To give some flavor of how microscopic theories are connected
to fluid dynamics, the present chapter is meant to offer a
minimal introduction (``flyby'') to 
four first-principles techniques, namely kinetic theory,
gauge/gravity duality, finite temperature field theory and lattice gauge theory.

For more than the minimalistic treatment offered here,
the interested reader is referred to the excellent
textbook literature on these topics, such as \cite{DeGroot:1980dk,CasalderreySolana:2011us,Laine:2016hma,Montvay:270707}.

\section{Kinetic Theory Flyby}
\label{sec:kin}

For many systems, the relevant degrees of freedom can be approximated as well-localized classical particles (or at least long-lived quasi-particles). Then, higher order correlations between particles may be neglected and the relevant quantity becomes the single particle distribution function $f(x^\mu,p^\mu)$, which has the interpretation of describing the number of particles $N$ with momentum $p^\mu$ at space-time point $x^\mu$. Thus for $d=4$, $f\propto \frac{d N}{d^{3}{\bf x} d^{3}{\bf p}}$. If the interaction between these particles is short-range, they may be thought of as propagating without collision for some period of time along geodesics characterized by an affine parameter $\cal T$.

Thus, $f$ will not change along the geodesic, and hence
\begin{equation}
  \label{eq:fs}
  \frac{d f}{d{\cal T}}=\frac{d x^\mu}{d{\cal T}}\partial_\mu f
+\frac{d p^{\mu}}{d{\cal T}} \frac{\partial f}{\partial p^{\mu}}=0\,.
\end{equation}
If the particles have mass $m$, then $m\frac{d x^\mu}{d{\cal T}}=p^\mu$ can be recognized as the momentum of a relativistic particle. The term $m \frac{d p^{\mu}}{d{\cal T}}=F^\mu$ describes the force acting on the particle arising for instance from electromagnetic or gravitational fields. 

If collisions can not be neglected, Eq.~(\ref{eq:fs}) will receive a correction term ${\cal C}[f]$ (``collision term'') such that Eq.~(\ref{eq:fs}) generalizes to 
\begin{equation}
  \label{eq:Boltzmann}
  p^\mu \partial_\mu f+F^\mu \partial_\mu^{(p)} f=-{\cal C}[f]\,,
\end{equation}
where the minus sign on the rhs is convention and $\partial_\mu^{(p)}\equiv \frac{\partial}{\partial p^\mu}$. Eq.~(\ref{eq:Boltzmann}) is the relativistic Boltzmann equation including mean-field interactions \cite{Boltzmann:1872}. The mean-field interactions are encoded in the force term $F^\mu$, which for gravity takes the form $F^\mu=-\Gamma^\mu_{\lambda \sigma} p^\lambda p^\sigma$ where $\Gamma^\mu_{\lambda\sigma}$ are the Christoffel symbols. For electromagnetic interactions of particles with charge $q$, the force becomes $F^\mu=q F^{\mu\nu}p_\nu$, where $F^{\mu\nu}$ is the electromagnetic field strength tensor.
\index{Boltzmann equation}
\index{Kinetic theory|see {Boltzmann equation}}

\subsection{Classical Fluid Dynamics from Kinetic Theory}
\label{sec:fluidfromkin}

Let us consider the particles to be uncharged $q=0$ and classical
in the sense that they should fulfill the mass shell condition $g_{\mu\nu}p^\mu p^\nu=-m^2$ and have positive energy ($p^0>0$). Integrating the single-particle
distribution $f$ over momenta $\int d^3p f$ has the interpretation of being proportional to the number of particles per volume, or number density. With the mentioned constraints this momentum integration can be cast in a Lorentz-covariant form as
\begin{equation}
\label{eq:intmeas}
  \int d\chi \equiv \int \frac{d^4p}{(2\pi)^4} \sqrt{-{\rm det}g_{\mu\nu}}(2\pi)\delta\left(g_{\mu\nu}p^\mu p^\nu+m^2\right) 2 \theta(p^0)\,,
\end{equation}
such that for instance the number current $n^\mu$ in kinetic theory is given by
\begin{equation}
\label{eq:numcurr}
  n^\mu \equiv \int d\chi\, p^\mu f\,.
\end{equation}
For Minkowski space-time, the time-like component of $N^\mu$ then defines the particle number density $n^0=\int \frac{d^3p}{(2\pi)^3}f$.

If the momentum integration is weighted by a factor of energy $p^0$, the result has units of number density times energy, or energy density. More rigorously, one can define the energy-momentum tensor in kinetic theory as
\begin{equation}
  \label{eq:KT}
  T^{\mu\nu} \equiv \int d\chi\, p^\mu p^\nu f\,.
\end{equation}

Dynamic equations for $T^{\mu\nu}$ can be obtained by acting with $d\chi p^\nu$ on the Boltzmann equation (\ref{eq:Boltzmann}), finding after some algebra \cite{Romatschke:2011qp}
\begin{equation}
 \int d\chi\, p^\nu  p^\mu \partial_\mu f=  \nabla_\mu \int d\chi\, p^\mu p^\nu f =\nabla_\mu T^{\mu\nu}=-\int d\chi\, p^\nu {\cal C}[f]\,,
  \end{equation}
where $\nabla_\mu$ is the geometric covariant derivative for the metric $g_{\mu\nu}$. Thus the energy-momentum tensor is covariantly conserved if the collision term fulfills the condition $\int d\chi p^\nu {\cal C}[f]=0$ for any $f$.

There may be special realizations $f=f_{(0)}(x^\mu,p^\mu)$ where ${\cal C}[f_{(0)}]=0$. These are traditionally associated with equilibrium configurations by employing Boltzmann's H-theory and a global entropy maximum cf. Ref.~\cite{DeGroot:1980dk}. However, it is conceivable that ${\cal C}[f_{(0)}]=0$ for non-equilibrium examples of $f_{(0)}$, for instance those corresponding to a local maximum of some non-equilibrium entropy functional. In the following, we will use the more traditional assumption that ${\cal C}[f_{(0)}]=0$ selects equilibrium particle distributions $f_{(0)}$.

In the case of global equilibrium $f=f_{(0)}$, the kinetic energy-momentum tensor (\ref{eq:KT}) should match that of equilibrium fluid dynamics, Eq.~(\ref{eq:hydro0f}). The particular choice
\begin{equation}
\label{eq:f0}
  f_{(0)}(x^\mu,p^\mu)=f_{(0)}\left(\frac{p^\mu u_\mu(x)}{T(x)}\right)\,,
\end{equation}
leads to the form Eq.~(\ref{eq:hydro0f}) with energy density $\epsilon$ and pressure $P$ given by
\begin{equation}
  \label{eq:defs1KT}
  \epsilon=\int d\chi \left(p^\mu u_\mu\right)^2 f_{(0)}\left(\frac{p^\mu u_\mu}{T}\right)\,,\quad
  P = \int d\chi \frac{p^\mu p^\nu \Delta_{\mu\nu}}{3} f_{(0)}\left(\frac{p^\mu u_\mu}{T}\right)\,.
  \end{equation}

Thus the functional form of the equilibrium distribution function $f_{(0)}$ is
given by a vector $u^\mu$ and a scalar $T$, which can be determined through the time-like eigenvector and eigenvalues of the energy momentum tensor. For simplicity, let us consider particles obeying classical statistics, noting that a generalization to include Bose-Einstein or Fermi-Dirac statistics is straightforward. For classical statistics, the relativistic equilibrium distribution $f_{(0)}$ is known to be given by the Maxwell-J\"uttner distribution \cite{DeGroot:1980dk}
\begin{equation}
  \label{eq:maxjut}
  f_{(0)}\left(\frac{p^\mu u_\mu}{T}\right)={\rm dof}\times e^{\frac{p^\mu u_\mu}{T}}\,,
\end{equation}
where ${\rm dof}$ is proportional to the number of degrees of freedom in the system. For instance, for a SU($N_c$) gauge theory with $N_f$ fundamental Dirac fermions,
\begin{equation}
\label{eq:KTdof}
{\rm dof}_{QCD}=\frac{\pi^4}{180}\left(4 (N_c^2-1)+7 N_c N_f\right)\,.
\end{equation}
For QCD, $N_c=3,N_f\simeq 2$, while for ${\cal N}=4$ SYM with $8$ bosonic and 8 adjoint fermionic degrees of freedom 
\begin{equation}
\label{eq:KTdofsym}
{\rm dof}_{SYM}=\frac{\pi^4 (N_c^2-1)}{90}\left(1 +\frac{7}{8}\right)8=\frac{\pi^4 (N_c^2-1)}{6}\,.
\end{equation}

Unlike $f_{(0)}$, the form of the microscopic collision kernel ${\cal C}$ depends on the particular form of particle interactions considered. However, it will be acceptable for the following discussion to approximate ${\cal C}$ using the single relaxation-time (or BGK \cite{PhysRev.94.511}) approximation, \index{BGK}
\begin{equation}
  \label{eq:BGK}
  {\cal C}=p^\mu u_\mu\frac{f_{(0)}-f}{\tau_R}\,,
\end{equation}
where all the details of the particle interactions have been absorbed into the single (energy-dependent) relaxation time $\tau_R$. A curious feature of the collision term (valid also outside the BGK approximation) is that ${\cal C}$ vanishes both in equilibrium ($f=f_{(0)}$) as well as in the limit of non-interacting particles ($\tau_R\rightarrow \infty$). This can be understood through the fact that in both of these cases, no entropy is produced, albeit for very different reasons. In equilibrium, the system is in a state of maximum entropy, forbidding further entropy production, while in the absence of interactions, phase space trajectories remain unperturbed and no entropy is generated.

The BGK approximation (\ref{eq:BGK}) has the nice feature that it automatically satisfies the condition $\int d\chi p^\nu {\cal C}[f]=0$ because
\begin{equation}
\label{eq:matching}
  \epsilon u^\nu=-u_\mu  \int d\chi p^\mu p^\nu f_{(0)}=-u_\mu  \int d\chi p^\mu p^\nu f\,,
\end{equation}
since $\epsilon, u^\mu$ are defined as the time-like eigenvalue and eigenvector of $T^{\mu\nu}=\int d\chi p^\mu p^\nu f$.

To summarize, the Boltzmann equation (\ref{eq:Boltzmann}) with the BGK collision kernel (\ref{eq:BGK}) given by
\begin{equation}
  \label{eq:BBGK}
  p^\mu \partial_\mu f + F^\mu \partial_{\mu}^{(p)}f=\frac{p^\mu u_\mu}{\tau_R}\left(f-f_{(0)}\right)\,,
\end{equation}
for energy momentum conserving forces $F^\mu$ leads to a covariantly conserved energy-momentum tensor $\nabla_\mu T^{\mu\nu}=0$ with time-like eigenvector and eigenvalue $u^\mu,\epsilon$, respectively. In equilibrium, $f_{(0)}$ is given by Eq.~(\ref{eq:maxjut}) and $\nabla_\mu T^{\mu\nu}=0$ reduces to the relativistic Euler equations (\ref{eq:eulerf2}).

\subsection{Free-Streaming Solutions}

Kinetic theory becomes particularly simple in the case when particle interactions can be neglected, e.g. (\ref{eq:BBGK}) with $\tau_R\rightarrow \infty$. In this limit the kinetic equations can be solved exactly by employing the method of characteristics. In the case of free-streaming in $d=4$ dimensional Minkowski space-time where $F^\mu=0$, (\ref{eq:BBGK}) becomes
\begin{equation}
\label{eq:FSmink}
\partial_t f+\frac{{\bf p}}{p^0}\cdot \nabla f=0\,,\quad p^0=\sqrt{{\bf p}^2+m^2}\,,
\end{equation}
which gives rise to the characteristic equations
\begin{equation}
\frac{d t}{ds}=1\,,\quad
\frac{d {\bf x}}{ds}=\frac{\bf p}{p^0}\,,
\end{equation}
with constant $f$. The solution to the characteristic equations is
\begin{equation}
{\bf x}(t)=\frac{\bf p}{p^0} t+{\rm const.}\,,
\end{equation}
such that all free-streaming solutions to (\ref{eq:FSmink}) are given by
\begin{equation}
f=f_{FS}\left({\bf p},{\bf x}-\frac{\bf p}{p^0} t\right)\,.
\end{equation}

Somewhat more interesting are free-streaming solutions to non-trivial space-times, such as the $d=4$ Milne universe (see section \ref{sec:idealbjork}) given by coordinates $\tau,x,y,\xi$ with $\tau=\sqrt{t^2-z^2}$ and $\xi={\rm arctanh}(z/t)$. In Milne coordinates, (\ref{eq:BBGK}) becomes
\begin{equation}
\label{eq:FSBjork}
\partial_\tau f+\frac{\bf p}{p^\tau}\cdot \nabla f-\frac{2 p^\xi}{\tau} \partial_\xi^{(p)}f=0\,,\quad p^\tau=\sqrt{(p^x)^2+(p^y)^2+\tau^2 (p^\xi)^2+m^2}\,,
\end{equation}
and the characteristic equations are
\begin{equation}
\frac{d {\bf x_\perp}}{d\tau}=\frac{{\bf p}_\perp}{p^\tau}\,,\quad
\frac{d \xi}{d\tau}=\frac{p^\xi}{p^\tau}\,,\quad
\frac{d p^\xi}{d\tau}=-\frac{2 p^\xi}{\tau}\,,
\end{equation}
where ${\bf x}_\perp=(x,y)$, ${\bf p}_\perp=(p_x,p_y)$. Again, the characteristic equations can be readily solved, finding the free-streaming solution to Eq.~(\ref{eq:FSBjork})  \cite{Romatschke:2015dha}
\begin{equation}
\label{eq:FSsolBj}
f=f_{FS}\left({\bf p}_\perp,p_\xi,{\bf x}_\perp-\frac{\tau {\bf p}_\perp p^\tau}{p_\perp^2+m^2},\xi+\ln\left[\frac{p^\tau}{p_\xi}+\frac{1}{\tau}\right]\right)\,.
\end{equation}

\subsubsection{Application: Bjorken flow}
\index{Bjorken flow! Kinetic Theory}

As a particular application of Eq.~(\ref{eq:FSsolBj}) let us consider Bjorken flow (see section \ref{sec:idealbjork}) where the system is assumed to be homogeneous in ${\bf x_\perp},\xi$. Then the free streaming solution is simply given by $f=f({\bf p}_\perp,p_\xi)$. If the system is prepared such that at some initial time $\tau=\tau_0$ the particles are distributed according to Eq.~(\ref{eq:maxjut}) with $T=T_0$, the free-streaming solution for all times is given by
\begin{equation}
\label{eq:BjorFSsol}
f_{FS}={\rm dof}\times \exp{\left[-\frac{\sqrt{p_\perp^2+p_\xi^2/\tau_0^2+m^2}}{T_0}\right]}\,.
\end{equation}
Note that it possible to rewrite the argument of the exponential such that
\begin{equation}
\sqrt{p_\perp^2+p_\xi^2/\tau_0^2+m^2}=\sqrt{(p^\tau)^2+\tau^2 (p^\xi)^2\left(\frac{\tau^2}{\tau_0^2}-1\right)}\,.
\end{equation}
Once the solution for the particle distribution function is known, the kinetic energy-momentum tensor can be calculated from (\ref{eq:KT}), which in Milne coordinates becomes
\begin{equation}
T^{\mu\nu}_{FS}=\tau \int \frac{d^2p_\perp dp^\xi}{(2\pi)^3}\frac{p^\mu p^\nu}{p^\tau} f_{FS}\,.
\end{equation}
Because of the explicit dependence of $f_{FS}$ on the momenta, it is straightforward to see that $T^{\mu\nu}_{FS}$ only has diagonal components. Hence the normalized time-like eigenvector of $T^{\mu\nu}_{FS}$ is $u^\mu=\left(1,0,0,0\right)$ and the corresponding eigenvalue is
\begin{equation}
\label{eq:epsFSbjor}
\epsilon_{FS}=\tau \int \frac{d^2 p_\perp dp^\xi}{(2\pi)^3} p^\tau f_{FS}=\int \frac{d^2 p_\perp dp^z}{(2\pi)^3} p^\tau f_{FS}\,,
\end{equation}
which has been formally rewritten by considering the variable substitution \hbox{$p^z=\tau p^\xi$}. Introducing the new set of coordinates $x^a=\left(\tau,x,y,z\right)$ and momenta $p^a=\left(p^\tau,p^x,p^y,p^z\right)$ with $p^\tau=\sqrt{{\bf p}^2+m^2}$, one notes that 
the free-streaming solution (\ref{eq:BjorFSsol}) is formally equivalent to an anisotropic deformation of the equilibrium distribution function (\ref{eq:maxjut}),
\begin{equation}
\label{eq:faniso}
f_{FS}=f_{(0)}\left(-\frac{\sqrt{p^a \Xi_{ab} p^b}}{T_0}\right)\,,\quad
\Xi_{ab}=u_{a}u_b+n_a n_b \Xi(\tau)\,,
\end{equation}
where $u_a=\left(1,0,0,0\right)$, $n_a=\left(0,0,0,1\right)$ and $\Xi(\tau)=\left(\frac{\tau^2}{\tau_0^2}-1\right)$. Such anisotropic deformations have been considered in the context of plasma instabilities in systems with momentum anisotropies \cite{Romatschke:2003ms} and are the basis for so-called anisotropic hydrodynamics.

\subsection{Eremitic Expansions}

In the limit of large but finite relaxation time $\tau_R$, systematic corrections to the free-streaming solution may be calculated \cite{Borghini:2010hy,Romatschke:2018wgi}. Expanding
\begin{equation}
f=f_{FS}+f_{\rm Hermit, (1)}+\ldots\,,
\end{equation}
with $f_{\rm Hermit, (1)}$ suppressed by a power of $\tau_R$ relative to $f_{FS}$ one finds for $F^\mu=0$
\begin{equation}
p^\mu \partial_\mu f_{\rm Hermit,(1)}=\frac{p^\mu u_\mu}{\tau_R}\left(f_{FS}-f_{(0)}\right)\,,
\end{equation}
where $T,u^\mu$ appearing in the argument of $f_{(0)}$ on the right-hand-side are to be evaluated from the time-like eigenvalue and eigenvector of $T^{\mu\nu}_{FS}=\int d\chi p^\mu p^\nu f_{FS}$. The above equation can once again be solved by the method of characteristics, finding
\begin{equation}
f_{\rm Hermit,(1)}(t,{\bf x},{\bf p})=\int_0^t dt^\prime \left.\frac{p^\mu u_\mu(t^\prime, {\bf x}^\prime)}{\tau_R p^0}\left(f_{FS}(t^\prime, {\bf x}^\prime,{\bf p})-f_{(0)}\right)\right|_{{\bf x}^\prime={\bf x}-\frac{{\bf p}}{p^0} t}\,.
\end{equation}
Higher order corrections may be obtained by repeating this procedure.

\subsection{Transport Coefficients from Kinetic Theory}
\index{Transport coefficients! Kinetic Theory}

Let us now calculate some of the transport coefficients for fluid dynamics
by considering kinetic theory in $d=4$ dimensional Minkowski space-time.

\subsubsection{Equilibrium}

With the form of the equilibrium distribution $f_{(0)}$ for particles
obeying classical statistics specified in (\ref{eq:maxjut}), it
is straightforward to evaluate the equilibrium pressure and energy density
for kinetic theory (\ref{eq:defs1KT}), finding 
\begin{eqnarray}
  \epsilon&=&{\rm dof}\times \int \frac{d^4p}{(2\pi)^3} (p^0)^2 \delta\left(-(p^0)^2+{\bf p}^2+m^2\right)2 \theta\left(p^0\right) e^{-p^0/T}\nonumber\\
  &=&{\rm dof}\times \int \left.\frac{d^3p}{(2\pi)^3} p^0 e^{-p^0/T}\right|_{p^0=\sqrt{{\bf p}^2+m^2}}\nonumber\\
  &=&{\rm dof}\times \frac{ T^4}{2 \pi^2} z^2 \left(3 K_2(z)+z K_1(z)\right)\,,\nonumber\\
  P&=&{\rm dof}\times \frac{ T^4}{2 \pi^2} z^2 K_2(z)\,,
  \label{eq:KTthermo}
\end{eqnarray}
where $z\equiv \frac{m}{T}$ and $K_n(z)$ are  modified Bessel functions. As a result, one finds the speed of sound in kinetic theory to be given by \cite{Romatschke:2011qp} \index{Equation of State (EoS)!Kinetic Theory}
\begin{equation}
  c_s\equiv \sqrt{\frac{dP}{d\epsilon}}=\left(3+\frac{z K_2(z)}{K_3(z)}\right)^{-1/2}\,.
\end{equation}
The results for kinetic theory in equilibrium (\ref{eq:KTthermo}) fulfill the usual thermodynamic identities such as
\begin{equation}
\label{eq:basicthermo}
  \frac{dP}{dT}=s\,,\quad \epsilon+P=s T\,,\quad \frac{d\epsilon}{ds}=T\,,
\end{equation}
with $s$ the equilibrium entropy density given by
\begin{equation}
  s\equiv {\rm dof}\times \frac{ T^3}{2 \pi^2} z^3 K_3(z)\,.
  \end{equation}

\subsubsection{Near Equilibrium}

Near equilibrium one expects the particle distribution function to be close to $f_{(0)}$, e.g.
\begin{equation}
  \label{eq:CE}
f=f_{(0)}+f_{(1)}+f_{(2)}+\ldots\,,
\end{equation}
where similar to the hydrodynamic gradient series (\ref{eq:tmunuseries})
one can expect the expansion to be in powers of space-time gradients, so
that $f_{(0)}$ is zeroth order in gradients, $f_{(1)}$ is first order in gradients
and so on. The expansion (\ref{eq:CE}) basically corresponds to the Chapman-Enskog expansion \cite{CE}.
\index{Chapman-Enskog expansion}

Inserting (\ref{eq:CE}) into Eq.~(\ref{eq:BBGK}) for $F^\mu=0$ one finds to first order in gradients
\begin{equation}
f_{(1)}=\frac{\tau_R}{p^\lambda u_\lambda}  p^\mu \partial_\mu f_{(0)}=\tau_R f_{(0)} \frac{p^\mu p^\nu}{p^\lambda u_\lambda} \left(\frac{\partial_\mu u_\nu}{T}-\frac{u_\nu\partial_\mu T}{T^2}\right)\,.
\end{equation}
Using the comoving time-like and space-like derivatives (\ref{eq:derivs}) one can write \hbox{$\partial_\mu = \nabla_\mu^\perp-u_\mu D$}. Similar to the case of near-equilibrium fluid dynamics, $D u_\mu$ and $D T$ are already determined by the relativistic Euler equations (\ref{eq:eulerf2}). Combining (\ref{eq:eulerf2}) with the thermodynamic relations (\ref{eq:KTthermo}), one finds for instance $D u_\mu=-\frac{\nabla^\perp_\mu P}{\epsilon+P}=-\frac{dP}{dT}\frac{\nabla^\perp_\mu T}{\epsilon+P}=-\nabla^\perp_\mu \ln T$. Collecting all terms, and noting that the term $p^\mu p^\nu$ demands symmetrization with respect to $\mu,\nu$, one finds \cite{Romatschke:2011qp}
\begin{equation}
\label{eq:KTf1}
f_{(1)}=\tau_R f_{(0)} \frac{p^\mu p^\nu}{T p^\lambda u_\lambda} \left[\frac{\sigma_{\mu\nu}}{2}+\left(\frac{\Delta_{\mu\nu}}{3}-c_s^2u_\mu u_\nu\right)\nabla^\perp_\lambda u^\lambda\right]
\end{equation}
The resulting kinetic energy-momentum tensor $T^{\mu\nu}$ to first order in gradients is given by
\begin{eqnarray}
\int d\chi p^\mu p^\nu \left(f_{(0)}+f_{(1)}\right)&=&T_{(0)}^{\mu\nu}-\left[\frac{\sigma_{\sigma \kappa}}{2}+\left(\frac{\Delta_{\sigma\kappa}}{3}-c_s^2u_\sigma u_\kappa\right)\nabla^\perp_\lambda u^\lambda\right]\frac{\tau_R}{T} I^{\mu\nu\sigma \kappa}_{(1)}\nonumber\\
I^{\mu\nu\sigma\kappa}_{(n)}&\equiv&\int d\chi \frac{p^\mu p^\nu p^\sigma p^\kappa}{(-p^\lambda u_\lambda)^n} f_{(0)}\left(\frac{p^\gamma u_\gamma}{T}\right)\,.
\label{eq:kkk}
\end{eqnarray}
The first-order gradient term can almost be recognized to match the fluid dynamic form (\ref{eq:hydro1f}). To make the match explicit, note that $I_{(n)}^{\mu\nu\sigma \kappa}$ only depends on one vector ($u^\gamma$). Because of relativistic covariance,  it must be possible to decompose $I_{(n)}^{\mu\nu\sigma \kappa}$ in terms of this vector and the metric tensor (or, equivalently, $\Delta^{\mu\nu}=g^{\mu\nu}+u^\mu u^\nu$),
\begin{eqnarray}
\label{eq:Imnsk}
I_{(n)}^{\mu\nu\sigma \kappa}&=&i_{40}^{(n)}u^\mu u^\nu u^\sigma u^\kappa + i_{41}^{(n)}\left(u^{\mu} u^\nu \Delta^{\sigma \kappa}+{\rm perm.}\right)\nonumber\\
&&+i_{42}^{(n)} \left(\Delta^{\mu \nu}\Delta^{\sigma \kappa}+\Delta^{\mu \sigma}\Delta^{\nu \kappa}+\Delta^{\mu \kappa}\Delta^{\nu \sigma}\right)\,,
\end{eqnarray}
where $i^{(n)}_{40},i^{(n)}_{41},i^{(n)}_{42}$ are functions of $T$ and ``${\rm perm}.$'' denotes all
symmetric permutations of indices. To find the coefficient functions $i_{40}^{(n)},i^{(n)}_{41},i^{(n)}_{42}$ one considers contractions of $I_{(n)}^{\mu\nu\sigma \kappa}$ with $u^\mu$ and $\Delta^{\mu\nu}$, respectively. One finds
$$
i_{40}^{(n)}=\int d\chi \left(p^0\right)^{4-n} f_{(0)}\,,\quad i_{41}^{(n)}=\int d\chi \frac{(p^0)^{2-n} {\bf p}^2}{3} f_{(0)}\,,\quad i_{42}^{(n)}=\int d\chi \frac{{\bf p}^4}{15 (p^0)^n} f_{(0)}\,,
$$
where we note that $i_{41}^{(1)}=c_s^2 i_{40}^{(1)}$ because Eq.~(\ref{eq:matching}) implies $\int d\chi f_{(1)}\left(p^\lambda u_\lambda\right)^2=0$.
 Plugging (\ref{eq:Imnsk}) into Eq.~(\ref{eq:kkk}), the kinetic energy-momentum tensor to first order gradients reduces to (\ref{eq:hydro1f}) with the transport coefficients $\eta,\zeta$ given by
\begin{eqnarray}
\label{eq:etazetaKT}
\eta&=&\frac{\tau_R}{15 T}\int\frac{d^3p}{(2\pi)^3} \frac{{\bf p}^4}{{\bf p}^2+m^2} e^{-\sqrt{{\bf p}^2+m^2}/T}=\frac{\tau_R}{T}\int_0^T dT^\prime \left(\epsilon+P\right) \,,\\
\zeta&=&\frac{\tau_R}{9 T}\int \frac{d^3p}{(2\pi)^3}\frac{{\bf p}^4-3 c_s^2({\bf p}^2+m^2){\bf p}^2}{{\bf p}^2+m^2} e^{-\sqrt{{\bf p}^2+m^2}/T}=\frac{5 \eta}{3}-\tau_R c_s^2 \left(\epsilon+P\right)\,,\nonumber
\end{eqnarray}
where the rhs expressions were obtained by manipulation of the thermodynamic relations (\ref{eq:KTthermo}).

Pushing the Chapman-Enskog expansion (\ref{eq:CE}) to second order in gradients, it is easy to show that $\tau_\pi=\tau_R$, \cite{Romatschke:2011qp}. Other transport coefficients require more extensive calculations, cf. Ref.~\cite{Jaiswal:2014isa}. 

\subsection{Conformal Kinetic Theory}
\label{sec:cKT}

A particularly simple example is the case when the particle mass is set to zero. Since there are no scales in the problem, this corresponds to the case of conformal kinetic theory.

In the case of conformal kinetic theory in $d=4$ Minkowski space-time, the transport coefficients from the preceding section reduce to
\begin{equation}
\label{eq:KTconfthermo}
P={\rm dof}\times \frac{T^4}{\pi^2}\,, \quad
\epsilon=3 P\,,\quad
s=\frac{4 P}{T}\,,\quad c_s=\frac{1}{\sqrt{3}}\,,\quad
\frac{\eta}{s}=\frac{\tau_R T}{5}\,,\quad \zeta=0\,,
\end{equation}
together with $\tau_\pi=\tau_R$. These results imply $C_\pi=5 C_\eta$, suggesting that the stability bound (\ref{eq:stability}) is easily satisfied for conformal kinetic theory.
Also, since $g_{\mu\nu}p^\mu p^\nu=0$ for massless on-shell particles, the first-order correction term (\ref{eq:KTf1}) to the equilibrium distribution function becomes\footnote{Note that the second equality in Eq.~(\ref{eq:KTconfFO}) only is true to first gradient order where $\pi^{\mu\nu}=-\eta \sigma^{\mu\nu}$.}
\begin{equation}
\label{eq:KTconfFO}
f_{(1)}=f_{(0)}\frac{\tau_R}{T} \frac{p^\mu p^\nu \sigma_{\mu\nu}}{2 p^\lambda u_\lambda}\simeq- f_{(0)} \frac{5 p^\mu p^\nu \pi_{\mu\nu}}{2 T (\epsilon+P) p^\lambda u_\lambda}\,.
\end{equation}
 The distribution function including first order corrections around equilibrium is thus given by
\begin{equation}
\label{eq:KTrec}
f\simeq f_{(0)}+f_{(1)}\simeq \left(1-\frac{5 p^\mu p^\nu \pi_{\mu\nu}}{2T (\epsilon+P) p^\lambda u_\lambda}\right)f_{(0)}=f_{\rm quadratic\ ansatz}\,,
\end{equation}
where $f_{(1)}/f_{(0)}$ is sometimes referred to as ``$\delta f$ correction''. Eq.~(\ref{eq:KTrec}) is known as ``quadratic ansatz'' in the literature \cite{Dusling:2009df}.
\index{Cooper-Frye prescription! Quadratic ansatz}
Apart from being accurate only to first order gradients, $f_{(0)}+f_{(1)}$ is not manifestly positive  for large momenta for any $\pi^{\mu\nu}\neq 0$, which creates considerable problems when using (\ref{eq:KTrec}) as a model to reconstruct a particle distribution function from fluid dynamic variables.

A simple \textit{model} of a distribution function that reproduces the exact energy-momentum tensor and is manifestly positive for non-vanishing $\pi^{\mu\nu}$ is
\begin{equation}
\label{eq:KTmodelexact}
f_{\rm model\, 1}=f_{(0)}\left(\frac{p^\lambda u_\lambda}{T}\left(1+\frac{10 p^\mu p^\nu \pi_{\mu\nu}}{(\epsilon+P)(p^\sigma u_\sigma)^2}\right)^{-1/4}\right)\,.
\end{equation}
Performing a variable substitution ${\bf p}^2=\tilde p^2\left(1+\frac{10 p^\mu p^\nu \pi_{\mu\nu}}{(\epsilon+P)(p^\sigma u_\sigma)^2}\right)^{1/4}$ (which --- despite the appearance --- is explicit since $|{\bf p}|$ cancels in the ratio $p^\mu p^\nu \pi_{\mu\nu}/(p^\sigma u_\sigma)^2$ for conformal systems) the energy-momentum tensor (\ref{eq:KT}) becomes
\begin{eqnarray}
\int d\chi p^\mu p^\nu f_{\rm model\, 1}&=&\int d\tilde \chi \tilde p^\mu \tilde p^\nu \left(1+\frac{10 p^\lambda p^\kappa \pi_{\lambda \kappa}}{(\epsilon+P)(p^\sigma u_\sigma)^2}\right)f_{(0)}\left(\frac{\tilde p^\lambda u_\lambda}{T}\right)\,,\nonumber\\
&=&T_{(0)}^{\mu\nu}+\frac{10}{\epsilon+P} \pi_{\lambda \kappa}I^{\mu\nu\lambda\kappa}_{(2)}=T_{(0)}^{\mu\nu}+\pi^{\mu\nu}\,,
\end{eqnarray}
where $I^{\mu\nu\lambda\kappa}_{(2)}$ is given by Eq.~(\ref{eq:Imnsk}) and the relevant coefficient $i_{42}^{(2)}=\frac{\epsilon}{15}$ has been used. Using the explicit form of the equilibrium distribution function (\ref{eq:maxjut}), it is also straightforward to show that $f_{\rm model\, 1}$ agrees with (\ref{eq:KTrec}) when expanding in powers of $\pi^{\mu\nu}$, but contains an infinite number of higher order terms. Also, it is straightforward to see that 
$f_{\rm model\, 1}$ is real for any value of the momenta ${\bf p}$ as long as
\begin{equation}
\label{eq:KTcondition}
{\rm max} \left(\frac{-p^\mu p^\nu\pi_{\mu\nu}}{(\epsilon+P)(p^\sigma u_\sigma)^2}\right)<\frac{1}{10}\,.
\end{equation}
Despite this attractive features, $f_{\rm model\, 1}$ will differ strongly from solutions of the kinetic equations (\ref{eq:Boltzmann}) for out-of-equilibrium situations, so it should be considered a near-equilibrium model for particle distribution functions.

For the free-streaming solution (\ref{eq:BjorFSsol}) in the conformal Bjorken flow case, it is convenient to introduce the variable substitution ${\bf p}^2=\tilde p^2\left(1+\Xi(\tau)\frac{(p^z)^2}{{\bf p}^2}\right)^{-1/2}$ \cite{Romatschke:2003ms} with $p^z=|{\bf p}| \cos \theta$ and again $\Xi(\tau)=\left(\frac{\tau^2}{\tau_0^2}-1\right)$. As a result, the energy-density (\ref{eq:epsFSbjor}) for the free-streaming solution becomes  \cite{Strickland:2014pga}
\begin{eqnarray}
\epsilon_{FS}(\tau)&=&\int_0^\infty \frac{d\tilde p}{2 \pi^2} \tilde p^3 f_{(0)}\left(-\frac{\tilde p}{T_0}\right)\int\frac{d(\cos\theta)}{2}\frac{1}{\left(1+\Xi(\tau)\cos^2\theta\right)^2}\,,\nonumber\\
&=& \frac{\epsilon(T_0)}{2}\left(\frac{1}{1+\Xi(\tau)}+\frac{\arctan(\sqrt{\Xi(\tau)})}{\sqrt{\Xi(\tau)}}\right) \,,
\end{eqnarray}
where $\epsilon(T_0)$ is the equilibrium energy density at temperature $T=T_0$ given by Eq.~(\ref{eq:KTconfthermo}). Note that in the late-time limit,
\begin{equation}
\lim_{\tau \rightarrow \infty} \epsilon_{FS}(\tau)=\frac{\epsilon(T_0) \pi \tau_0 }{4\tau}\,,
\end{equation}
which clearly differs from the behavior $\epsilon\propto \tau^{-4/3}$ obtained for equilibrium fluid dynamics in Eq.~(\ref{eq:ebjor0}).

The shear stress tensor $\pi^{\mu\nu}$ can similarly be calculated for the free-streaming Bjorken solution. Since $u_\mu \pi^{\mu\nu}=0=\pi^\mu_\mu$, and $T^{\mu\nu}$ is diagonal because $u^\mu=\left(1,0,0,0\right)$, we have $\pi^{\mu\nu}={\rm diag}\left(0,\pi^{xx},\pi^{xx},-2 \pi^{xx}/\tau^2\right)$. The only independent component of $\pi^{\mu\nu}$ may be calculated from $\pi^{xx}=T^{xx}-P$, where $P=\epsilon_{FS}/3$, finding
\begin{equation}
\pi^{xx}=\epsilon(T_0)\left(\frac{\sqrt{\Xi}(3+\Xi)-(3+2\Xi-\Xi^2)\arctan(\sqrt{\Xi})}{12\Xi^{3/2}\left(1+\Xi\right)}\right)\,.
\end{equation}

For this free-streaming system, it is straightforward to verify that the condition (\ref{eq:KTcondition}) for the $f_{\rm model 1}$ is violated for $\Xi\gtrsim 1.35$. Also, for this free-streaming system, $f_{\rm quadratic\ ansatz}$
 turns negative for $p^z\gtrsim 3.5 T_0$ for this value of $\Xi$ and ${\rm dof}=\frac{\pi^2}{3}$. These issues motivate the search for yet another non-equilibrium model ansatz for $f$.

Taking inspiration from the anisotropic free-streaming solution (\ref{eq:faniso}), 
a third model for the out-of-equilibrium distribution function is the ``exponential ansatz''
\begin{equation}
\label{eq:KTmodel2}
f_{\rm exponential\ ansatz}=f_{(0)}\left(\frac{p^\lambda u_\lambda}{T}\left(1-\frac{5 p^\mu p^\nu \pi_{\mu\nu}}{(\epsilon+P)(p^\sigma u_\sigma)^2}\right)^{1/2}\right)\,,
\end{equation}
which can be recognized to be of the form (\ref{eq:faniso}) with
\begin{equation}
\Xi_{ab}=u_a u_b-\frac{5\pi_{ab}}{\epsilon+P}\,.
\end{equation}
The model (\ref{eq:KTmodel2}) is better behaved than (\ref{eq:KTmodelexact}) because for the above free-streaming solution it is free from pathologies for \textit{arbitrary} anisotropies larger $\Xi\gtrsim -0.75$. Within this regime, $f_{\rm exponential\ ansatz}$ is manifestly positive for all momenta. However, unlike $f_{\rm model\, 1}$ or the quadratic ansatz, Eq.~(\ref{eq:KTmodel2}) does not reproduce the exact energy momentum tensor upon momentum integration
\index{Cooper-Frye prescription! Exponential ansatz}


 section \ref{sec:ads}.

\subsection{Collective Modes in Kinetic Theory}
\label{sec:KTcollmodes}

To study the collective mode structure of kinetic theory it is necessary to calculate the retarded two-point correlator for the energy-momentum tensor (\ref{eq:2ptcorr}) in complete analogy to section \ref{sec:hydrocoll}. To this end, consider Eq.~(\ref{eq:Boltzmann}) in the BGK approximation in a non-trivial metric background such that $F^\mu=-\Gamma^\mu_{\lambda \sigma}p^\lambda p^\sigma$. The two-point correlator may be obtained by considering small perturbations $\delta f$ in the particle distribution function around some background configuration $f_0$, which in analogy to 
section \ref{sec:hydrocoll} will be taken to be the equilibrium particle distribution $f_{0}=e^{-p_0/T_0}$. Note that --- unlike for the Chapman-Enskog expansion --- no restrictions on the gradient strength in $\delta f$ are imposed. As a consequence, genuine off-equilibrium situations where gradients are large can be studied, albeit only for the \textit{linear response} dispersion relations defined by the two-point correlator\footnote{Non-linear response corrections to the dispersion relations can in principle be studied by calculating higher-point correlation functions.}. With these choices, and focusing on the case of conformal kinetic theory (massless particles), the Boltzmann equation in linear response becomes
\begin{equation}
\label{eq:KTlinresp}
p^\mu \partial_\mu \delta f(t,{\bf x},{\bf p})-\Gamma^\nu_{\lambda \sigma} p^\lambda p^\sigma \partial_\nu^{(p)} f_0 = -\frac{p^0}{\tau_R}\left(\delta f-\delta f_{0}\right)\,,
\end{equation}
where $\delta f_{0}$ denotes the equilibrium distribution expanded to first order in fluctuations. Care must be taken to properly suppress contact terms in $\delta f_0$ (see section \ref{sec:NScoll}), which can be achieved by demanding that the fluctuations of the equilibrium energy density $\delta \epsilon$ and fluid velocity $\delta u^i$ fulfill $\delta \epsilon=\delta T^{00}, \delta u^i=\frac{\delta T^{0i}}{\epsilon_0+P_0}$. In particular, a coordinate change from the original momenta $\tilde p^i=r^i_{\ j}p^j$ to new momenta $p^i$ such that $r^i_{\ m} r^{j}_{\ n}g_{ij}=\delta_{ij}$ allows to re-write the solution for the on-shell condition $\tilde p^\mu \tilde p^\nu g_{\mu\nu}=0$ as $\tilde p^0=|{\bf p}|\left(1+\frac{\delta g_{00}}{2}+\frac{p^i}{|{\bf p}|} \delta g_{0i}\right)$, where ${\bf p}^2\equiv p^i p^j \delta_{ij}$. To linear order in perturbations, the Jacobian for the transformation leads to $d^3\tilde p\rightarrow d^3p \left({\rm det}g_{ij}\right)^{-1/2}$, canceling part of the determinant in the integration measure (\ref{eq:intmeas}). Furthermore, using $u_0=-1+\delta g_{00}/2$ one finds for the fluctuation of the equilibrium distribution function \cite{Romatschke:2015gic,Kurkela:2017xis}
\begin{equation}
\delta f_0(t,{\bf x},{\bf p})= \frac{f_0({\bf p})}{T_0}\left({\bf p}\cdot \delta {\bf u}(t,{\bf x})+\frac{p^0\delta T(t,{\bf x})}{T_0}\right)\,,
\end{equation}
which obeys $\int d\chi p^0 p^0 \delta f_0=\delta \epsilon, \int d\chi p^0 p^i \delta f_0 = (\epsilon_0+P_0) \delta u^i$. It is then convenient to Fourier transform (\ref{eq:KTlinresp}) from $t,{\bf x}$ to $\omega,{\bf k}$, finding the linear response of the energy-momentum tensor (\ref{eq:KT}) to be given by\footnote{Note that $\Gamma^i_{\mu \nu}p^\mu p^\nu p_i=\Gamma^{0}_{\mu \nu}p^\mu p^\nu p_0$ because $\Gamma_{\mu \nu\lambda}p^\mu p^\nu p^\lambda=p^\mu \partial_\mu \left(p^2\right)=0$ for on-shell particles.}
\begin{equation}
\delta T^{\mu\nu}(\omega, {\bf k})=\int \frac{dp}{2 \pi^2 T_0} f_0({\bf p}) \int\frac{d \Omega}{4 \pi} v^\mu v^\nu \frac{{\bf v}\cdot {\delta u}+\delta T/T_0-\tau_R \Gamma^0_{\lambda \sigma}v^\lambda v^\sigma}{1+\tau_R(-i \omega+i {\bf k}\cdot {\bf v})}\,,
\end{equation}
where $v^\mu=\left(1,{\bf p}/|{\bf p}|\right)$. This equation corresponds to self-consistency conditions on the fluctuations of energy density $\delta \epsilon=\delta T^{00}$ and fluid velocity $\delta u^i=\frac{\delta T^{0i}}{\epsilon_0+P_0}$. Choosing a specific dependence of the metric on the coordinates such as $\delta g^{\mu\nu}=\delta g^{\mu\nu}(t,x_3)$ as in section \ref{sec:NScoll}, one obtains for the retarded two-point correlators in momentum space \cite{Romatschke:2015gic}:
\begin{eqnarray}
\label{eq:allhydrores}
G_T^{12,12}(\omega,k)&=&\frac{3 i \omega \tau_R H_0}{16}
\left[\frac{10}{3}\frac{1-i \tau_R \omega}{k^2 \tau_R^2}+\frac{2 (1-i \tau_R \omega)^3}{k^4 \tau_R^4}+i\frac{\left((1-i \tau_R \omega)^2+k^2 \tau_R^2\right)^2}{k^5 \tau_R^5} L \right]\,,\nonumber\\
G_T^{01,01}(\omega,k)&=&-H_0\frac{2 k \tau_R\left(2 k^2 \tau_R^2+3 (1-i \tau_R \omega)^2\right)+3 i (1-i\tau_R \omega)\left(k^2 \tau_R^2+(1-i \tau_R \omega)^2\right) L}{2 k \tau_R (3+2 k^2 \tau_R^2-3 i \tau_R \omega)+3 i \left(k^2 \tau_R^2+(1-i \tau_R \omega)^2\right) L}\,,\nonumber\\
G_T^{13,13}(\omega,k)&=&\frac{\omega^2}{k^2}G_T^{01,01}(\omega,k)\,,\nonumber\\
G_T^{00,00}(\omega,k)&=&-3H_0\left[1+k^2 \tau_R \frac{2 k \tau_R+i(1-i \tau_R\omega) L}{2 k \tau_R(k^2 \tau_R +3 i \omega)+i \left(k^2 \tau_R+3\omega(i+\tau_R \omega)\right) L}\right]\,,\nonumber\\
G_T^{03,00}(\omega,k)&=&-3H_0\frac{\omega}{k}\left[k^2 \tau_R \frac{2 k \tau_R+i(1-i \tau_R\omega) L}{2 k \tau_R(k^2 \tau_R +3 i \omega)+i \left(k^2 \tau_R+3\omega(i+\tau_R \omega)\right)  L}\right]\,,\nonumber\\
G_T^{03,03}(\omega,k)&=&-3H_0\left[\frac{1}{3}+\omega^2 \tau_R \frac{2 k \tau_R+i(1-i \tau_R\omega) L}{2 k \tau_R(k^2 \tau_R +3 i \omega)+i \left(k^2 \tau_R+3\omega(i+\tau_R \omega)\right) L}\right]\,,\nonumber\\
\end{eqnarray}
where the shorthand notations
$$
H_0=\epsilon_0+P_0,\quad L=\ln \left(\frac{\omega-k+\frac{i}{\tau_R}}{\omega+k+\frac{i}{\tau_R}}\right)\,,
$$
were used.

It is easy to verify that these results for the correlators reduce to the form dictated by rBRSSS (\ref{eq:rBRSSS2pt}) in the limit of small gradients $\omega/T \ll 1$, ${\bf k}/T \ll 1$. The generalized Kubo relations from Eq.~(\ref{eq:rBRSSS2pt}) then determine the value of the transport coefficients in kinetic theory as
\begin{equation}
\frac{\eta}{s}=\frac{\tau_R T}{5}\,, \quad \tau_\pi=\tau_R\,,\quad \kappa=0\,,
\end{equation}
which in the case of massless particles considered here is in full agreement with Eq.~(\ref{eq:etazetaKT}) calculated using the Chapman-Enskog expansion, but additionally provides the kinetic theory value for the second-order transport coefficient $\kappa$.

As in section \ref{sec:NScoll}, the singularities of $G^{\mu\nu,\gamma\delta}(\omega,k)$ determine the collective modes of kinetic theory in linear response. In the limit of small gradients, one finds that Eqns.~(\ref{eq:allhydrores}) have simple poles corresponding to the hydrodynamic sound  and shear mode dispersion relations given in Eqns.~(\ref{eq:soundmode}) and (\ref{eq:dispNSshear}), respectively. This is consistent with the expectation that near equilibrium, kinetic theory can be matched with a near-equilibrium fluid description. However, in addition to the familiar hydrodynamic modes near equilibrium, Eqns.~(\ref{eq:allhydrores}) contain additional singularities that will be discussed below.

We close by noting that the kinetic theory correlation functions do not contain terms proportional to $\omega^4$ which are known to exist from free field theory calculations \cite{Teaney:2006nc}. The reason for this apparent discrepancy is that the derivation of the Boltzmann equation from field theory involves discarding higher order gradients in the Wigner transform, cf. Ref.~\cite{Blaizot:2001nr}. Therefore, kinetic theory correlation functions are not accurate approximations of quantum field theory when gradients are strong, even if the field theory coupling is weak.

\subsection{Non-Hydrodynamic Modes in Kinetic Theory}
\label{sec:ktnonhydro}
\index{Non-hydrodynamic mode! Kinetic Theory}

In almost all treatise on the subject, the connection between kinetic theory and fluid dynamics is based on the near-equilibrium Chapman-Enskog expansion (\ref{eq:CE}). However, kinetic theory is well defined also far away from equilibrium, whereas the Chapman-Enskog expansion diverges \cite{Denicol:2016bjh} similar to the gradient expansion in fluid dynamics, cf. the discussion in section \ref{sec:offeq}. For fluid dynamics, the Central Lemma outlined in section \ref{sec:offeq} suggests a quantitative description for off-equilibrium situations as long as generalized ``Borel-resummed'' hydrodynamic degrees of freedom dominate over all other, non-hydrodynamic modes. Thus, the question arises on how the Borel-resummed hydrodynamic modes (and the competing non-hydrodynamic modes) manifest themselves in the case of off-equilibrium kinetic theory. 

\begin{figure}[t]
  \begin{center}    
     \includegraphics[width=.49\linewidth]{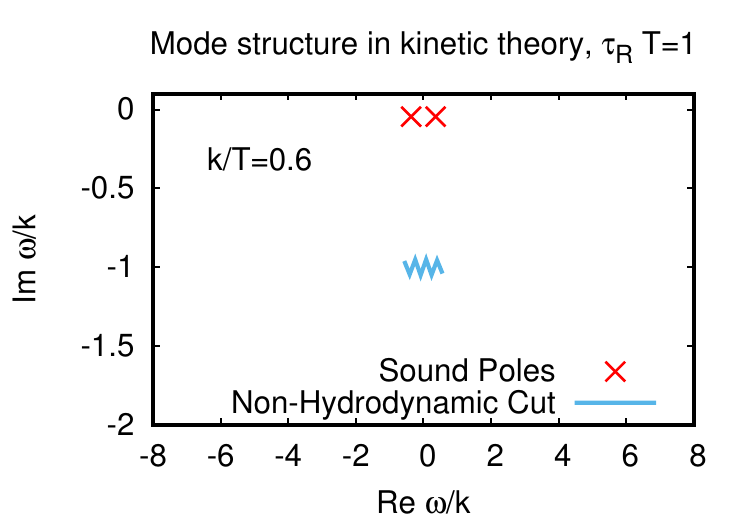}
     \hfill
     \includegraphics[width=.49\linewidth]{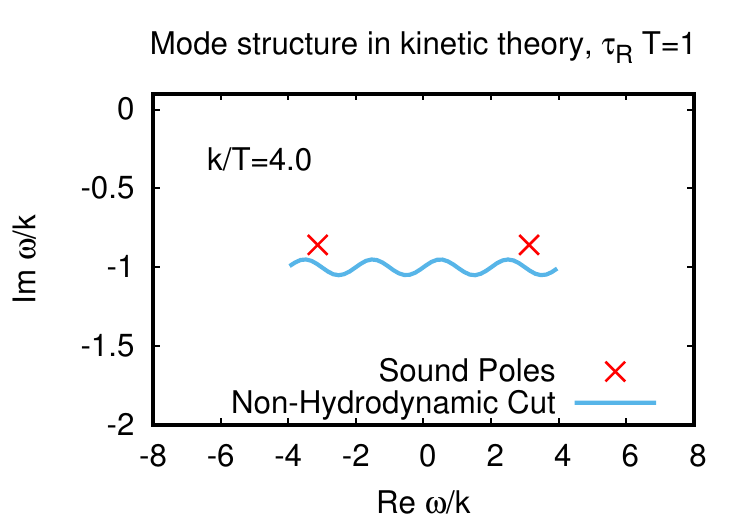}
     \end{center}
  \caption{\label{fig3:nonhy} Kinetic theory collective modes in the sound channel, corresponding to singularities of $G^{00,00}(\omega,k)$ for fixed coupling strength $\tau_R T=1$ in the complex frequency plane. Shown are results near equilibrium (left) and off-equilibrium (right). Increasing the gradient strength beyond $k \tau_R\gtrsim 4.531$, the hydrodynamic sound poles cease to exist and only the non-hydrodynamic branch cut remains.}
\end{figure}

To this end, consider \textit{all} the singularities of $G^{\mu\nu,\gamma\delta}(\omega,k)$ in Eqns.~(\ref{eq:allhydrores}). Let us focus\footnote{The singularities in the shear channel $G^{01,01}(\omega,k)$ and scalar channel $G^{12,12}(\omega,k)$ of kinetic theory are discussed in Ref.~\cite{Romatschke:2015gic}.} on the sound channel given by the singularities of $G^{00,00}(\omega,k)$. As outlined above, for small spatial gradients\footnote{Note that the kinetic theory correlators (\ref{eq:allhydrores}) only depend on the combinations $\omega \tau_R,k \tau_R$. For this reason, collective modes in the strong coupling limit $\tau_R T\rightarrow 0$ and the weak gradient limit $k/T\rightarrow 0$ are identical as long as $k \tau_R$ for the two cases coincides, cf. Ref.~\cite{Romatschke:2015gic}.} $|{\bf k}|/T\ll 1$, there are two simple poles corresponding to the hydrodynamic sound modes Eqns.~(\ref{eq:soundmode}). In addition, $G^{00,00}(\omega,k)$ has a logarithmic branch cut emanating from the branch points located at $\omega_{nh}=\pm k -\frac{i}{\tau_R}$ (see Fig.~\ref{fig3:nonhy}), which may be taken to run from $\omega_{nh}=-k -\frac{i}{\tau_R}$ to $\omega_{nh}=k -\frac{i}{\tau_R}$. Since this branch cut does not correspond to any mode contained in the Navier-Stokes equations, it is a non-hydrodynamic mode according to the definition of section \ref{sec:nonhydro}. At weak coupling $\tau_R T\rightarrow \infty$, this non-hydrodynamic branch cut moves closer to the real axis and becomes the branch cut familiar from Landau-damping \cite{Blaizot:2001nr}, with branch points at $\omega=\pm k$ corresponding to single-particle excitations.

Starting near equilibrium and increasing the gradient strength $|{\bf k}|/T$, the location of the sound poles contained in $G^{00,00}(\omega,k)$ move in the complex frequency plane, and their dispersion relation starts to differ appreciably from (\ref{eq:soundmode}) for $k \tau_R>1.5$. For instance for $k \tau_R=4$ (cf Fig.~\ref{fig3:nonhy}), the Navier-Stokes result Eq.~(\ref{eq:soundmode}) predicts $\omega \tau_R \simeq \pm 2.31-2.13 i$ whereas the hydrodynamic sound poles in the kinetic theory correlator are located at $\omega \tau_R \simeq \pm 3.12-0.85 i$. However, while the Navier-Stokes result, derived under the assumption of near-equilibrium, can not be trusted in off-equilibrium situations where $k \tau_R=4$, the kinetic theory result is on solid footing and implies that hydrodynamic sound modes exist off-equilibrium. Since the hydrodynamic sound poles are located closer to the real axis than the non-hydrodynamic branch cut, they have smaller relative damping (cf. Eq.~(\ref{eq:sound})) and as a consequence can be expected to dominate the evolution of perturbations $\delta T^{\mu\nu}$. Hence off-equilibrium kinetic theory evolution will be quantitatively described in terms of a kind of fluid dynamical evolution with modes that are qualitatively similar to, but quantitatively different from, those contained in Navier-Stokes or BRSSS. This offers a concrete example of how the Central Lemma of fluid dynamics is realized in a particular microscopic theory setting.

Increasing the gradient strength $|{\bf k}|/T$ further, the collective mode structure changes qualitatively from the results shown in Fig.~\ref{fig3:nonhy}. For $k \tau_R>k_c \tau_R\simeq 4.5313912\ldots$ the kinetic theory correlator $G^{00,00}(\omega,k)$ no longer contains any hydrodynamic sound poles\footnote{Note that a study for kinetic theory  with momentum-dependent relaxation time in Ref.~\cite{Kurkela:2017xis} suggests that the disappearance of sound poles may be an artifact of the BGK approximation used here.}, whereas the non-hydrodynamic branch cut remains for arbitrary $k\tau_R$. The mathematical reason is that the sound poles present for $k<k_c$ move through the logarithmic branch cut onto the next Riemann sheet as $k$ is increased above $k_c$. A physics interpretation of this behavior is that there is a hydrodynamic onset transition present at $k=k_c$ which marks the onset (or turn-off) of sound in kinetic theory. Above $k>k_c$, sound modes are absent, and transport occurs exclusively via non-hydrodynamic degrees of freedom (in this case damped single particle excitations). For kinetic theory at gradient strengths that exceed $k>k_c$, Borel fluid dynamics has broken down.
\index{Fluid Dynamics! Break-down of}

\section{Gauge/Gravity Duality Flyby}
\label{sec:ads}

Gauge gravity duality\footnote{The terms ``gauge/gravity duality'' and ``holography'' will be used synonymously, while the term ``AdS/CFT correspondence'' will be reserved for the gravity duality of conformal field theories.} denotes the conjectured correspondence between gauge theories in Minkowski space-time on the one hand and gravity (or more generally string theory) on the other hand. A particular realization of the conjecture is the correspondence between ${\cal N}=4$ SYM  and string theory in space-times with a negative cosmological constant (Anti-de-Sitter or ``AdS'' space-times) \cite{Maldacena:1997re}. While this duality has been conjectured to hold for arbitrary coupling $g$, and arbitrary number of colors $N_c$ on the gauge theory side, in the following we shall be concerned with the particular limit of large $N_c$ and large 't Hooft coupling $g^2 N_c\rightarrow \infty$. In this limit, gauge/gravity duality states that ${\cal N}=4$ SYM in four-dimensional Minkowski space-time is dual to classical (Einstein) gravity in five-dimensional AdS (times an additional S$_5$ space-time which will be unimportant for the following discussion).
\index{Gauge/Gravity duality}
\index{Holography|see {Gauge/Gravity duality}}
\index{AdS/CFT|see {Gauge/Gravity duality}}

In the absence of matter, the Einstein equations (\ref{eq:einstein}) with a cosmological constant $\Lambda$ may be solved by a metric ansatz for the line element
\begin{equation}
\label{eq:m1}
ds^2=f^2(u)\left(-dt^2+dx^2+dy^2+dz^2+du^2\right)\,,
\end{equation}
with $x^\mu=(t,x,y,z,u)$ and $u$ the coordinate along the ``extra'' fifth dimension. In these coordinates, the form of Eq.~(\ref{eq:m1}), and in particular the requirement that $f(u)$ is a function of $u$ only, is dictated by requiring Poincar\'e invariance in every four dimensional space-time slice of $u$.

Plugging (\ref{eq:m1}) into the Einstein equations with cosmological constant $\Lambda$ and solving for $f(u)$ leads to $f(u)=\sqrt{-\frac{6}{\Lambda u^2}}$. Choosing units where the cosmological constant $\Lambda=-6/L^2$ is expressed in terms of an AdS length scale $L$, this implies
\begin{equation}
\label{eq:m2}
ds^2=\frac{L^2}{u^2}\left(-dt^2+dx^2+dy^2+dz^2+du^2\right)=\frac{L^2}{u^2}\left(-dt^2+d{\bf x}^2+du^2\right)\,,
\end{equation}
which is the line-element for empty AdS$_5$ space-time. In addition to Poincar\'e invariance in four dimensions, the AdS$_5$ line element (\ref{eq:m2}) is also invariant under scaling transformations
\begin{equation}
\label{eq:conft1}
(t,{\bf x})\rightarrow (w t,w {\bf x})\,,\quad u\rightarrow w u\,.
\end{equation}
Since four-dimensional conformal field theories are also invariant under the reparametrizations $(t,{\bf x})\rightarrow (w t,w {\bf x})$, the symmetry (\ref{eq:conft1}) of gravity in AdS$_5$ is the first indication for a relation between conformal field theories and gravity in anti-de-Sitter space-times.

The AdS$_5$ line-element (\ref{eq:m2}) diverges for $u=0$. More careful investigations imply that the AdS$_5$ space-time has a ``conformal'' boundary at $u=0$ in the sense that $d\tilde s^2=ds^2 u^2$ has a $d=4$ dimensional Minkowski space-time boundary for this value of $u$ \cite{Zaffaroni:2000vh}\footnote{In addition to the singularity at $u=0$, the line element (\ref{eq:m2}) also is ill-defined for $u\rightarrow \infty$. It turns out that we will not need the $u\rightarrow \infty$ limit of the line element in the following, and hence we will skip the discussion of the physics interpretation of $u\rightarrow \infty$ here.}. In the following, we will use the interpretation that the gauge theory ``lives'' on this conformal boundary at $u=0$, while the full AdS$_5$ space-time provides a ``hologram'' of this field theory in the fifth dimension for $u>0$. For the case described by (\ref{eq:m2}), this hologram is not very interesting from a fluid dynamics perspective: empty AdS$_5$ simply is a hologram of vacuum ${\cal N}=4$ SYM in $d=4$ Minkowski space-time.

In order to describe field theory with matter, it is mandatory to study solutions to the Einstein equations (\ref{eq:einstein}) which do not correspond to empty AdS, but are nevertheless close to (\ref{eq:m2}) near the conformal boundary (``asymptotically AdS''). In many cases, the next simplest case to consider besides empty space-time is a space-time with a black hole.

\subsection{Black Brane Metric}

In order to find black hole solutions to the Einstein equations in asymptotic AdS space-times, let us make an ansatz based on a modification of (\ref{eq:m2}), which is inspired by the Schwarzschild metric in usual Minkowski space-time:
\begin{equation}
\label{eq:m3}
ds^2=\frac{L^2}{u^2}\left(-f(u) dt^2+d{\bf x}^2+f^{-1}(u)du^2\right)\,,
\end{equation}
with $f(u)$ again being a function of $u$ only. Solving Einstein equations (\ref{eq:einstein}) with this ansatz leads to the solution
\begin{equation}
f(u)=1-\frac{u^4}{u_s^4}
\end{equation}
with $u_s$ an undetermined constant. Close to the conformal boundary at $u=0$, the line element (\ref{eq:m3}) matches (\ref{eq:m2}), so the space-time described by (\ref{eq:m3}) is asymptotically AdS. However, in addition to $u=0$, Eq.~(\ref{eq:m3}) has a singularity at $u=u_s$. As will be shown below, this singularity indicates the presence of a marginally trapped surface. Since $u=u_s$ is a singular point for all ${\bf x}$, the geometry of this trapped surface is that of a plane rather than that of a sphere as would have been expected for a spherical black hole. As a consequence, Eq.~(\ref{eq:m3}) is referred to as the black brane metric.

Let us now investigate the space-time structure close to the singular point of Eq.~(\ref{eq:m3}) given by
\hbox{$u=u_s\left(1-\rho^2\right)$}. Expanding (\ref{eq:m3}) to leading order in $\rho\ll 1$ leads to 
\begin{equation}
\label{eq:m4}
ds^2\simeq L^2\left(-\frac{4 \rho^2}{u_s^2}dt^2+\frac{d {\bf x}^2}{u_s^2}+d\rho^2\right)\,.
\end{equation}
The trapped surface at constant time $t$ can be parametrized by \hbox{$\rho=-\psi$}, where $\psi$ could in principle depend on ${\bf x}$ but does not in this case because of the symmetries involved. There are two normal vectors on the trapped surface which can be parametrized as $l_\mu^{\pm}dx^\mu=c_1 dt+c_2 d\rho$ (since $d\psi=0$). These vectors are null (``light-rays'') if they fulfill the condition $l_\mu l^\mu=0$, which leads to $c_2=\pm \frac{u_s}{2 \psi} c_1$. Thus, up to an unimportant overall constant,
\begin{equation}
l_\mu^+=\left(-1,0,\frac{u_s}{2 \psi}\right)\,,\quad
l_\mu^-=\left(1,0,\frac{u_s}{2 \psi}\right)\,.
\end{equation}
It is easy to check that the expansion of the outgoing null vector $l_\mu^+$ vanishes which indicates the presence of an apparent horizon. The induced metric $g_{\mu\nu}^{\rm ind}$ on the horizon surface is given by (\ref{eq:m4}) with $dt=0,d\rho=0$, or $ds^2_{\rm ind}=L^2\frac{d{\bf x}^2}{u_s^2}$. The area of the horizon is thus given by
\begin{equation}
A_h=\int d^3x \sqrt{\rm det{\, g_{\mu\nu}^{\rm ind}}}=\frac{V_3 L^3}{u_s^3}\,,
\end{equation}
where $V_3=\int d^3x$ is the volume of space on the conformal boundary at $u=0$. The horizon area is related to the black brane entropy via the Bekenstein-Hawking formula
\begin{equation}
\label{eq:BH}
S=\frac{A_h}{4 G_5}\,,
\end{equation}
where $G_5$ is the effective 5-dimensional gravitational constant.

Black holes are classically stable, but emit particles quantum-mechanically through Hawking radiation. The associated Hawking temperature $T$ of the black brane can be calculated from (\ref{eq:m3}) as follows: first recall that field theory at finite temperature $T$ is obtained by considering time to be imaginary $t=i\tau$ with periodic boundary conditions identifying $\tau=0$ and $\tau=\frac{1}{T}$ (see section \ref{sec:fieldtheory}). Considering the variable change $t=i\tau$ in (\ref{eq:m4}) leads to 
\begin{equation}
\label{eq:m5}
ds^2\simeq L^2\left(\frac{4 \rho^2}{u_s^2}d\tau^2+\frac{d {\bf x}^2}{u_s^2}+d\rho^2\right)\,.
\end{equation}
However, the space-time described by (\ref{eq:m5}) has a conical singularity at $\rho=0$ unless the ``angle'' $\frac{2 \tau}{u_s}$ is periodic with a period of $2\pi$. Identifying the periodicity of $\tau$ with the inverse temperature leads to $\frac{2}{u_s}\frac{1}{T}=2\pi$ or 
\begin{equation}
\label{eq:us2}
u_s=\frac{1}{\pi T}\,,
\end{equation}
which connects the black brane geometry with the Hawking temperature of the black brane.

In the gauge/gravity setup, the Hawking temperature of the black brane is identified with the temperature of matter of the ${\cal N}=4$ SYM field theory on the conformal boundary. This provides the first entry in the dictionary between gravity and gauge theory.

\subsection{Transport Coefficients from Gauge/Gravity Duality}
\label{sec:ggtrans}
\index{Transport coefficients! Holography}

The previous black brane setup in asymptotic AdS space-times
allows the calculation of transport coefficients for strongly
coupled ${\cal N}=4$ SYM.

\subsubsection{Equilibrium}

The black brane metric (\ref{eq:m3}) corresponds to an equilibrium gravity solution. The Hawking temperature $T$ of the black brane corresponds to the temperature of the associated equilibrium field theory in $d=4$ Minkowski space-time. In order to connect gravity to the field theory, we mention without proof that the five-dimensional gravitational constant is related to the AdS length scale $L$ as\footnote{This result can be gleaned e.g. from Ref.~\cite{Policastro:2002tn} by writing the gravitational action as \hbox{$S=\frac{\pi^3 L^5}{2 \kappa_{10}^2}\int d^5x \sqrt{-g} \left(R-2 \Lambda\right)$} with $\kappa_{10}^2=\frac{4 \pi^5 L^8}{N^2}$ in terms of five-dimensional Newton's constant, $\frac{\pi^3 L^5}{2 \kappa_{10}^2}=\frac{1}{16 \pi G_5}$, cf. Eq.~(\ref{eq:gravityaction}).} \cite{CasalderreySolana:2011us}
\begin{equation}
G_5=\frac{\pi L^3}{2 N^2}\,.
\end{equation}
As a consequence, the entropy density in the field theory (defined as entropy over volume $S/V_3$) can be obtained from the Bekenstein-Hawking result (\ref{eq:BH}) as
\begin{equation}
\label{eq:SBH}
s=\frac{N^2 \pi^2 T^3}{2}\,.
\end{equation}
Since the entropy density scales as $T^{d-1}$, Eq.~(\ref{eq:SBH}) implies conformality for the boundary field theory, and as a consequence the thermodynamic relations (\ref{eq:basicthermo}) imply
\begin{equation}
\label{eq:ThermBH}
P=\frac{N^2 \pi^2 T^4}{8}\,,\quad \epsilon=3 P\,,\quad c_s=\frac{1}{\sqrt{3}}\,.
\end{equation}
Nevertheless, the gauge/gravity relations (\ref{eq:SBH}), (\ref{eq:ThermBH}) are curiously different from the conformal weak-coupling (kinetic) results (\ref{eq:KTconfthermo}), (\ref{eq:KTdofsym}) in the large $N_c$ limit in that the ratio
\begin{equation}
\frac{P_{\rm gauge/gravity}}{P_{\rm kinetic}}=\frac{\frac{N^2 \pi^2 T^3}{8}}{\frac{N^2 \pi^2 T^3}{6}}=\frac{3}{4}\,,
\end{equation}
is not equal to unity. This result implies that even for conformal theories the effective number of degrees of freedom will depend on the effective coupling constant.

\subsubsection{Near Equilibrium}

Away from equilibrium, fluctuations around the black brane metric (\ref{eq:m3}) have to be considered. For instance, we might be interested in fluctuations in the background metric $\delta g_{\mu\nu}(t,{\bf x},u)$ that will generate fluctuations in the boundary stress tensor $T^{\mu\nu}$. Similar to the calculation of correlation functions in fluid dynamics (\ref{eq:2ptcorr}) and kinetic theory (\ref{eq:allhydrores}), this will allow the calculation of retarded correlators for strongly coupled ${\cal N}=4$ SYM.

For simplicity, let us consider a particular form of the metric fluctuation, namely only $\delta g_{12}(t,z,u)$ to be non-vanishing (for a more complete treatment, the interested reader is referred to Ref.~\cite{Kovtun:2005ev}). To simplify computations it is convenient (but not necessary) to change coordinates $u=u_s \sqrt{v}$ such that the black brane horizon is located at $v=1$.
Studying an individual Fourier component by inserting $g_{\mu\nu}=\frac{L^2}{v u_s^2}{\rm diag}\left(-(1-v^2),{\bf 1},\frac{u_s^2}{4 v (1-v^2)}\right)+\frac{L^2}{u_s^2 v}\delta g_{\mu\nu}$ with
\begin{equation}
\delta g_{12}(t,z,v)=\delta g_{21}(t,z,v)=\phi(v) e^{-i \omega t+i k z}\,,
\end{equation}
into the Einstein equations (\ref{eq:einstein}) with $\Lambda=-\frac{6}{L^2}$ and linearizing in $\delta g_{12}$ leads to
\begin{equation}
\label{eq:minscal}
\phi^{\prime\prime}(v)-\phi^\prime \frac{(1+v^2)}{v (1-v^2)}+\phi(v)\frac{u_s^2\left(\omega^2 -k^2(1-v^2)\right)}{4 v(1-v^2)^2}=0\,.
\end{equation}
Near the boundary $v=0$, Eq.~(\ref{eq:minscal}) admits solutions of the form
$\phi(v)\propto v^\alpha$ with characteristic exponents $\alpha=0,2$. Therefore, the asymptotic behavior of $\phi(v)$ near the boundary can be written as
\begin{equation}
\label{eq:phinearboundary}
\phi(v)={\cal A}(\omega,k) Z_A(\omega,k,u)+{\cal B}(\omega,k) Z_B(\omega,k,u)\,,\
\end{equation}
with
\begin{eqnarray}
Z_A(\omega,k,v)&=&1+z_{a1}v+z_{ab}\ln v\,  Z_B(\omega,k,v)+z_{a3} v^3+\ldots\,,\nonumber\\
Z_B(\omega,k,v)&=&v^2\left(1+z_{b1}v+z_{b2}v^2\ldots\right)\,,
\end{eqnarray}
and constant coefficients $z_{a1},z_{ab},z_{a3},z_{b1},z_{b2},\ldots$. It is customary to introduce the dimensionless frequency and wave-vector
\begin{equation}
{\mathfrak w}=\frac{u_s \omega}{2}=\frac{\omega}{2 \pi T}\,,\quad
{\mathfrak K}=\frac{u_s k}{2}=\frac{k}{2 \pi T}\,.
\end{equation}
In particular, using these definitions one finds $z_{ab}=-\frac{({\mathfrak w}^2-{\mathfrak K}^2)^2}{2}$. 
If the coefficient functions ${\cal A},{\cal B}$ in Eq.~(\ref{eq:phinearboundary}) were known, the corresponding ${\cal N}=4$ SYM retarded energy-momentum tensor correlation function would be given by the AdS/CFT dictionary relation \cite{Kovtun:2005ev}
\begin{equation}
\label{eq:ggG1212}
G^{12,12}(\omega,k)=-\frac{\pi^2 N^2 T^4}{2}\frac{{\cal B}(\omega,k)}{{\cal A}(\omega,k)}\,.
\end{equation}
To calculate the coefficients ${\cal A},{\cal B}$, it is necessary to solve (\ref{eq:minscal}) near the black brane horizon with retarded (in-falling) boundary conditions at $v=1$. In analogy with (\ref{eq:phinearboundary}), Eq.~(\ref{eq:minscal}) admits solutions of the form $\phi(v)\propto (1-v)^\beta$ near the horizon with characteristic exponents $\beta=\pm\frac{i {\mathfrak w}}{2}$. The choice $\beta=-\frac{i {\mathfrak w}}{2}$ corresponds to in-falling boundary conditions and one can solve Eq.~(\ref{eq:minscal}) using a generalized power series ansatz, cf.~\cite{Starinets:2002br}
\begin{equation}
\label{eq:phinearhorizon}
\phi(v)=\left(1-v\right)^{-i {\mathfrak w}/2}\sum_{n=0}^\infty f_n({\mathfrak w},{\mathfrak K})(1-v)^{n}\,.
\end{equation}
The coefficients $f_n$ may be obtained order-by-order by inserting this ansatz in Eq.~(\ref{eq:minscal}), where $f_0=1$ without loss of generality. Once the near-horizon solution (\ref{eq:phinearhorizon}) has been found, it can be used to obtain the coefficients ${\cal A},{\cal B}$ through
\begin{equation}
{\cal A}=\phi(0)=\sum_{n=0}^\infty f_n({\mathfrak w},{\mathfrak K})\,,\quad
{\cal B}=\frac{1}{2}\left(\phi^{\prime\prime}(0)-\lim_{v\rightarrow 0}{\cal A} z_{ab}\left(3+2 \ln(v)\right)\right)\,.
\end{equation}
Truncating the sum in (\ref{eq:phinearhorizon}) at finite order $n=M$, the coefficients become
\begin{eqnarray}
{\cal A}&=&\sum_{n=0}^M f_n({\mathfrak w},{\mathfrak K})\,,\\
{\cal B}&=&\frac{1}{2}\sum_{n=0}^M f_n({\mathfrak w},{\mathfrak K})\left[\left(n-\frac{i{\mathfrak w}}{2}\right)\left(n-\frac{i{\mathfrak w}}{2}-1\right)+\frac{({\mathfrak w}^2-{\mathfrak K}^2)^2}{2}\left(3-2 H_{N}\right)\right]\nonumber\,,
\end{eqnarray}
where the harmonic number $H_N$ arises from considering the near-boundary approximation $\lim_{v\rightarrow 0}\ln(v)=-\lim_{v\rightarrow 0}\sum_{n=0}^N \frac{(1-v)^n}{n}=H_N$.

The resulting retarded correlator $G^{12,12}(\omega,k)$ has simple poles in the complex $\omega$ plane, which can be found as the zeros of ${\cal A}({\mathfrak w},{\mathfrak K})$. For instance, for $k=0$ the poles closest to the origin in the complex $\omega$ plane are located at
\begin{equation}
\label{eq:scalarqnm}
\omega^{(1)}_{\pm}(k=0)\simeq 2 \pi T \left(\pm 1.56-1.37i\right)\,.
\end{equation}
Since the Navier-Stokes equation result for $G^{12,12}$ does not exhibit any poles, the excitations corresponding to Eq.~(\ref{eq:scalarqnm}) must be non-hydrodynamic modes, see section \ref{sec:nonhydro}. In fact, they correspond to the non-hydrodynamic quasi-normal modes of the black brane \cite{Berti:2009kk}. 

To obtain transport parameters such as the shear viscosity for strongly coupled ${\cal N}=4$ SYM, it is most economical to solve Eq.~(\ref{eq:minscal}) for small $\omega,k$, while keeping the correct in-falling boundary conditions, e.g. $\phi(v)=(1-v)^{-i {\mathfrak w}/2}\left(1+{\cal O}(\omega,q)\right)$. Normalizing
$\lim_{v\rightarrow 1}\phi(v) (1-v)^{i {\mathfrak w}/2}=1$, one finds for instance
\begin{eqnarray}
\phi(v)&=&\left(1-v\right)^{-i {\mathfrak w}/2}\left[1-\frac{i {\mathfrak w}}{2}\ln \left(\frac{1+v}{2}\right)-
{\mathfrak K}^2\ln \frac{1+v}{2}\right.\nonumber\\
&&\left.+\frac{{\mathfrak w^2}}{2}\left({\rm Li}_2\left(\frac{1+v}{2}\right)-\frac{\pi^2}{6}+\frac{\ln\left(\frac{1+v}{2}\right)}{4}\ln\left(\frac{e^8 (1+v)(1-v)^4}{32}\right)\right)\right.\nonumber\\
&&\left.+{\cal O}\left({\mathfrak w}^3,{\mathfrak K}^3\right)\right]\,.
\end{eqnarray}
To this order of accuracy for small frequencies and wavenumbers $\omega,k\ll 1$, ${\cal B}=\frac{\phi^{\prime \prime}(0)}{2}$ and ${\cal A}=\phi(0)$ such that
\begin{equation}
\label{eq:2ptN4}
G^{12,12}(\omega,k)=s\left[-\frac{i\omega}{4 \pi}-\frac{k^2}{8 \pi^2 T}+\frac{\omega^2(1-\ln 2)}{8 \pi^2 T}+{\cal O}\left(\omega^3,k^3\right)\right]\,.
\end{equation}
Comparing this expression to the Navier-Stokes result $G^{12,12}=-i \eta \omega$ from Eq.~(\ref{eq:NS2pt}) where the overall constant (contact term) has been ignored, one finds \cite{Policastro:2001yc}
\begin{equation}
\label{eq:etaos}
\frac{\eta}{s}=\frac{1}{4\pi}\,,
\end{equation}
for strongly coupled ${\cal N}=4$ SYM. \index{Shear viscosity! Lower bound (string-theory)}
It has been conjectured by Kovtun, Son and Starinets (KSS) in Ref.~\cite{Kovtun:2004de}, that $\frac{\eta}{s}\geq\frac{1}{4\pi}$ serves as a lower bound for a wide class of systems, making earlier qualitative arguments for the existence of such a bound \cite{Danielewicz:1984ww} precise  (however, see Refs.~\cite{Kats:2007mq,Brigante:2008gz} for counter-examples). 
The higher order terms in (\ref{eq:2ptN4}) can be matched to rBRSSS theory (\ref{eq:rBRSSS2pt}), finding \cite{Baier:2007ix}
\begin{equation}
\label{eq:taupi}
\kappa=\frac{s}{4\pi^2 T}\,,\quad \tau_\pi=\frac{2-\ln 2}{2 \pi T}\,.
\end{equation}

\subsection{Non-Hydrodynamic Modes in Gauge/Gravity Duality}
\label{sec:ggnonhydro}
\index{Non-hydrodynamic mode! Holography}

\begin{figure}[t]
  \begin{center}    
     \includegraphics[width=.49\linewidth]{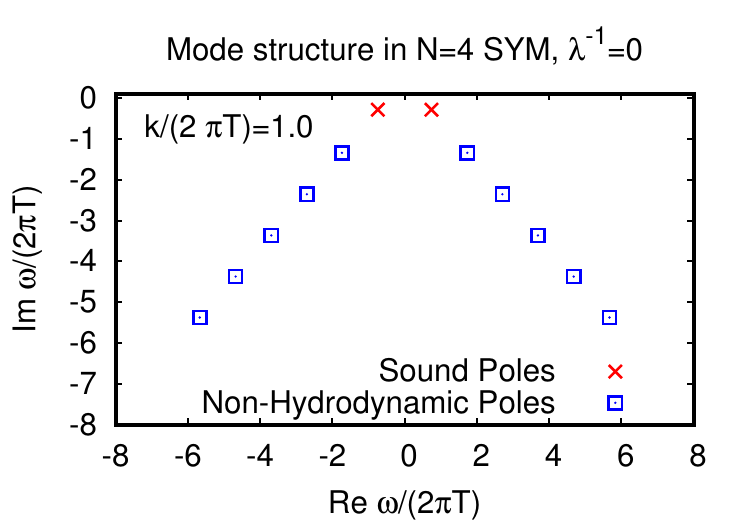}
     \hfill
     \includegraphics[width=.49\linewidth]{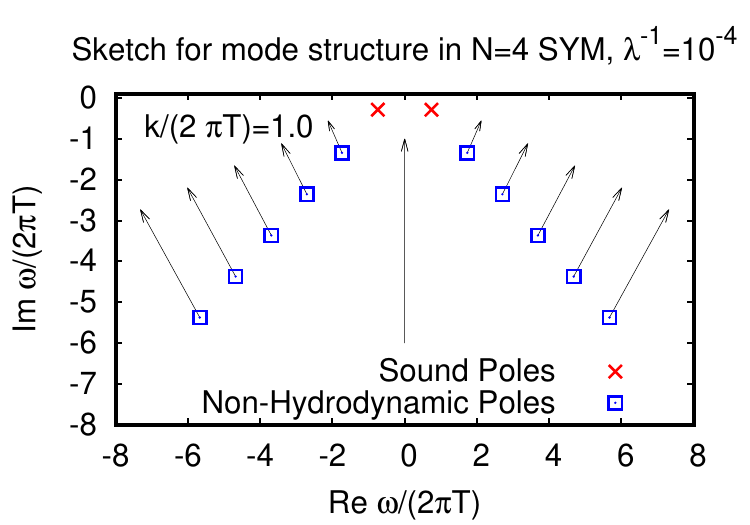}
     \end{center}
  \caption{\label{fig3:ggnonhy} Collective modes for strongly coupled ${\cal N}=4$ SYM in the sound channel at fixed spatial gradient $k/T=2 \pi$ from Ref.~\cite{Kovtun:2005ev}, corresponding to singularities of $G^{00,00}(\omega,k)$ in the complex frequency plane. Shown are results at infinitely large 't Hooft coupling ${\lambda}$ and a sketch for the behavior at finite 't Hooft coupling based on the results from Ref.~\cite{Grozdanov:2016vgg}. In this sketch, as the coupling is decreased, the movement of poles  has been indicated by arrows. Decreasing 't Hooft coupling or increasing the gradient strength above $k > k_c$, the hydrodynamic sound poles become sub-dominant to the non-hydrodynamic poles.}
\end{figure}

 Similar to the discussion in \ref{sec:KTcollmodes}, it is interesting to study more broadly the collective structure in gauge/gravity duality defined as the 
singularities of the retarded two-point correlator of the energy-momentum tensor. The sound channel (defined by the singularities of $G^{00,00}(\omega,k)$) can be obtained in complete analogy to the scalar channel $G^{12,12}(\omega,k)$ given in Eq.~(\ref{eq:ggG1212}). For small spatial gradients $k/T\ll 1$, there are two simple poles corresponding to the hydrodynamic sound modes Eqns.~(\ref{eq:soundmode}). In addition, $G^{00,00}(\omega,k)$ has an infinite number of simple poles as shown in Fig.~\ref{fig3:ggnonhy}. The additional, non-hydrodynamic, poles correspond to the non-hydrodynamic quasi-normal modes of the black brane in the dual gravitational description\footnote{Note that non-hydrodynamic quasi-normal modes of astrophysical black holes have recently been observed in the gravitational wave detection of a binary black hole  merger by LIGO, cf. Refs.~\cite{TheLIGOScientific:2016wfe,TheLIGOScientific:2016src}.}

Similar to the findings for kinetic theory in section \ref{sec:ktnonhydro}, the hydrodynamic poles in $G^{00,00}(\omega,k)$ are present also in non-equilibrium situations even though their associated dispersion relation no longer quantitatively matches the Navier-Stokes result Eq.~(\ref{eq:soundmode}). This is another concrete example of how the Central Lemma of fluid dynamics is realized in a particular microscopic theory setup.

However, decreasing the 't Hooft coupling $\lambda$ or alternatively increasing the spatial gradient strength beyond a critical, coupling-dependent value of $k>k_c({\lambda})$, the hydrodynamic sound poles are no longer the singularities of $G^{00,00}(\omega,k)$ closest to the real axis. In particular, when decreasing $\lambda$, additional poles not present at $\lambda^{-1}=0$ start to enter the complex $\omega$ plane from below as sketched in Fig.~\ref{fig3:ggnonhy}, and replace the hydrodynamic sound poles as the dominant singularities for $k>k_c(\lambda)$ \cite{Spalinski:2016fnj}. A fit to $k_c(\lambda)$ for ${\cal N}=4$ SYM obtained numerically in Ref.~\cite{Grozdanov:2016vgg} is given by \cite{Romatschke:2016hle}
\begin{equation}
k_c(\lambda)\simeq \frac{\pi T}{4}\ln\left(\frac{6.65 \lambda^{3/2}}{1+105 \lambda^{-3/2}}\right)\,.
\end{equation}

For $k>k_c$, thus, sound modes are sub-dominant in ${\cal N}=4$ SYM, reminiscent of the onset transition discussed in the context of kinetic theory in section \ref{sec:ktnonhydro}, and transport occurs via non-hydrodynamic degrees of freedom (in this case, non-hydrodynamic quasi-normal modes). For strongly coupled ${\cal N}=4$ SYM at gradient strengths $k>k_c$, Borel fluid dynamics has broken down.
\index{Fluid Dynamics! Break-down of}

Finally, it should be pointed out that the non-hydrodynamic mode spectrum of strongly coupled ${\cal N}=4$ SYM at $N_c\rightarrow \infty$ can be qualitatively reconciled with weak-coupling kinetic theory results. At first glance, the non-hydrodynamic mode structure between the two regimes (strong coupling and weak coupling) appears to be qualitatively different, with simple poles corresponding to non-hydrodynamic black hole quasi-normal modes on the one hand and logarithmic branch cuts corresponding to Landau damping on the other hand. At second glance, however, one may adopt a non-standard convention for the logarithmic branch cuts in kinetic theory such that they stretch from $\omega=(-\infty,-k-i/\tau_R]$ and $\omega=[k-i/\tau_R,\infty)$. Approximating the branch cuts as a series of poles, it is plausible that the poles thin out and move in the complex $\omega$ plane as to resemble the structure sketched in Fig.~\ref{fig3:ggnonhy} as the coupling is increased (cf. the discussion and evidence presented in Refs.~\cite{Romatschke:2016hle,Grozdanov:2016vgg}). In this interpretation of two-point retarded energy-momentum correlators for $N_c\rightarrow \infty$, branch cuts would exist only at \textit{vanishing} coupling $\lambda=0$, while for any nonzero $\lambda>0$ the singularities of $G^{\mu\nu,\lambda\kappa}(\omega,k)$ would be simple poles which approximate the branch cuts for $\lambda\rightarrow 0$. It would be interesting to test this interpretation by a first-principles calculation of a retarded energy-momentum correlator at fixed (but arbitrary) 't Hooft coupling $\lambda\neq 0$.

\subsection{Holographic Renormalization}
\label{sec:holoren}

As the last section of this gauge/gravity flyby, let us mention a useful relation between the metric and the boundary energy-momentum tensor. This is part of a much more complete renormalization program in holography, cf. Ref.~\cite{deHaro:2000vlm}. Let us again consider the black brane metric (\ref{eq:m3}), and now consider a coordinate transformation
\begin{equation}
z=\sqrt{2  u_s^2} \sqrt{\frac{u_s^2}{u^2}-\sqrt{\frac{u_s^4}{u^4}-1}}\,,
\end{equation}
which brings the black brane line element into the form
\begin{equation}
\label{eq:mylinele}
ds^2=\frac{L^2}{z^2}\left(-\frac{\left(1-\frac{z^4}{4 u_s^4}\right)^2}{\left(1+\frac{z^4}{4 u_s^4}\right)}dt^2+\left(1+\frac{z^4}{4 u_s^4}\right)d{\bf x}^2+dz^2\right)\,,
\end{equation}
where the boundary is located at $z=0$. Note that in this form, mixed terms such as $dt dz$ are absent and the coefficient of the $dz^2$ term has been fixed to $\frac{L^2}{z^2}$. It can be shown that there is enough freedom in diffeomorphism transformations to bring any line element in a similar form, and the corresponding coordinates are referred to as Fefferman-Graham coordinates \cite{Fefferman}. Expanding the line-element (\ref{eq:mylinele}) in a power series close to the boundary leads to
\begin{equation}
\label{eq:mylinele2}
ds^2=\frac{L^2}{z^2}\left(-dt^2+d{\bf x}^2+dz^2\right)+L^2 z^2\left(\frac{3}{4 u_s^4} dt^2+\frac{1}{4 u_s^4}d{\bf x}^2\right)+{\cal O}(z^4)\,,
\end{equation}
where the first term in the expansion can easily be recognized to be the Minkowski metric at the conformal boundary. Using (\ref{eq:us2}) the terms proportional to ${\cal O}(z^2)$ in Eq.~(\ref{eq:mylinele2}) are  proportional to $T^4$, similar to the energy density and pressure of a conformal system at temperature $T$ in $d=4$. More general considerations lead to the conclusion that -- in Fefferman-Graham coordinates -- the line element can be written as
\begin{equation}
\label{eq:fg}
ds^2=\frac{L^2\left(g_{\mu\nu}(t,{\bf x},z)dx^\mu dx^\nu+dz^2\right)}{z^2}
\end{equation}
where the near boundary expansion of $g_{\mu\nu}(t,{\bf x},z)$ is given by
\begin{equation}
g_{\mu\nu}(t,{\bf x},z)=g_{\mu\nu}^{(0)}(t,{\bf x})+z^2 g_{\mu\nu}^{(2)}(t,{\bf x})+z^4 g_{\mu\nu}^{(4)}(t,{\bf x})+\ldots\,.
\end{equation}
In this expansion, $g_{\mu\nu}^{(0)}$ can be identified with the metric of the $d=4$ dimensional boundary theory and $g_{\mu\nu}^{(4)}$ is proportional to the boundary-theory energy-momentum tensor,
\begin{equation}
g_{\mu\nu}^{(4)}(t,{\bf x})\propto T_{\mu\nu}(t,{\bf x})\,.
\end{equation}
This implies that the near-boundary expansion of the line-element in Fefferman-Graham coordinates is a practical means of obtaining the energy-momentum tensor of the boundary theory.

\section{Finite Temperature QFT Flyby}
\label{sec:fieldtheory}

\subsection{Basics of Finite Temperature Field Theory}

Consider a free scalar field $\phi$ in Minkowski space with Lagrangian
$${\cal L}=-\frac{1}{2}\partial_\mu \phi \partial^\mu \phi-\frac{1}{2}m^2 \phi^2\,,$$
and an associated action $S=\int dt d^3 x{\cal L}$. Demanding that $S$ be invariant under symmetry transformations $x^\mu\rightarrow x^{\prime \mu}=x^\mu+\delta x^\mu$ leads to an associated Noether current
\begin{equation}
\label{eq:tmunufreeqft}
T^{\mu\nu}=\partial^\mu \phi \partial^\nu \phi+g^{\mu\nu}{\cal L}\,,
\end{equation}
where $T^{00}=\frac{1}{2}(\partial_t \phi)^2+\frac{1}{2}(\partial \phi)^2+\frac{1}{2}m^2 \phi^2={\cal H}$ can be recognized as the Hamiltonian density of the system. For a time-independent Hamiltonian $H=\int d^3x {\cal H}$, the time evolution from $t=0$ to $t=t_f$ can be calculated using the evolution operator $\hat U=e^{-i \hat H t_f}$. At finite temperature $T$, the partition function of the system is given by
\begin{equation}
\label{eq:partitionfunc}
Z={\rm Tr}e^{-\hat H/T}\,,
\end{equation}
where the trace is over the full vector space of the theory. One can formally identify the finite-temperature density matrix $e^{-\hat H/T}$ with the time-evolution operator $e^{-i \hat H t_f}$ by associating
\begin{equation}
t_f=-i/T=-i \beta\,,
\end{equation}
with $\beta\equiv \frac{1}{T}$ the inverse temperature. This is most easily accomplished by introducing imaginary time
\begin{equation}
\tau=i t\,,\quad \tau \in [0,\beta]\,,
\end{equation}
with $\phi(\tau,{\bf x})$ fulfilling periodic boundary conditions on $\tau\in [0,\beta]$ because of the trace in (\ref{eq:partitionfunc}). In terms of imaginary time $\tau$, the partition function $Z(\beta)$ can be written as a path-integral
\begin{equation}
Z(\beta)=\int {\cal D}\phi e^{-S_E}\,,\quad S_E=\int_0^{\beta}d\tau \int d^3x {\cal L}_E\,,
\end{equation}
where ${\cal L}_E$ is the \textit{Euclidean} Lagrangian density related to the Lagragian density ${\cal L}$ in Minkowski space-time as
\begin{equation}
{\cal L}_E=-{\cal L}(t\rightarrow -i \tau)=\frac{1}{2}(\partial_\tau \phi)^2+\frac{1}{2}(\partial \phi)^2+\frac{1}{2}m^2 \phi^2
\end{equation}
Because of the periodicity of $\phi$, its time dependence can be expressed in terms of a Fourier series
\begin{equation}
\label{eq:phiexpress}
\phi(\tau,{\bf x})=T \sum_{n=-\infty}^\infty \tilde \phi(\omega_n,{\bf x}) e^{i \omega_n \tau}\,,
\end{equation}
where $\omega_n\equiv 2 \pi T n$ are the so-called Matsubara frequencies and the sum in Eq.~(\ref{eq:phiexpress}) is referred to as a thermal sum.

  \subsection{Energy-momentum tensor for free scalar field}
 \label{sec:fieldemt}

Equipped with the basic setup of finite temperature quantum field theory, let us calculate the expectation value of the energy-momentum tensor (\ref{eq:tmunufreeqft}) for a free scalar field in the local rest frame at equilibrium. The expectation value $\langle T^{\mu\nu}\rangle$ can be calculated as
\begin{equation}
\langle T^{\mu\nu}\rangle_{\rm LRF}=\frac{\int {\cal D}\phi e^{-S_E} T^{\mu\nu}}{Z(\beta)}\,,
\end{equation}
where on the rhs $T^{\mu\nu}$ is given by (\ref{eq:tmunufreeqft}) with the replacement $t\rightarrow -i \tau$. Transforming $\phi$ into Fourier space, one encounters integrals such as
\begin{equation}
\langle (\partial_\tau \phi)^2 \rangle=\frac{\int {\cal D}\phi e^{-S_E} (\partial_\tau \phi)^2}{Z(\beta)}=T \sum_n \int \frac{d^3k}{(2\pi)^3}\frac{\omega_n^2}{\omega_n^2+m^2+{\bf k}^2}\,,
\end{equation}
where $\sum_n$ denotes a thermal sum as in (\ref{eq:phiexpress}). Using standard techniques to evaluate these thermal sums \cite{Laine:2016hma}, and subtracting the zero-temperature (vacuum) contribution one finds
\begin{equation}
\label{eq:tsumex1}
\langle (\partial_\tau \phi)^2 \rangle=- \int \frac{d^3k}{(2 \pi)^3} \frac{\sqrt{m^2+{\bf k}^2}}{e^{\sqrt{m^2+{\bf k}^2}\beta}-1}\simeq
-\frac{3 T^4 \pi^2}{90}\,,
\end{equation}
where the approximation is the leading contribution in the high temperature limit $\frac{m}{T}\rightarrow 0$. Using results such as these one finds for the energy-momentum tensor of a free scalar field in the high temperature limit
\begin{equation}
\langle T^{\mu\nu}\rangle_{\rm LRF}=\frac{T^4 \pi^2}{90}\left(\begin{array}{cccc}
3 & 0 & 0 & 0\\
0 & 1 & 0 & 0 \\
0 & 0 & 1 & 0\\
0 & 0 & 0 & 1
\end{array}\right)\,.
\end{equation}
Since $T^{00}$ is the Hamiltonian density of the system, the corresponding expectation value is the local energy density,
\begin{equation}
\langle T^{00}\rangle_{\rm LRF}=\epsilon\,.
\end{equation}
On the other hand, one may recognize the spatial entries $\langle T^{xx}\rangle_{\rm LRF}$,  $\langle T^{yy}\rangle_{\rm LRF}$, $\langle T^{zz}\rangle_{\rm LRF}$ to correspond to the local thermodynamic pressure $P$ of the system by explicitly calculating \cite{Laine:2016hma}
\begin{equation}
P=\lim_{V\rightarrow \infty}\frac{\ln Z}{\beta V}=-T \int \frac{d^3 k}{(2 \pi)^3}\ln \left(1-e^{-\sqrt{m^2+{\bf k}^2} \beta}\right)\simeq \frac{\pi^2 T^4}{90}\,,
\end{equation}
where only the leading term in the high temperature limit $\frac{m}{T}\rightarrow 0$ has been kept. This shows that the expectation value of the energy-momentum tensor in the local rest frame is given by
$$\langle T^{\mu\nu}\rangle_{\rm LRF}={\rm diag}\left(\epsilon,P,P,P\right)\,,$$
in free quantum field theory \cite{HabichBach:2011}. These results can readily be generalized to other field theories, such as SU($N_c$) gauge theory where the Stefan-Boltzmann pressure is given by\cite{Laine:2016hma}
\begin{equation}
\label{eq:freepress}
P=\frac{2 (N_c^2-1)\pi^2 T^4}{90}\,.
\end{equation}
\index{Equation of State (EoS)!Bose-Einstein}

\subsection{Transport Coefficients from Free Field Theory}
\index{Transport coefficients! Free field theory}

While the calculation of most transport coefficients in field theory requires elaborate techniques to treat the interactions, the value of some transport coefficients can be obtained in free field theory. These transport coefficients have been dubbed ``thermodynamic'' transport coefficients \cite{Moore:2012tc} because they are not associated with entropy production, unlike the more familiar shear and bulk viscosity coefficients \cite{Romatschke:2009kr}.

Let us consider calculating the energy-momentum tensor correlation function $G^{12,12}(\omega,{\bf k})$ for the massless free scalar field at finite temperature. With the energy-momentum tensor given by (\ref{eq:tmunufreeqft}), the Euclidean correlator (equal to \textit{minus} the retarded correlator upon analytic continuation) is given by
\begin{equation}
\langle T^{12}(\tau,{\bf x}) T^{12}(0,0)\rangle=
\langle\partial_{x_1} \phi(\tau,{\bf x})\partial_{x_2}\phi(\tau,{\bf x})\partial_{x_1} \phi(0,0)\partial_{x_2}\phi(0,0) \rangle\,.\nonumber
\end{equation}
Fourier-transforming the field $\phi$, performing the Wick-contractions and using $\langle \tilde \phi(\omega_n,{\bf k})\tilde \phi(\omega_m,{\bf p})\rangle=\frac{\beta \delta_{\omega_n,\omega_m} \delta({\bf k}+{\bf p})}{\omega_n^2+{\bf k}^2}$ leads to
\begin{equation}
G^{12,12}(\omega,k)=- \lim_{\epsilon\rightarrow 0} 2 T \sum_n \int \frac{d^3p}{(2 \pi)^3}\frac{p_{x_1}^2 p_{x_2}^2}{\left(\omega_n^2+{\bf p}^2\right)\left((\omega_n+i \omega-\epsilon)^2+({\bf p}-{\bf k})^2\right)} \,,
\end{equation}
where contact terms (proportional to $\delta(\omega)\delta(k)$) were dropped and ${\bf k}$ was assumed to point along the $x_3$ direction. Restricting our consideration to the term proportional to ${\cal O}(k^2)$ in $G^{12,12}(\omega,k)$ we may directly access this coefficient as \cite{Moore:2012tc}
\begin{equation}
\partial_{k}^2 G^{12,12}(\omega=0,k)=4 T \sum_n \int \frac{d^3p}{(2 \pi)^3} \left[\frac{p_{x_1}^2p_{x_2}^2}{\left(\omega_n^2+{\bf p}^2\right)^3}-\frac{4 p_{x_1}^2p_{x_2}^2 p_{x_3}^2}{\left(\omega_n^2+{\bf p}^2\right)^4}\right]=\frac{T^2}{72}\,,
\end{equation}
using $T \sum_n \frac{d^3p}{(2 \pi)^3} \frac{({\bf p})^{2n}}{\left(\omega_n^2+{\bf p}^2\right)^{n+1}}=\frac{(2n+1)!}{2^{2n}(n!)^2} \frac{T^2}{12}$ where all zero-temperature contributions have been ignored. Matching this result to rBRSSS theory  (\ref{eq:rBRSSS2pt}) leads to
\begin{equation}
\label{eq:kappaweak}
\kappa=-\frac{T^2}{72}
\end{equation}
for the free scalar field \cite{Moore:2012tc}. A similar calculation for an SU($N_c)$ gauge theory leads to $\kappa=\frac{N_c^2-1}{18}T^2$ \cite{Romatschke:2009kr}. Note that no non-trivial viscosity coefficient arises in the free theory, cf. Ref.~\cite{Aarts:2002cc}.

\section{Lattice Gauge Theory Flyby}
\label{sec:lattice}

The primary applications of lattice gauge theory for the purpose of this work are non-perturbative 
calculations of thermodynamic equilibrium quantities, such as the equation of state and the speed of sound.
With this aim in mind, let us give an introduction to lattice gauge theory by considering the case of pure SU($N_c$) Yang-Mills theory
in four space-time dimensions, without discussing  the more advanced treatment of fermions and scalars on the lattice, cf.~\cite{Karsch:2001cy}.

To study properties of the gauge theory in thermodynamic equilibrium, the starting point is the partition function $Z(\beta)$
of the theory (see section \ref{sec:fieldtheory}). It is possible to write the partition function for a gauge theory at finite temperature as a Euclidean path integral \cite{Kapusta:2006pm}
\begin{equation}
\label{eq:parti}
Z(\beta,V)=\int {\cal D} A_\mu e^{-S_E(\beta,V)}\,,
\end{equation}
where $A_\mu$ are SU($N_c$) gauge fields, and $S_E[\beta,V]$ is the Euclidean action for SU($N_c$) Yang-Mills theory at finite volume $V$. The Euclidean action is given by
\begin{equation}
\label{eq:FTaction}
S_E[\beta]=\frac{1}{2g^2}\int_0^\beta d\tau \int d^3x {\rm Tr}F_{\mu\nu}F_{\mu\nu}\,,
\end{equation}
where $g$ is the Yang-Mills coupling, and $F_{\mu\nu}=\partial_\mu A_\nu-\partial_\nu A_\mu+\left[A_\mu,A_\nu\right]$ is the Yang-Mills field strength tensor\footnote{Note that because space-time is Euclidean, there is no difference between covariant and contravariant indices.}.

Discretizing Euclidean space-time on a square lattice with lattice spacing $a$ with $N_\sigma$ sites in the three ``spatial'' directions each, and $N_\tau$ sites in the ``time'' direction, the lattice volume and temperature are given by
\begin{equation}
V=(N_\sigma a)^3\,, \quad T=\frac{1}{N_\tau a}=\beta^{-1}\,,
\end{equation}
respectively. To discretize the lattice action (\ref{eq:FTaction}) it is useful to consider the so-called link variables 
\begin{equation}
U_{x,\mu}\equiv {\rm P} e^{\int_x^{x+\hat \mu a} dx^\mu A_\mu(x)}\,,
\end{equation}
which are associated with the link between two neighboring sites $x,x+\hat \mu a$ on the lattice\footnote{Here the symbol ${\rm P}$ denotes path-ordering.}. 
In terms of the link variables, a standard regularization of the lattice action (\ref{eq:FTaction})
is given by
\begin{equation}
\label{eq:lateffac}
S_E(\beta)\simeq S_L(N_\tau,N_\sigma,\beta_L)=\beta_L\sum_{x}\sum_{\Box}\left(1-\frac{1}{N_c}{\rm Re}{\rm Tr} U_\Box\right)+{\cal O}(a^2)\,,
\end{equation}
where $\beta_L\equiv \frac{2 N}{g^2}$ is the lattice coupling (not to be confused with the inverse temperature $\beta$) and $U_\Box$ is the plaquette operator expressed in terms of link variables as \cite{Montvay:270707}
\begin{equation}
U_{\rm \Box,\mu\nu}\equiv U_\mu(x)U_\nu(x+\hat \mu a)U_\mu^\dagger(x+\hat \nu a)U_\nu^\dagger(x)\,.
\end{equation}
The plaquette operator corresponds to a simple loop on the lattice, and $\sum_{\Box}$ in (\ref{eq:lateffac}) denotes the sum over all loops with only one orientation
\begin{equation}
\sum_{\rm \Box}\equiv \sum_{1\leq \mu < \nu \leq 4}\,.
\end{equation}

Lattice gauge theory calculations consist of two steps. Step number one is to obtain configurations of link variables (gauge configurations) consistent with an importance sampling of the probability distribution $e^{-S_E(\beta)}$. Step number two then consists of using these gauge configurations to calculate expectation values of relevant observables. These steps will be discussed in reverse order below.

\subsection{Observables from Lattice Gauge Theory}

If the partition function (\ref{eq:parti}) could be effectively evaluated on the lattice, this would allow calculation of thermodynamic properties such as the pressure through the well-known thermodynamic properties $P=\frac{1}{\beta V}\ln Z$. In practice, $\ln Z$ is typically
not accessible in lattice calculations, but its derivative is
\begin{equation}
\frac{\partial \ln Z}{\partial \beta_L}=-\langle \frac{\partial S_L}{\partial \beta_L}\rangle=-\langle \sum_{x}\sum_{\Box}\left(1-\frac{1}{N_c}{\rm Re}{\rm Tr} U_\Box\right)\rangle\,,
\end{equation}
where $\langle \cdot \rangle$ denotes the expectation value of a quantity ${\cal O}$ as
\begin{equation}
\label{eq:expva}
\langle {\cal O} \rangle \equiv \frac{\int d{\cal A}_\mu e^{-S_E} {\cal O}}{Z}\,.
\end{equation}
If one can calculate $\frac{\partial \ln Z}{\partial \beta_L}$ on the lattice, then the pressure may be obtained by integration as
$P=-(N_\sigma^3 N_\tau a^4)^{-1}\int d\bar\beta_L \frac{S_L}{\bar\beta_L}$. The integration is best performed starting from a $\beta_L=\beta_L^{0}$ value where the system is in a confined state such that $\left. P\right|_{\beta_L^{0}}$ is small. Also, it is typically necessary to subtract the zero-temperature (vacuum) contribution from the pressure. This can effectively be done by subtracting $S_L/\beta_L$ for $N_\tau=N_\sigma$ from the integrand. With these steps, the pressure over the fourth power of temperature at lattice coupling $\beta_L$ can be obtained as
\begin{equation}
\frac{P}{T^4}(\beta_L)=-\frac{N_\tau^3}{N_\sigma^3}\int_{\beta_L^{0}}^{\beta_L} d\bar\beta_L \frac{S_L(N_\tau,N_\sigma,\bar \beta_L)-S_L(N_\sigma,N_\sigma,\bar \beta_L)}{\bar\beta_L}\,.
\end{equation}
Note that Renormalization Group (RG) flow techniques \cite{Luscher:2009eq} have recently been presented that allow more direct access to thermodynamic quantities from the lattice, cf. Ref.\cite{Makino:2014taa}.

In order to know the physical temperature for a given lattice simulation, it is necessary to relate the lattice coupling $\beta_L$, or rather the lattice spacing at the lattice coupling $a(\beta_L)$ to a physical scale (``scale setting''). Scale setting in lattice QCD is done by calculating (dimensionless) quantities on the lattice and comparing these to the actually observed values \cite{Sommer:2014mea}. For instance, one could proceed to calculate the $0^{++}$ glueball mass $m_g$ on the lattice for a given lattice coupling $\beta_L$ as $m_L=m_g a(\beta_L)=1.2345$\footnote{Note that the particular numerical value for $m_L$ was invented for this example.}. Knowledge of the \textit{physical} glueball mass $m_g\simeq 1.7$ GeV \cite{Morningstar:1999rf} would then imply $a(\beta_L)=\frac{m_L}{m_g}\simeq 0.15$ fm. Clearly, $m_L$ will have lattice artifacts at finite lattice spacing $a$. Therefore, it is necessary to perform a \textit{continuum limit} $a\rightarrow 0$ by repeating the calculation at increased lattice volumes. \index{Continuum limit}

Note that more modern versions of scale setting exist that employ RG techniques from Wilson flow, cf. Refs.~\cite{Luscher:2009eq,Bornyakov:2015eaa}.

Once the relation between $\beta_L$ and the physical lattice spacing $a(\beta_L)$ is known, it is possible to relate $N_\tau$ and the physical temperature as $T=\frac{1}{N_\tau a(\beta_L)}$, which in turn allows to express $\frac{P}{T^4}(\beta_L)$ as a function of temperature. 

\subsection{Generating Gauge Configurations using HMC}

Let us now discuss the issue of generating gauge configurations that will allow evaluation of the expectation values of observables discussed above. There are many ways to generate configurations, and in the following only one such algorithm, the Hybrid Monte Carlo (HMC) algorithm, will be described \cite{Montvay:270707}. The HMC algorithm consists of a standard heat-bath algorithm combined with classical gauge field evolution and is straightforward to implement. Consider a fictitious classical action (\ref{eq:FTaction}) in \textit{five} dimensional \textit{Minkowski} space-time (consisting of four dimensional Euclidean space-time plus one additional fictitious time direction), and employ the temporal gauge $A_0=0$. In this case, conjugate momenta to the remaining gauge field degrees of freedom are given by $E_\mu=\partial_0 A_\mu=F_{0\mu}$, which may be interpreted as fictitious electric fields. The resulting equations of motion for the gauge fields can be taken to arise from a five-dimensional classical Hamiltonian density ${\cal H}_5$ in addition to a Gauss-law constraint $D_\mu E_\mu=0$. Expressing the gauge fields in terms of a set of generators $\lambda^a$ (traceless hermitian complex $N_c\times N_c$ matrices) as
\begin{equation}
A_\mu=-i g A^a_\mu \lambda^a\,,
\end{equation}
where $g$ is again the gauge coupling and the coefficients $A^a_\mu$ are real, the Hamiltonian density is then given by 
\begin{equation}
\label{eq:h5}
{\cal H}_5=\frac{1}{2}E_\mu^a E_\mu^a+\frac{1}{4}F_{\mu\nu}^aF^a_{\mu\nu}\,,
\end{equation}
where $F_{\mu\nu}^a=\partial_\mu A_\nu^a-\partial_\nu A_\mu^a+g f^{abc}A_\mu^b A_\nu^c$ and $f^{abc}$ are the SU($N_c$) structure constants that obey $[\lambda^a,\lambda^b]=\sum_c i f^{abc}\lambda^c$. The Hamiltonian density may be discretized on a lattice as
\begin{equation}
\frac{1}{4}F_{\mu\nu}^aF^a_{\mu\nu}\rightarrow\beta_L a^{-4} \sum_{\Box}\left(1-\frac{1}{N_c}{\rm Re}{\rm Tr} U_\Box\right)\,.
\end{equation}
The classical Hamiltonian equations then give rise to the equations of motion for the link variables and fictitious momenta \cite{Hanada:2016qbz}
\begin{eqnarray}\label{eq:classeoms}
U_\mu(x,t+\Delta t)&=&e^{i \Delta t E_\mu}U_\mu(x,t)\,,\\
E_\mu\left(x,t+\frac{1}{2}\Delta t\right)&=&E_\mu\left(x,t-\frac{1}{2}\Delta t\right)-\Delta t \sum_{|\nu|\neq \mu} {\rm Adj}\left[U_\mu(x,t) S_{\mu\nu}^\dagger(x,t)\right]\,,\nonumber\\
S_{\mu\nu}(x,t)&=&U_{\nu}(x,t)U_\mu(x+\hat \nu,t)U_\nu^\dagger(x+\hat \mu,t)\,,\nonumber
\end{eqnarray}
where ${\rm Adj}[M]=-\frac{i}{2}\left[M-M^\dagger-\frac{1}{N_c}{\rm Tr}\left(M-M^\dagger\right)\right]$ and $\Delta t$ is the time step in the fictitious time direction. These classical equations of motion can readily be solved in the fictitious time on the lattice using well-established techniques \cite{Hanada:2016qbz,Gottlieb:1987mq,Bodeker:1999gx,Rebhan:2004ur,Berges:2008zt}. Starting from some initial conditions for $E_\mu,U_\mu$, solutions of the classical equations of motion evolve the gauge link configurations $U_\mu$ in the fictitious time $t$, where it should be noted that $U_\mu(x,t)$ at fixed fictitious time $t$ correspond to gauge configurations of \textit{four-dimensional} Euclidean lattice action (\ref{eq:lateffac}). Thus, generating different gauge field configurations $U_\mu$ implies a Monte-Carlo sampling procedure for the path integral (\ref{eq:parti}) with probability distribution $e^{-S_L(N_\tau,N_\sigma,\beta_L)}$. It turns out that the classical dynamics governed by the equations (\ref{eq:classeoms}) correspond to an efficient algorithm for generating new four-dimensional gauge configurations where every field variable $U_\mu$ is changed, but the Hamiltonian corresponding to (\ref{eq:h5}) is conserved up to discretization errors of order ${\cal O}(\Delta t^2)$ \cite{Montvay:270707}.

Starting with trivial gauge field configurations $U_\mu(x,t=0)={\bf 1}$, a complete HMC algorithm \cite{Duane:1987de} can be formulated as follows \cite{Hanada:2016qbz}:
\begin{enumerate}
\item
Throw fictitious momenta $\tilde E_\mu^a$ from a Gaussian random ensemble
\item
Scale momenta as $\tilde E_\mu^a\rightarrow E_\mu^a=\tilde E_\mu^a \sqrt{2N/\beta_L}$
\item
Calculate initial total ``energy'' of the system $H_i=a^4\sum_x {\cal H}_5$
\item
Evolve $E_\mu,U_\mu$ according to the classical equations of motion (\ref{eq:classeoms}) up to fictitious time $t=1$ (as $\Delta t$ is decreased, this will imply a higher number of evolution steps). Calculate the final total energy $H_f=a^4\sum_x {\cal H}_5$ of the system. 
\item
At the end of the evolution, a new configuration $U_\mu(x)$ is accepted with probability $P={\rm min}\left(1,e^{H_i-H_f}\right)$. If the new configuration is rejected, load previous gauge field configuration
\item
Repeat
\end{enumerate}

For a given lattice size, if $\Delta t$ is chosen too large then energy conservation during the classical evolution will be poor, and as a consequence the acceptance probability of new configurations will be low. On the other hand, if $\Delta t$ is chosen too small, then acceptance probability will approach unity and successive gauge field configurations will exhibit strong auto-correlations \cite{DeGrand:2006zz}. Ideally, $\Delta t$ should be chosen such that the acceptance rate is somewhere between 50 and 80 percent.

Having generated a sufficiently large number $N$ of uncorrelated gauge field configurations $U_\mu^{(1)},U_\mu^{(2)},\ldots,U_\mu^{(n)}$, expectation values of observables (\ref{eq:expva}) can be evaluated as
\begin{equation}
\langle {\cal O}\rangle = \frac{1}{N}\sum_{n=1}^{N} {\cal O}^{(n)}\,,
\end{equation}
where ${\cal O}^{(n)}$ corresponds to the observable evaluated using gauge configuration $U_\mu^{(n)}$.

\subsection{Results for the QCD Equation of State}
\label{sec:latticeresults}
\index{Equation of State (EoS)!Lattice QCD}

The previous subsection focused on elucidating how lattice gauge theory is used to obtain results for observables in thermodynamic equilibrium for pure gauge theories. In order to obtain results for QCD, fermions need to be included in the description, which requires techniques in addition to the ones discussed here. However, full QCD lattice calculations are qualitatively similar to pure gauge calculations, albeit computationally much more expensive, in particular for physical (small) quark masses. For this reason, we permit ourselves to highlight some of the major results obtained from full lattice QCD calculations without discussion of how fermionic degrees of freedom are implemented (cf. Refs.~\cite{Montvay:270707,DeGrand:2006zz}).

\begin{figure}[t]
  \begin{center}    
     \includegraphics[width=.49\linewidth]{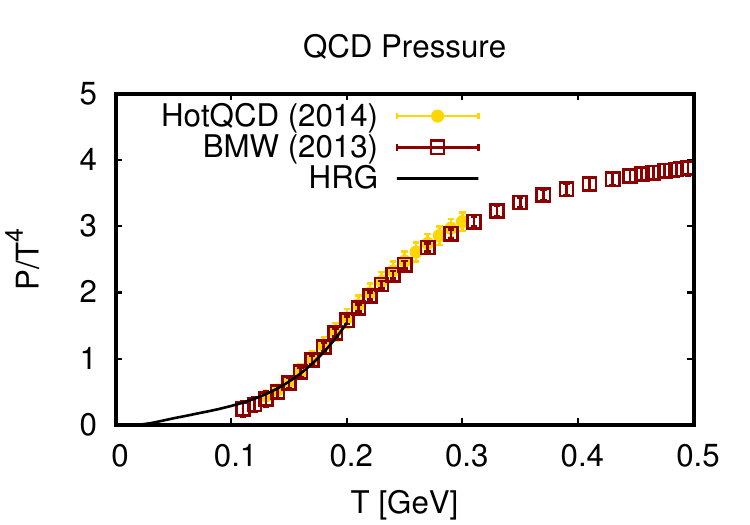}
     \hfill
     \includegraphics[width=.49\linewidth]{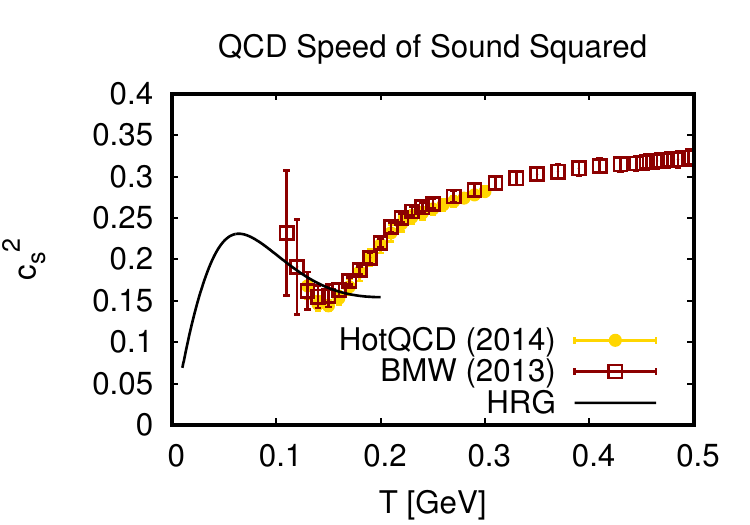}
     \end{center}
  \caption{\label{fig3:P} Normalized QCD pressure (left) and speed of sound squared (right) versus temperature in $d=4$ obtained from lattice QCD calculations at physical quark masses (BMW collaboration \cite{Borsanyi:2010cj,Borsanyi:2013bia} and HotQCD collaboration \cite{Bazavov:2014pvz}). All results are compared to a gas of non-interacting hadron resonances (``HRG'') with masses less than $2.5$ GeV \cite{Olive:2016xmw}.
  }
\end{figure}
\index{Speed of sound}

In Fig.~\ref{fig3:P}, results for the QCD pressure and speed of sound squared are shown from full lattice QCD calculations for physical quark masses as obtained from two independent lattice QCD groups, the BMW collaboration \cite{Borsanyi:2010cj,Borsanyi:2013bia} and the HotQCD collaboration \cite{Bazavov:2014pvz}. As is evident from Fig.~\ref{fig3:P}, historical differences between two collaborations resulting from lattice artifacts have been resolved, and the 2014 HotQCD collaboration results are in full agreement with the results obtained by the BMW collaboration.

The results shown in Fig.~\ref{fig3:P} indicate a significant rise in the normalized QCD pressure, suggesting a rapid increase in the number of degrees of freedom of the system. This behavior is associated with the confinement-deconfinement transition in QCD, where a low-temperature hadronic phase in which QCD color degrees of freedom are confined gives rise to a high-temperature quark-gluon plasma phase with unconfined color degrees of freedom. As suggested by the first-principles lattice QCD results for the pressure shown in Fig.~\ref{fig3:P}, the QCD deconfinement ``transition'' is not an actual transition at all, but has been shown to be an analytic crossover \cite{Aoki:2009sc}. Because the phenomenon of deconfinement does not correspond to an actual transition, there is no unambiguous definition for the ``critical'' temperature $T_c$ at which it occurs. However, one can define a ``critical'' temperature $T_c$ for the confinement-deconfinement transition through the peak of the order parameter (Polyakov loop) associated with the confinement-deconfinement transition in theories that \textit{do} exhibit a phase transition, finding
\begin{equation}
\label{eq:tc}
T_c=170\pm4\ {\rm MeV}\,,
\end{equation}
for QCD at physical quark masses \cite{Aoki:2009sc}. This result for $T_c$ will be referred to as ``critical temperature'' for QCD in the following, even though it should be understood that the QCD deconfinement transition is a broad analytic cross-over.

At low temperatures, the relevant degrees of freedom for QCD are known to be hadronic in nature. Thus, one can expect that a description  in terms of kinetic theory of hadrons should be in reasonable quantitative agreement with lattice QCD results. To this end, results from a gas of non-interacting hadrons and hadron resonances in thermal equilibrium (``hadron resonance gas'' or ``HRG'') are shown in Fig.~\ref{fig3:P} for comparison, which simply correspond to summing Eq.~(\ref{eq:KTthermo}) over all known hadron resonances in the particle data book\footnote{It is straightforward to generalize (\ref{eq:KTthermo}) to bosonic and/or fermionic statistics. However, for the purpose of comparison in Fig.~\ref{fig3:P} these modifications are minor.} \cite{Karsch:2003vd}. One observes that the HRG results shown in  Fig.~\ref{fig3:P} do seem to offer a good quantitative description of lattice QCD results at low temperature \cite{Borsanyi:2011sw,Bazavov:2012jq}. The success of the HRG model can be understood through the fact that it emerges from lattice QCD in the strong coupling limit \cite{Langelage:2010yn}.

At very high temperatures (not shown in Fig.~\ref{fig3:P}),  one observes good agreement between lattice QCD results and resummed perturbative QCD \cite{Blaizot:2000fc,Vuorinen:2003fs,Blaizot:2003iq,Laine:2006cp,Haque:2013sja}.

 \begin{figure}[t]
  \begin{center}    
     \includegraphics[width=.7\linewidth]{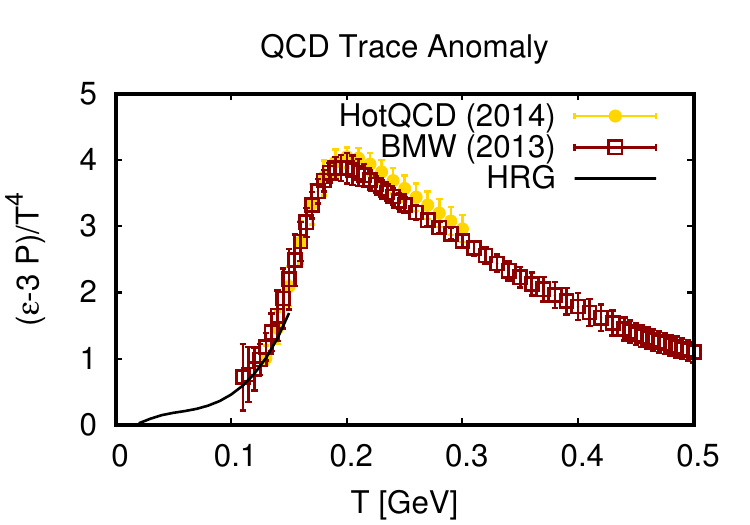}
     \end{center}
  \caption{\label{fig3:trac} QCD trace anomaly $T^\mu_\mu$ in $d=4$ obtained from lattice QCD calculations (BMW collaboration \cite{Borsanyi:2010cj,Borsanyi:2013bia} and HotQCD collaboration \cite{Bazavov:2014pvz}) compared to a gas of non-interacting hadron resonances ('HRG') with masses less than $2.5$ GeV \cite{Olive:2016xmw}.
  }
\end{figure}

While it is possible to use the tabulated lattice QCD results shown in Fig.~\ref{fig3:P} directly, easy-to-use functional representations have been provided by the lattice collaborations. For the BMW collaboration, this functional representation is based on the equilibrium QCD trace anomaly $T^\mu_\mu=\epsilon-3 P$ (see Fig.~\ref{fig3:trac}), and parametrized as
\begin{equation}
  \label{eq:traceanomaly}
 I_{\rm BMW}(T)\equiv \frac{\epsilon-3 P}{T^4}=e^{-h_1/t-h_2/t^2}\left(h_0+\frac{f_0\left({\rm tanh}(f_1 t+f_2)+1\right)}{1+g_1 t+g_2 t^2}\right)\,,
  \end{equation}
  where $t\equiv T/(0.2\, {\rm GeV})$, $h_0=0.1396$, $h_1=-0.18$, $h_2=0.035$, $f_0=1.05$, $f_1=6.39$, $f_2=-4.72$, $g_1=-0.92$, $g_2=0.57$ \cite{Borsanyi:2013bia}. The QCD pressure may be obtained from the trace anomaly by integration,
\begin{equation}
\label{eq:latticeeos1}
\frac{P}{T^4}= \int_{0}^T \frac{dT^\prime}{T^{\prime}}I_{\rm BMW}(T^\prime)\,.
\end{equation}

For the 2014 HotQCD collaboration, the pressure itself has been parametrized as
  \begin{equation}
   \label{eq:latticeeos2}
  \frac{P}{T^4}=
  \frac{1}{2}\left(1+{\rm tanh}(c_t(\bar t-t_0))\right) \frac{p_{\rm id}+a_n/\bar t+b_n/\bar t^2+d_n/\bar t^4}{1+a_d/\bar t+b_d/\bar t^2+d_d/\bar t^4}\,,
  \end{equation}
  where $\bar t=T/(0.154\, {\rm GeV})$, $p_{\rm id}=\frac{95 \pi^2}{180}$, $c_t=3.8706$, $t_0=0.9761$, $a_n=-8.7704$, $b_n=3.9200$, $d_n=0.3419$, $a_d=-1.26$, $b_d=0.8425$, $d_d=-0.0475$  \cite{Bazavov:2014pvz}.
  From the pressure, other thermodynamic quantities may be found through the equilibrium thermodynamic relations (\ref{eq:basicthermo}).

\section{Regime of Applicability of Microscopic Approaches}

The first-principles techniques discussed in the preceding sections each possess their own set of assumptions. These assumptions in turn restrict the regime where each first-principle method can be expected to deliver quantitatively accurate results. Since the regimes of applicability of kinetic theory, gauge/gravity duality, free field theory and lattice gauge theory differ among each other, we summarize this information in this subsection and put it in context with the regime of applicability of fluid dynamics.

 \begin{figure}[t]
  \begin{center}    
     \includegraphics[width=.9\linewidth]{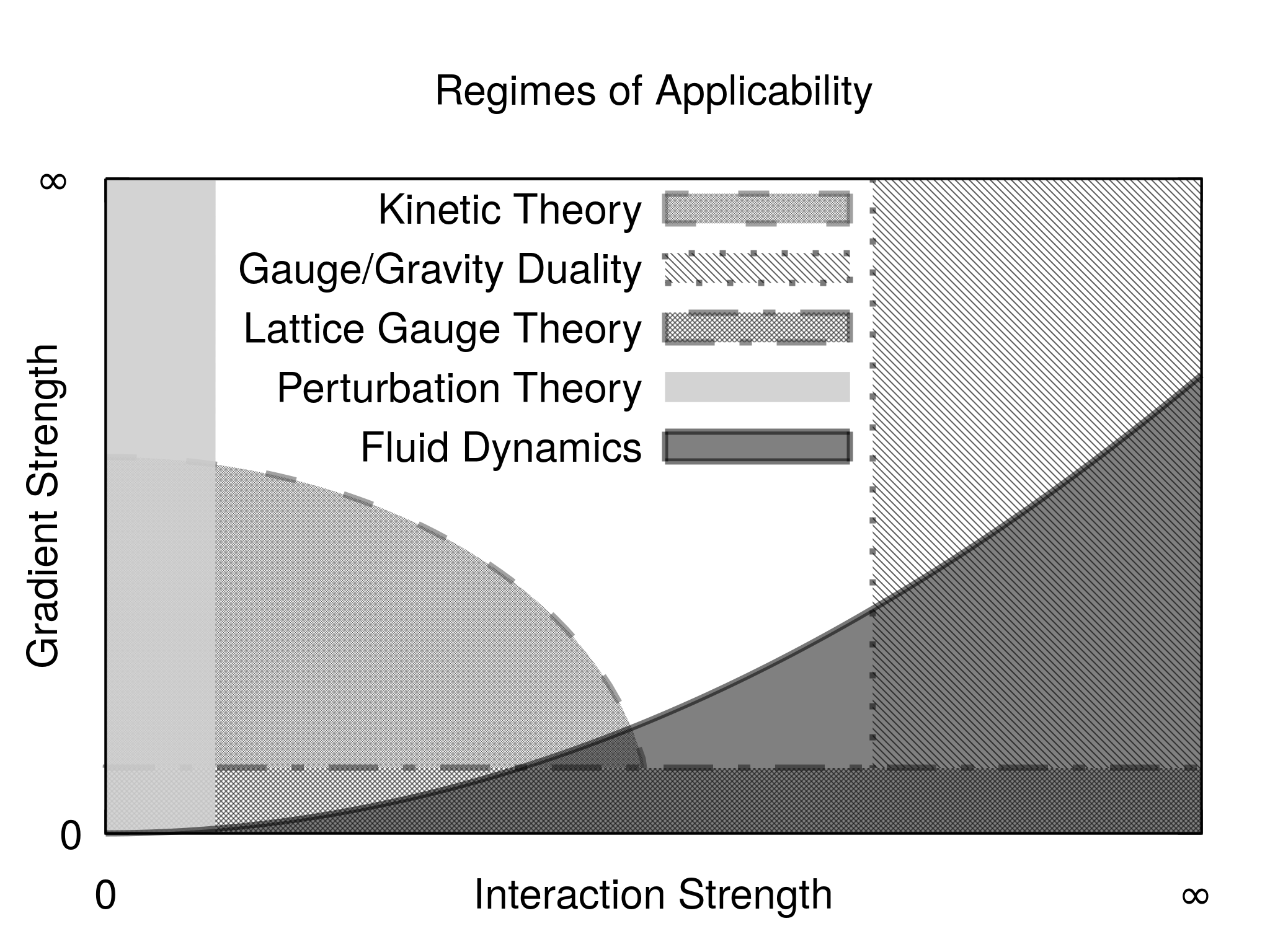}
     \end{center}
  \caption{\label{fig3:apre} Sketch of regime of validity of different approaches in terms of strength of interaction and gradients: kinetic theory, gauge/gravity duality, weak coupling perturbation theory, lattice gauge theory and fluid dynamics. Note that vanishing gradient strength implies equilibrium. Overlapping regimes of validity imply that more than one approach can be used to reliably calculate observables there. Boundaries should be understood to be gradual rather than sharp.
  }
\end{figure}

\begin{itemize}
\item
Kinetic theory assumes the system to be made out of weakly coupled (quasi-)particle degrees of freedom. This limits the applicability of kinetic theory to weak coupling. Furthermore, when deriving kinetic theory from quantum field theory, higher order gradients are neglected, cf. Ref.~\cite{Blaizot:2001nr}. For this reason, kinetic theory is limited to a regime where gradients are not too strong. 
\item
Gauge gravity duality assumes the system to be strongly coupled, while no assumptions about the strength of gradients are made.
\item
Free field theory assumes the coupling of the theory to be exactly zero, but allows arbitrarily strong gradients. Weak coupling perturbation theory allows to extend the regime of applicability to non-zero, but small coupling. 
\item
Lattice gauge theory assumes the system to be in thermal equilibrium, but can provide information about system properties at any interaction strength.
\item
Fluid dynamics applies whenever hydrodynamic modes dominate over non-hydrodynamic modes. For weak coupling, this implies that fluid dynamics can only access regimes close to equilibrium. However, as the coupling strength is increased,  the regime of applicability of fluid dynamics extends to increasingly stronger gradients and thus to systems out of equilibrium.
\end{itemize}

Fig.~\ref{fig3:apre} contains a qualitative sketch of the various regimes of applicability of these approaches as a function of interaction strength and gradient strength. Overlapping regions in Fig.~\ref{fig3:apre} indicate that more than one approach can be used to reliably calculate observables in this region. For instance, at not-to-strong coupling and close to equilibrium, both kinetic theory and lattice gauge theory could reasonably be expected to provide input to fluid dynamics, such as by calculations of transport coefficients. Also, Fig.~\ref{fig3:apre} highlights the existence of a regime where fluid dynamics is expected to be quantitatively applicable, but kinetic theory is not.

\chapter{Simulating Relativistic Nuclear Collisions}
\label{sec:numsim}

The preceding chapters of this work were concerned with
the setup of fluid dynamics theory from first principles,
including its relation to microscopic theories
and calculation of transport coefficients.

This chapter will give an introduction on how to apply the knowledge
thus gained to simulating real-world high-energy nuclear collisions.
As is often the case when aiming for a description of experimental
measurements, some ingredients for the simulations cannot
(yet) be derived from first principles, thus making it necessary
to introduce {\it models}. Not surprisingly, models for individual
simulation components vary greatly in number, sophistication
and reliability. In some cases, two extreme-case models exist that
can be used to bound real-world behavior from opposite sides, such
as for the initial conditions models. In other cases, such as
the hydrodynamic to particle decoupling stage, only one model
with unknown systematic uncertainty exists. Clearly, there is
ample opportunity for improvement.

It should be emphasized that final results obtained
from present-day simulations of nuclear collisions come with
two sets of systematic uncertainties. On the one hand,
there are uncertainties from applying fluid dynamics
itself to an out-of-equilibrium system using a numerical
algorithm. These uncertainties can in practice be at least
estimated using knowledge on non-hydrodynamic modes gained
in the previous chapters, as well as numerical convergence
tests. On the other hand, there are uncertainties
originating from using various model components, which
tend to dwarf the fluid theory uncertainties just mentioned.
Unfortunately, these model component uncertainties
are essentially ignored in present-day comparisons to experimental data.

As will be pointed out below, the biggest uncertainties in
present day nuclear-collision fluid simulations arise from
initial condition models as well as fluid-particle
decoupling models. At present, these uncertainties constitute
a road-block in the current effort of making high-energy nuclear
physics a precision endeavor. However, the situation also constitutes
an opportunity in the sense that new ideas for first-principle
calculations to replace model components, or alternatively
data-driven model validation, would guarantee instant
success and fame in the high-energy nuclear physics community.
We hope that some readers of this work therefore would consider
taking on the challenges.

\section{Brief Introduction to Relativistic Nuclear Collisions}
\label{sec:hicintro}

Collisions of nuclei are said to be relativistic if the center-of-mass
collision energy is larger than the rest mass of the nuclei,
implying that the nuclei move with speeds that are a considerable
fraction of the speed of light. For historical reasons,
one distinguishes between heavy-ion\footnote{A heavy ion is an atom heavier than helium stripped of its electrons.} collisions,
asymmetric collisions (such as heavy-on-light-ion
collisions) and collisions of protons. Considering nuclei
 to be made up of individual nucleons (protons and neutrons),
 the relevant relativistic nuclear collision parameter is the center-of-mass collision energy per nucleon pair: $\sqrt{s}$. In terms of this parameter,
 the typical Lorentz contraction factor $\gamma$ for the nucleons can be approximated as
 \begin{equation}
   \gamma\simeq \frac{\sqrt{s}}{2 m_p}\simeq \frac{\sqrt{s}}{2\, {\rm GeV}}\,,
 \end{equation}
 where $m_p$ is the proton mass and $1\, {\rm GeV}=10^9\, {\rm eV}$ is the
 energy unit of choice for high energy physics. Present-day experiments
 achieve collision energies in the TeV range, such that $\gamma \gg 1$. This implies that nucleons have speeds very close to the speed of light, and the collision energy is dominated by kinetic energy, not rest mass energy.

 The high Lorentz contraction factors imply that nuclei in the center-of-mass frame are severely distorted as compared to their rest configuration. A picture often employed thus compares observing a nucleus moving at relativistic speed to a ``pancake'' rather than an approximately spherical object.

 \begin{figure}[t]
  \begin{center}    
     \includegraphics[width=.7\linewidth]{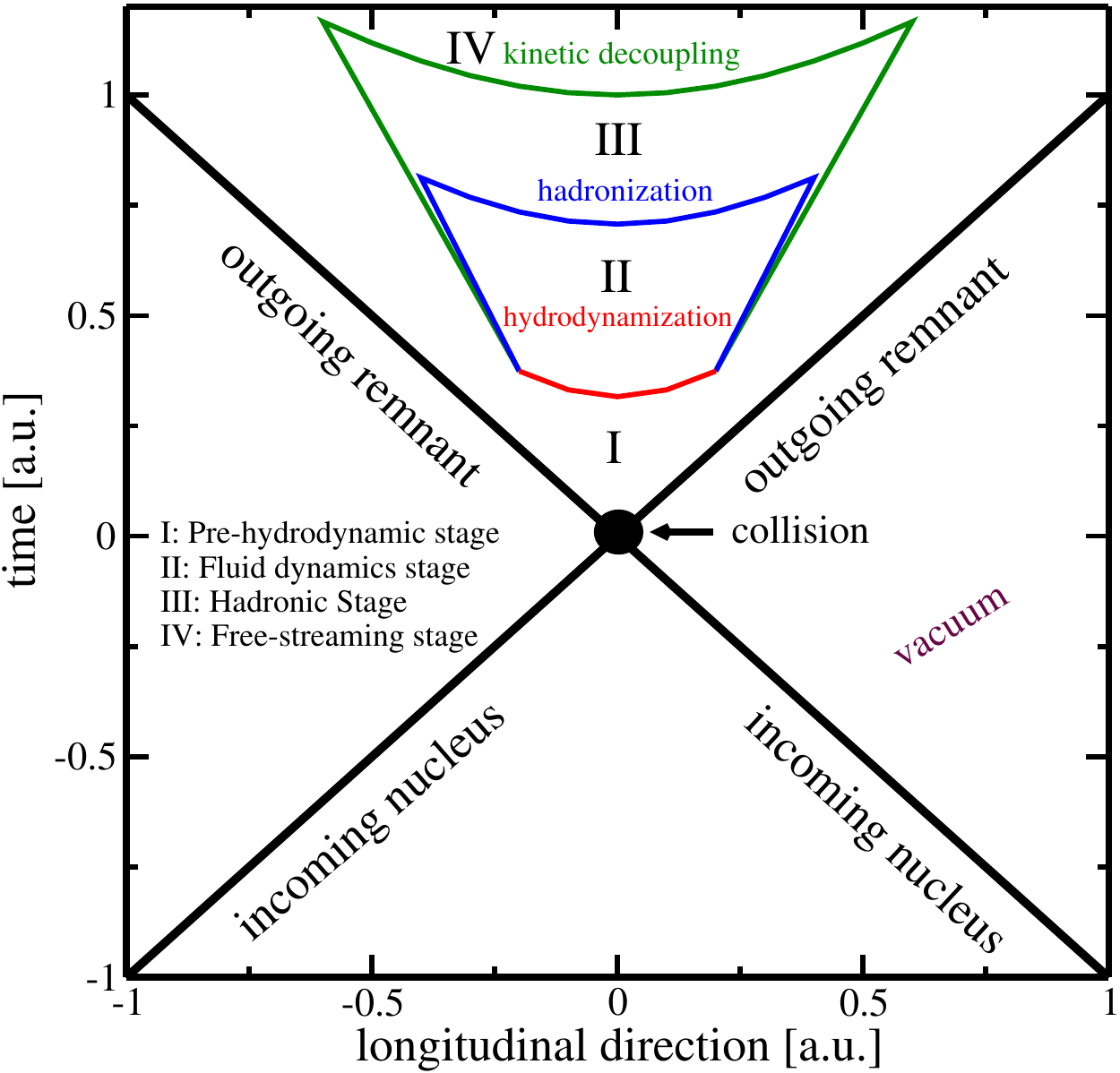}
     \end{center}
  \caption{\label{fig4:one} Schematic sketch of a relativistic nuclear collision. Lines represent boundaries of various stages for the evolution of matter, characterized by a far-from-equilibrium region ("pre-hydrodynamics"), the onset of fluid dynamics (``hydrodynamization''), fluid-particle conversion (``hadronization'') and kinetic decoupling.}
\end{figure}
 
 A schematic view of a nuclear collision is shown in Fig.~\ref{fig4:one}. Before the collision, nuclei approach each other on the light-cone (defined by positions \hbox{$z=\pm t$}). The collision is taking place at $t=z=0$, and any remnants of the nuclei that do not take part in the collision process leave the collision region on the forward light-cone. Interacting matter will lead to energy deposition in the center of the collision region, close to $z\simeq 0$. Initially, this deposited energy will be far from equilibrium, denoted as stage I in figure \ref{fig4:one}. However, as the matter spreads into the vacuum, it expands, cools and interacts, which leads to a decrease of gradients. Even if the matter does not come close to equilibrium, its bulk evolution can be described by fluid dynamics after this ``hydrodynamization'' time, cf. the discussion in section \ref{sec:faraway}. In stage II of Fig.~\ref{fig4:one}, the system continues to expand and cool according to the laws of relativistic fluid dynamics, until the local energy density will come close to the transition energy density of the QCD cross-over (\ref{eq:tc}). At energy densities below the QCD cross-over transition, the relevant degrees of freedom are hadrons, predominantly pions, kaons and protons. QCD interactions will confine any colored degrees of freedom into hadrons at energy densities below the critical energy density (``hadronization''). It is in principle possible that fluid dynamics could be applicable in the hadronic stage, but this is unlikely because of the phenomenon of cavitation, see section \ref{sec:hadro1}. 


Even after hadronization, hadrons will interact significantly through scattering. The corresponding dynamics can be captured by particle-cascade simulations, and constitutes stage III in Fig.~\ref{fig4:one}. Since the matter keeps expanding, particle densities drop, and eventually all particle interactions cease, signaling what is known as kinetic decoupling. In the final stage IV in Fig.`\ref{fig4:one}, particles travel on straight-line trajectories (``free-streaming'') until they are registered by experimental detectors.

\section{Overview of Simulation Components}
\label{sec:overview}

Modern simulations of relativistic nuclear collisions consist of a host of interconnected modules designed to describe the various physics phenomena as realistically and accurately as possible.

The main components of a simulation package are:
\begin{itemize}
\item
Model for the geometry of the nuclei before the collision, either on the level of nucleons or including sub-nucleonic degrees of freedom. See for instance sections \ref{sec:geo1},\ref{sec:geo2}. Modern geometry models implement geometry fluctuations through event-by-event Monte-Carlo sampling, cf. section \ref{sec:MC}.
\item
Model for the energy distribution deposited in a nuclear collision, such as the Glauber model discussed in section \ref{sec:MCGlauber}. First-principles theory guidance for the energy deposition is available at extremely weak and extremely strong interaction (see sections \ref{sec:weak},\ref{sec:strong}). 
\item
Model for the evolution of the system in the pre-hydrodynamic stage. Examples include the zero-flow assumption (fluid velocities vanish at the start of hydrodynamics), extremely weak coupling (see Eq.~(\ref{eq:preeq1}) for IP-Glasma) or extremely strong coupling (see Eq.~(\ref{eq:preeq2}) for AdS/CFT).
\item
Numerical algorithm to solve the relativistic viscous hydrodynamics equations of motion, see section \ref{sec:numalgo}.
\item
Inputs for the transport coefficients (and their temperature dependence) in the simulation, such as speed of sound (equation of state), shear and bulk viscosity and relaxation times, see sections \ref{sec:workhorse}, \ref{sec:nonconfwork}.
\item
Prescription of how to convert fluid degrees of freedom into hadrons, see section \ref{sec:hadro}.
\item
Numerical algorithm to simulate hadron interactions until kinetic freeze-out, see section \ref{sec:cascade}
\item
Data analysis of hadrons including proper averaging over Monte-Carlo generated events. Calculation of observables comparable to experimental data, see section \ref{sec:obs}
\end{itemize}
Many different implementations of these components exist, with different levels of assumptions and accuracy. Since the prospect of implementing all of these components from scratch could be intimidating for newcomers, it should be noted that several complete packages are publicly available, such as MUSIC \cite{MUSIC}, superSONIC \cite{SONIC} and VISHNU \cite{VISHNU}. Different modules will be discussed in detail below.

\section{Initial Condition Models}
\label{sec:Igeo}

\subsection{Geometry Models for Nuclei}
\label{sec:geo1}

In order to study nuclear collisions, it is necessary to have
a baseline model of the nuclear geometry. Fortunately,
elastic electron scattering provides an excellent experimental
probe on the electric charge density distribution $\rho({\bf x})$
of nuclei. For large spherically symmetric nuclei, experimental
electron scattering data is well described by a so-called
Woods-Saxon (or two parameter Fermi) distribution \cite{WoodsSaxon}
\begin{equation}
  \label{eq:WS} \index{Woods-Saxon distribution}
  \rho_A({\bf x})=\frac{\rho_0}{1+e^{(|{\bf x}|-r_0)/a_0}}\,,\quad A\gg 1
\end{equation} 
where the coordinate system ${\bf x}$ is centered around the nucleus. Here $r_0$ is the nuclear radius parameter, $a_0$ is the nuclear skin thickness parameter, $A$ denotes the atomic weight of the nucleus and $\rho_0$ is an overall constant that is conveniently fixed  by requiring $\int\, d^3{\bf x} \rho_A({\bf x})=1$.
Tabulated results for $r_0,a_0$ for different nuclei can e.g. be found in Refs.~\cite{DeJager1974479,DeVries1987495}, cf. Tab.~\ref{tab4:one}.
In the Woods-Saxon model, the charge distribution $\rho({\bf x})$ is taken to describe the distribution of \textit{nucleons} (both protons and neutrons) inside the nucleus, cf. Fig.~\ref{fig4:two}.

For a nucleus moving with relativistic speeds in the laboratory frame, the Woods-Saxon nucleon distribution (\ref{eq:WS}) will appear highly Lorentz contracted. The direction of the nucleus in the center-of-mass frame is referred to as the beam axis, and is conventionally taken to be the longitudinal $z$ axis, whereas the transverse directions spanned by ${\bf x}_\perp=(x,y)$ are referred to as the transverse plane. \index{Transverse plane}
For relativistic nuclei,  it is useful to define the nuclear thickness function
\begin{equation}
  \label{eq:TA} \index{Nuclear thickness function}
  T_A({\bf x}_\perp)=\int_{-\infty}^\infty dz\, \rho_A({\bf x})\,,
\end{equation}
which corresponds to squeezing the nuclear density distribution into a two-dimensional sheet, which is a good approximation for $\gamma \gg 1$. The nuclear thickness function has the interpretation of probability per unit area of finding a nucleon at a position ${\bf x}_\perp$ away from the center of the nucleus.

\begin{table}
  \centering
\begin{tabular}{cccc}
  \hline
 Isotope  & $r_0$ [fm] & $a_0$ [fm]\\
 \hline
 $^{27}$Al & 3.07(9) & 0.519(26)\\
 $^{63}$Cu & 4.163(27) & 0.606(11)\\
  $^{197}$Au & 6.38(6) & 0.535(27)\\
  $^{208}$Pb & 6.62(6) & 0.546(10)\\
  \hline
\end{tabular}
\caption{\label{tab4:one} Table of Wood-Saxon parameters for some nuclear isotopes, cf.~\cite{DeVries1987495}.}
\end{table}

For light ions, the distribution (\ref{eq:WS}) no longer offers a good description of electron scattering data or nuclear positions. In the case of the deuteron, nucleon positions can be sampled from the analytically known deuteron wave-function. If one nucleon of the deuteron is located at the origin of the coordinate system, the probability distribution of finding the second nucleon a distance $r_{np}$ away from the first nucleon is is given by the Hulth\'en wave function \cite{Hulten:1957} squared,
\begin{equation}
  \rho_{A=2}({\bf x})=\rho_0
  \frac{\left(e^{-\alpha r}-e^{-\beta r}\right)^2}{r^2}\,,
\end{equation}
with $\alpha=0.228$ fm$^{-1}$ and $\beta=1.18$ fm$^{-1}$ \cite{Miller:2007ri}. 

In the case of $^3$He, numerical results for $\rho_A({\bf x})$ from modern nuclear structure calculations can be employed, cf.~\cite{Carlson:1997qn,Nagle:2013lja}.

\subsection{Nuclear Geometries from Monte-Carlo Sampling}
\label{sec:MC}
\index{Monte-Carlo sampling}

Once a probability distribution function is known, one can generate individual realizations of nucleon positions (``events'') by performing a Monte-Carlo sample of this distribution. But how does one do this in practice?

As an example, let us consider the Woods-Saxon distribution function Eq.~(\ref{eq:WS}) for a gold nucleus $A=197$. We want to interpret $\rho_{A}$ as a probability distribution function for a nucleon at a distance $|{\bf x}|=r$ away from the center of the nucleus. The requirement $\int d^3{\bf x}\rho_A({\bf x})=1$ implies $\rho_0=8.6\times 10^{-4}$ with values for $a_0,r_0$ for $^{197}$Au given in Tab.~\ref{tab4:one}. To perform the Monte-Carlo sampling, we need a (pseudo-) random number generator which produces random numbers $u$ uniformly in the interval $u\in[0,1]$.

\begin{figure}[t]
  \begin{center}
    \hspace*{-1cm}
    \includegraphics[width=.64\linewidth]{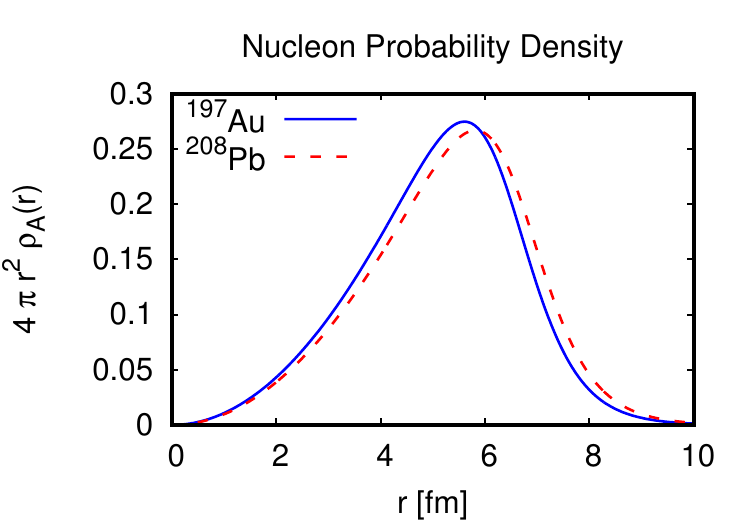}\hspace*{-1.5cm}
      \includegraphics[width=.64\linewidth]{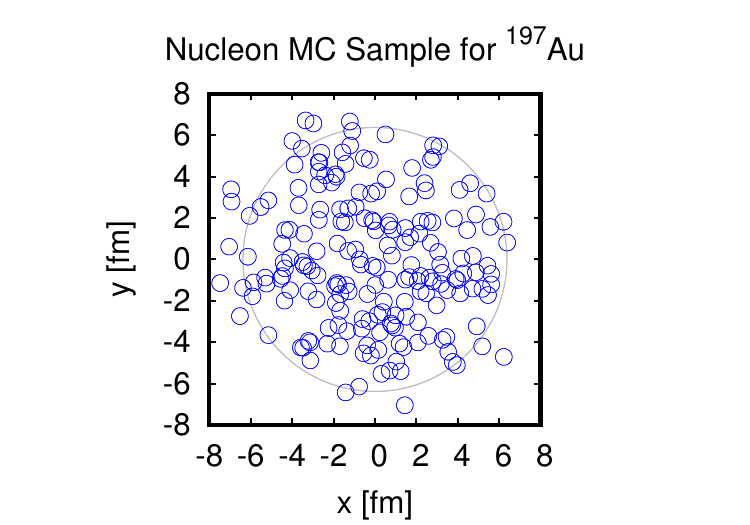}\hfill
     \end{center}
  \caption{\label{fig4:two} Left: Nucleon probability density from Woods-Saxon profile for gold and lead nuclei. Right: a single Monte-Carlo (MC) sample for nucleon positions for a $^{197}$Au nucleus. Nucleon positions have been projected to the transverse plane and are shown as circles with radius of $R=0.4$ fm. As a guide to the eye, the gold radius $r=r_0$ from Tab.~\ref{tab4:one} is indicated by a large circle.}
\end{figure}

To draw a sample from the probability distribution function density $\rho_A({\bf x})$ in Eq.~(\ref{eq:WS}), throw four random numbers $u_1,u_2,u_3,u_4$ using the uniform random number generator. The first three random numbers are used to pick a random position for a candidate nucleon in the nucleus as
\begin{equation}
  \label{eq:Mc1}
r=R u_1,\quad \theta={\rm arccos} (2 u_2-1),\quad \phi=2 \pi u_3\,,
\end{equation}
where spherical coordinates $r,\theta,\phi$ were used. In (\ref{eq:Mc1}), $R$ is the cutoff scale for the radial position of a nucleon beyond which no nucleon position will be allowed; in principle, this value should be set to infinity, but this would make the sampling procedure inefficient. Inspecting the probability density distribution (\ref{eq:WS}), it is straightforward to see that --- while not exactly zero --- the probability of having a nucleon at position $R\gg r_0$ is exponentially small. For instance, $R\simeq 3r_0$ is a good choice in practice. With the candidate nucleon position given by (\ref{eq:Mc1}), Monte-Carlo sampling is performed by accepting the candidate nucleon if
\begin{equation}
  u_4\leq r^2\rho_A(r)\,,
\end{equation}
and rejecting the candidate nucleon otherwise. If the candidate nucleon is rejected, one throws again four random numbers and repeats the above steps. If the candidate nucleon is accepted, its position is stored and one proceeds to repeat the above procedure to generate the next candidate nucleon until a total number of $A$ nucleons have been accepted. The nucleon positions thus generated will in general not correspond to the center of mass frame. One can re-center the coordinate system by subtracting ${\bf x}_\perp^{\rm center}=\frac{1}{A}\sum_{i=1}^A (x_i,y_i)$ from each of the $i$ nucleons. However, for $A\gg 1$, in practice this will only involve a modest change in nucleon positions.

The set of $A$ nucleons then corresponds to a single Monte-Carlo sample of a gold nucleus from the probability density distribution $\rho_A(|{\bf x}|)$. A representative $^{197}$Au  Monte-Carlo sample obtained by the above procedure is shown in Fig.~\ref{fig4:two}. Similarly, the nuclear thickness function $T_A({\bf x}_\perp)$ may for instance be obtained by summing over all nucleon positions in the transverse plane with a Gaussian weight, i.e.
\begin{equation}
\label{eq:glauberta}
T_A({\bf x}_\perp)\propto \sum_{i=1}^A e^{-\frac{({\bf x}_\perp-{\bf x}_\perp^{(i)})^2}{2 R^2}}\,.
\end{equation}

\subsection{Sub-Nucleonic Geometry Models}
\label{sec:geo2}

The discussion above focused on nuclear geometry models based on constituent nucleons. However, especially for light nuclei, in particular protons, it will turn out to be essential to allow for the description of geometry on sub-nucleonic scales. 

The study of nuclear geometries at sub-nucleonic scales is still in its infancy, but at a qualitative level one expects partonic excitations such as quarks and gluons to lead to local distortions from a mean spherical nucleon geometry. But how many partons should be included in the description, and what constitutes a good model for their distribution inside the nucleon?

\begin{figure}[t]
  \begin{center}
    \hspace*{-1cm}
    \includegraphics[width=.64\linewidth]{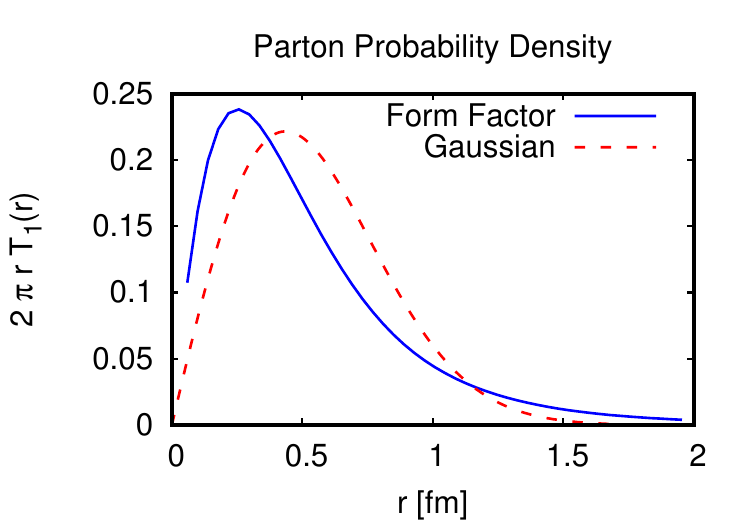}\hspace*{-1cm}
      \includegraphics[width=.64\linewidth]{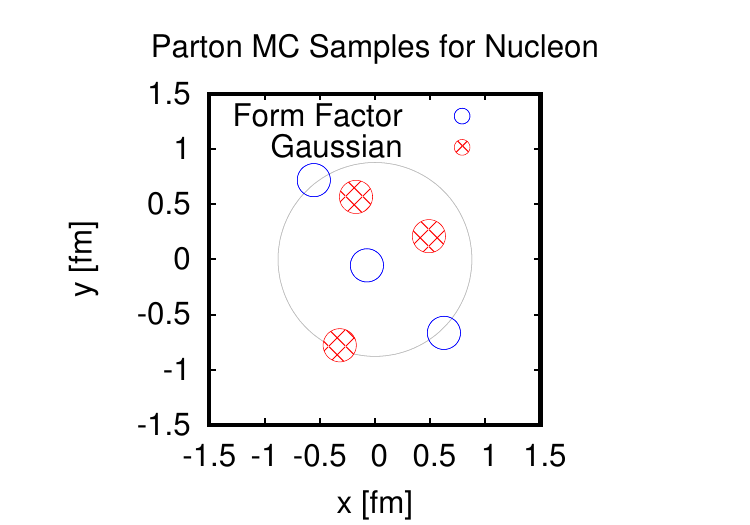}\hfill
     \end{center}
  \caption{\label{fig4:three} Left: Parton transverse probability density from the proton form factor and a simple Gaussian distribution, respectively. 
    Right: a single Monte-Carlo (MC) sample for the positions of three partons in a nucleon, again for the form factor and simple Gaussian distribution. Parton positions are shown as circles with radius of $R=0.15$ fm. As a guide to the eye, the proton radius $r=0.878$ fm is indicated by a large circle.}
\end{figure}

One could try to copy the approach for nuclei from above by using the measured elastic (electric plus magnetic) form factor $F(Q^2)$ of the proton to define a nuclear thickness function
\begin{equation}
  \label{eq:protonTA}
  T_{A=1}^{{\rm form\ factor}}({\bf x}_\perp)=\rho_0\int \frac{d^2q}{(2\pi)^2} e^{-i {\bf q}\cdot {\bf x_\perp}} F\left(Q^2={\bf q}^2\right)\,,
  \end{equation}
where a convenient parametrization of $F(Q^2)$ can e.g. be found in Ref.~\cite{Venkat:2010by}. A reasonable fit to Eq.~(\ref{eq:protonTA}) is provided by
an exponential parametrization,
\begin{equation}
  T_{A=1}^{\rm form\ factor}({\bf x}_\perp)\simeq \rho_0 e^{-|{\bf x}_\perp|/\alpha}\,,
\end{equation}
where $\alpha\simeq 0.263$ fm. However, often a much simpler Gaussian model
\begin{equation}
  \label{eq:protonTAG}
  T_{A=1}({\bf x}_\perp)=\rho_0 e^{-\frac{{\bf x}_\perp^2}{2 R^2}}\,,
\end{equation}
is considered where $R$ corresponds to the width of the nucleon. In order to retain some constraint from electron scattering data, one can consider setting $R\simeq 0.436$ fm such that the mean square transverse radius from Eqns.~(\ref{eq:protonTA}) and (\ref{eq:protonTAG}) agree:
\begin{equation}
  \int d^2{\bf x}_\perp \left(T_{A=1}^{{\rm form\ factor}}-T_{A=1}\right) {\bf x}_\perp^2=0\,.
  \end{equation}
Using this choice, the form factor and simple Gaussian parametrization for the thickness function of the nucleon are shown in Fig.~\ref{fig4:three}. See e.g. Ref.~\cite{Heinz:2011mh} for other choices of $R$.

As an example for sub-nucleonic geometries, let us consider nucleons to be made out of three partons (e.g. constituent quarks)\footnote{There is no fundamental reason why the constituent number in such a model has to be three, and in fact it could equally well be taken to be two, four or 27.}. To generate a sample configuration of such a nucleon, one can perform a Monte-Carlo sampling of the simple Gaussian $T_{A=1}$ distribution with three partons. Once a sample of three parton positions has been generated, it is important to re-center the coordinate system unlike what was the case for sampling nucleon configurations for a nucleus with $A\gg 1$.

Monte-Carlo samples of a single nucleon configuration obtained from (\ref{eq:protonTA}) and (\ref{eq:protonTAG}) are shown in Fig.~\ref{fig4:three}. One finds that parton distributions look generically similar despite the qualitatively different form of the distributions (\ref{eq:protonTA}) and (\ref{eq:protonTAG}), because the transverse probability density $r T_{A=1}(r)$ only differs by the location of the peak probability, see the left panel of Fig.~\ref{fig4:three}. Compared to nucleon position distributions shown in Fig.~\ref{fig4:two}, the parton distributions in the nucleon in Fig.~\ref{fig4:three} exhibit considerably stronger fluctuations.

It is of course possible to combine the parton and nucleon distributions and e.g. obtain a sampling of parton positions in a $^{197}$Au nucleus, thus providing a unified framework of nuclear geometry from protons to heavy ions, cf.~\cite{Welsh:2016siu,Weller:2017tsr}.

\section{The Glauber Model for Nuclear Collisions}
\label{sec:MCGlauber}
\index{Glauber Model}

The Glauber model \cite{Glauber,Glauber:2006gd} for relativistic nuclear collisions is a phenomenological model for the energy (or entropy) deposited in the collision process. The model does not attempt to describe the collision process itself, but instead provides initial conditions for the subsequent fluid-dynamic evolution. 

The two main variants of the Glauber model are the ``optical'' Glauber model, where the nucleus is treated as uniform, and the Monte-Carlo Glauber model, where the nucleus is treated as a collection of nucleons with positions sampled by a Monte-Carlo procedure \cite{Miller:2007ri}. While much of the simulation modeling of relativistic nucleus collisions has been based on the optical Glauber model, it is now considered somewhat outdated, which is why only the Monte-Carlo Glauber model will be described in the following.

\subsection{Glauber Model with Disk-Like Nucleons}

Let us consider the collision process of two nuclei with atomic weights $A$ and $B$ at a center-of-mass energy $\sqrt{s}$ and at a given impact parameter ${\bf b}_\perp$ in the transverse plane. The position of nucleons inside each nucleus can be sampled from the Woods-Saxon distribution (\ref{eq:WS}) via Monte-Carlo sampling. In the simplest version of the Glauber model, a collision between individual nucleons takes place if their distance $|{\bf x}_\perp|$ in the transverse plane satisfies
\begin{equation}
\label{eq:glaubercrit}
|{\bf x}_\perp|<\sqrt{\frac{\sigma_{NN}(\sqrt{s})}{\pi}}
\end{equation}
where $\sigma_{NN}(\sqrt{s})$ is the total inelastic nucleon-nucleon cross section at the respective center-of-mass collision energy (cf. Tab. \ref{tab:sigma}). The condition (\ref{eq:glaubercrit}) is a purely geometrical constraint, with nucleons modeled as uniform ``disks'', as opposed to modeling nucleons as having a radius-dependent probability profile for scattering, cf. section \ref{sec:ppglauber}.

\begin{table}
\begin{center}
\begin{tabular}{|c|ccccccc|}
\hline
$\sqrt{s}$ [GeV] & 19.6 & 62.4 & 130 & 200 & 2760 & 7000 & 13000\\
\hline
$\sigma_{NN}$ [mb] & 32.3 & 35.6 & 40 & 42 & 60 & 72 & 78.1 \\
\hline
$\lambda_{NN}$ [mb] & 5  &  5.5     &   6.5    & 7 & 13 & 18.2 & 22.5\\
\hline
\end{tabular}
\end{center}
\caption{\label{tab:sigma} Values for the total inelastic nucleon-nucleon cross section $\sigma_{NN}$ in millibarns as a function of center of mass collision energy $\sqrt{s}$ from Ref.~\cite{Miller:2007ri,Aad:2011eu,Aaboud:2016mmw}. Also shown are inferred values for the parameter $\lambda_{NN}$ in Eq.~(\ref{eq:g2prob}) for $b_{\rm cutoff}=8$ fm, $R=0.52$ fm and $\sigma_g=0.46$ fm. }
\end{table}

\begin{figure}[t]
  \begin{center}    
     \includegraphics[width=.7\linewidth]{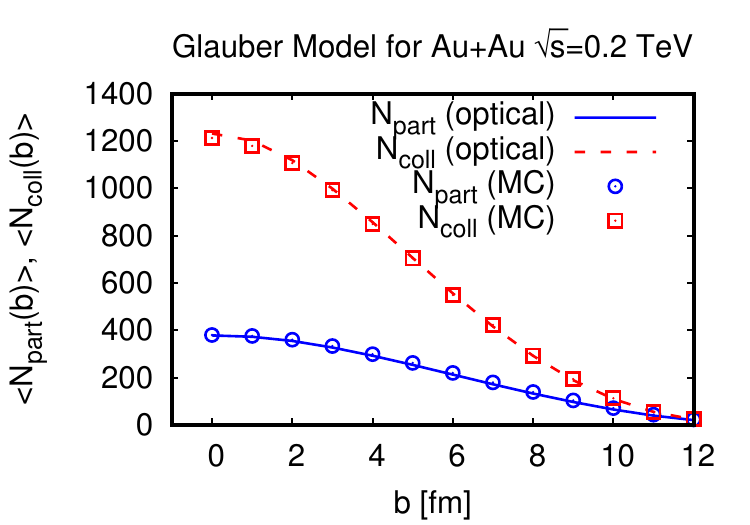}
     \end{center}
  \caption{\label{fig4:Npart} Number of participants and number of collisions as a function of impact parameter in the Glauber model for a Au+Au collision at $\sqrt{s}=0.2$ TeV.}
\end{figure}

Two quantities that can be calculated within the Monte-Carlo Glauber model are the number of participants $N_{\rm part}$ and the number of collisions $N_{\rm coll}$. These quantities are not directly measurable, but are often quoted as reference in experimental studies. The number of participants is simply the number of nucleons which have undergone at least one collision according to the criterion (\ref{eq:glaubercrit}), whereas the number of collisions is the total number of nucleon-nucleon binary collisions that have occurred (recall that in a nucleus Eq.~(\ref{eq:glaubercrit}) may allow a single nucleon to have multiple collisions). It turns out that for large nuclei $A\gg 1, B\gg1$, $N_{\rm part}$ and $N_{\rm coll}$ for a collision at impact parameter ${\bf b}_\perp$ are well approximated by the (optical Glauber) relations \cite{Miller:2007ri}
\begin{eqnarray}
\label{eq:optGlauber}
N_{\rm part}({\bf b}_\perp)&\simeq& A \int d^2{\bf x}_\perp T_A({\bf x}_\perp)\left[1-\left(1-T_B({\bf x}_\perp-{\bf b}_\perp)\sigma_{NN}\right)^B\right]\nonumber\\
&&+B \int d^2{\bf x}_\perp T_B({\bf x}_\perp-{\bf b}_\perp)\left[1-\left(1-T_A({\bf x}_\perp)\sigma_{NN}\right)^A\right]\,,\nonumber\\
N_{\rm coll}({\bf b}_\perp)&\simeq& A B \int d^2{\bf x}_\perp T_A({\bf x}_\perp)T_B({\bf x}_\perp-{\bf b}_\perp) \sigma_{NN}\,.
\end{eqnarray}
This similarity is demonstrated in Fig.~\ref{fig4:Npart} where results from Eq.~(\ref{eq:optGlauber}) at various impact parameter values $b=|{\bf b}_\perp|$ are compared to $\langle N_{\rm part}(b)\rangle ,\langle N_{\rm coll}(b)\rangle$ \textit{event-averaged} over 100 Monte-Carlo events at each value of $b$ for a Au+Au collision at $\sqrt{s}=0.2$ TeV.

In the Monte-Carlo Glauber model, it is also possible to keep track of participants and number of collisions locally, corresponding to participant and collision density
\begin{equation}
n_{\rm part}({\bf x_\perp})\equiv \frac{dN_{\rm part}}{d^2{\bf x_\perp}},\quad
n_{\rm coll}({\bf x_\perp})\equiv \frac{dN_{\rm coll}}{d^2{\bf x_\perp}}\,.
\end{equation}
In the Monte-Carlo approach, this is most easily achieved by tracking the location of nucleons undergoing a collision (or those that participate) and then depositing a Gaussian (\ref{eq:protonTAG}) with adjustable width $R$ at each of those nucleon locations. The participant and collision density then can be calculated by summing up the individual Gaussian contributions. (Alternatively, variants of the model have been proposed where deposition occurs at the mid-point in between participants/collisions).

Given these quantities, the Glauber model for relativistic nuclear collisions then consists of assuming that the energy density $\epsilon$ (or, in a different variant, the pseudo-entropy density $s$) deposited in the collision process is proportional to a combination of $n_{\rm part},n_{\rm coll}$, e.g. through a two-component model \cite{Kharzeev:2000ph,Kolb:2001qz}
\begin{equation}
\label{eq:2comp}
\epsilon({\bf x_\perp})\ {\rm or}\ s({\bf x_\perp}) \propto n_{\rm part}(1-\kappa)({\bf x_\perp})+\kappa\, n_{\rm coll}({\bf x_\perp})\,,
\end{equation}
where $\kappa\in [0,1]$ is a parameter controlling the relative amount of participant and binary collision contributions, respectively\footnote{This purely phenomenological parameter should not be confused with the second order transport coefficient also denoted as $\kappa$, see e.g. Tab.~\ref{tab:one2}.}. In the Glauber model, there is no obvious way to access information about the longitudinal (rapidity) dependence of quantities, so the model is generically used in conjunction with Bjorken's boost-invariance assumption (see section \ref{sec:idealbjork}). As an example, the energy-density distribution $\epsilon({\bf x}_\perp)$ in the Monte-Carlo Glauber model is shown in Fig.~\ref{fig4:ED} for a single Au+Au collision event at $\sqrt{s}=0.2$ TeV and $b=0$ using $n_{\rm coll}$ scaling ($\kappa=1$) by placing Gaussians (\ref{eq:protonTAG}) with width $R=0.4$ fm at the position of each nucleon that experiences a collision. As can be seen in Fig.~\ref{fig4:ED}, this results in an energy-density distribution that is not azimuthally symmetric even at zero impact parameter. One recovers azimuthal symmetry at zero impact parameter when averaging $\epsilon({\bf x}_\perp)$ over many events. However, observables that are sensitive to azimuthal asymmetries will in general be sensitive to such event-by-event fluctuations (see sections \ref{sec:eccs},\ref{sec:obs}).

The phenomenological Glauber model (\ref{eq:2comp}) has been tremendously successful in describing relativistic nuclear collision data, which is why it is considered as a baseline in many more elaborate models of initial conditions. Some of this success will be reviewed in chapter \ref{chap:experiment}. However, given its phenomenological nature, considerable efforts have been taken to put the collision process and the mechanism of energy deposition on more solid theoretical footing, which will be discussed in the following sections.

\subsection{Glauber Model for Sub-Nucleonic Constituents}
\label{sec:ppglauber}
\index{Glauber Model! with partons}

Despite the phenomenological success of the nucleon-based Glauber model described in the previous section, recent experimental data in proton-proton collisions suggests that sub-nucleonic degrees of freedom should be included in the description, cf. section \ref{sec:geo2}. In addition, it has been noted that the simple Glauber model with disk-like nucleons does not accurately capture experimental results for rare fluctuations. As a remedy for both, let us thus discuss a modified version of the nucleon-based Glauber model.

Let us consider again the collision process of two nuclei with atomic weights $A$ and $B$ at a center-of-mass energy $\sqrt{s}$. The impact parameter ${\bf b}_\perp$ of the collision in the transverse plane is obtained by sampling ${\bf b}_\perp^2$ from a uniform probability distribution in the interval $[0,b_{\rm cutoff}^2]$, where $b_{\rm cutoff}$ is chosen such that collisions with $b>b_{\rm cutoff}$ have negligible probability. For each nucleus, the position of its nucleons are obtained from Monte-Carlo sampling Eq.~(\ref{eq:WS}). In a departure from the standard collision criterion (\ref{eq:glaubercrit}), however, let us now pose that a collision between two nucleons $i\in A$ and $j\in B$ occurs with probability
\begin{equation}
\label{eq:g2prob}
P_{\rm collision}^{i,j}=1-e^{-\lambda_{NN} C^{i,j}_{NN}}\,,
\end{equation}
where $C_{NN}^{i,j}$ is similar to an overlap function of nucleons $i$ and $j$. Here $\lambda_{NN}$ is a parameter that is fixed by requiring that the expectation value of the collision probability for proton-proton collisions satisfies $\langle P_{\rm collision}\rangle=\frac{\sigma_{NN}}{\pi b_{\rm cutoff}^2}$, where $\sigma_{NN}$ is the inelastic nucleon-nucleon cross section, cf. Table \ref{tab:sigma}.

In addition, let us assume that individual nucleons consist of three partons (cf. the discussion in section \ref{sec:geo2}). The collision function $C_{NN}^{i,j}$ will then be taken to be given by
\begin{equation}
\label{eq:cnn}
C_{NN}^{i,j}=\sum_{k,l=1}^3 \frac{1}{4\pi \sigma_g^2} e^{-|{\bf x}_{\perp,k}^{(i)}-{\bf x}_{\perp,l}^{(j)}|^2/(4 \sigma_g^2)}\,,
\end{equation}
where ${\bf x}_{\perp,k}^{(i)}$ are the transverse plane positions\footnote{Note that the longitudinal component of the positions will not enter the Glauber model description and is ignored in the following.} of the three partons of nucleon $i$, obtained from Monte-Carlo sampling the probability distribution (\ref{eq:protonTAG}). The parameter $\sigma_g\simeq 0.46$ fm in Eq.~(\ref{eq:cnn}) corresponds to the transverse size of the individual partons, inspired by the gluon cloud surrounding a valence quark \cite{Welsh:2016siu}.
Random sampling of the collision probability $P_{\rm collision}$ determines if a collision occurs for any pair of nucleons $i\in A$, $j\in B$, replacing the standard collision criterion (\ref{eq:glaubercrit}). Once all binary collision pairs are known, the number of participants and number of binary collisions are obtained as in the standard nucleon-based Glauber model described above.

\section{Theory of Collisions at Extremely Weak Coupling}
\label{sec:weak}

With the geometry of a single boosted nucleus known from section \ref{sec:Igeo}, let us now
consider a theoretical description of the actual collision process. The relevant interaction
strength of the collision process will be given by the
QCD coupling
\begin{equation}
\label{eq:alphadef}
\alpha_s=\frac{g^2}{4\pi}\,.
\end{equation}
The value of this coupling
depends on the energy scale $Q$ of the interaction process, and for
QCD, $\alpha_s(Q^2)$ is a monotonically decreasing function of $Q$,
because of QCD's property of asymptotic freedom.
Thus, for sufficiently large $Q$, $\alpha_s$ will be small,
and its value can be calculated from first principles via
perturbative QCD. To lowest order, one has \cite{Olive:2016xmw}
\begin{equation}
\label{eq:oneloop}
\alpha_s(Q^2)\simeq \frac{4\pi}{\beta_0 \ln \left(\frac{Q^2}{\Lambda_{\overline{MS}}^2}\right)}\,,\quad \beta_0=\frac{33-2 N_f}{3}\,,
\end{equation}
where $\Lambda_{\overline{MS}}$ is the renormalization scale in the $\overline{MS}$ scheme and $N_f$ is the number of fermion flavors with masses that are light compared to the scale $Q$. Rather than quoting a value of $\Lambda_{\overline{MS}}$, it is customary to quote a value of $\alpha_s(Q^2)$ at a particular scale, e.g. $\alpha_s(Q^2=91.18^2\,{\rm GeV}^2)\simeq 0.118$ \cite{Olive:2016xmw}.

Let us now consider a theorist's universe where the relevant energy scale $Q$ is very large and hence $\alpha_s(Q^2)$ is very small. In this limit, it is possible to treat the collision process from first-principles. The results thus found may then be extrapolated to values of $\alpha_s$ which are expected for present-day nuclear collision experiments. However, it will be important to recall that while calculated from first principles at weak coupling, the extrapolated results again constitute a model for describing real-world collision systems.

As a first approximation, let us ignore the presence of fermions in QCD, thus setting $N_f=0$. This is of course a drastic approximation, but well within the limits of what is allowed in a theorist's universe. Once fermions are ignored, and the coupling is very weak, the equations of motion for the gluon fields $A^\mu$ become the ordinary classical Yang-Mills equations
\begin{equation}
\label{eq:YM}
D_\mu F^{\mu\nu}=j^\nu\,,\quad F^{\mu\nu}=\partial^\mu A^\nu-\partial^\nu A^\mu+i g \left[A^\mu, A^\nu\right]\,,
\end{equation}
where $D_\mu=\partial_\mu + i g \left[A_\mu,.\right]$ is the gauge-covariant derivative and $[.,.]$ denotes the commutator. Note that $g$ is related to $\alpha_s$ via Eq.~(\ref{eq:alphadef}), and that in this section we have normalized the gauge fields $A_\mu$ differently than in section \ref{sec:kin} in order to match the common notation in the literature. (The conventions matching Eq.~(\ref{eq:FTaction}) are recovered when letting $A_\mu\rightarrow  -i g^{-1} A_\mu$, $F_{\mu\nu}\rightarrow -i g^{-1} F_{\mu\nu}$ in Eq.~(\ref{eq:YM}).)
In the above expression, $j^\mu$ is the color current density which provides the source for the Yang-Mills equations (\ref{eq:YM}). For vanishing $j^\mu$, Eqns.~(\ref{eq:YM}) just describe the evolution of pure gauge fields, similar to the case of electromagnetism without matter. Similarly, the color current should describe the matter content in the case of a nuclear collision, which is given by the color current of the nuclei. Consistent with the assumption of weak coupling made above, the collision energy scale has to be very high, so effectively we are dealing with nuclei with very large Lorentz boost factors in the center-of-mass frame. Because of Lorentz contraction, the color current density for a single boosted nucleus will thus be highly localized at $z(t)=\pm t$, for the left- and right-moving nuclei, respectively. Furthermore, it is plausible that the transverse profile of the color current density for a boosted nucleus is proportional to the nuclear thickness function $T_A({\bf x}_\perp)$ discussed in sections \ref{sec:geo1}, \ref{sec:geo2}. Using the generators of SU($N_c$) gauge theory $\lambda^a$ with $a=1,2,\ldots N_c^2-1$ normalized as ${\rm Tr} \lambda^a \lambda^b=\frac{1}{2}\delta^{ab}$, the color current density may be modeled as
\begin{equation}
\label{eq:colorcurr}
j^\mu=j^{\mu,b} \lambda^b=g^{\mu-}T_{A_1}^a({\bf x}_\perp) \lambda^a \delta\left(x^-\right)+g^{\mu+}T_{A_2}^a({\bf x}_\perp)\lambda^a\delta\left(x^+\right)\,,
\end{equation}
where $T_{A_1}^a,T_{A_2}^a$ can be thought of as ``colored'' versions of the nuclear thickness functions of nucleus one and two, respectively, and new coordinates $x^{\pm}=\frac{t\pm z}{\sqrt{2}}$ have been introduced for convenience. A specific way to obtain $T_{A_1}^a,T_{A_2}^a$ will be discussed as ``IP-Glasma model'' below. \index{IP-Glasma model}
\index{Color Glass Condensate|see {IP-Glasma model}}

Before the collision occurs ($t<0$), a solution to the Yang-Mills equations (\ref{eq:YM}) can be found by choosing the gauge $A^\pm=0$. For this gauge choice, (\ref{eq:YM}) reduce to
\begin{equation}
\label{eq:YM2}
D_i F^{i+}=\partial_- D_i A^i=-T_{A_1}\delta(x^-)\,, \quad D_i F^{i-}=\partial_+ D_i A^i=-T_{A_2}\delta(x^+)\,.
\end{equation}
Because of the delta functions in $j^\pm$, the first of these two equations suggests solutions of the form $A^i\propto \theta(x^-)$ whereas the second suggests $A^i\propto \theta(x^+)$. For $t<0$, one can combine these as \cite{McLerran:1993ni,McLerran:1993ka}
\begin{equation}
\label{eq:apm}
A^\pm=0\,, A^i=\theta(x^-)\theta(-x^+)\alpha_1^i({\bf x}_\perp)+\theta(-x^-)\theta(x^+)\alpha_2^i({\bf x}_\perp)\,,
\end{equation}
with
\begin{equation}
\label{eq:YM3}
D_i \alpha_{1,2}^i({\bf x}_\perp)=-T_{A_{1,2}}({\bf x}_\perp)\,,
\end{equation}
subject to the constraint $F^{ij}=0$. 
The functions $\alpha_{1,2}({\bf x}_\perp)$ can be found from (\ref{eq:YM3}) as \cite{JalilianMarian:1996xn,Kovchegov:1996ty}
\begin{equation}
\label{eq:single}
\alpha^i_{1,2}=\frac{i}{g} e^{i \Lambda_{1,2}}\partial^i e^{-i \Lambda_{1,2}}\,,\quad
\partial_\perp^2 \Lambda_{1,2}=-g T_{A_{1,2}}({\bf x}_\perp)\,.
\end{equation}

The above results are solutions to the Yang-Mills equation everywhere except when both $x^+\geq 0$ and $x^-\geq 0$, or equivalently $2 x^+ x^-=\tau^2\geq 0$ using a Milne proper time coordinate (cf appendix \ref{chap:coor}). In order to obtain the solutions to the Yang-Mills equations in the forward light-cone $\tau>0$, it is convenient to guess an ansatz close to the light-cone at $x^+=0\,,x^-=0$ or $\tau=0$. Following Ref.~\cite{Kovner:1995ts}, such an ansatz is given by
\begin{eqnarray}
\label{eq:apm2}
A^i&=&\theta(x^-)\theta(-x^+)\alpha_1^i({\bf x}_\perp)+\theta(-x^-)\theta(x^+)\alpha_2^i({\bf x}_\perp)+\theta(x^-)\theta(x^+)\alpha_3^i({\bf x}_\perp,\tau)\,,\nonumber\\
A^\pm&=&\pm x^\pm \theta(x^-)\theta(x^+) \beta({\bf x}_\perp,\tau)\,,
\end{eqnarray}
where the Fock-Schwinger gauge $x^+ A^-+x^- A^+=0$ has been chosen. In principle, the functions $\alpha_3,\beta$ can depend on $x^+,x^-$ (or equivalently $\tau,\eta$) independently.
Plugging (\ref{eq:apm2}) into the equations of motion (\ref{eq:YM}) results in many terms. Of these, the most singular will be terms proportional to $\delta(x^-),\delta(x^+)$ and their product. It turns out that all the delta-functions in the equations of motion cancel exactly if \cite{Kovner:1995ts}
\begin{equation}
\label{eq:approx}
\alpha_3^i({\bf x}_\perp,\tau=0)=\alpha_1^i({\bf x}_\perp)+\alpha_2^i({\bf x}_\perp)\,,
\quad
\beta({\bf x}_\perp,\tau=0)=-\frac{i g}{2}\left[\alpha_{1,i}({\bf x}_\perp),\alpha_{2,i}({\bf x}_\perp)\right]\,,
\end{equation}
where the single nucleus fields $\alpha_{1,2}^i$ are still determined by Eq.~(\ref{eq:single}). However, recall that other terms in (\ref{eq:YM})  (e.g. terms not proportional to delta functions) will \textit{not} cancel, which implies that (\ref{eq:approx}) is only an approximate solution to the equations of motion. Along the same lines, note that if nuclei with finite (as opposed to infinite) boost factor $\gamma$ were to be considered, the delta-functions in (\ref{eq:colorcurr}) would be ``smeared-out'' by a finite width proportional to $\gamma^{-1}$, and the analytic treatment leading to (\ref{eq:approx}) breaks down (see Ref.~\cite{Ipp:2017lho} for a numerical simulations for finite-width collisions).

For any classical gauge field configuration, the energy-momentum tensor is given by
\begin{equation}
\label{eq:YMemt}
T^{\mu\nu}=-2 {\rm Tr}\left[F^{\mu}_{\ \lambda}F^{\lambda \nu}+\frac{1}{4}g^{\mu\nu} F^{\kappa \lambda}F_{\kappa \lambda}\right]\,.
\end{equation}
Note that $T^\mu_\mu=0$ for classical Yang-Mills, which implies that the solution (\ref{eq:apm2}) has conformal symmetry. There is another symmetry (boost-invariance) that can be gleaned by calculating the non-vanishing components of the energy-momentum tensor before the collision ($t<0$) as
\begin{equation}
\label{eq:delta2}
T^{++}=2 {\rm Tr}\left[\alpha_{1}^i({\bf x}_\perp)\alpha_{1}^i({\bf x}_\perp)\right] \delta^2(x^-)\,,\quad
T^{--}=2 {\rm Tr}\left[\alpha_{2}^i({\bf x}_\perp)\alpha_{2}^i({\bf x}_\perp)\right] \delta^2(x^+)\,,
\end{equation}
where the trace is over color degrees of freedom. Clearly, the presence of the square of delta functions suggests this naive result is not well defined even in a distributional sense. This point will be revisited below.
Using $x^{\pm}=\frac{\tau}{\sqrt{2}} e^{\pm\xi}$, the center-of-mass energy-density $T^{\tau\tau}$ is found from (\ref{eq:delta2}), (\ref{eq:tmntrafo}) as
\begin{equation}
\label{eq:biweak}
T^{\tau\tau}=\frac{1}{2}\left( T^{++}e^{-2 \xi}+2 T^{+-}+T^{--}e^{2 \xi}\right)=2 \sum_{m=1}^2{\rm Tr}\left[\alpha_{m}^i({\bf x}_\perp)\alpha_{m}^i({\bf x}_\perp)\right] \delta^{2}(\tau)\,,
\end{equation}
which is manifestly independent of space-time rapidity $\xi$. While only derived from the result (\ref{eq:delta2}) valid for $t<0$, this symmetry is not destroyed by the collision process. As a consequence, the coefficients $\alpha_3,\beta$ in (\ref{eq:apm})  are indeed independent of rapidity (only dependent on $\tau$) in the limit of infinite boost-factors $\gamma$ and in the absence of quantum (finite coupling) corrections.

Neglecting the rapidity dependence in the matching conditions for $\alpha_3,\beta$ implies that the resulting dynamics in the forward light-cone will remain boost-invariant. Since the sources $j^\mu\propto \delta(\tau)$ remain localized on the light-cone, the solution for $\tau>0$ can be obtained by directly solving the vacuum Yang-Mills equations $D_\mu F^{\mu\nu}=0$ with initial conditions at $\tau=0$ given by (\ref{eq:approx}), for instance through numerical techniques \cite{Krasnitz:1999wc, Lappi:2003bi}. Re-introducing rapidity-fluctuations (which are expected to be sourced by quantum corrections or nuclei of finite width) qualitatively changes the behavior of the solutions because of the presence of plasma instabilities \cite{Romatschke:2005pm,Romatschke:2006nk,Fukushima:2011nq,Berges:2012cj,Gelis:2013rba}.

\subsection{The (Monte-Carlo) IP-Glasma Model}
\index{IP-Glasma model}

\begin{figure}[t]
  \begin{center}    
     \includegraphics[width=.7\linewidth]{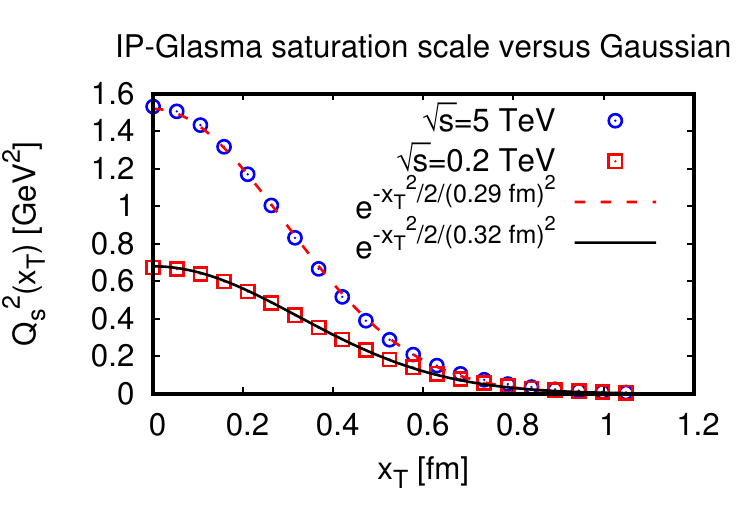}
     \end{center}
  \caption{\label{fig4:four} Transverse coordinate dependence of the saturation scale in the IP-Glasma model from Eq.~(\ref{eq:qs2}) for two representative center-of-mass collision energies $\sqrt{s}$. For comparison, a simple Gaussian parametrization is shown.}
\end{figure}

\begin{figure}[t]
  \begin{center}    
\includegraphics[width=.49\linewidth]{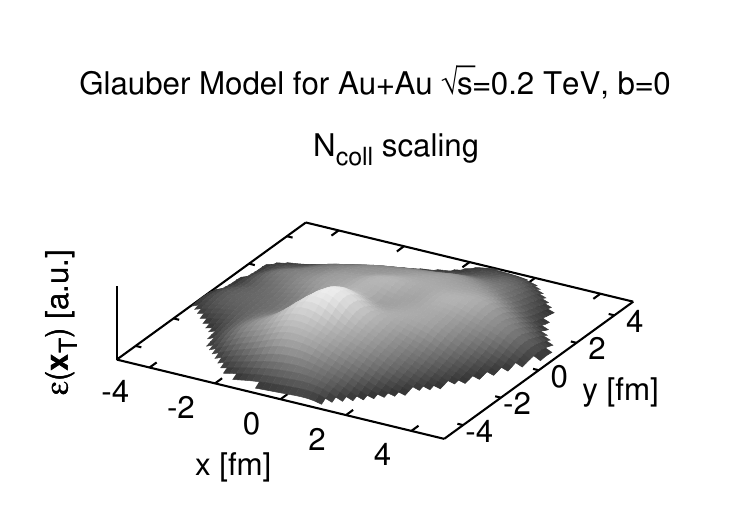}
\includegraphics[width=.49\linewidth]{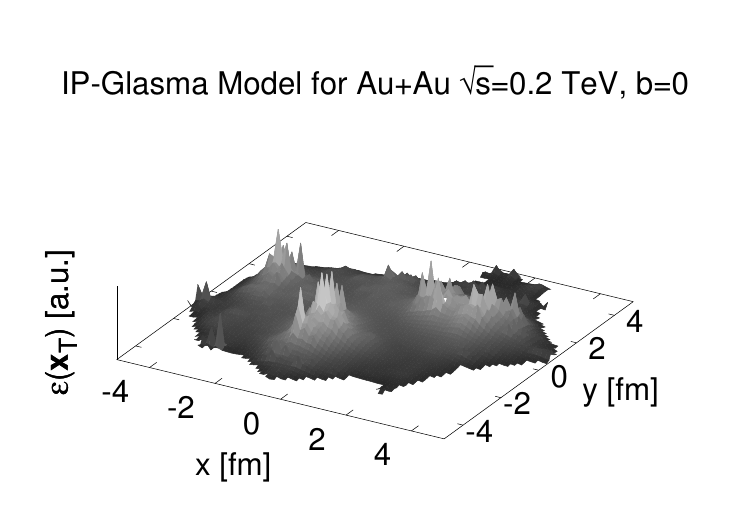}
     \end{center}
  \caption{\label{fig4:ED} Transverse energy density $\epsilon({\bf x}_\perp)$ for a single Au+Au collision event at $b=0$ and $\sqrt{s}=0.2$ TeV using the  Monte-Carlo Glauber model with Gaussian width parameter $R=0.4$ fm (left) and (for the same nucleon positions as the Glauber model) the IP-Glasma model with $R=0.32$ fm (right), cf. Fig.~\ref{fig4:four}. While individual nucleon contributions cannot be easily discerned, $\epsilon({\bf x}_\perp)$ is not azimuthally symmetric even at zero impact parameter. Note that the color fluctuations seen in the IP-Glasma model are on the scale of the lattice spacing used to discretize (\ref{eq:colorcorr}), and are therefore not entirely physical.}
\end{figure}

With the framework of treating collisions at the classical level (extremely weak coupling) laid out above, let us now provide details on implementing the sources $T^a_{A_1,A_2}$ for the color current (\ref{eq:colorcurr}) in a particular model, the so-called IP-Glasma model \cite{Schenke:2012wb}. It will be assumed that the color structure is Gaussian distributed such that for a single nucleus $A$ the correlators between $T_A^a$'s are given by
\begin{equation}
\label{eq:colorcorr}
\langle T_A^a({\bf x}_\perp) T_A^b({\bf y}_\perp) \rangle_{\rm cf} = \delta^{ab}\delta^2({\bf x}_\perp-{\bf y}_\perp) \chi_A({\bf x}_\perp)\,,
\end{equation}
where $\langle\rangle_{\rm cf}$ denotes an average over color fluctuations. Not surprisingly, $\chi_A({\bf x}_\perp)\propto T_A({\bf x}_\perp)$ on average \cite{Krasnitz:2002mn}. To obtain a more realistic nuclear geometry, it is again useful to consider Monte-Carlo sampling of nucleon positions from $\chi_A({\bf x}_\perp)$. Once found, the IP-Glasma model essentially is a prescription on the contribution of each individual nucleon to $\chi_A({\bf x}_\perp)$ for a given event. To this end, some concepts from perturbative QCD such as the DGLAP \index{DGLAP} evolution of the gluon structure function $f(x,Q^2)$ will be needed \cite{Martin:2008cn}:
\begin{eqnarray}
\label{eq:DGLAP}
\frac{\partial f\left(x,Q^2\right)}{\partial{\ln Q^2}} &=& \frac{\alpha_s(Q^2)}{2 \pi}\left\{\int_x^1 dz \left[6\left(\frac{1-z}{z}+z(1-z)\right)f\left(\frac{x}{z},Q^2\right)\right.\right.\\
&&\left.\left.+6\frac{z f\left(\frac{x}{z},Q^2\right)-f\left(x,Q^2\right)}{1-z}\right]+\left(\frac{11}{2}-\frac{N_f}{3}\right)f\left(x,Q^2\right)\right\}\nonumber\,.
\end{eqnarray}
To start the evolution, a seed function $f\left(x,Q^2\right)$ at some fiducial scale $Q=Q_0$ is needed, which is obtained by a fit to data from the HERA experiment using the so-called IP-Sat \index{IP-Sat model} model \cite{Kowalski:2003hm}:
\begin{equation}
f\left(x,Q_0^2\right)=A_g x^{-\lambda_g}(1-x)^{5.6}\,,
\end{equation}
with the fit parameters $A_g=2.308$, $\lambda_g=0.058$ and $Q_0^2=1.51$ GeV$^2$ from Ref.~\cite{Rezaeian:2012ji}. As in Ref.~\cite{Schenke:2012fw}, we use $\alpha_s(Q^2)$ from Eq.~(\ref{eq:oneloop}) with $N_f=3$ in Eq.~(\ref{eq:DGLAP}) to evolve $f(x,Q^2)$ from $Q_0$ to arbitrary $Q$ and the empirical relation
\begin{equation}
x=\frac{0.425-0.0197\ln(s/{\rm GeV}^2)+0.00156 \ln^2(s/{\rm GeV}^2)}{\sqrt{s}/{\rm GeV}}
\end{equation}
from the mean hadronic transverse momentum $\langle p_T\rangle$ in proton-proton collisions at center-of-mass collision energy $\sqrt{s}$ \cite{Khachatryan:2010xs} to set the momentum fraction $x$. In a coordinate system centered around a nucleon, we define $Q_s^2(x,{\bf x}_\perp)$ by self-consistently solving
\begin{equation}
\label{eq:qs2}
Q_s^2(x,{\bf x}_\perp)=\frac{\pi}{3 R^2}\alpha_s\left(Q_0^2+2 Q_s^2\right) f\left(x,Q_0^2+2 Q_s^2\right)e^{-\frac{{\bf x}_\perp^2}{2 R^2}}\,,
\end{equation}
with $R\simeq 0.35$ fm a particular choice for another phenomenological constant \cite{Schenke:2012fw} (see also the discussion around Eq.~(\ref{eq:protonTAG})). Performing a Monte-Carlo sampling procedure to obtain nucleon positions $x_\perp^{(i)}$ in a given nucleus with atomic number $A$ (cf. Fig.~\ref{fig:two}) one thus obtains $\chi_A({\bf x}_\perp)$ as \cite{Schenke:2012fw}
\begin{equation}
\label{eq:nucleonweight}
\chi_A({\bf x}_\perp)\propto\sum_{i=1}^AQ_s^2(x,{\bf x}_\perp^{(i)})\,,
\end{equation}
with a proportionality factor that is close to unity \cite{Schenke:2012fw}. It turns out that the shape function $Q_s(x,{\bf x}_\perp)$ for an individual nucleon thus obtained from Eq.~(\ref{eq:qs2}) is very similar to that of a Gaussian, cf. Fig.~\ref{fig4:four}, with a width that diminishes from $R\simeq 0.35$ in Eq.~(\ref{eq:qs2}) as $\sqrt{s}$ is increased.

The IP-Glasma model is constructed by first obtaining nucleon positions inside nucleus one by performing a Monte-Carlo sampling of the Woods-Saxon potential (\ref{eq:WS}). This fixes $\chi_{A_1}({\bf x}_\perp)$ via Eq.~(\ref{eq:nucleonweight}), which essentially corresponds to summing up Gaussian contributions (\ref{eq:protonTAG}) centered around the nucleon positions with an energy-dependent width $R(\sqrt{s})$. Discretizing the transverse coordinates on a lattice with lattice spacing $a_\perp$, on each lattice site ${\bf x}_\perp$, $N_c^2-1$ numbers $r_a=\left\{r_1,r_2,\ldots r_{N_c^2-1}\right\}$ are Monte-Carlo sampled from a Gaussian probability distribution $e^{-r^2/2}/\sqrt{2 \pi}$ to obtain a single event color current source $T_{A_1}^a({\bf x}_\perp)=\sqrt{\chi_A({\bf x}_\perp) a_\perp^{-2}}r^a$. In general, this will result in nucleus configurations that are not exactly color neutral, e.g. $\int d^2{\bf x}_\perp T_{A_1}^a({\bf x}_\perp)\neq 0$, and hence lead to problems with the zero mode when attempting to solve for $\Lambda({\bf x}_\perp)$ in (\ref{eq:single}). Thus, to avoid problems with the zero mode, it is important to ensure overall color neutrality of the nucleus by subtracting $\int d^2{\bf x}_\perp T_{A_1}^a({\bf x}_\perp)/\int d^2{\bf x}_\perp$ from $T_{A_1}^a({\bf x}_\perp)$ \cite{Lappi:2003bi}. The procedure is repeated for the second nucleus. With the color current sources thus fixed, gauge field configurations may be obtained by solving (\ref{eq:single}) on the lattice.

The spatial scale of the color fluctuations from (\ref{eq:colorcorr}) as implemented above is the lattice spacing $a_\perp$. Thus the detailed structure of fluctuations is not entirely physical, and hence it is useful to consider the color average over gauge fields (\ref{eq:single}) for fixed nucleon positions or given $\chi_A({\bf x}_\perp)$.
To lowest order in the coupling $g$, the color average over the fields becomes
$$
\langle \alpha_1^{i,a}({\bf x}_\perp) \alpha_1^{j,b}({\bf x}_\perp^\prime)\rangle_{\rm cf}=\delta^{ab}\partial^i \partial^{j} \frac{1}{\partial_\perp^4}\delta\left({\bf x}_\perp-{\bf x}_\perp^\prime\right) \chi_A({\bf x}_\perp)\,,
$$
where $\frac{1}{\partial_\perp^2}$ denotes the (suitably regularized, cf. Ref.~\cite{Schenke:2012fw}) inverse of the Laplacian in two dimensions (an equivalent relation holds for the fields $\alpha_2$). To arbitrary order in $g$, the above relation generalizes to \cite{JalilianMarian:1996xn}
\begin{equation}
\label{eq:kernel}
\langle \alpha_1^{i,a}({\bf x}_\perp) \alpha_1^{j,b}({\bf x}_\perp^\prime)\rangle_{\rm cf}=\delta^{ab}\partial^i \partial^{j} \frac{1}{\partial_\perp^4}\delta\left({\bf x}_\perp-{\bf x}_\perp^\prime\right) C_A({\bf x}_\perp)\,,
\end{equation}
with $C_A({\bf x}_\perp)$ the kernel of the gauge field correlator.

\subsection{Energy Deposition and Pre-Hydrodynamic Flow}
\index{Pre-hydrodynamic flow}

Within the weak-coupling (IP-Glasma) approach, there are two sources of fluctuations for individual nuclei: on the one hand, the positions of nucleons within nuclei will vary, and secondly, for fixed nucleon positions, color fluctuations in Eq.~(\ref{eq:colorcorr}) will lead to color fluctuations of the current densities (\ref{eq:colorcurr}). As was pointed out above, the color currents fluctuate on the scale of the lattice spacing in the IP-Glasma model. Let us thus consider averaging quantities over the color fluctuations (\ref{eq:colorcorr}) such that they may be directly compared to models that do not employ color fluctuations, such as the Glauber model in section \ref{sec:MCGlauber}.

To this end, let us consider the energy-momentum tensor (\ref{eq:YMemt}) after the collision which is most easily accessed using Milne coordinates $\tau,\xi$ (cf. appendix \ref{chap:coor}). For instance, we have
\begin{equation}
T^{--}=2 {\rm Tr}\left(F^{-i}\right)^2\,,\ 
T^{+-}=2 {\rm Tr}\left[\frac{1}{2}\left(F^{+-}\right)^2+\frac{1}{2}\left(F^{12}\right)^2\right]\,,\ 
T^{++}=2 {\rm Tr}\left(F^{+i}\right)^2\,,
\end{equation}
where for $\tau>0$ the solutions (\ref{eq:apm2}) imply
\begin{equation}
F^{+i}=-D_i A^+={\cal O}(\tau)\,, F^{-i}=-D_i A^-={\cal O}(\tau)\,, F^{+-}=2 \beta+{\cal O}(\tau^2)\,,
\end{equation}
such that for early times after the collision $\tau \ll 1$
\begin{equation}
\label{eq:ttt}
T^{\tau\tau}=T^{+-}+{\cal O}(\tau^2)=2 {\rm Tr}\left[2\left(\beta({\bf x}_\perp,0)\right)^2+\frac{1}{2}\left(F^{12}\right)^2\right]+{\cal O}(\tau^2)\,.
\end{equation}
To calculate $T^{\tau\tau}$ averaged over color but not nucleon position fluctuations, we follow Ref.~\cite{Lappi:2006hq} and integrate (\ref{eq:ttt}) over an area $A_\perp$ in the transverse plane in which the nucleon weight (\ref{eq:nucleonweight}) can be assumed constant,
\begin{equation}
\label{eq:tttredef}
T^{\tau\tau}=\frac{1}{A_\perp}\int d^2{\bf x_\perp} 2 {\rm Tr}\left[2\left(\beta({\bf x}_\perp,0)\right)^2+\frac{1}{2}\left(F^{12}\right)^2\right]+{\cal O}(\tau^2)\,.
\end{equation}
Inserting the result for $\alpha_3,\beta$ from (\ref{eq:approx}) and performing the color average (\ref{eq:colorcorr}) for nucleus $1$ and $2$ at impact parameter ${\bf b}_\perp$, respectively, yields \cite{Lappi:2006hq}
\begin{equation}
\langle T^{\tau\tau}\rangle_{\rm cf}=\frac{g^2}{2}N_c\left(N_c^2-1\right)C_{A_1}(0)C_{A_2}(0)+{\cal O}(\tau^2)\,,
\end{equation}
where $C_{A_1},C_{A_2}$ are the kernels of the gauge field correlator (\ref{eq:kernel}) for nucleus 1 and 2, respectively, that up to a logarithmic regulator are given by
\begin{equation}
C_A({\bf x}_\perp)\propto \chi_A({\bf x_\perp})\propto \sum_{i=1}^A Q_s^2(x,{\bf x}_\perp^{(i)})\,.
\end{equation}
Note that while the ${\bf x}_\perp$ dependence of $C_A$ has been averaged over by the integration procedure introduced in (\ref{eq:tttredef}), it is possible to sneak in the ${\bf x}_\perp$ dependence of the \textit{nucleons} through $\chi_A({\bf x}_\perp)$ into this equation.
Since the functional shape of $Q_s^2(x,{\bf x}_\perp^{(i)})$ for each individual nucleon $i$  is very close to that of a Gaussian (cf. Fig.~\ref{fig4:four}), this implies that $C_A({\bf x}_\perp)$ is proportional to the sum over Gaussians at all nucleon positions in the nucleus, or the nuclear thickness function $T_A({\bf x}_\perp)$ as defined in (\ref{eq:glauberta}).
 The final result for the energy density in the weak-coupling approximation then reads
\begin{equation}
\label{eq:emtweak1}
\langle T^{\tau\tau}\rangle_{\rm cf}\propto g^2 T_{A_1}({\bf x}_\perp) T_{A_2}({\bf x}_\perp+{\bf b}_\perp)+{\cal O}(\tau^2)\,.
\end{equation}
Note that it is possible to extend the analytic analysis by formally expanding in a power series in proper time $\tau$, finding e.g. near mid-rapidity $|\xi|\ll 1$ \cite{Chen:2015wia}
\begin{eqnarray}
\label{eq:emtweak2}
\langle T^{\tau \perp}\rangle_{\rm cf}&=& -\frac{\tau}{2}\frac{\partial}{\partial {\bf x}_\perp} \langle T^{\tau\tau}\rangle_{\rm cf} +{\cal O}(\tau^3)+{\cal O}(\xi)\,,\nonumber\\
\langle T^{ij}\rangle_{\rm cf}&=& \delta^{ij} \langle T^{\tau\tau}\rangle_{\rm cf}+{\cal O}(\tau^2)\,,\nonumber\\
\langle T^{\xi\xi}\rangle_{\rm cf} &=& -\tau^{-2}\langle T^{\tau\tau}\rangle_{\rm cf}+{\cal O}(\tau^0)\,.
\end{eqnarray}
It is straightforward to calculate the time-like eigenvector $u^\mu=\left(u^0,u^0{\bf v}\right)$ and associated eigenvalue $\epsilon$ from this energy-momentum tensor using Eq.~(\ref{eq:umudef}), 
\begin{equation}
\label{eq:preeq1}
\epsilon = \langle T^{\tau\tau}\rangle_{\rm cf}+{\cal O}(\tau)\,,\quad
{\bf v}^\perp=-\frac{\tau}{4} \frac{\partial}{\partial {\bf x}_\perp} \ln\epsilon \,.
\end{equation}

A few remarks are in order:
\begin{itemize}
\item
The weak-coupling treatment of nuclear collisions at the leading (classical Yang-Mills) level corresponds to a conformal theory, and as a consequence the energy-momentum tensor obeys $T^\mu_\mu=0$
\item
The weak-coupling treatment of nuclear collisions implies boost-invariance unless finite width or higher-order coupling corrections are considered, cf. Eq.~(\ref{eq:biweak}).
\item
In the weak-coupling framework, the energy deposited  in the transverse plane at early times after the collision (\ref{eq:emtweak1}) is essentially  the same as the binary collision scaling variant of the Glauber model, e.g. Eq.~(\ref{eq:2comp}) with $\kappa=1$. Time evolution (governed either by Yang-Mills \cite{Schenke:2012fw} or far from equilibrium fluid dynamics) will change the energy density profile
\item
In the weak-coupling framework, the energy deposited in the transverse plane after the collision is proportional to the strong coupling constant $\alpha_s(Q^2)$, cf.~(\ref{eq:alphadef}). The coupling $\alpha_s$ can either be taken as a constant (``fixed-coupling case'') or dependent on the local energy scale $Q$ (``running coupling case''). In the running coupling case, a prescription for the energy scale $Q$ needs to be supplied. In the IP-Glasma framework, a phenomenologically successful prescription is to take $Q=\langle {\rm max}\left( T_{A_1}^{1/2}({\bf x}_\perp),T_{A_2}^{1/2}({\bf x}_\perp)\right)\rangle$, where $\langle\cdot \rangle$ denote the average over the transverse ${\bf x}_\perp$ plane \cite{Schenke:2012fw}.
\item
From Eq.~(\ref{eq:emtweak2}), the negative effective pressure $P_{L}=\langle T^\xi_\xi\rangle_{\rm cf}=-\epsilon$ (cf. Eq.~(\ref{eq:effpress})) indicates that the system does not admit a description in terms of low-order hydrodynamics
\item
Since Eq.~(\ref{eq:emtweak1}) implies $\tau \partial_\tau \ln \epsilon\rightarrow 0$, the weak-coupling result happens to fall on the start of the hydrodynamic attractor solution for kinetic theory, cf. the panel labeled 'Boltzmann equation' in Fig.~\ref{fig:one}. This potentially suggests that a far-from equilibrium fluid description could be applicable immediately after the collision.
\item
The pre-hydrodynamic transverse flow profile ${\bf v}_\perp$ can be compared to the relativistic Euler equation (\ref{eq:eulerf2}) or $\partial_\tau {\bf v}^\perp=-c_s^2 \frac{\partial_\perp \epsilon}{\epsilon+P}$ if ${\bf v}_\perp \ll 1$. For ideal fluids, the thermodynamic relations $\partial_\perp \epsilon=T \partial_\perp s$ and $\epsilon+P=s T$ apply, leading to $\partial_\tau {\bf v}^\perp=-\frac{1}{3} \partial_\perp \ln s$ for a conformal theory. Therefore ${\bf v}_\perp$ from (\ref{eq:preeq1}) formerly agrees with the expectations from ideal fluid dynamics if $s({\bf x}_\perp)\propto \epsilon({\bf x}_\perp)^{3/4}\propto\left(T_{A_1}({\bf x}_\perp) T_{A_2}({\bf x}_\perp+{\bf b}_\perp) \right)^{3/4}$
\end{itemize}

\section{Theory of Collisions at Extremely Strong Coupling} 
\label{sec:strong}

\subsection{Digression: Energy-Momentum Tensor for Boosted Charge in QED}

To get started, a simple calculation of the energy-momentum tensor for a boosted electric charge in QED will be instructive. First consider a static point charge $q$ in QED. If this point charge is boosted along the z-axis with constant velocity $v$, finding the corresponding field strength tensor in light-cone coordinates $x^\mu=\left(x^+,x,y,x^-\right)$ is a text-book problem in classical electrodynamics \cite{Jackson:1998nia,Steinbauer:1996fv}:
\begin{equation}
\label{eq:qed}
F^{\mu\nu}=\frac{\gamma q (1+v)}{4\pi\sqrt{2}\left[\gamma^2 (z-v t)^2+{\bf x}_\perp^2\right]^{3/2}}
\left(
\begin{array}{cccc}
0 & -x & -y &\frac{\sqrt{2} (z-v t)}{1+v} \\
x & 0 & 0 & \frac{x(1-v)}{1+v} \\
y & 0 & 0 & \frac{y(1-v)}{1+v} \\
-\frac{\sqrt{2} (z-v t)}{1+v} & -\frac{x(1-v)}{1+v} & -\frac{y(1-v)}{1+v} &0
\end{array}
\right)
\end{equation}
For sufficiently large velocities $v\rightarrow 1$, the above field strength tensor becomes a distribution when realizing that
\begin{equation}
\lim_{v\rightarrow 1} \frac{\gamma/\sqrt{2}}{\left(2 \gamma^2 (x^-)^2 + {\bf x}_\perp^2\right)^{3/2}}=\frac{1}{{\bf x}_\perp^2}\delta(x^-)\,,
\end{equation}
such that the only non-vanishing component of $F^{\mu\nu}$ becomes
\begin{equation}
F^{+i}\propto \frac{{\bf x}_\perp^i}{{\bf x}_\perp^2}\delta(x^-)\,,
\end{equation}
matching the structure from (\ref{eq:apm}). Using Eq.~(\ref{eq:qed}), the relevant energy-momentum tensor entry is given by
\begin{equation}
\label{eq:tppqed}
T^{++}=\frac{\gamma^2 q^2 {\bf x}_\perp^2}{8 \pi^2\left[2 \gamma^2 (x^-)^2+{\bf x}_\perp^2\right]^{3}}+{\rm subleading}\,,
\end{equation}
where terms originating from $F^{+-}$ have been dropped since they are subleading corrections for $v\rightarrow 1$. Eq.~(\ref{eq:tppqed}) does not have a well-defined limit $v\rightarrow 1$ even in the distributional sense, just like Eq.~(\ref{eq:delta2}).

For this reason, following \cite{Steinbauer:1996fv,Lousto:1988ua} one takes the charge $q$ to decrease as
\begin{equation}
\label{eq:chargedec}
q^2\rightarrow q^2/\gamma\,,
\end{equation}
such that the energy-momentum tensor with this redefinition obeys
\begin{equation}
\label{eq:tppqed2}
\lim_{v\rightarrow 1}T^{++}\rightarrow \frac{3 q^2 \sqrt{2}}{64 \pi |{\bf x}_\perp|^3}\delta(x^-)=T_{--}\,,
\end{equation}
whereas all the other components vanish in this limit. Thus the $T_{--}$ component of a boosted charge is proportional to $\delta(x^-)$ and the transverse charge density $q/|{\bf x}_\perp|^2$ to the power 3/2. Therefore, for a nucleus with charge $A$, we may expect
\begin{equation}
\label{eq:modelta}
T_{--}\propto T_A^{p}({\bf x}_\perp) \delta (x^-)\,,
\end{equation}
where $T_A({\bf x}_\perp)$ is the nuclear thickness function (\ref{eq:TA}) and $p\simeq 1.5$ a numerical coefficient that would parametrize how transverse charge density generalizes to baryon density $T_A$. In the following we will set $p=1$ for simplicity.

\subsection{Collisions of ``Nuclei'' at Extremely Strong Coupling}

Let us now consider a theorist's universe where the relevant coupling constant is extremely large, in an attempt to tackle the opposite limit that was considered in section \ref{sec:weak}. It will turn out that in order to keep the problem tractable, we will also have to switch our consideration from actual nuclear matter (described by QCD) to more exotic matter in the from of ${\cal N}=4$ SYM in the limit of a large number of colors. In this limit, it is possible to consider a gravitational dual of matter with a given energy-momentum tensor via the holographic renormalization prescription outlined in section \ref{sec:holoren}.  Specifically, for an energy-momentum tensor of a boosted charge such as (\ref{eq:modelta}), the holographic line-element in light-cone coordinates $x^{\pm}=\frac{t\pm z}{\sqrt{2}}$ can be written using Eq.~(\ref{eq:fg}) as
\cite{Grumiller:2008va,Romatschke:2013re}
\begin{equation}
\label{eq:shock}
ds^2=\frac{-2 dx^+ dx^-+d{\bf x}_\perp^2+dz^2+ \Phi({\bf x}_\perp,z)\delta(x^-)dx^{- 2}}{z^2}\,,
\end{equation}
where the near-boundary expansion of $\Phi({\bf x}_\perp,z)$ should correspond to the transverse charge density of interest, e.g.
\begin{equation}
\label{eq:nearboundaryphi}
\lim_{z\rightarrow 0}\frac{\Phi({\bf x}_\perp,z)}{z^2}\propto T_A({\bf x}_\perp)\,.
\end{equation}
Line elements \index{Gravitational shock-wave collisions} with delta-distributions such as (\ref{eq:shock}) are referred to as gravitational shock-waves in the literature on general relativity. In analogy to the weak coupling case, the shock-wave line element (\ref{eq:shock}) should be an exact solution to Einstein's equations which require $\Phi$ to fulfill \cite{Taliotis:2010pi,Romatschke:2013re} 
\begin{equation}
\label{eq:constrainedeq}
\left[\partial_z^2-\frac{3}{z}\partial_z+\partial_\perp^2\right] \Phi({\bf x}_\perp,z)=0\,.
\end{equation}
The solution to this equation is most easily expressed in Fourier-space in the transverse coordinates ${\bf x}_\perp$ as \cite{Avsar:2009xf}
\begin{equation}
\label{eq:phisol}
\Phi(k_\perp,z)=c_2(k_\perp)z^2 I_2\left(z k_\perp\right)\,,
\end{equation}
where $I_2$ is a modified Bessel function and $c_2(k_\perp)$ is an arbitrary function of the transverse momenta $k_\perp$. (Note that there is a second solution to (\ref{eq:constrainedeq}), which, however, does not allow the conformal boundary metric to be Minkowski and has therefore been discarded.) As an example for a solution, one can take $c_2(k_\perp)=K_2 (z_0 k_\perp)$ with $K_2$ a modified Bessel function of the second kind, such that
$$
\Phi(|{\bf x}|_\perp,z)\propto z q^{-3}\ _2F_1(3,5/2,3,-1/q)\,,
$$
with $q=\frac{x_\perp^2+(z-z_0)^2}{4 z z_0}$ and $_2F_1$ a hypergeometric function which is particularly simple \cite{Gubser:2008pc,Avsar:2009xf}. This solution has the property that the near-boundary expansion implies
$$
\lim_{z\rightarrow 0}\frac{\Phi({\bf x}_\perp,z)}{z^2}\propto T_A({\bf x}_\perp)\propto \frac{1}{(x_\perp^2+z_0^2)^3}\,,
$$
which corresponds to a simple choice for the profile function $T_A$.

It should be stressed that any physical profile of a nucleus in the transverse plane, including a Monte-Carlo generated event-by-event sampling of nucleon positions, can be implemented through (\ref{eq:phisol}). Once $\Phi$ for a single ``nucleus'' has been fixed, one can proceed to consider collisions of two ``nuclei'' by calculating the outcome of the collision of two gravitational shock waves. This is a classic problem in general relativity in Minkowski space (cf. Refs.~\cite{Khan:1971vh,DEath:1992plq,DEath:1992mef,DEath:1992nmz}), which for the purpose of holography is generalized to space-times that are asymptotically AdS.

Similar to the case of weak-coupling \cite{Ipp:2017lho}, techniques have been developed to simulate the entire collision by numerically solving Einstein equations \cite{Chesler:2010bi,Casalderrey-Solana:2013aba,Fernandez:2014fua, Chesler:2015wra,Grozdanov:2016zjj}. However, similar to the weak-coupling case, considerable insight can be gained by developing analytic techniques applicable to times before and shortly after the collision process itself.

To this end, let us consider the following coordinate transformation for a single shock wave \cite{Yoshino:2002br,Romatschke:2013re}
\begin{equation}
\label{eq:rosen}
x^+=u\,,\quad
x^-=v+\frac{1}{2}\Phi \theta(u)+\frac{u \theta^2(u)}{8} \partial_i
\Phi\,\delta^{ij}\, \partial_j \Phi\,,\quad
x^i=\tilde x^i + \frac{1}{2} u \theta(u) \delta^{ij} \partial_j \Phi\,,
\end{equation}
where here $x^i=(x,y,z)$. It is straightforward to verify that after the transformation (\ref{eq:rosen}), the line element (\ref{eq:shock}) no longer contains any $\delta$ functions, implying that the metric is continuous across the light-like hypersurface at $u=0$ (``Rosen'' form). A similar transformation may be performed on a second shock wave traveling in the opposite direction, and as a consequence the pre-collision line element will be given as a simple superposition of the individual shock waves \cite{Romatschke:2013re}:
\begin{equation}
\label{ds2pre}
ds^2_{\rm pre} = \frac{-2 du dv + d\tilde x^i d\tilde x^j \left(\delta^{kl}
H_{ik}^{(1)} H_{jl}^{(1)}+\delta^{kl} H_{ik}^{(2)} H_{jl}^{(2)}-\delta_{ij}\right)
}{\left[\tilde z+\frac{1}{2} \left(u \theta(u) \partial_z \Phi_{(1)} +
v \theta(v) \partial_z \Phi_{(2)}\right)\right]^2}\,,\quad u<0,v<0\,,
\end{equation}
and
$$
H^{(1)}_{ij}=\delta_{ij} + \frac{u \theta(u)}{2}\partial_i \partial_j \Phi_{(1)}
\,,\quad
H^{(2)}_{ij}=\delta_{ij} + \frac{v \theta(v)}{2}\partial_i \partial_j \Phi_{(2)}\,.
$$
(Note that this is operationally equivalent to the weak-coupling case (\ref{eq:apm}) where the pre-collision gauge fields were given as a simple superposition of the gauge configurations of individual nuclei.)

It should be pointed out that the pre-collision line-element corresponds to an energy-momentum tensor in the boundary theory which is the simple superposition of $T^{++}=T_{A_1}({\bf x}_\perp)\delta(x^-)$ and $T^{--}=T_{A_2}({\bf x}_\perp+{\bf b}_\perp)\delta(x^+)$, cf. Eq.~(\ref{eq:tppqed2}) where ${\bf b}_\perp$ is the impact parameter of the collision. Thus, according to Eq.~(\ref{eq:tmntrafo}) the energy-density $T^{\tau\tau}$ before the collision is given by
\begin{equation}
\label{eq:boostbroken}
T^{\tau\tau}\propto \delta(\tau)\left(T_{A_1}({\bf x}_\perp)e^{-\xi}+T_{A_2}({\bf x}_\perp+{\bf b}_\perp)e^{\xi}\right)\,,
\end{equation}
where again $x^\pm=\frac{\tau}{\sqrt{2}}e^{\pm \xi}$ has been used to rewrite the $\delta$-functions. In contrast to the result found for \textit{weak coupling} collisions (\ref{eq:biweak}), the energy-density for \textit{strong coupling} collisions is \textit{not} boost-invariant. This is a direct consequence of having replaced the $\delta^2(x^-)$ in Eq.~(\ref{eq:delta2}) by $\delta(x^-)$ through the procedure (\ref{eq:chargedec}). 

Since the metric in Rosen form is continuous across the light-cone, one can proceed to make an ansatz for the \textit{full} line element as
\begin{equation}
ds^2=ds_{\rm pre}^2+\theta(u) \theta(v) ds^2_{\rm int}\,,
\end{equation}
similar to what has been done in the weak-coupling case for gauge fields. The interaction piece $ds^2_{\rm int}$ has to be determined by solving the equations of motion (the Einstein equations) across the light-cone.

Unlike the case of weak-coupling (\ref{eq:approx}), where only the most singular terms (delta-functions) in the Yang-Mills equations are required to match, in the gravitational case this information is not sufficient to fix $ds^2_{\rm int}$ unambiguously. Rather, it is necessary to also match the Einstein equations exactly from pre- to post-collision line element, e.g. order-by-order in an early-time (or near boundary) expansion. This program has been carried through explicitly for highly symmetric situations where $\Phi({\bf x}_\perp,z)$ either does not depend on ${\bf x}_\perp$ (homogeneous shock waves \cite{Grumiller:2008va}) or only depends on the absolute value of $|{\bf x}|_\perp$ (radially symmetric shock waves \cite{Romatschke:2013re}).

\subsection{Energy Deposition and Pre-Hydrodynamic Flow}
\index{Pre-hydrodynamic flow}

Once the post-collision line element has been found, it is possible to obtain the energy-momentum tensor \textit{after the collision} by again performing the holographic renormalization procedure outlined in section \ref{sec:holoren}. This leads for instance to the results near mid-rapidity $|\xi|\ll 1$ \cite{Romatschke:2013re}
\begin{eqnarray}
\label{eq:emtstrong}
T^{\tau\tau}&\propto& T_{A_1}({\bf x}_\perp) T_{A_2}({\bf x}_\perp+{\bf b}_\perp) \tau^2+{\cal O}(\tau^4)\,,\nonumber\\
T^{\tau\perp}&=& -\frac{\tau}{2}\frac{\partial}{\partial {\bf x}_\perp} T^{\tau\tau}+{\cal O}(\tau^5)\,,\nonumber\\
T^{ij}&=& 2 \delta^{ij} T^{\tau\tau} +{\cal O}(\tau^4)\,,\nonumber\\
T^{\xi\xi}&=& -3 \tau^{-2} T^{\tau\tau} +{\cal O}(\tau^2)\,,
\end{eqnarray}
using Milne coordinates (see appendix \ref{chap:coor}).
The proportionality sign in Eqns.~(\ref{eq:emtstrong}) indicates the freedom (\ref{eq:nearboundaryphi})  of multiplying by an arbitrary (dimensionful) constant. Note that while the results for the energy-momentum tensor are independent of space-time rapidity to this order in the small time expansion, higher orders do depend on rapidity, as was expected because of Eq.(\ref{eq:boostbroken}).  For the case of symmetric homogeneous shockwaves (where $T_{A_1}({\bf x}_\perp)=T_{A_1}=T_{A_2}=T_A={\rm const}$), this has been explicitly calculated in Ref.~\cite{Grumiller:2008va}, finding e.g.
\begin{equation}
\label{eq:negvel}
T^{\tau\xi}\propto - 3 T_A^3 \sinh\xi \tau^6+{\cal O}(\tau^9)\,.
\end{equation}
Calculating the time-like eigenvector $u^\mu=\left(u^0,u^0{\bf v}\right)$ and associated eigenvalue $\epsilon$ from this energy-momentum tensor using Eq.~(\ref{eq:umudef}) one finds
\begin{equation}
\label{eq:preeqp2}
\epsilon = T^{\tau\tau}+{\cal O}(\tau^3)\,,\quad
{\bf v}^\perp=-\frac{\tau}{6} \frac{\partial}{\partial {\bf x}_\perp} \ln \epsilon \,.
\end{equation}

A few remarks are in order:
\begin{itemize}
\item
The strong-coupling treatment of nuclear collisions at the leading (classical gravity) level corresponds to a conformal theory, and as a consequence the energy-momentum tensor obeys $T^\mu_\mu=0$
\item
The strong-coupling treatment of nuclear collisions explicitly breaks boost-invariance, cf. Eq.~(\ref{eq:boostbroken})
\item
The longitudinal velocity $v^\xi\propto T^{\tau \xi}$ from Eq.~(\ref{eq:negvel}) can be seen to indicate flow \textit{towards} mid-rapidity $\xi=0$ rather than outward. This could have implications for phenomenology of nuclear collisions, cf. Ref.~\cite{Stephanov:2014hfa}
\item
In the strong-coupling framework, the energy deposited in the transverse plane at early times after the collision (\ref{eq:emtstrong}) is essentially the same as the collision-scaling variant of the Glauber model, e.g. Eq.~(\ref{eq:2comp}) with $\kappa=1$. Time evolution (governed either by gravitational \cite{vanderSchee:2015rta} or far from equilibrium fluid dynamics) will change the energy density profile. However, note that this scaling hinges on assuming $p=1$ in the scaling of the single-nucleus energy-momentum tensor (\ref{eq:modelta}) with the thickness function
\item
From (\ref{eq:emtstrong}), the negative effective pressure $P_{L}=T^\xi_\xi=-3\epsilon$ (cf. Eq.~(\ref{eq:effpress})) indicates that the system does not admit a description in terms of low-order hydrodynamics
\item
Unlike the case of the weak-coupling framework, (\ref{eq:emtstrong}) implies $\tau \partial_\tau \ln \epsilon\rightarrow 2$, and therefore the strong-coupling result does not seem to fall onto the start of the hydrodynamic attractor solution, cf. the panel labeled 'AdS/CFT' in Fig.~\ref{fig:one}. (Recall, however, that the attractor solution was calculated assuming boost-invariance while the above treatment of collisions at strong coupling violates boost-invariance, see above)
\item
The pre-hydrodynamic transverse flow profile ${\bf v}_\perp$ can be compared to the relativistic Euler equation (\ref{eq:eulerf2}) or $\partial_\tau {\bf v}^\perp=-c_s^2 \frac{\partial_\perp \epsilon}{\epsilon+P}$ if ${\bf v}_\perp \ll 1$. For ideal fluids, the thermodynamic relations $\partial_\perp \epsilon=T \partial_\perp s$ and $\epsilon+P=s T$ apply, leading to $\partial_\tau {\bf v}^\perp=-\frac{1}{3} \partial_\perp \ln s$ for a conformal theory. Therefore ${\bf v}_\perp$ from (\ref{eq:preeq2}) formally agrees with the expectations from ideal fluid dynamics if $s({\bf x}_\perp)\propto \left(T_{A_1}({\bf x}_\perp) T_{A_2}({\bf x}_\perp+{\bf b}_\perp) \right)^{1/2}$
\item
Non-linear evolution of the system changes the flow profile ${\bf v}_\perp$ from Eq.~(\ref{eq:preeq2}). However, it has been found in Ref.~\cite{Habich:2014jna} that the flow profile can be very well approximated by
\begin{equation}
\label{eq:preeq2}
{\bf v}^\perp\simeq-\frac{\tau}{3.0} \frac{\partial}{\partial {\bf x}_\perp} \ln\left(T_{A_1}({\bf x}_\perp) T_{A_2}({\bf x}_\perp+{\bf b}_\perp)\right)\,.
\end{equation}
\end{itemize}

\section{Initial State Eccentricities}
\label{sec:eccs}
\index{Initial-state eccentricities}

The previous chapters contain three models for the energy deposition in relativistic nuclear collisions: the Glauber model, the weak-coupling (IP-Glasma) model and the strong-coupling (AdS/CFT) model. All of these models have event-by-event fluctuations of the positions of nucleons inside the nuclei in common, such that the energy deposited in an individual event will typically involve a ``lumpy'' energy-density profile in the transverse plane.

To characterize the properties of these lumpy energy-density profiles, it is customary to consider the so-called initial state eccentricities $e_n$ and participant plane angles $\Phi_n$ defined as \cite{Alver:2010gr,Teaney:2010vd,Bhalerao:2011yg}
\begin{equation}
\label{eq:eccsdef}
e_n e^{i n \Phi_n}\equiv - \frac{\int dr_\perp d\phi\, r_\perp^{n+1} e^{i n \phi}\bar\epsilon(r_\perp,\phi)}{\int dr_\perp d\phi\, r_\perp^{n+1} \bar\epsilon(r_\perp,\phi)}\,,\quad n\geq 2\,
\end{equation}
where $r_\perp,\phi$ are polar coordinates in the transverse plane, and $\bar\epsilon(r_\perp,\phi)\equiv \epsilon\left({\bf x}_\perp-\bar {\bf x}_\perp\right)$ with $\bar {\bf x}_\perp\equiv \frac{\int d^2{\bf x}_\perp \epsilon({\bf x}_\perp) {\bf x}_\perp}{\int d^2{\bf x}_\perp \epsilon({\bf x}_\perp)}$.

\begin{figure}[t]
  \begin{center}    
     \includegraphics[width=\linewidth]{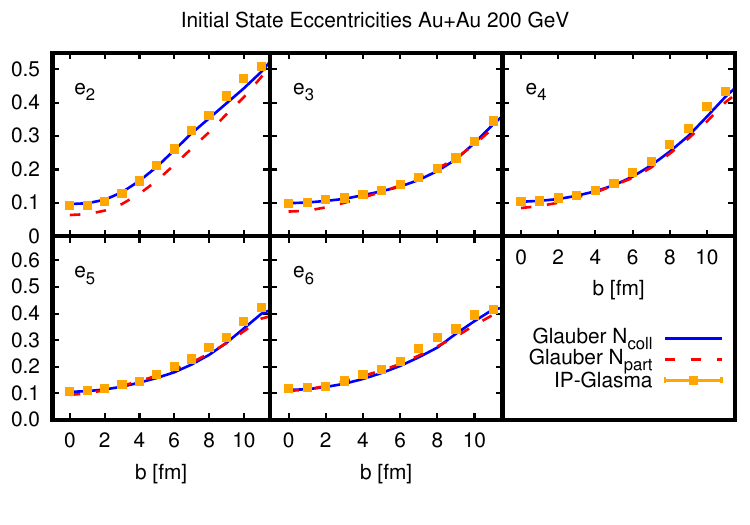}
     \end{center}
  \caption{\label{fig4:six} Initial state eccentricities from the Glauber model using $N_{\rm coll}$ and $N_{\rm part}$ scaling as a function of impact parameter with $R=0.4$ fm for a Au+Au collision at $\sqrt{s}=200$ GeV compared to eccentricities from the IP-Glasma model (data points from Ref.~\cite{Schenke:2012fw}). }
\end{figure}

In the Glauber model defined in section \ref{sec:MCGlauber}, it is particularly easy to calculate the even  eccentricities $e_{2n}$ because $\epsilon({\bf x}_\perp)$ is just a sum of Gaussians with width $R$ and hence all integrals can be evaluated analytically to give
\begin{eqnarray}
\left.e_{2} e^{i 2 \Phi_2}\right|_{\rm Glauber}&=&\sum_{{\rm nucleon}\  m} \frac{(x_m+i y_m)^2}{2 R^2+x_m^2+y_m^2}\,,\nonumber\\
\left.e_{4} e^{i 4 \Phi_4}\right|_{\rm Glauber}&=&\sum_{{\rm nucleon}\  m} \frac{(x_m+i y_m)^4}{8 R^4+8 R^2 (x_m^2+y_m^2)+(x_m^2+y_m^2)^2}\,,\nonumber\\
\left.e_{6} e^{i 6 \Phi_6}\right|_{\rm Glauber}&=&\sum_{{\rm nucleon}\  m} \frac{(x_m+i y_m)^6}{48 R^6+72 R^4 (x_m^2+y_m^2)+18 R^2 (x_m^2+y_m^2)^2+(x_m^2+y_m^2)^3}\,,\nonumber
\end{eqnarray}
where the sum is either over all participant nucleons or over all binary collision nucleons ($N_{\rm part}$ or $N_{\rm coll}$ scaling). For the odd $n$, the denominator of $e_n$ is not easily expressed in terms of simple functions, but we find that
\begin{eqnarray}
\left.e_{3} e^{i 3 \Phi_3}\right|_{\rm Glauber}&\simeq &\sum_{{\rm nucleon}\  m} \frac{(x_m+i y_m)^3}{\left((x_m^{2}+y_m^{2})^{a_3}+(9 \pi/2)^{a_3/3}R^{2 a_3}\right)^{3/(2a_3)}}\,,\quad a_3=\frac{14}{15}\nonumber\\
\left.e_{5} e^{i 5 \Phi_5}\right|_{\rm Glauber}&=&\sum_{{\rm nucleon}\  m} \frac{(x_m+i y_m)^5}{\left((x_m^{2}+y_m^{2})^{a_5}+(225 \pi/2)^{a_5/5}R^{2 a_5}\right)^{5/(2a_5)}}\,,\quad a_5=\frac{25}{29}\,,\nonumber
\end{eqnarray}
offer approximations that are accurate to within three percent. The results for the eccentricities in the Glauber model with number of collision and number of participant scaling with $R=0.4$ fm are shown in Fig.~\ref{fig4:six}. Overall, 
all model results give very similar trends for the initial state eccentricities.
However, one observes a clear difference for $e_2$ between the Glauber $N_{\rm part}$ scaling and the other two models for all impact parameters shown, as well as for $e_{3},e_{4}$ for $b\simeq 0$. This difference causes a corresponding difference in experimentally measurable observables, favoring the Glauber $N_{\rm coll}$/IP-Glasma values over the Glauber $N_{\rm part}$ results \cite{Retinskaya:2013gca}.

Also, Fig.~\ref{fig4:six} demonstrates that for all impact parameters shown, 
there is almost perfect agreement between the IP-Glasma results and Glauber $N_{\rm coll}$ scaling, which was first pointed out for $e_2$ in Ref.~\cite{Lappi:2006xc} and later for $e_3$ in Ref.~\cite{Schenke:2012wb}. It is likely that this apparent matching can be understood from the fact that the IP-Glasma initial conditions obey an $N_{\rm coll}$ scaling when performing the averaging over color fluctuations of the energy-momentum tensor as derived in Eqns.~(\ref{eq:emtweak2}). For the IP-Glasma results from Ref.~\cite{Schenke:2012fw} shown in Fig.~\ref{fig4:six}, the color average and nucleon position average have been performed on equal footing, yet the final result agrees with performing a nucleon position average of Eqns.~(\ref{eq:emtweak2}).

\section{Numerical Algorithms for Fluid Dynamics}
\label{sec:numalgo}
\index{Numerical Algorithms}

Relativistic viscous hydrodynamic evolution equations such as (\ref{eq:rBRSSS}),(\ref{eq:BRSSSf}) form a set of coupled partial different equations for the fluid variables. Since finding analytic solutions to these equations is not feasible except in a limited number of special cases, numerical techniques will be needed to obtain solutions.

Several different algorithms to obtain solutions to relativistic viscous hydrodynamics exist \cite{Romatschke:2007mq,Song:2007ux,Chaudhuri:2007qp,Dusling:2007gi,Molnar:2009tx,Mendoza:2009gm,Schenke:2010rr,Bozek:2011ua,DelZanna:2013eua}, three of which will be discussed in more detail below. For simplicity, the case of conformal fluid dynamics (see section \ref{sec:confsym}) will be discussed, since algorithms readily generalize to the case of non-conformal systems.

\subsection{Naive Discretization}
\label{sec:naivedisc}

Possibly the most straightforward approach is to use naive discretization of derivatives, employed in \cite{Romatschke:2007mq,Luzum:2008cw}. The hydrodynamic evolution equations (\ref{eq:rBRSSS}), (\ref{eq:BRSSSf}) for the fluid variables $\epsilon,u^\mu,\pi^{\mu\nu}$ are first-order in time. Furthermore, the evolution equation for the shear stress (\ref{eq:rBRSSS}) may be expressed in terms of first-order time-derivatives of $\epsilon,u^\mu$. 
This allows rewriting the d-dimensional evolution equations in matrix form as\footnote{Because of the constraints $u_\mu u^\mu=-1$, not all of the components of the fluid variables are independent. Here, spatial components $u^{i}$ were chosen as the independent components of $u^\mu$.}
\begin{equation}
\left(\begin{array}{cccc}
a_{00} & a_{01} & \ldots & a_{0d}\\
a_{10} & a_{11} & \ldots & a_{1d}\\
\ldots\\
a_{d0} & a_{d1} & \ldots & a_{dd}\\
\end{array}\right)\cdot
\left(\begin{array}{c}
\partial_\tau \epsilon\\
\partial_\tau u^1\\
\ldots\\
\partial_\tau u^{d-1}\\
\end{array}\right)
=\left(\begin{array}{c}
b_0\\
b_1\\
\ldots\\
b_{d-1}
\end{array}\right)\,,
\end{equation}
where the matrix and vector in the above equation will be denoted as ${\bf a},{\bf b}$ in the following. Using an equation of state $P=P(\epsilon)$, the quantities ${\bf a},{\bf b}$ can readily be expressed in terms of $\epsilon,u^i$ as well as the first and second order spatial derivatives of these. A straightforward way to formally solve the hydrodynamic evolution equations is then given by
\begin{equation}
\label{eq:disc}
\left(\begin{array}{c}
\partial_\tau \epsilon\\
\partial_\tau u^1\\
\ldots\\
\partial_\tau u^{d-1}\\
\end{array}\right) = {\bf a}^{-1}\cdot {\bf b}\,.
\end{equation}
Using a naive discretization scheme for space and time by replacing differentials of a quantity $X(\tau,x)$ as
\begin{equation}
\partial_\tau X(\tau,x)=\frac{X(\tau+\delta \tau,x)-X(\tau,x)}{\delta \tau}\,,
\quad
\partial_x X(\tau,x)=\frac{X(\tau,x+\delta x)-X(\tau,x-\delta x)}{2 \delta x}\,,\nonumber\end{equation}
and similarly for $\partial_x^2 X(\tau,x)$, the quantities ${\bf a},{\bf b}$ as well as the inverse ${\bf a}^{-1}$ can be evaluated on a spatial grid with lattice spacing $\delta x$. Knowledge of the independent fluid variable components $\epsilon,u^i$ at some time $\tau$ thus can be used to step the variables forward in time by an increment $\delta \tau$ through solving (\ref{eq:disc}). Knowledge of $\epsilon,u^i$ including their derivatives in space and time then can be used to step the shear stress forward in time by similarly solving Eq.~(\ref{eq:rBRSSS}). Repetition of the above steps constructs a numerical solution to the relativistic viscous hydrodynamic evolution equations with given initial conditions $\epsilon(\tau_0,{\bf x})$, $u^\mu(\tau_0,{\bf x})$, $\pi^\mu(\tau_0,{\bf x})$ at $\tau=\tau_0$ on a space-time grid with grid spacings $\delta \tau,\delta x$, respectively.

A few remarks are in order:
\begin{itemize}
\item
The above discretization is first-order accurate in the time-increment $\delta \tau$, but second-order accurate in the lattice spacing $\delta x$. This means that numerical errors accumulated by executing the above algorithm are expected to scale as ${\cal O}(\delta \tau)$, ${\cal O}\left((\delta x)^2\right)$ for small $\delta \tau,\delta x$. This discretization scheme has the advantage of being easy to program, but has the disadvantage that in practice $\delta \tau \ll \delta x$ is needed in order to obtain accurate and stable solutions. Upgrading the time-discretization scheme to second order accuracy is possible by employing an additional correction step when obtaining the time-incremented values for $\epsilon,u^i$.
\item
In the case of ideal fluid dynamics (zero viscosity), the above naive discretization algorithm would fail dramatically, see e.g. the discussion in Ref.~\cite{Press:2007:NRE:1403886}. The reason for this behavior is that ideal hydrodynamics is inherently unstable with respect to the growth of small scale fluctuations\footnote{Note that in numerical simulations, small fluctuations are always present at all scales because of finite machine accuracy.}. Interestingly, these instabilities are not just numerical artifacts, but rather correspond to the turbulent instabilities that occur at high Reynolds numbers. In ideal fluid dynamics, the Reynolds number is infinite, hence it is not surprising to encounter the equivalent of turbulent instabilities in numerical simulations of ideal fluid dynamics. In order to control these turbulent instabilities in numerical simulations of ideal fluid dynamics, a different discretization scheme is used that introduces a non-vanishing \textit{numerical} viscosity to stabilize the evolution algorithm. As pointed out in Ref.~\cite{Luzum:2008cw}, the naive discretization discussed above leads to stable evolution for sufficiently small $\delta \tau$ as long as the \textit{real} viscosity coefficient is non-vanishing.
\item
While the above algorithm allows stable evolution of sufficient smooth initial data giving rise to laminar flows, it is not suitable to deal with shock propagation or strong gradients that are typically encountered when simulating initial conditions obtained from Monte-Carlo sampling of nuclear geometries, cf. Fig.~\ref{fig4:four}. As long as the overall flow remains laminar, it is possible to deal with large gradients by patching the above algorithm with a smearing step, first discussed in Ref.~\cite{Nagle:2013lja}. The smearing step is implemented through a condition on the local energy density in the evolution. If the local energy density $\epsilon$ falls below a set minimum energy density at the same time-step, then it is replaced by an average over the nearest-neighbor values of $\epsilon$. While aesthetically displeasing, it has been shown  that the overall flow pattern remains unaffected by the smearing in the case of laminar flows \cite{Nagle:2013lja}. Furthermore, the correspondingly patched algorithm allows robust evolution even in strong-gradient situations encountered for instance in the simulation of proton+nucleus collisions, cf. chapter \ref{chap:experiment}.
\end{itemize}

\subsection{Kurganov-Tadmor}

A more sophisticated discretization method employed in Ref.~\cite{Schenke:2010rr} uses the Kurganov-Tadmor method \cite{KT:2000}. The hydrodynamic evolution equations (\ref{eq:rBRSSS}), (\ref{eq:BRSSSf}) are first-order in time in terms of the ideal fluid energy momentum tensor $T^{\mu\nu}_{(0)}$ given in (\ref{eq:hydro0f}) and $\pi^{\mu\nu}$; thus, they can be rewritten as
\begin{equation}
\label{eq:KThydroother}
\partial_\tau T^{\tau\mu}_{(0)}=-\partial_\tau \pi^{\tau \mu}+b_1^\mu\,,\quad
\partial_\tau \pi^{\mu\nu}=b_2^{\mu\nu}\,,
\end{equation}
where $b_1^\mu,b_2^{\mu\nu}$ are functions of $T^{\mu\nu}_{(0)},\pi^{\mu\nu}$ and their first spatial derivatives.

Let us first explain the Kurganov-Tadmor scheme for a simple, one-dimensional model equation $\partial_t \rho(t,x)+\partial_x J(t,x)=0$. To solve this equation, replace $\rho(t,x)$ by a cell-average $\rho(t,x)\rightarrow \bar \rho(t,x)=\frac{1}{\delta x}\int_{x-\delta x/2}^{x+\delta x/2} dx \rho(t,x)$. The time-average of this quantity will be given by
\begin{equation}
\label{eq:example}
\partial_t \bar \rho=-\frac{J(t,x+\delta x/2)-J(t,x-\delta x/2)}{\delta x}\,.
\end{equation}
This discretization leads to discontinuities at the points $x\pm \delta x/2$. The discontinuities do not spread through the grid instantaneously, but rather possess a maximal local propagation speed $a(t,x)=\left|\frac{\partial J}{\partial \rho}\right|$. The Kurganov-Tadmor scheme exploits this finite propagation speed and replaces Eq.~(\ref{eq:example}) by \cite{KT:2000,Schenke:2010nt}
\begin{equation}
\partial_t \bar \rho=-\frac{H(t,x+\delta x/2)-H(t,x-\delta x/2)}{\delta x}\,.
\end{equation}
where
\begin{eqnarray}
\label{eq:KTh}
H(t,x\pm\delta x/2)&=&\frac{J^+\left(t,x\pm\delta x/2\right)+J^-\left(t,x\pm\delta x/2\right)}{2}\\
&&+\frac{a(t,x\pm \delta x/2)}{2}\left(\bar \rho^+(t,x\pm\delta x/2)-\bar \rho^-(t,x\pm\delta x/2)\right)\nonumber\\
J^\pm(t,x)&=&J(t,x\pm a(t,x) \delta t)\,, \bar\rho^\pm(t,x)=\bar\rho(t,x\pm\delta x/2)\mp \frac{\delta x}{2}\partial_x \bar \rho(t,x\pm \delta x/2)\,.\nonumber
\end{eqnarray}
To limit oscillatory behavior in the evolution, a minmod flux limiter is employed in calculating the spatial derivatives $\partial_x \bar\rho(t,x)$:
\begin{eqnarray}
\label{eq:minmod}
\partial_x \rho(t,x) = {\rm minmod}&&\left(\theta \frac{\rho(t,x+\delta x)-\rho(t,x)}{\delta x},\frac{\rho(t,x+\delta x)-\rho(t,x-\delta x)}{2 \delta x},\right.\nonumber\\
&&\left.\quad\theta \frac{\rho(t,x)-\rho(t,x-\delta x)}{\delta x}\right)\,,\nonumber
\end{eqnarray}
with $1\leq \theta \leq 2$ a numerical control parameter and 
\begin{equation}
{\rm minmod} (x_1,x_2,\ldots)=\left\{\begin{array}{c}
{\rm min}(x_i)\,,\ {\rm if}\ x_i>0\ \forall x_i\,, \\
{\rm max}(x_i)\,,\ {\rm if}\ x_i<0\ \forall x_i\,,\\
0\,, {\rm else}\,.
\end{array}
\right.
\end{equation}
The scheme can readily be upgraded to higher dimensions and non-vanishing source terms. For instance, the equation
\begin{equation}
\label{eq:KTmodel}
\partial_t \rho(t,x,y)+\partial_x J^x(t,x,y)+\partial_y J^y(t,x,y)=q\left(\rho(t,x,y)\right)
\end{equation}
can be recast into \cite{KT:2000,Naidoo:2004}
\begin{eqnarray}
\label{eq:KT3}
\partial_t \bar \rho
&=&-\frac{H^x(t,x+\delta x/2,y)-H^x(t,x-\delta x/2,y)}{\delta x}\nonumber\\
&&-\frac{H^y(t,x,y+\delta y/2)-H^y(t,x,y-\delta y/2)}{\delta y}+q\left(\bar\rho(t,x,y)\right)\,,
\end{eqnarray}
where $H^{x,y}$ are analogous to Eq.~(\ref{eq:KTh}), see Ref.~\cite{KT:2000}.

Equations such as (\ref{eq:KT3}) may be solved either by naive first-order time discretization or by a second-order discretization scheme, sometimes referred to as Heun's rule (see e.g. Ref.~\cite{Schenke:2010nt}). Since the relativistic viscous hydrodynamics equations (\ref{eq:KThydroother}) for $T^{\tau\mu}_{(0)},\pi^{\mu\nu}$ are similar to the model equation (\ref{eq:KTmodel}), the Kurganov-Tadmor scheme can be used for these equations, cf. Ref.~\cite{Schenke:2010rr}. With $T^{\tau\nu}_{(0)}$ known, one can calculate $\epsilon,u^\mu$ by matching to the ideal fluid dynamic form (\ref{eq:hydro0f}). From $\epsilon,u^\mu$, one can reconstruct the full ideal energy-momentum tensor $T^{\mu\nu}_{(0)}$. Summing $T^{\mu\nu}_{(0)}$ and $\pi^{\mu\nu}$ gives the full energy momentum tensor at the next time-step.

A few remarks are in order:
\begin{itemize}
\item
The Kurganov-Tadmor scheme is a flux-conserving scheme that is most powerful for ideal (non-viscous) hydrodynamics which cannot be treated using a naive discretization.
\item
The time-derivative of the shear-stress in Eq.~(\ref{eq:KThydroother}) can be handled by using a naive first-order integration of $\partial_\tau \pi^{\mu \nu}$ as a source in the equation for $\partial_\tau T^{\tau \mu}$ as a first step, and later correcting this estimate in a second step.
\item
In the case of strong gradients, the Kurganov-Tadmor scheme suffers from the same kind of numerical instabilities as the naive discretization scheme discussed above. The patch for the algorithm proposed in Ref.~\cite{Schenke:2010rr} was to set the local viscosity coefficient to zero as well as decreasing the pressure by five percent whenever any of the local effective pressures $P_{\rm eff}^{(i)}$ in Eq.~(\ref{eq:effpress}) become negative. Once viscosity has been set to zero, the flux-conserving nature including numerical viscosity of the Kurganov-Tadmor scheme guarantees stability of the algorithm in this region. Similar to the patched naive discretization algorithm, Ref.~\cite{Schenke:2010rr} reported no strong effects on observables of interest after applying this patch.
\end{itemize}

\subsection{Lattice Boltzmann}

Both the naive discretization and the Kurganov-Tadmor scheme discussed above suffer from the fact that they need patches in order to allow to solve the relativistic viscous hydrodynamics equations without encountering numerical instabilities whenever gradients become strong. While the proposed patches for these algorithms seem reasonably benign in terms of not affecting the bulk of the hydrodynamic evolution, they are nevertheless ad-hoc patches that are at the very least aesthetically displeasing. This motivates the search for other algorithms that do not suffer from this problem, such as the Lattice Boltzmann approach as employed in Ref.~\cite{Mendoza:2009gm}.

Despite its name, the Lattice Boltzmann approach is an algorithm to solve fluid dynamics, not kinetic theory. However, from the discussion in section \ref{sec:fluidfromkin} it is known that for near equilibrium situations, the Boltzmann equation gives rise to fluid dynamics. Thus the equations of fluid dynamics (albeit only with certain fixed values of transport coefficients, cf. Eq.~(\ref{eq:etazetaKT})) are contained within the Boltzmann equation. Therefore, solving the Boltzmann equation for near-equilibrium situations is akin to solving the equations of fluid dynamics.

Given that a solution of the Boltzmann equation (\ref{eq:Boltzmann}) involves solving an integro-differential equation in $2d-1$ dimensional phase-space\footnote{The on-shell particle distribution function $f$ depends on $d$ space-time coordinates and $d-1$ momenta.}, this seems a rather impractical way of performing fluid dynamics simulations. The crucial insight of the Lattice-Boltzmann approach is that it is possible to obtain exact results for fluid dynamics if the particle momenta are discretized on a (rather coarse) momentum grid. This reduces the dimensionality of the problem.

Recall from section \ref{sec:fluidfromkin} that fluid dynamics emerges from the Boltzmann equation by taking momentum moments of the particle distribution function. For illustration, let us choose $d=4$ Minkowski space-time and massless particles. The central object for fluid dynamics is the energy momentum tensor (\ref{eq:KT}), which becomes
\begin{equation}
\label{eq:LBtmunu}
T^{\mu\nu}=\int \frac{d^3 p}{(2\pi)^3} \frac{p^\mu p^\nu}{p^0} f(t,{\bf x},{\bf p})\,,
\end{equation}
where the on-shell particle energy obeys $p^0=|{\bf p}|$.

Changing the particle distribution function $f$ will leave fluid-dynamics unchanged if the above integral over the distribution function remains the same for all components of $T^{\mu\nu}$. Let us expand $f(t,{\bf x},{\bf p})$ in a suitable basis to make this freedom explicit \cite{Romatschke:2011hm}:
\begin{equation}
\label{eq:LBansatz}
f(t,{\bf x},{\bf p})=e^{-p^0/T^0}\sum_{k=0}^\infty \sum_{n=0}^\infty L_k^{(\alpha)}(p^0/T^0) P_{i_1 i_2\ldots i_n}^{(n)}({\bf p}/p^0) a^{(n,k)}_{i_1 i_2\ldots i_n}(t,{\bf x})\,,
\end{equation}
where $L_k^{(\alpha)}$ are the generalized Laguerre polynomials and $P_{i_1 i_2\ldots i_n}^{(n)}({\bf v})$ with ${\bf p}/p^0\equiv {\bf v}$ are tensor polynomials of order $n$ that are orthogonal with respect to the solid angle integration of the unit sphere\footnote{Note that since particles are assumed massless, ${\bf v}$ is a unit vector.}. Plugging (\ref{eq:LBansatz}) into (\ref{eq:LBtmunu}) one finds that the particular choice of $\alpha=3$ for the generalized Laguerre polynomials implies that of the series $\sum_{k=0}^\infty$ in (\ref{eq:LBansatz}) only the $k=0$ term contributes to $T^{\mu\nu}$. Similarly, one finds that of the series $\sum_{n=0}^\infty$ in (\ref{eq:LBansatz}) only terms $n\leq 2$ contribute. Therefore, specifying the coefficients $a^{(n,0)}(t,{\bf x})$ with $n\leq 2$ in (\ref{eq:LBansatz}) will result in an energy-momentum tensor that \textit{exactly} matches $T^{\mu\nu}$ calculated with the full distribution function $f(t,{\bf x},{\bf p})$.

Based on (\ref{eq:LBansatz}), the energy-momentum tensor (\ref{eq:LBtmunu}) is given by an integral over orthogonal polynomials of low degree. It is well known that integrals over polynomials can be represented \textit{exactly} as sums over the roots of the associated polynomials (cf. the Gauss-Hermite quadrature rule). For the case at hand, defining for simplicity $\bar{p}\equiv p^0/T^0$, the quadrature rule implies that for a polynomial $g(\bar{p})$ of degree less than $2 N_p$, the result
\begin{equation}
\int d\bar p\, e^{-\bar p}\bar p^\alpha g(\bar p) =\sum_{i=0}^{N_p-1} w_i g(\bar p_i)\,,
\end{equation}
is exact if the nodes $\bar{p}_i=\bar{p}_0,\bar p_1,\ldots, \bar p_{N_p-1}$ are the roots of the generalized Laguerre polynomials $L_{N_p}^{(\alpha)}(\bar p_i)=0$ and the weights $w_i$ are given by \cite{Romatschke:2011hm}
\begin{equation}
w_i^{(p)}=\frac{(N_p+\alpha)!}{N_p!}\frac{\bar p_i}{(N_p+1)^2 \left(L_{N_p+1}^{(\alpha)}(\bar p_i)\right)^2}\,.
\end{equation}

\begin{figure}[t]
  \begin{center}    
     \includegraphics[width=.5\linewidth]{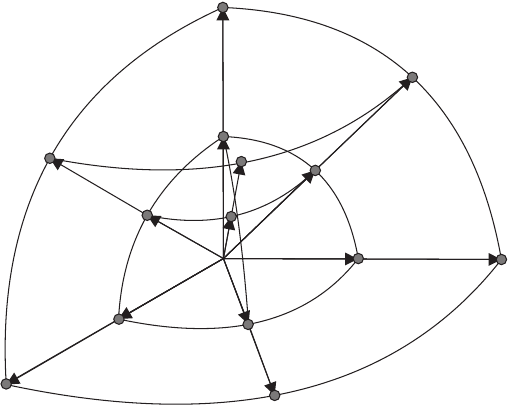}
     \end{center}
  \caption{\label{fig4:eight} Sketch of a possible momentum node $\bar {\bf p}_i$ configuration for a relativistic Lattice Boltzmann  algorithm. Figure from Ref.~\cite{Romatschke:2011hm}.}
\end{figure}

As a consequence, once the weights $w_i^{(p)}$ are calculated, it is sufficient to know the distribution function at the node points $\bar p_i$ in order to calculate the momentum integral in Eq.~(\ref{eq:LBtmunu}) exactly. Similar arguments apply to the solid angle integral in in Eq.~(\ref{eq:LBtmunu}) such that one may express \cite{Romatschke:2011hm,Mendoza:2013km}
\begin{equation}
\label{eq:LBtmunu2}
T^{\mu\nu}=\sum_{\bar {\bf p}_i} w_i \bar p_i^\mu \bar p_i^\nu f_i(t,{\bf x})\,,
\end{equation}
with weights $w_i$ a combination of $w_i^{(p)}$ as well as the angular integration weights. $\bar {\bf p}_i$ are the three-dimensional quadrature nodes and $f_i(t,{\bf x})$ the particle distribution function evaluated at nodes $\bar {\bf p}=\bar {\bf p}_i$. The nodes $\bar {\bf p}_i$ form a lattice in momentum space, see for instance Fig.~\ref{fig4:eight}. Effectively, the possibility of representing integrals by sum over node points implies that knowledge of the particle distribution function on this momentum lattice is sufficient to allow for an exact representation of the energy-momentum tensor. This reduces the effective dimensionality of the computational problem from $2d-1=7$ to $d=4$ in four space-time dimensions.

After the reduction from continuous momenta to the momentum lattice, let us now discuss the Lattice Boltzmann algorithm. For simplicity, consider the Boltzmann equation in the BGK approximation in 4-dimensional Minkowski space-time without any additional forces (see Refs.~\cite{Romatschke:2011hm,Romatschke:2011qp,Ambrus:2015rnx,Succi:2014,2017arXiv170501801V} for examples with different metrics and non-vanishing forces). Discretizing space-time on a square grid with \textit{equal} temporal and spatial lattice spacings $\delta t$, the BGK Boltzmann equation Eq.~(\ref{eq:BBGK}) can be written as
\begin{equation}
\label{eq:LBalgo}
f_i(t+\delta t,{\bf x}+{\bf v}_i \delta t)-f_i(t,{\bf x})=-\frac{v_i^\mu u_\mu}{\tau_R}\left(f_i-f_i^{(0)}\right)\,,
\end{equation}
where ${\bf v}_i=\bar {\bf p}_i/|\bar {\bf p}_i|$, $v^\mu_i=\left(1,{\bf v}_i\right)$ and $\tau_R$ is the (possibly temperature-dependent) relaxation time. Here $f_i^{(0)}$ is the lattice equilibrium distribution function, cf. Eq. (\ref{eq:maxjut}), which is obtained by calculating the coefficients $a_{i_1i_2\ldots i_n}^{(n,k)}$ in (\ref{eq:LBansatz}) through integration of the continuum distribution function $f^{(0)}(t,{\bf x},{\bf p})$. Because $f^{(0)}$ depends on $x^\mu$ only implicitly through $T(t,{\bf x}),u^\mu(t,{\bf x})$, the equilibrium coefficients $a_{i_1i_2\ldots i_n}^{(n,k)}$ will similarly be simple functions of $T(t,{\bf x}),u^\mu(t,{\bf x})$.

Initial conditions for the Lattice-Boltzmann algorithm consist of providing $f_i(t_0,{\bf x})$ at some initial time $t=t_0$. A simple choice is to use the lattice equilibrium distribution function $f_i=f_i^{(0)}$ with a given temperature and velocity profile. Once initial conditions are provided, an algorithm to solve the discretized Lattice-Boltzmann equation (\ref{eq:LBalgo}) consists of a streaming step in configuration space followed by collisional relaxation \cite{Romatschke:2011hm}. The streaming step consists of updating $f_i(t+\delta t,{\bf x})=f_i(t,{\bf x}-{\bf v}_i \delta t)$. Note that in the case of the momentum lattice above, ${\bf x}-{\bf v}_i \delta t$ will generically not end up on a node of the spatial lattice. A simple solution is to use linear interpolation to obtain the updated $f_i$'s on the spatial grid nodes. Another solution is to express ${\bf v}\cdot \nabla f$ on the lhs of (\ref{eq:BBGK}) by using a minmod flux limiter (\ref{eq:minmod}) (see Refs.~\cite{Romatschke:2011hm,PhysRevE.58.R4124,2003PhRvE..68a6701U,Mendoza:2013km} for other solutions). Using the updated $f_i's$, the updated energy-momentum tensor is calculated from Eq.~\ref{eq:LBtmunu2} and $\epsilon,u^\mu$ are obtained as the time-like eigenvalue and eigenvalue of $T^{\mu\nu}$. Using the equation of state, the pseudo-temperature $T$ corresponding to $\epsilon$ is obtained, which together with $u^\mu$ can be used to construct the lattice equilibrium distribution $f_i^{(0)}$ from (\ref{eq:LBansatz}) using the known equilibrium coefficients $a_{i_1i_2,\ldots i_n}^{(n,k)}$. With $u^\mu,f_i^{(0)}$ known, collisional relaxation is performed by adding the rhs of Eq.~(\ref{eq:LBalgo}) to $f_i(t+\delta t,{\bf x})$. Repetition of the above steps constructs a numerical solution for $\epsilon,u^\mu$ with given initial conditions on a space-time grid with uniform lattice spacing $\delta t$.

A few remarks are in order:
\begin{itemize}
\item
The above discretization is first-order accurate in the space-time increment $\delta t$. This means that numerical errors accumulated by executing the above algorithm are expected to scale as ${\cal O}(\delta t)$ for small $\delta t$. Upgrading the algorithm to second order accuracy is possible, cf. Ref.~\cite{Brewer:2015hua}.
\item
The streaming formulation (\ref{eq:LBalgo}), or alternatively the flux-limited version of (\ref{eq:BBGK}), are numerically stable as long as $f_i\geq 0$ \cite{Ansumali2002}. This allows simulations for initial conditions with strong gradients or shock waves without the necessity of the patches needed in the naive discretization or Kurganov-Tadmor schemes discussed above. However, supersonic flows typically violate $f_i\geq 0$ and lead to instabilities in the Lattice Boltzmann algorithm. While the streaming step in (\ref{eq:LBalgo}) maintains the sign of $f_i$, the equilibrium distribution $f_i^{(0)}$ will typically have negative entries for $|{\bf v}|>c_s$. Various ways to control these instabilities have been suggested, such as entropic stabilization \cite{PhysRevE.92.061301} and bulk viscosity \cite{PhysRevD.87.083003}.
\item
While relativistic Lattice-Boltzmann algorithms have been successfully employed in a variety of of applications \cite{Romatschke:2011hm,2014PhRvD..90l5028M,Bantilan:2014sra}, a full implementation for simulating nuclear collisions is currently lacking.
\end{itemize}

\section{Hadronization and Hadronic Cascade}
\label{sec:hadro1}
\index{Hadronization|see {Cooper-Frye prescription}}

The hydrodynamic algorithms discussed in section \ref{sec:numalgo} allow for the numerical evolution of initial conditions for the hydrodynamic fields. However, hydrodynamic field variables are not directly observable in relativistic nuclear collision experiments. Rather, the experiments observe particles (hadrons, leptons, photons) that are emitted before, during and after the hydrodynamic evolution (cf. Fig.~\ref{fig4:one}).

Little is known about particle emission in the pre-hydrodynamic stage, partly because the onset of hydrodynamic behavior itself is only now starting to be understood in terms of non-hydrodynamic mode decay to attractor solutions (see the discussion in section \ref{sec:offeq}). One notable exception is the pre-hydrodynamic emission of photons and dileptons, see for instance Refs.~\cite{Gelis:2002ki,Arleo:2004gn,Martinez:2008di,Martinez:2008mc,Rebhan:2011ke,Oliva:2017pri,Berges:2017eom}.

Let us now focus on hadrons generated during the hydrodynamic evolution, and hold off discussing the emission of particles after the hydrodynamic evolution has ended until section \ref{sec:cascade}. 
One expects that a fluid description of QCD matter is applicable as long as the effective fluid pressure is larger than the hadron gas pressure. The fluid pressure gets reduced in particular by bulk viscous effects (\ref{eq:peff}), with the QCD bulk viscosity coefficient being sourced by the QCD trace anomaly \cite{Jeon:1995zm,Arnold:2006fz,Lu:2011df,Dobado:2011qu}. The QCD trace anomaly in equilibrium has been calculated using lattice QCD \cite{Borsanyi:2010cj,Borsanyi:2013bia,Bazavov:2014pvz}, cf. Fig.~\ref{fig3:trac}. However, the precise relation between the peak in the trace anomaly and the maximal bulk viscosity coefficient is complicated, cf. the discussion in Ref.~\cite{Romatschke:2009ng}. 
In the following, it will be assumed that the temperature at which the QCD bulk viscosity is maximal coincides with the QCD deconfinement temperature (\ref{eq:tc}) of $T_c\simeq 0.17$ GeV \cite{Aoki:2009sc} (see Ref.~\cite{Li:2009by} for model studies supporting this assumption). 
If bulk viscosity becomes large close to the QCD cross-over transition, this will make the bulk stress $\Pi$ large and negative, and hence decrease the effective local pressure (\ref{eq:effpress}). This is a known phenomenon from ordinary (non-relativistic) fluids such as water, where e.g. near fast-moving turbines a decrease of the effective fluid pressure below the vapor pressure leads to the formation of bubbles (``cavitation''). It has been argued that an analogous phenomenon should exist in QCD fluids, where the rise in bulk viscosity close to the QCD cross-over transition would lead to a decrease of the fluid pressure below the hadron gas pressure, effectively forcing a change in the degrees of freedom from deconfined quark-gluon fluid to hadrons ("hadronization") \cite{Rajagopal:2009yw,Klimek:2011by,Bhatt:2011kr,Habich:2014tpa,Sanches:2015vra,Fogaca:2016eat,Monnai:2016kud} (see the discussion in section \ref{sec:confborel}). It should be pointed out that it is not known if hadronization is indeed driven by cavitation, hence these assumptions introduce a significant systematic error in final observables.
\index{Cavitation}

Because the temperature distribution of the fluid will not be uniform,  hadronization happens dynamically in regions where the temperature in the local fluid rest frame is close to $T_c$, until there are no longer any fluid cells with temperatures exceeding the critical temperature.

The dynamics of hadrons (``hadronic phase'') will in general not be well captured by fluid dynamics. This is because the gas is weakly coupled, and as a consequence non-hydrodynamic modes are expected to dominate hadron dynamics (see section \ref{sec:ktnonhydro}). However, kinetic theory is expected to provide a reasonable description of the evolution of the system in the hadronic phase. Some evidence supporting this expectation is that thermodynamic lattice QCD results are very well approximated at low temperatures by an non-interacting  hadron resonance gas, section \ref{sec:latticeresults}.

However, for practical applications to model relativistic nuclear collisions, two main problems remain. First, as pointed out above, the non-uniform temperature profile in ion collisions will lead to a hadronization of some fluid cells, while others have not yet hadronized. Currently, no framework exists that allows simultaneous treatment of both the hadron gas and fluid dynamic phase. Instead, the location of fluid cells that hadronize are stored during the hydrodynamic evolution, giving rise to a so-called switching hypersurface delineating the border between fluid phase and gas phase. After the last fluid cell has hadronized, and the switching hypersurface has been calculated, the values of fluid variables on the hypersurface are used to initialize a kinetic description (``hadron cascade code'') that is used to simulate the hadronic phase. A consequence of this sequential treatment is that it does not capture events where hadrons from a hadronized fluid cell are re-absorbed in a neighboring fluid cell that has not yet hadronized. It has been estimated that such events are not very common in the simulation of heavy-ion collisions, yet they require some ad-hoc procedure which will introduce a systematic error in final observables.

Secondly, while the dynamics of e.g. a pion gas is very well understood, the scattering cross-section of heavier hadronic resonances at temperatures above $100$ MeV are not at all well measured experimentally. For this reason, hadron scatterings implemented in cascade codes rely on audacious extrapolations from low energy data or ``educated guesswork'', introducing another significant systematic error in final observables.

Keeping these limitations in mind, we now turn to review how hadronization and hadron cascades are currently implement in order to model relativistic ion collisions. 

\subsection{Hadronization}
\label{sec:hadro}
\index{Switching Hypersurface}

As discussed above, it will be assumed in the following that hadronization of a QCD fluid cell occurs whenever the local pseudo-temperature\footnote{Recall that for out-of-equilibrium situations, the word pseudo-temperature merely refers to the energy scale corresponding to the local energy-density $\epsilon(T)$ via the equation of state, cf. the discussion in sections \ref{sec:neqeos},\ref{sec:confborel}.} drops below the QCD confinement temperature $T_c\simeq 0.17$ GeV \cite{Aoki:2009sc}. This is known as isothermal decoupling. Given a numerical scheme that simulates the time-evolution of an initial condition in relativistic viscous hydrodynamics on a space-time grid (see section \ref{sec:numalgo} above), one locates the location of the switching hypersurface by comparing the local pseudo-temperature in nearest neighbor cells.

In the following, an algorithm for constructing a ``block-element'' approximation to the hypersurface will be discussed, which allows for unconditionally stable hypersurface construction even in the case of highly non-uniform temperature profiles (see e.g. Ref.~\cite{Romatschke:2009im,Huovinen:2012is} for other algorithms). Let us consider the example of a simulation using Milne coordinates (\ref{eq:milne}), with a local temperature profile given by $T(\tau_i,x_j,y_k,\xi_l)$ on grid points $\tau_i,x_j,y_k,\xi_l$ with grid spacing $\delta \tau,\delta x,\delta y,\delta \xi$. If $T(\tau_i,x_j,y_k,\xi_l)>T_c$ and $T(\tau_i,x_{j}+\delta x,y_k,\xi_l)<T_c$, record $x^\mu=\left(\tau_i,x_j+\frac{\delta x}{2},y_k,\xi_l\right)$ to be a point on the switching hypersurface $\Sigma$ with the corresponding oriented hypersurface element given by \cite{Ruuskanen:1986py,Rischke:1996em}
$$
d\Sigma^\mu=\left(d\Sigma^\tau,d\Sigma^x,d\Sigma^y,d\Sigma^\xi\right)=\left(0,1,0,0\right)\tau \delta \tau\delta y\delta \xi\,.
$$
On the other hand, if $T(\tau_i,x_j,y_k,\xi_l)<T_c$ and $T(\tau_i,x_{j}+\delta x,y_k,\xi_l)>T_c$, then $x^\mu=\left(\tau_i,x_j+\frac{\delta x}{2},y_k,\xi_l\right)\in \Sigma$ but the oriented element would be given by $d\Sigma^\mu=\left(0,-1,0,0\right)\tau \delta \tau\delta y\delta \xi$.  Once it has been determined that $x^\mu\in \Sigma$, the fluid variables $\epsilon(\Sigma),u^\mu(\Sigma),\pi^{\mu\nu}(\Sigma),\Pi(\Sigma)$ at that point are also recorded.
Proceeding in this fashion by comparing local cell pseudo-temperatures of nearest-neighbor cells (including in the temporal direction) leads to the construction of an approximation to the hypersurface $\Sigma$ along with the associated oriented surface elements $d\Sigma^\mu$. The approximation may be systematically improved by using smaller grid-spacings in the respective directions.

Equipped with the switching hypersurface $\Sigma$ and oriented elements $d\Sigma^\mu$ at which the fluid variable values are known, hadronization is now implemented via a generalized Cooper-Frye prescription \cite{Cooper:1974mv}. The hadronization procedure should conserve energy and momentum, so the energy-momentum tensor is required to be continuous across $\Sigma$:
\begin{equation}
\label{eq:CF}
T^{\mu\nu}_{\rm fluid}=T^{\mu\nu}_{\rm particles}\,.
\end{equation}
The particle energy-momentum tensor is a generalization of Eqns.~ (\ref{eq:KT},\ref{eq:intmeas}) to multiple particles, e.g.
\begin{equation}
T^{\mu\nu}_{\rm particles}=\sum_{r\in\, {\rm resonances}}\int d\chi_r p^\mu p^\nu f_r(x^\mu,p^\mu)\,,
\end{equation}
where the integration measure $d\chi_r$ for a given resonance with mass $m_r$, spin and isospin degeneracies $s_r,g_r$ is given by
$$
 \int d\chi_r \equiv (2s_r+1)(2g_i+1)\int \frac{d^4p}{(2\pi)^4} \sqrt{-{\rm det}g_{\mu\nu}}(2\pi)\delta\left(g_{\mu\nu}p^\mu p^\nu+m_r^2\right) 2 \theta(p^0)\,.
$$

If the fluid cells on $\Sigma$ were all in equilibrium ($\pi^{\mu\nu}=0,\Pi=0$) then the on-shell distribution function $f_r$ for each resonance would be given by the equilibrium distribution function (\ref{eq:f0}) with $u^\mu=u^\mu(\Sigma)$, $T=T(\Sigma)$. This corresponds to the original Cooper-Frye prescription \cite{Cooper:1974mv}. Since for viscous hydrodynamics, the cells on $\Sigma$ will typically not be in equilibrium, a generalization of the Cooper-Frye prescription to non-equilibrium systems is needed. This is an ill-defined problem, since it corresponds to reconstructing the three-dimensional momentum information of every resonance using only the constraint (\ref{eq:CF}). For this reason, various models for the out-of-equilibrium distribution function $f_r$ have been considered.

\index{Cooper-Frye prescription! in simulations}
For conformal theories out-of-equilibrium, the ``quadratic ansatz'' (\ref{eq:KTrec}) is one of the most popular models for $f_r$, cf.~\cite{Teaney:2003kp,Baier:2006um}. However, while enjoying wide popularity and ease of implementation, the quadratic ansatz is problematic for at least two reasons: it is disfavored by comparisons with experimental data \cite{Luzum:2010ad} and it leads to negative particle distribution functions $f_r<0$ for large momenta for any non-vanishing $\pi^{\mu\nu}$. A viable alternative enjoying phenomenologic success is the ``exponential ansatz'' (\ref{eq:KTmodel2}) \cite{Pratt:2010jt,vanderSchee:2013pia,Nopoush:2015yga}, which also guarantees $f_r>0$ for all momenta if $\pi^{\mu\nu}$ is not too large, cf. the discussion in section \ref{sec:cKT}. A significant drawback of the ``exponential ansatz'' is that it only fulfills the matching condition (\ref{eq:CF}) approximately, introducing a significant systematic error in final observables.

Unfortunately, non-conformal systems with non-vanishing bulk stress $\Pi\neq 0$ still pose major challenges for the hadronization procedure. In particular, it is known that a simple generalization of the ``quadratic ansatz'' (\ref{eq:KTrec}) to non-conformal systems leads to unacceptably large distortions of the particle distribution function  even if bulk viscosity is small \cite{Monnai:2009ad}. The strategy adopted in some recent works  is to perform the switching at considerably lower temperatures than $T_c$, where the bulk viscosity coefficient can be expected to be small and non-equilibrium contributions to $f_r$ therefore manageable. However, it is our expectation that the cavitation instability triggered by the large bulk stress near $T_c$ initializes hadronization in the fluid, and correspondingly should not be ignored. It is possible that future work using a two-phase (fluid plus hadron gas) simulation could help resolve the problem of bulk viscous corrections to particle spectra. In the absence of such a two-phase simulation, we will entirely neglect bulk viscous corrections to $f_r$ in the following. Not surprisingly, this introduces a sizable systematic error for final observables.

With the single particle distribution function given by the ``exponential ansatz'' (\ref{eq:KTmodel2}) for every hadron resonance under consideration, it would be possible to directly calculate observables (see section \ref{sec:obs}). While this approach has been successful historically, it completely ignores hadron-hadron interactions after hadronization. For this reason, modern simulations employ a so-called hybrid approach where the particle distribution functions only serve as initial condition for the subsequent hadron cascade simulation described in the following section.

\subsection{Hadron Cascade}
\label{sec:cascade}
\index{Hadron Cascade}

With the particle distribution functions for hadron resonances determined on a switching hypersurface $\Sigma$, one performs a Monte-Carlo sampling\footnote{Because $f_r$ falls strongly for large momenta, a single Monte-Carlo sampling will typically not generate a significant number of high momentum hadrons. In order to boost statistics of high momentum observables in the simulation, a frequently employed trick is to sample many hadron cascade events from a single hydrodynamic $\Sigma$ (``oversampling'').} of hadron resonances interpreting $f_r(x^\mu,p^\mu)$ as probability distribution functions (cf. Refs.~\cite{Pratt:2010jt,Novak:2013bqa}) and section \ref{sec:MC} for details). A delicate issue encountered in this procedure is that, occasionally, the Monte-Carlo procedure generates particles that move inward into a space-like hypersurface, rather than outward. The patch applied in the following is to reject these particles in the Monte-Carlo procedure \cite{Pratt:2010jt} (see Refs.~\cite{Bugaev:1999uy,Molnar:2005gx} for further discussion of this issue).

Once hadrons have been generated from the Monte-Carlo procedure, their dynamics can be simulated by standard particle cascade codes \cite{Zhang:1997ej,Bass:1998ca,Molnar:2000jh,Xu:2004mz,Novak:2013bqa}. In these approaches, particles follow straight-line trajectories until particles come sufficiently close to undergo a collision event, where 'sufficiently close' can e.g. be determined by a scattering cross-section $\sigma$, cf. Eq.~(\ref{eq:glaubercrit}).

One particular cascade code, B3D \cite{Novak:2013bqa}, simulates interactions of hadron resonances identified in the particle data book \cite{Olive:2016xmw} via s-wave scattering with a constant cross section of $\sigma=10$ mb as well as scattering through resonances modeled as Breit-Wigner forms. The constant cross-section assumed in B3D is unlikely to correspond to the true hadron cross-sections, which are not known at the relevant energy densities. A different cascade code, URQMD \cite{Bass:1998ca}, instead employs parametrizations of cross-sections extrapolated from low-energy experimental results. As mentioned above, the unknown hadron cross sections introduce a sizable systematic error in final observables.

Once hadrons have stopped interacting, the system reaches kinetic freeze-out, and hadrons follow straight-line trajectories until registered by experimental detectors.

\section{Observables and Event Averaging}
\label{sec:obs}

The result from the hadron cascade simulation described in the previous section is a list of particles and their momenta. In principle, in the simulation also the particles' position information would be available, but this information will be ignored since it is not directly accessible by experiment. It is customary in high energy nuclear collisions to express momenta consistent with the Milne coordinate system (see section \ref{sec:milnecoo}), that is $p^\mu=\left(p^\tau,p^x,p^y,p^\xi\right)$ where for a particle of mass $m$ 
\begin{equation}
\label{eq:momdefs}
p^\tau=\sqrt{m^2+p_\perp^2}\cosh\left(Y-\xi\right)\,,\quad
p^\xi=\tau^{-1}\sqrt{m^2+p_\perp^2}\sinh\left(Y-\xi\right)\,,
\end{equation}
where $p_\perp\equiv \sqrt{p_x^2+p_y^2}$ and
\begin{equation}
\label{eq:rap}
Y\equiv {\rm arctanh}\left(p^z/p^t\right)\,, \index{Rapidity}
\end{equation}
is the (momentum) rapidity. Furthermore, the momentum polar angle $\phi$ is introduced through
\begin{equation}
p^x=p_\perp \cos\phi\,,\quad p^y=p_\perp \sin\phi\,.
\end{equation}
Note that in experiment, a frequently used quantity is pseudo-rapidity $\eta$ (not to be confused with the shear viscosity coefficient $\eta$), defined as
\begin{equation}
\eta\equiv  {\rm arctanh}\left(p^z/\sqrt{p_\perp^2+p_z^2}\right)\,,
\end{equation}
which coincides with rapidity $Y$ in the limit of ultrarelativistic particles \hbox{$m/|{\bf p}|\ll 1$}. Unlike $Y$, pseudo-rapidity does not require knowledge of the particle's mass, making it the experimental quantity of choice when presenting results for unidentified particles. 

\subsection{Multiplicity and Momentum Spectra}
\index{Multiplicity}
\index{Momentum spectra}
For a single simulation event, the number of hadrons $N_r$ of species $r$ from the hadron cascade is binned in rapidity intervals $dY$, momentum angles $d\phi$ and total transverse momentum $dp_\perp$, resulting in $\frac{dN_r}{dY d^2{\bf p}_\perp}$ where  $d^2{\bf p}_\perp\equiv p_\perp dp_\perp d\phi$. Summing particles over all angles and transverse momenta results in the so-called multiplicity per unity rapidity $\frac{dN_r}{dY}$. One can choose to report the multiplicity of a particular observable hadron species such as $\pi^{\pm},K^\pm$, or alternatively sum over hadron species to obtain the unidentified multiplicity $\frac{dN}{dY}=\sum_r \frac{dN_r}{dY}$. In experiment, uncharged particles are harder to observe, so often the multiplicity for unidentified \textit{charged} hadrons
\begin{equation}
\label{eq:chargedmult}
\frac{dN_{\rm ch}}{dY}=\sum_{r\in{\rm charged}} \frac{dN_r}{dY}\,,
\end{equation}
is reported instead. As a rule of thumb, roughly 2/3 of hadrons and hadron resonances carry electric charge. As another rule of thumb, one can approximate the pseudo-rapidity distribution as
\begin{equation}
\label{eq:pseudomult}
\frac{dN_{\rm ch}}{d \eta}\simeq \frac{1}{1.1} \frac{dN_{\rm ch}}{dY}\,.
\end{equation}

\index{Centrality}
In a theory simulation, the impact parameter of the simulated collision (``centrality'') is known, but that quantity is not directly accessible in experiment. However, for larger impact parameter, the theory results (\ref{eq:2comp}), (\ref{eq:emtweak2}), (\ref{eq:emtstrong}) imply that the deposited energy will tend to decrease. As a consequence, one typically expects fewer particles (smaller multiplicity) for larger impact parameters. Therefore, the multiplicity is a key quantity in experiment because it serves as a proxy for the impact parameter of a given collision. 

Modern simulations of nuclear collisions start with Monte-Carlo generated initial conditions (see section \ref{sec:overview}), requiring the use of \textit{event-by-event} averaging\index{Event-by-event averaging} in order to obtain results that can be compared to experimentally determined quantities.  For the case of the multiplicity, the event-averaged multiplicity is simply defined as
\begin{equation}
\label{eq:entav1}
\langle\langle \frac{dN_{\rm ch}}{dY} \rangle\rangle = \frac{1}{\rm events} \sum_{\rm events}\frac{dN_{\rm ch}}{dY}\,,
\end{equation}
where here and in the following $\langle\langle \cdot \rangle\rangle$ denotes an event-averaged quantity.

A more differential observable than the multiplicity is the identified particle transverse momentum spectrum obtained by integration over polar angles, 
\begin{equation}
\label{eq:momspectrum}
\frac{d N_r}{2\pi p_\perp dp_\perp dY}=\int \frac{d\phi}{2\pi}\frac{d N_r}{p_\perp dp_\perp d\phi dY}\,.
\end{equation}
The transverse momentum spectrum carries information on the amount of radial expansion (``radial flow'') of the system (see section \ref{sec:radialflow} below). The event-averaged momentum spectrum is given by an expression analogous to (\ref{eq:entav1}).
\index{Flow! Radial}

\subsection{Anisotropic Flow}
\index{Flow! Anisotropic}

Of particular interest is the angular information carried by particles. This information may be expressed in terms of the Fourier series \cite{Poskanzer:1998yz}
\begin{equation}
\label{eq:fourierdec}
\frac{d N_r}{p_\perp dp_\perp d\phi dY}=\frac{d N_r}{2\pi p_\perp dp_\perp dY}\left(1+2 \sum_{n=1}^\infty v_n(p_\perp,Y) \cos\left[n(\phi-\Psi_n(p_\perp,Y))\right]\right)\,,
\end{equation}
where $v_n(p_\perp,Y)$ are the (differential) flow coefficients and $\Psi_n(p_\perp,Y)$ are the associated flow angles\footnote{Note that $\Psi_n(p_\perp,Y)$ seem to exhibit little $p_T$ dependence for $n=2,3,4$ in the interval $p_\perp\in [0.5,2]$ GeV \cite{Heinz:2013bua}.}  of hadron species $r$ \cite{Heinz:2013th}. The coefficients $v_1,v_2,v_3$ are known as directed flow, elliptic flow and triangular flow \cite{Ollitrault:1992bk,Voloshin:1994mz,Alver:2010gr,Alver:2010dn,Snellings:2011sz}. 
If the momentum information of flow coefficients is not of interest, one integrates (\ref{eq:fourierdec}) over momenta to find 
\begin{equation}
\label{eq:fourierdec2}
\frac{d N_r}{d\phi dY}=\frac{d N_r}{2\pi dY}\left(1+2 \sum_{n=1}^\infty v_n(Y) \cos\left[n(\phi-\Psi_n(Y))\right]\right)\,,
\end{equation}
where $v_n^{\rm int}(Y)$ are the ``$p_\perp$-integrated'' flow coefficients of hadron species $r$\footnote{When comparing to experimental data, it is worth pointing out that experimental detectors typically have a low momentum cut-off $p_\perp^{\rm cut}$ below which particles are not measured. For ``integrated'' flow coefficients with a lower integration limit of $p_\perp^{\rm cut}\gg 0.1$ GeV, the resulting values are considerably larger than the ``true'' integrated results without cutoff \cite{Luzum:2010ag}.}

Note that in the definitions (\ref{eq:fourierdec}), (\ref{eq:fourierdec2}), the flow angles $\Psi_n$ are not known. For this reason, the flow coefficients $v_n$ have to be estimated using azimuthal particle correlations. A simple algorithm estimating the flow coefficients for a single simulation event from 2-particle correlations is given by
\begin{equation}
\langle v_n^2(p_\perp,Y)\rangle \equiv \frac{\left(\sum_i \cos(n \phi_i)\right)^2+\left(\sum_i \sin(n \phi_i) \right)^2}{M^2}\,,
\end{equation}
where $\phi_i$ is the polar angle for particle $i$ and the sums $\sum_{i=1}^M$ are over all $M$ hadrons of species $r$ with transverse momentum $p_T$. An event-averaged flow coefficient similar to the rms definition in Ref.~\cite{Alver:2008zza} can then be obtained as \cite{Heinz:2013th,Romatschke:2015gxa}
\begin{equation}
\label{eq:vnabsdef}
v_n^{\{abs\}}(p_\perp,Y)=\frac{1}{\rm events}\sum_{\rm events}\sqrt{\langle v_n^2(p_\perp,Y)\rangle}\,.
\end{equation}

Note that this algorithm does not correct for autocorrelations and that 2-particle correlations in experiment are known to be affected by so-called non-flow contributions \cite{Snellings:2011sz}. For this reason, higher order particle correlations are often employed. A standard way to  define the average 2- and 4-particle azimuthal correlations for a single simulation event is given by \cite{Bilandzic:2010jr}
\begin{eqnarray}
\langle 2 \rangle&\equiv& \langle e^{i n (\phi_1-\phi_2)}\rangle=\frac{1}{P_{M,2}} \sum_{i\neq j} e^{i n (\phi_i-\phi_j)}\,,\\
\langle 4 \rangle&\equiv& \langle e^{i n (\phi_1+\phi_2-\phi_3-\phi_4)}\rangle=\frac{1}{P_{M,4}} \sum_{i\neq j\neq k\neq l} e^{i n (\phi_i+\phi_j-\phi_k-\phi_l)}\,,
\end{eqnarray}
where $P_{n,m}=n!/(n-m)!$ and for identified particle observables the sums run over all $M$ hadrons of a particular species (multiplicity of hadron species $r$). An efficient algorithm to calculate these 2- and 4-particle correlations employs the concepts of Q-cumulants \cite{Borghini:2000sa,Borghini:2001vi,Bilandzic:2010jr}
\begin{equation}
Q_n\equiv \langle e^{i n \phi}\rangle = \sum_{i=1}^M e^{i n \phi_i}\,,
\end{equation}
where again the sum is over all $M$ hadrons of species $r$ in a single event. In terms of the Q-cumulants, one finds \cite{Bilandzic:2010jr}
\begin{eqnarray}
\langle 2\rangle &=& \frac{|Q_n|^2-M}{M(M-1)}\,,\\
\langle 4\rangle &=& \frac{|Q_n|^4+|Q_{2n}|^2-2 {\rm Re}\left(Q_{2n} (Q_n^*)^2\right)-4 (M-2)|Q_n|^2+2 M (M-3)}{M (M-1)(M-2)(M-3)}\,,\nonumber \index{Multi-particle correlations}
\end{eqnarray}
where ${\rm Re}(a),z^*$ denote the real part and complex conjugate of a complex number $z$. With single-event 2- and 4-particle correlations determined, a weighted event-average
\begin{equation}
\langle\langle m\rangle\rangle\equiv \frac{\sum_{\rm events}W_{m} \langle m \rangle}{\sum_{\rm events}W_{m}}
\end{equation}
with $W_{m}=M!/(M-m)!$ and $M$ the hadron multiplicity of a single event is employed to minimize the effect of multiplicity variations. The event-averaged 2- and 4- particle correlations can be used to calculate the ``genuine'' 2- and 4-particle correlations
\begin{equation}
c_n^{\{2\}}=\langle \langle 2\rangle\rangle\,,\quad
c_n^{\{4\}}=\langle \langle 4\rangle\rangle-2 \langle \langle 2\rangle\rangle^2\,,
\end{equation}
which in turn can be used to calculate the event-averaged $p_\perp$-integrated 2- and 4-particle flow coefficients
\begin{equation}
v_n^{\rm int,\{2\}}(Y)=\sqrt{c_n^{\{2\}}}\,,\quad
v_n^{\rm int,\{4\}}(Y)=^4\sqrt{-c_n^{\{4\}}}\,.
\end{equation}
Note that in low-multiplicity events (such as in very peripheral nucleus-nucleus collisions or proton-proton collisions), $c_n^{\{4\}}$ turns out to be positive, preventing interpretation of $v_n^{\rm int,\{4\}}(Y)$ as a real valued four-particle flow coefficient. Nevertheless, comparison to experimental data directly for $c_n^{\{4\}}$ is possible.

Similar to $p_\perp$-integrated flow coefficients, differential event-averaged flow  coefficients can be calculated using the Q-cumulants as \cite{Bilandzic:2010jr}
\begin{equation}
v_n^{\{2\}}(p_\perp,Y)=\frac{\langle\langle 2^\prime\rangle\rangle}{\sqrt{c_n^{\{2\}}}}\,,\quad
\langle\langle 2^\prime\rangle\rangle = \frac{\sum_{\rm events} {\rm Re}(p_n Q_n^*-m_q)}{\sum_{\rm events}  m_p M- m_q}\,,
\end{equation}
where $p_n=\sum_i^{m_p}e^{i n \phi_i}$ is the differential cumulant for the $m_p$ particles of interest calculated in reference to $m_q$ particles of reference. (See Ref.~\cite{Bilandzic:2012wva} for the corresponding expressions for 4-particle and 6-particle flow coefficients.) Note that in practical applications, $v_n^{\{2\}}(p_\perp,Y)$ often turns out to be indistinguishable from $v_n^{\{abs\}}(p_\perp,Y)$ defined in (\ref{eq:vnabsdef}).

\subsection{HBT Radii}
\index{HBT Radii}
\index{Femtoscopy|see {HBT Radii}}

Hanbury-Brown-Twiss (HBT) radii are a concept originally invented in astronomy to determine sizes of distant stars \cite{1956Natur.177...27B}. In the context of nuclear collisions, the normalized two particle correlation function $C(K^\mu,q^\mu)$ is constructed as the ratio between two-particle inclusive and single particle inclusive momentum spectra  where
\begin{equation}
K^\mu=\frac{p_1^\mu+p_2^\mu}{2}\,,\quad
q^\mu=\frac{p_1^\mu-p_2^\mu}{2}\,,
\end{equation}
are the average and difference between the two particle momenta $p_1^\mu,p_2^\mu$. Under certain assumptions concerning the emission source \cite{Lisa:2005dd}, the two-particle correlation function at small ${\bf q}$ may be written as \cite{Novak:2013bqa},
\begin{equation}
C(K^\mu,q^\mu)=\int d^3x S(K,x) \left(1+\cos(2 {\bf q}\cdot {\bf x})\right)\,,
\end{equation}
where $S(K,x)$ is the emission source function determined by the single-particle distribution $f$ 
\begin{equation}
S(K,x)=\frac{\int d^3x_1 d^3x_2 f(K,x_1) f(K,x_2)\delta^3\left({\bf x}-{\bf x}_1+{\bf x}_2\right)}{\int d^3x_1 d^3x_2 f(K,x_1) f(K,x_2)}\,.
\end{equation}
A detailed construction for obtaining $S(K,x)$ from the output of a hadron cascade calculation can be found in Ref.~\cite{Novak:2013bqa}.

Assuming the emission source function to be Gaussian $S(K,x)\propto e^{-{\bf x}\cdot {\bf M} \cdot {\bf x}}$ with ${\bf M}$ a diagonal matrix depending on $K$, one obtains
\begin{equation}
\label{eq:HBT}
C(K^\mu,q^\mu)\simeq 1+e^{-R_{\rm out}^2(K) q_{\rm out}^2-R_{\rm side}^2(K) q_{\rm side}^2-R_{\rm long}^2(K) q_{\rm long}^2}\,,
\end{equation}
where the ``outward'' direction is transverse to the beam axis and parallel to ${\bf K}$, ``longitudinal'' is along the beam axis and ``sideward'' is perpendicular to both. The radii $R_{\rm out}, R_{\rm side}, R_{\rm long}$ are commonly referred to as HBT radii and characterize the extent and lifetime of the emission source.

Historically, it had been challenging for hydrodynamic models of nuclear collisions to describe experimentally measured HBT radii \cite{Teaney:2000cw,Soff:2000eh,Hirano:2002hv}, which is known as the ``HBT Puzzle''. The puzzle was finally solved with the advent of modern hybrid models combining pre-equilibrium flow, viscous hydrodynamic evolution and hadronic rescatterings \cite{Pratt:2008qv,Habich:2014jna}.

\section{Physics Interpretation of Observables in a Simplified Model}
\label{sec:simplified}

In order to highlight the physics interpretation of the observables discussed in the previous section, let us consider a simplified model facilitating the corresponding calculations. First, let us consider neglecting hadron interactions and decays, such that hadrons emitted from the switching hypersurface are assumed to follow straight-line trajectories to the detector. While quantitatively inaccurate, neglecting the hadron cascade does not change the qualitative physics interpretation. Neglecting the hadron cascade, the switching hypersurface $\Sigma$ is identical with the kinetic freeze-out hypersurface discussed above. Second, let us make the simplifying assumption that the system freezes out at a given proper time $\tau=\tau_c$ (isochronous decoupling) rather than the isothermal decoupling discussed above. This assumption is justified if we are only interested in qualitative results. Third, let us neglect quantum statistics so that the equilibrium particle distribution function of hadron resonances will be given by the Maxwell-J\"uttner form (\ref{eq:maxjut}). Finally, we will make the assumption that the system obeys certain additional symmetries, such as Bjorken boost invariance and/or azimuthal symmetry, etc, which will allow for semi-analytic calculation of observables without qualitatively changing the physics.

For a freeze-out hypersurface $\Sigma$, the total number of hadrons of a given species $r$ will be given by
\begin{equation}
N_r=-\int n^\mu d\Sigma_\mu\,,\quad n^\mu=\int d\chi_r p^\mu f_r(x^\mu,p^\mu)\,,
\end{equation}
where $n^\mu$ is the kinetic theory expression for the particle current (\ref{eq:numcurr}). Since $\Sigma$ is parametrized by $\tau=\tau_c$, its oriented surface element is given by $d\Sigma^\mu=\left(1,{\bf 0}\right)\tau dx dy d\xi$.  Using (\ref{eq:momdefs}), one finds
\begin{equation}
\label{eq:hist1}
\frac{dN_r}{d^2p_\perp dY}=d_r \tau \sqrt{m_r^2+p_\perp^2} \int \frac{dx dy d\xi}{(2\pi)^3}  \cosh(Y-\xi) f_r\,,
\end{equation}
where $d_r\equiv (2s_r+1)(2g_r+1)$ is the degeneracy factor for the hadron species $r$. 

\subsection{Particle Yields}
\index{Particle yields}

Let us first consider the case of boost invariance and a spatially homogeneous system (Bjorken flow), implying $u^x=u^y=u^\xi=0$ and $\pi^x_x=\pi^y_y=-\frac{1}{2}\pi^\xi_\xi$. For illustrative purposes, let us consider the quadratic ansatz (\ref{eq:KTrec}) for $f_r$ to implement the non-equilibrium hadronization procedure. Then, the particle spectrum for hadron species $r$ is given by
\begin{equation}
\label{eq:fullyield}
\frac{dN_r}{d^2p_\perp dY}=d_r \tau \sqrt{m_r^2+p_\perp^2} \frac{2 A_\perp}{(2\pi)^3}   \left(K_1(z)+\frac{5\pi^\xi_\xi\left(- p_\perp^2 K_0(z)+2 z T^2 K_1(z)\right)}{4 T(\epsilon+P)\sqrt{m_r^2+p_\perp^2}} \right)\,,
\end{equation}
where $A_\perp=\int dx dy$ is the transverse area of the system, $K(z)$ are modified Bessel functions and the shorthand notation $z\equiv \frac{\sqrt{m_r^2+p_\perp^2}}{T}$ has been introduced for convenience. Total hadron yields per unit rapidity may be obtained from $\int d^2 p_\perp \frac{dN_r}{d^2p_\perp dY}$. In equilibrium (setting $\pi^\xi_\xi=0$), hadron yields are given by 
\begin{equation}
\label{eq:statmodel}
\frac{dN_r^{\rm eq}}{dY}=\frac{2 d_r \tau A_\perp}{(2\pi)^2}m_r^2 T K_2\left(\frac{m_r}{T}\right)\,.
\end{equation}

 \begin{figure}[t]
  \begin{center}    
     \includegraphics[width=.49\linewidth]{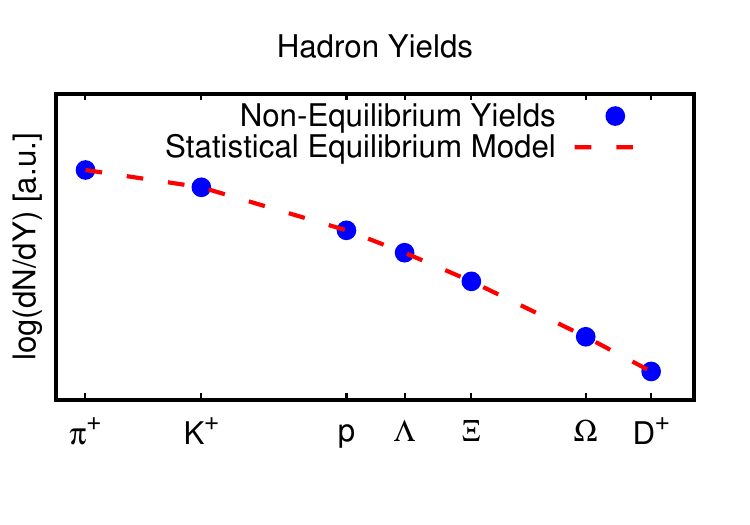}
     \hfill
     \includegraphics[width=.49\linewidth]{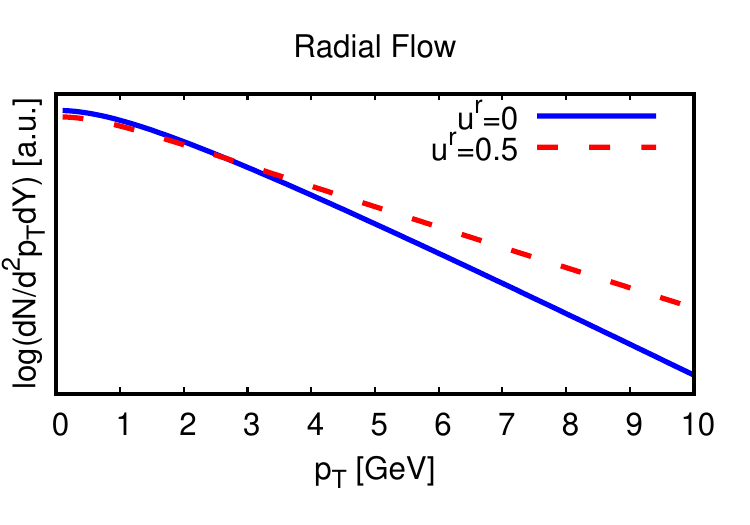}
     \end{center}
  \caption{\label{fig4:appli} Left: Non-equilibrium particle yields compared to statistical model results assuming equilibrium, Eq.~(\ref{eq:statmodel}), as a function of particle mass for $\pi^+,K^+,p,\Lambda,\Xi,\Omega,D^+$. Non-equilibrium contribution to particle yields are found to vanish upon momentum integration for arbitrarily strong non-equilibrium stresses. Right: Momentum spectra with vanishing and non-vanishing radial fluid velocity.}
\end{figure}

Distribution functions assuming thermal and chemical equilibrium have been successfully employed in statistical model fits to experimental data \cite{Becattini:2003wp,BraunMunzinger:1994xr}. In fact, the successes of the statistical models assuming equilibrium are commonly interpreted as indication that hadrons are emitted from an equilibrated source. Fig. ~\ref{fig4:appli} compares the ``statistical equilibrium model'' hadron yields (\ref{eq:statmodel}) to the full, non-equilibrium yields $\int d^2 p_\perp \frac{dN_r}{d^2p_\perp dY}$ using (\ref{eq:fullyield}). Results shown in Fig. ~\ref{fig4:appli} for equilibrium and non-equilibrium not only appear to be very close, they are in fact identical owing to the fact that the non-equilibrium correction to the hadron yields vanishes identically, 
\begin{equation}
\int d^2p_\perp \frac{5\pi^\xi_\xi\left(- p_\perp^2 K_0(z)+2 z T^2 K_1(z)\right)}{4 T(\epsilon+P)} =0\,,
\end{equation}
for \textit{arbitrary} values of the shear stress $\pi^\xi_\xi$. Cautioning that several simplifying assumptions have been made in deriving this result analytically, it nevertheless strongly suggests that total particle yields measured in experiment are not very sensitive to kinetic non-equilibrium contributions. Therefore, successful description of experimental data with equilibrium statistical models most likely can not be used as evidence for thermalization of the medium.

\subsection{Radial Flow}
\label{sec:radialflow}
\index{Flow! Radial}

Let us lift the assumption of spatial homogeneity in the transverse plane used in the previous subsection and consider a system that is boost-invariant (implying $u^\xi=0$) and azimuthally symmetric (implying $u^x=u^y=u^r/\sqrt{2}$) instead. Assuming an equilibrium distribution of the form (\ref{eq:maxjut}), Eq.~(\ref{eq:hist1}) then leads to transverse momentum spectra (\ref{eq:momspectrum}) of the form \cite{Baier:2006gy,Romatschke:2009im}
\begin{equation}
\frac{dN_r}{2 \pi p_\perp dp_\perp dY}=2 d_r \tau \sqrt{m_r^2+p_\perp^2}\int \frac{dx dy}{(2\pi)^3}  K_1\left(\frac{\sqrt{m_r^2+p_\perp^2} u^\tau}{T}\right)I_0\left(\frac{p_\perp u^r}{T}\right)\,,
\end{equation}
where $I_0(z)$ denotes a modified Bessel function. Transverse momentum spectra are shown in Fig.~\ref{fig4:appli} by plotting $\frac{N_r}{2\pi p_\perp dp_\perp dY dx dy}$ for $u^r=0$ and $u^r\neq 0$, respectively. As can be seen from this figure, non-vanishing radial flow $u^r\neq 0$ implies flatter momentum spectra (cf. Ref.~\cite{Huovinen:2006jp} for direct simulations using ideal fluid dynamics). Note that non-equilibrium corrections (\ref{eq:fullyield}) tend to have the same effect as radial flow, also resulting in flatter momentum spectra, \cite{Baier:2006gy,Romatschke:2009im}.

Approximating momentum spectra as $\frac{dN}{d^2p_\perp dY}\propto e^{-p_\perp/T_{\rm slope}}$ at $p_\perp\gg m_r$, flatter momentum spectra imply a larger slope parameter $T_{\rm slope}$. This is consistent with the expectation that a radially flowing source leads to an effective blue-shifted temperature \cite{Schnedermann:1993ws}.

\subsection{Elliptic Flow}
\index{Flow! Elliptic}

Let us lift the assumption of azimuthal symmetry used in the previous subsection and consider a system that is boost-invariant (implying $u^\xi=0$) instead. Assuming an equilibrium distribution function (\ref{eq:maxjut}), the particle spectra (\ref{eq:hist1}) for high momentum particles $p_\perp \gg T,m_r$ may be calculated by a saddle point approximation of the integral. The saddled point corresponds to a minimum of 
\begin{equation}
\label{eq:arg}
-p^\mu u_\mu=\sqrt{m_r^2+p_\perp^2}\cosh(Y-\xi) u^\tau-{\bf p} \cdot {\bf u}\,.
\end{equation}
The minimum may be conveniently located by employing the (local) velocity decomposition ${\bf u}={\bf u}_{||}+{\bf u}_{\perp}$ with ${\bf u}_{||}$ a component parallel to ${\bf p}_\perp$ and ${\bf u}_{\perp}$ perpendicular to ${\bf p}_\perp$ such that $u^\tau=\sqrt{1+u_{||}^2+u_\perp^2}$. In terms of these coordinates, the minimum of (\ref{eq:arg}) is found at \cite{Borghini:2005kd}
\begin{equation}
\xi=Y\,,\quad u_\perp({\bf x}_\perp)=0\,,\quad u_{||}({\bf x}_\perp)=u_{\rm max}({\bf x}_\perp)\,,
\end{equation}
where $u_{\rm max}({\bf x}_\perp)$ corresponds to the local maximum value of ${\bf u}$. Expanding (\ref{eq:arg}) around the saddle point leads to
the angular information of the particle spectra (\ref{eq:hist1}) \cite{Borghini:2005kd}
\begin{equation}
\frac{d N_r}{d^2{\bf p}_\perp dY}\propto \frac{1}{\sqrt{p_t-\sqrt{m_r^2+p_\perp^2} u_{\rm max}/u_{\rm max}^\tau}}e^{\frac{p_\perp u_{\rm max}-\sqrt{m_r^2+p_\perp^2} u_{\rm max}^\tau}{T}}\,,
\end{equation}
where $u_{\rm max}^\tau=\sqrt{1+u_{\rm max}^2}$. Let us now consider a velocity profile with the angular dependence
\begin{equation}
\label{eq:exflowprof}
u_{\rm max}=u^r\left(1+2 \sum_{n=1}^\infty V_n \cos(n\phi)+\ldots\right)\,.
\end{equation}
If $T$ is sufficiently small, the azimuthal dependence of $\frac{d N_r}{d^2{\bf p}_\perp dY}$ will be dominated by the exponential. Expanding the exponential to leading order in $V_n$, one obtains the flow coefficients (\ref{eq:vnabsdef}) \cite{Borghini:2005kd}
\begin{equation}
\label{eq:loflow}
v_n^{\rm abs}(p_\perp)=\frac{V_n u^r}{T}\left(p_\perp-\frac{\sqrt{m_r^2+p_\perp^2} u^r}{\sqrt{1+u_r^2}}\right)+{\cal O}(V_n^2)\,.
\end{equation}
The $n=2$ coefficient is called ``elliptic flow'', because it corresponds to an elliptic modulation of the flow profile (\ref{eq:exflowprof}).
Higher values of $n$ are referred to as ``triangular flow'' (n=3), ``quadrangular flow'' (n=4), etc.
\index{Flow! Triangular}
The result (\ref{eq:loflow}) suggests that heavier particles have smaller $v_n$ in ideal hydrodynamics, which has been given the name of ``mass ordering'' in the context of relativistic nuclear collisions. Non-linear contributions to the flow coefficients may also be calculated in the same manner \cite{Borghini:2005kd}. For instance, one finds that for the so-called ``quadrangular flow'' coefficient, the second-order contribution
\begin{equation}
v_4^{\rm abs}(p_\perp)\simeq \left(\frac{u^r}{T}\left(p_\perp-\sqrt{m_r^2+p_\perp^2} v^r\right)\right)^2\frac{1}{2}\left(V_2^2+2 V_1 V_3+2 V_1 V_5+2 V_2 V_6+\ldots\right)\,,
\end{equation}
dominates over (\ref{eq:loflow}) for sufficiently large $p_\perp$. These non-linear ``mode-mixing'' contributions have been discussed in more detail in Ref.~\cite{Teaney:2010vd}.

\chapter{Comparison to Experimental Data}
\label{chap:experiment}

The previous chapters contain a discussion of the theory setup of relativistic hydrodynamics,
as well as the necessary background to perform hydrodynamic simulations of relativistic
nuclear collisions. The topic of this chapter is the comparison of theory simulations to
experimental data. We start by a discussion of the regime of applicability of hydrodynamic simulations to nuclear collision systems, 
addressing the question of which set of experimental data can be expected to be quantitatively described by hydrodynamic simulations. 

This discussion is followed-up by a summary of the hydrodynamic simulation package superSONIC that will be used
for the data-theory comparisons in this chapter. The hydrodynamic simulations will be based on first-principles insight detailed in the preceding chapters, constraining the form of the energy-deposition following a nuclear collision, the presence of pre-hydrodynamic flow, transport coefficients, hadronization, hadron cascades and analysis of experimentally measured observables. In the subsequent data-theory comparison, we have deviated from the valuable traditional approach taken in the field to tune model ingredients to obtain optimal agreement with certain (sub-)sets of experimental data. Instead, the emphasis in this work will be on the overall broad data-theory agreement that can be achieved by using first-principles theory input without any fine-tuning at all. For this reason, we will employ a single (simple) choice for all the parameters of the fluid simulations for \textit{all} systems and \textit{all} observables shown. Because of the minimal number of adjustable parameters in the model, no precision agreement with experimental data can be expected. 

\section{Applicability of Hydrodynamic Simulations}

Historically, the application of hydrodynamic simulations was thought to be reserved for
collisions of sufficiently large nuclei (``heavy-ion collisions''). The rational behind
this limitation was that only for large nuclei, the traditional criterion (\ref{eq:landaucrit}) of system size $L$ being much larger than the local mean free path $\lambda_{\rm mfp}$ was expected to hold.
Estimating 
the system size for a heavy-ion collision by the radius of a lead ion $L\simeq 6$ fm (see Tab.~\ref{tab4:one}), and estimating the mean free path as $\lambda_{\rm mfp}\simeq \frac{\eta}{s T}$ at a temperature of $T=0.2\,{\rm GeV}\simeq 1\, {\rm fm}^{-1}$, with the shear viscosity over entropy ratio given by pQCD value of $\frac{\eta}{s}\simeq 1.5$ at a QCD coupling of $\alpha_s=0.3$, one finds that the traditional criterion (\ref{eq:landaucrit}) for the applicability of fluid dynamics is barely fulfilled\footnote{In fact, it is known that local equilibrium is \textit{not} reached even in heavy-ion collisions \cite{Romatschke:2016hle}.}. Moreover, based on these estimates, one would not expect hydrodynamic simulations to be applicable to smaller collision systems, such as those created in deuteron-gold or proton-proton collisions with system sizes $L\lesssim 2$ fm.
However, within the past decade, two developments have cast doubt on this historic estimate for the applicability of fluid dynamics to smaller systems. First, calculations in strongly coupled gauge theories have indicated that the shear viscosity over entropy ratio for nuclear matter may be considerably smaller than the pQCD estimate, e.g. $\frac{\eta}{s}\simeq 0.1$ rather than $\frac{\eta}{s}\simeq 1.5$, cf. Tab.~\ref{tab4:one}. Second, as outlined in section \ref{sec:offeq}, the modern theory of relativistic viscous hydrodynamics as presented in this work applies even in regimes where $L<\lambda_{\rm mfp}$  
as long as a hydrodynamic attractor is present, and the effects from non-hydrodynamic modes can be neglected\footnote{Note that present-day theory simulations based on rBRSSS can only be expected to be quantitatively applicable whenever second-order hydrodynamics is close to the hydrodynamic attractor, see section \ref{sec:unreasonable}.}. 

Taken together, these two developments open up the possibility of previously unimaginable applications of hydrodynamics to much smaller systems, effectively leading to a change of
paradigm for relativistic nuclear collisions \cite{Schukraft:2017nbn,Zajc:2017hbs,Florkowski:2017olj}. Nevertheless, at sufficiently small scales, non-hydrodynamic modes become dominant. Once this happens, there no longer is a hydrodynamic attractor, and fluid dynamics has broken down. Calculations on the critical momentum beyond which fluid dynamics breaks down have been performed in weak coupling kinetic theory (see section \ref{sec:ktnonhydro} and Ref.~\cite{Romatschke:2015gic}) and strong coupling gauge/gravity duality (see section \ref{sec:ggnonhydro} and Ref.~\cite{Grozdanov:2016vgg}). Results suggest fluid dynamics breaks down for momenta exceeding four to seven times the local pseudo-temperature \cite{Romatschke:2016hle}. 
The pseudo-temperature itself varies during the hydrodynamic evolution, typically in the range of $T\in[0.17,0.6]$ GeV, as the system expands into the vacuum and cools down. This suggests that hydrodynamics can offer a quantitatively reliable description for particles with low transverse momenta $p_\perp \leq 0.5$ GeV, and will break down completely at transverse momenta exceeding $p_\perp\geq 4$ GeV. 
Therefore, quantitative agreement of hydrodynamic simulations with experimental data is expected to gradually worsen for increasing particle momenta $p_\perp \in [0.5,4]$ GeV, and fail qualitatively for particles with momenta $p_\perp \gg 4$ GeV.




In this chapter, we present evidence that hydrodynamic simulations indeed are able to offer quantitatively accurate descriptions of experimental data for low-momentum particles, for large, medium and even the smallest collision systems that can be probed in relativistic nuclear collisions. To highlight the breadth of the model description, we have selected a subset of key observables and collision systems that in our opinion are representative of the overall quality of the description. We expect the model description to perform similarly well for a multitude of other collision systems and low momentum observables which are not discussed in this work.

\section{Model Simulation Package: superSONIC}

Many model packages for simulating the evolution of relativistic nuclear collisions exist,
among them MUSIC, VISHNU, superSONIC, EKRT, EPOS, ECHO-QGP and others \cite{MUSIC,SONIC,VISHNU,Niemi:2015qia,Werner:2010aa,DelZanna:2013eua,Bozek:2009dw,Alqahtani:2017jwl}. These model packages have respective
strengths and weaknesses, and have been optimized for certain aspects or certain collision systems.
Common to all the theoretical simulations of nuclear collisions is that all input parameters and initial
conditions need to be chosen, see the discussion in section \ref{sec:overview}. In order to make the
following data-theory comparison as transparent as possible, all theory simulation results presented in this chapter have been calculated using the same simulation package with
the same set of input parameters. Out of convenience, we have chosen our own superSONIC model
for this comparison, but we expect other model packages to obtain quantitatively similar results.

The choices for the superSONIC simulations for relativistic heavy-ions, heavy-on-light-ions and proton-proton collisions
shown in the remainder of this chapter are based on the following physics inputs:
\begin{itemize}
\item
  {\bf Initial Conditions:} In first-principles calculations both at weak and strong coupling discussed in sections \ref{sec:weak}, \ref{sec:strong}, it was found that the deposited energy scales as the product of the overlap functions of the target and projectile, respectively. Based on these results, in superSONIC the energy density at mid-rapidity deposited in a nuclear collision is taken to follow a Glauber-model binary collision scaling, cf. Eq.~(\ref{eq:2comp}). In order to describe proton-proton collisions, a Glauber model including three constituent partons with positions ${\bf x}_{\perp,k}$, $k=1,2,3$, will be employed, as described in section \ref{sec:ppglauber}, using $R=0.52$ fm for the width of the nucleon (\ref{eq:protonTAG}) and $\sigma_g=0.46$ fm for the width of the partons. The deposited energy-density profile in superSONIC is calculated as
\begin{equation}
  \label{eq:glauberdep}
  \epsilon({\bf x}_\perp)=\epsilon_0\sum_{i}^{\rm collisions} \sum_{k=1}^3 \gamma^{(k)} \frac{1}{2\pi \sigma_g^2}e^{-|{\bf x_\perp}-{\bf x}_{\perp,k}^{(i)}|^2/(2\sigma_g^2)}\,,
\end{equation}
where the parameter $\epsilon_0$ depends on the collision system and collision energy (but does not depend on the impact parameter of the collision). In practice it is taken as a free parameter that is fixed by matching to experimental data for the total multiplicity for central collisions. Note that this is the \textit{only} parameter of the model that will be tuned to reproduce experimental data in the various collision systems and energies!
Fluctuations in energy deposition are modeled by allowing $\gamma^{(k)}$ to fluctuate for individual partons according to a probability distribution \cite{Shen:2014vra,Welsh:2016siu}
\begin{equation}
  P_\gamma(\gamma)=\frac{\gamma^{-1+1/(3\theta)}e^{-\gamma/\theta}}{\theta^{1/(3\theta)} \Gamma(\frac{1}{3\theta})}\,,
\end{equation}
with $\theta$ a collision-energy parameter controlling the shape of the multiplicity fluctuations (we fix $\theta=2$ in the following). 
\item
  \textbf{Pre-Hydrodynamic Evolution}:
  In first-principles calculations both at weak and strong coupling it was found that the pre-hydrodynamic flow profile at early proper times $\tau$ is given by (\ref{eq:preeq1}), (\ref{eq:preeq2})
  \begin{equation}
    \label{eq:preeqsonic}
    {\bf v}_\perp=-\frac{\tau}{b}\frac{\partial}{\partial {\bf x}_\perp} \ln \epsilon({\bf x}_\perp)\,,
  \end{equation}
  where the range of $b\in [3,4]$ correspond to the limits of infinitely strong (weak) coupling (see also the corresponding discussions in Refs.~\cite{Vredevoogd:2008id,Keegan:2016cpi}). In superSONIC, Eq.~(\ref{eq:preeqsonic}) with $b=3$ is implemented for the flow profile $u^\mu$ at the start of the hydrodynamic evolution at $\tau=\tau_0$. For simplicity, and in accordance with the findings in Ref.~\cite{vanderSchee:2013pia}, $\epsilon({\bf x}_\perp)$ at $\tau=\tau_0$ is left unchanged from the initial distribution (\ref{eq:glauberdep}). Finally, the shear and bulk stress tensors $\pi^{\mu\nu},\Pi$ are set to zero at $\tau=\tau_0$, since their initial values do not appear to influence results even in proton-proton collisions, cf. Ref.~\cite{Weller:2017tsr}. As shown in Ref.~\cite{vanderSchee:2013pia}, hydrodynamic results with these starting profiles are essentially insensitive to the choice of the hydrodynamic starting time $\tau_0$. In the following, the choice $\tau_0=0.25$ fm/c is used in superSONIC calculations.
\item
  \textbf{Transport Coefficients}:
  For the QCD equation of state $\epsilon=\epsilon(P)$, the lattice QCD results (\ref{eq:latticeeos1}) are used. For the shear and bulk viscosity coefficients, absent first-principles calculations of their temperature dependence in QCD, constant values of the ratios $\frac{\eta}{s}$ and $\frac{\zeta}{s}$ are employed\footnote{Gradient-dependent effective viscosities as those resulting from Borel resummation shown in Fig.~ \ref{fig:two}, are not currently implemented in superSONIC.}. For the second-order transport coefficients, the bulk relaxation time is set to $\tau_\Pi=\tau_\pi$ (cf. Table \ref{tab:one3}) and the shear relaxation time is set to $\tau_\pi=C_\pi\frac{\eta}{s T}$ where $T$ is the local pseudo-temperature and $C_\pi\in [2.61,5.9]$ corresponding to the expected range from strong to weak coupling (cf. Table \ref{tab:one2}). The values of the other second-order transport coefficients are set to zero since they do not seem to affect final results, cf. Ref.~\cite{Luzum:2008cw}. Because of the relation of $\tau_\pi$ to the non-hydrodynamic mode damping rate in rBRSSS (\ref{eq:rbrsssnonhydro}), changing the value of $C_\pi$ also probes the sensitivity of final results to non-hydrodynamic modes. Small variations resulting from varying $C_\pi$ indicate insensitivity to non-hydrodynamic degrees of freedom, whereas large variations indicate a breakdown of fluid dynamics.
\item
  \textbf{Numerical Fluid Dynamics Algorithm}:
  A naive discretization scheme with a gradient patch described in section \ref{sec:naivedisc} is used to numerically solve the fluid dynamics equations $\nabla_\mu T^{\mu\nu}=0$ with $T^{\mu\nu}=T^{\mu\nu}_0+\pi^{\mu\nu}+\Delta^{\mu\nu} \Pi$, cf. Eqns. (\ref{eq:r2h}). For superSONIC simulations, boost invariance is assumed, such that the equations are effectively only 2+1 dimensional. Typical choices for the temporal and spatial lattice spacings are $\delta \tau=0.01 \delta x$ and $\delta x=0.5$ ${\rm GeV}^{-1}$ on a $250\times 250$ spatial grid for heavy-ion collisions. 
\item
  \textbf{Switching Conditions}:
  Isothermal switching at the QCD deconfinement transition temperature $T_c=0.17$ GeV is employed. Elements of the switching hypersurface are constructed using the ``block-element'' approximation discussed in section \ref{sec:hadro}. 
\item
  \textbf{Hadronic Cascade}:
  Initial hadron configurations are obtained by Monte-Carlo sampling off-equilibrium ``exponential-ansatz'' distribution functions (\ref{eq:KTmodel2}) on the switching hypersurface. Owing to the unknown form for the off-equilibrium hadronic distributions for non-conformal systems, bulk stress tensor contributions to hadronic spectra are ignored. Hadron interactions are then simulated in superSONIC using the hadron cascade code B3D (see section \ref{sec:cascade}) until particles have reached kinetic freeze-out and follow straight-line trajectories to the detectors.
\item
  \textbf{Particle Analysis}:
  For a single hadron cascade event, observables are calculated as described in section \ref{sec:obs}. In order to increase statistics, superSONIC employs an oversampling approach where typically 1,000 hadron cascade events are combined for every single hydrodynamic event into one ``super-event''. This procedure artificially suppresses decay correlations arising in the hadronic phase.
  \end{itemize}

 \begin{figure}[t]
  \begin{center}    
    \includegraphics[width=.7\linewidth]{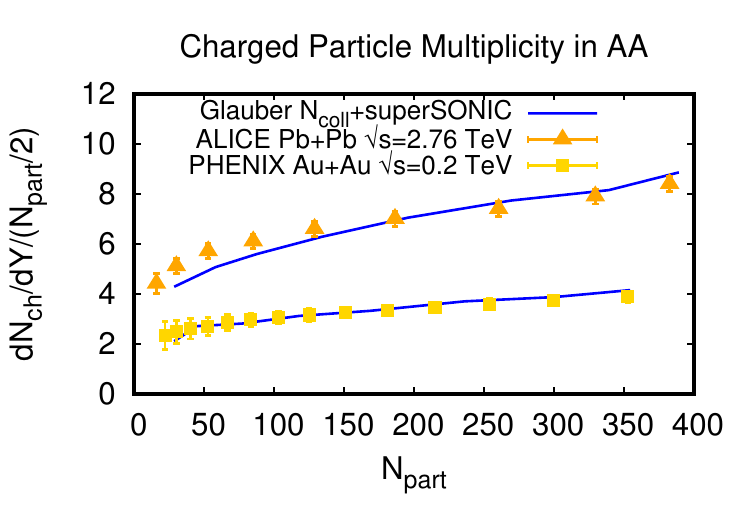}
    \includegraphics[width=.7\linewidth]{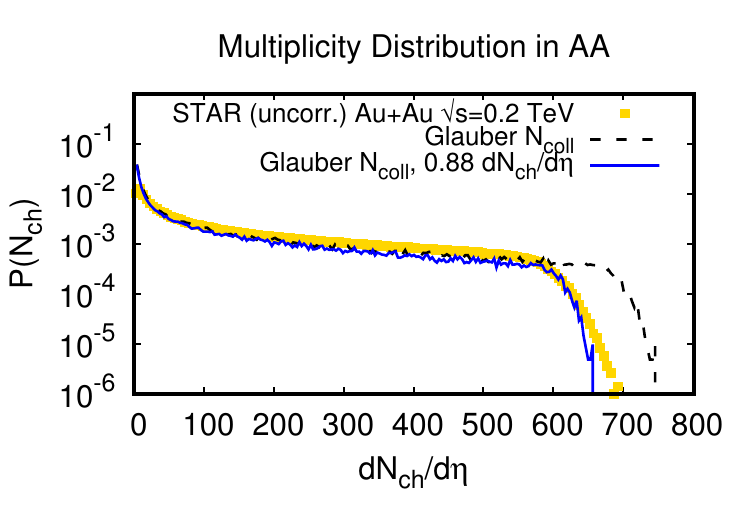}
     \end{center}
  \caption{\label{fig5:multAA} Upper panel: charged particle multiplicity from theory simulations (``superSONIC'', using $\eta/s=0.12$, $\zeta/s=0.02$) for heavy-ion collisions as a function of centrality (number of participants) compared to experimental data from Refs.~\cite{Adler:2004zn,Aamodt:2010cz}. Lower panel: multiplicity distribution from Glauber initial conditions (no hydrodynamic simulations) compared to uncorrected experimental data from Ref.~\cite{Abelev:2008ab} for Au+Au collisions at $\sqrt{s}=0.2$ TeV. To account for unknown experimental resolution effects, the theory curve with a multiplicity multiplication factor of $0.88$ is also shown.}
\end{figure}

\section{Heavy-Ion Collisions}


\begin{figure}[t]
  \begin{center}    
    \includegraphics[width=.7\linewidth]{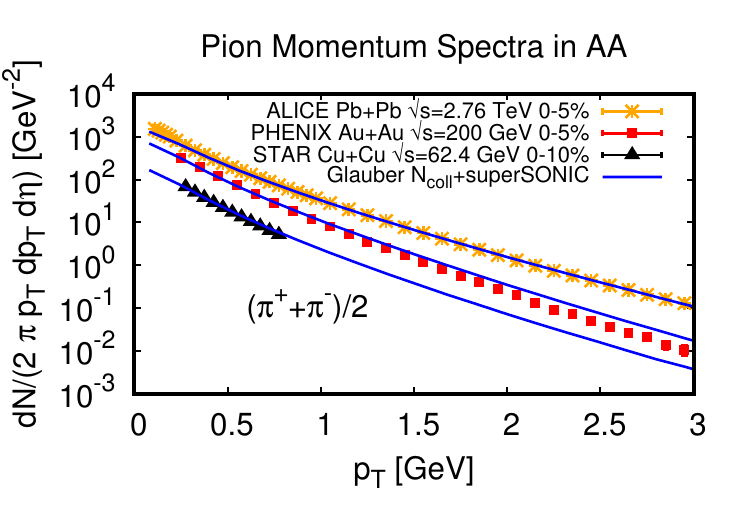}
    \hfill
     \end{center}
  \caption{\label{fig5:IdAA} Pion spectra as a function of transverse momentum from theory simulations (``superSONIC'', using $\eta/s=0.12$, $\zeta/s=0.02$) for heavy-ion collisions compared to experimental data from Refs.~\cite{Abelev:2013vea,Adler:2003cb,Aggarwal:2010pj}.
  }
\end{figure}

Fig.~\ref{fig5:multAA} shows a comparison of the multiplicity per unit rapidity from superSONIC compared to experimental data for Au+Au collisions at $\sqrt{s}=200$ GeV and Pb+Pb collisions at $\sqrt{s}=2.76$ TeV. Both simulation and experimental results are expressed in terms of number of participants $N_{\rm part}$ in a Glauber model, which controls the centrality of the collisions. One finds that the $N_{\rm coll}$ scaling suggested by first-principle theory calculations in sections \ref{sec:weak}, \ref{sec:strong} are able to capture the centrality dependence reasonably well, similar to what has been observed in ideal hydrodynamic simulations \cite{Kolb:2001qz}. It should be noted that to good approximation, the multiplicity resulting from a full superSONIC run is given by the total deposited entropy corresponding to (\ref{eq:glauberdep}) via the QCD equation of state (\ref{eq:latticeeos1}). This assumes that the viscosity coefficients used in the hydrodynamic simulation are not too large such that entropy is approximately conserved \cite{Romatschke:2007jx} and that the final number of particles can be obtained from the final entropy by dividing by a constant factor \cite{Gubser:2008pc,Romatschke:2009im}. Using this method as a proxy for the superSONIC multiplicity is considerably faster than performing full hydrodynamic simulations and therefore allows for efficient extraction of the multiplicity probability distribution $P(N_{\rm ch})$ for $\sqrt{s}=200$ GeV Au+Au collisions shown in Fig.~\ref{fig5:multAA} (note that rapidity has been converted into pseudo-rapidity using Eq.~(\ref{eq:pseudomult}) in order to compare to experiment). Again, one finds that the theory simulations are able to capture the probability distribution fairly accurately, in particular when recalling that the experimental data has not been corrected for detector inefficiencies.


Fig.~\ref{fig5:IdAA} shows a comparison of identified particle (pion) momentum spectra from superSONIC compared to experimental data from a variety of collision systems and energies (cf. Ref.~\cite{Habich:2014jna}). One finds good overall agreement between the theory simulations and experimental data when using constant (temperature and collision system/energy independent) values $\eta/s=0.12$, $\zeta/s=0.02$ for the shear and bulk viscosity coefficient. However, it should be noted that allowing in particular $\zeta/s$ to vary with collision energy (or allowing for temperature-dependent profiles) leads to improved fits with respect to the experimental data \cite{Ryu:2015vwa}.


\begin{figure}[t]
  \begin{center}    
    \includegraphics[width=.7\linewidth]{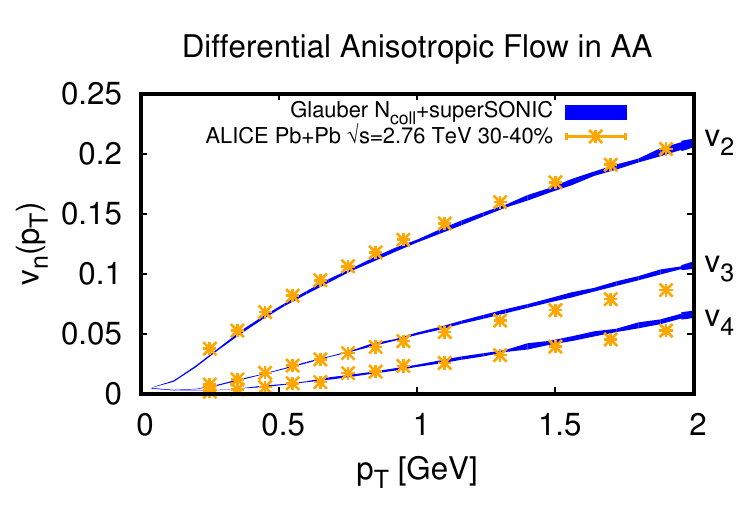}
    \includegraphics[width=.7\linewidth]{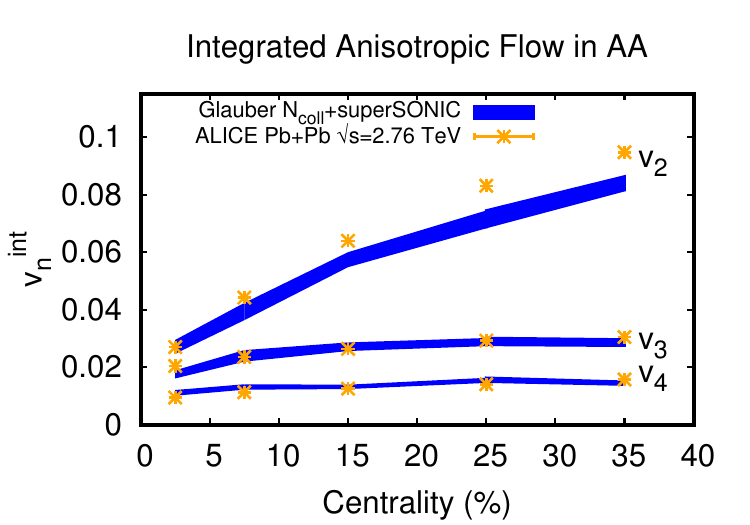}
     \end{center}
  \caption{\label{fig5:flowAA} Differential (upper panel) and integrated (lower panel) unidentified charged hadron flow coefficients $v_n$ for Pb+Pb collisions at $\sqrt{s}=2.76$ TeV from theory simulations (``superSONIC'', using $\eta/s=0.12$, $\zeta/s=0.02$) compared to experimental data from Refs.~\cite{Abelev:2012di,ALICE:2011ab}.  Theory uncertainty bands indicate sensitivity to non-hydrodynamic modes.
  }
\end{figure}

Fig.~\ref{fig5:flowAA} shows charged hadron anisotropic flow coefficients (\ref{eq:vnabsdef})  from superSONIC compared to experimental data for Pb+Pb collisions at $\sqrt{s}=2.76$ TeV. Overall, one finds good agreement between theory simulations and experimental data, even though the agreement can be considerably improved when allowing viscosity coefficients to be temperature-dependent functions \cite{Ryu:2015vwa,Niemi:2015qia}. This observation has given rise to an effort to determine the temperature dependent functions $\eta/s$, $\zeta/s$ by fitting parametrized trial functions to experimental data using Bayesian analysis, cf. Refs.~\cite{Novak:2013bqa,Bernhard:2016tnd}.

It should be pointed out that the flow results in heavy-ion collisions shown in Fig.~\ref{fig5:flowAA} (as well as those shown in Figs.~\ref{fig5:IdAA}, \ref{fig5:multAA}) 
show very little sensitivity to the choice of $C_\pi$. \index{Non-hydrodynamic mode! Simulations} In conjunction with the Central Lemma of fluid dynamics, the apparent insensitivity to the non-hydrodynamic mode damping rate suggests that simulation results are quantitatively reliable for heavy-ion collisions, even though fluid dynamics simulations are out of equilibrium \cite{Romatschke:2016hle}.

\begin{figure}[t]
  \begin{center}    
    \includegraphics[width=.8\linewidth]{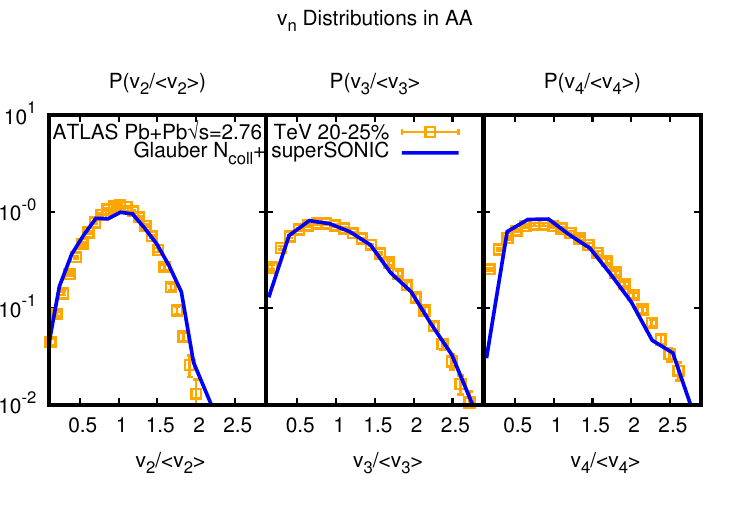}
     \end{center}
  \caption{\label{fig5:Pvn} Probability distributions for integrated flow coefficients $v_n$ from 1200 theory simulations events (``superSONIC'', using $\eta/s=0.12$, $\zeta/s=0.02$) for Pb+Pb collisions at $\sqrt{s}=2.76$ TeV in the 20-25 percent centrality class compared to experimental data from Ref.~\cite{Aad:2013xma}.  }
  \end{figure}

Another key observable to consider is the event-by-event probability distribution for anisotropic flow coefficients $P(v_n/\langle v_n\rangle)$. The corresponding data-theory comparison is shown in Fig.~\ref{fig5:Pvn}, exhibiting reasonable agreement as in Ref.~\cite{Gale:2012rq}. It should be pointed out that earlier calculations based on the Glauber-model with $N_{\rm part}$ scaling are typically not in agreement with the measured $P(v_n/\langle v_n\rangle)$ distributions, while for all practical purposes, the Glauber-model with $N_{\rm coll}$ scaling seems indistinguishable from results based on the IP-Glasma model \cite{Gale:2012rq}. This agreement is expected because of the results discussed in sections \ref{sec:weak}, \ref{sec:eccs}.


\begin{figure}[t]
  \begin{center}    
    \includegraphics[width=.8\linewidth]{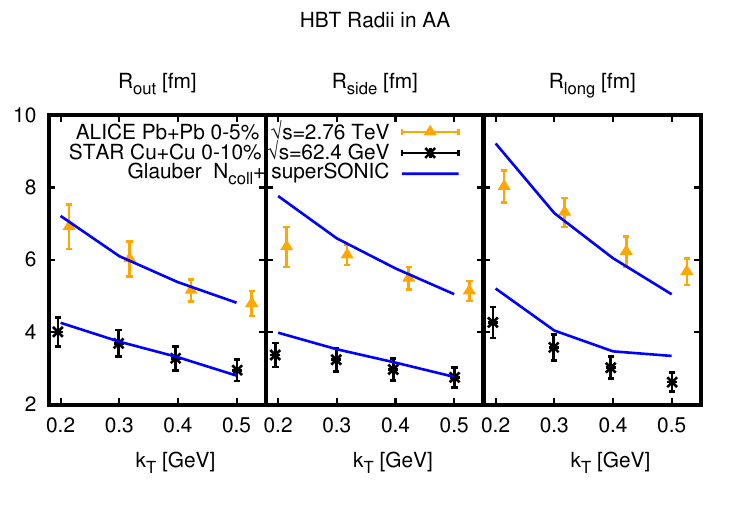}
     \end{center}
  \caption{\label{fig5:hbt} Pion HBT radii as a function of pair transverse momentum from theory simulations (``superSONIC'', using $\eta/s=0.12$, $\zeta/s=0.02$) for heavy-ion collisions compared to experimental data from
  Refs.~\cite{Aamodt:2011mr,Abelev:2009tp}.
  }
  \end{figure}

In Fig.~\ref{fig5:hbt}, pion HBT radii $R_{\rm out},R_{\rm side},R_{\rm long}$ (\ref{eq:HBT})  from superSONIC are compared to experimental data for heavy-ion collisions. There is reasonable agreement between experiment and simulation, similar to what has been observed before in Ref.~\cite{Habich:2014jna}. Note that the qualitative trend of HBT radii falling with increasing pair momentum $k_\perp$ by itself is not indicative of hydrodynamic behavior, because it arises also in simulations where a hydrodynamic phase is not present \cite{Romatschke:2015dha}.

\section{Heavy-on-Light Ion Collisions}

Historically, heavy-on-light ion collisions such as p+Pb and d+Au were conceived as control studies for heavy-ion collisions based on the assumption that flow effects were absent in small collision systems. By now, experimental evidence to the contrary is overwhelming, showing that flow patters (radial flow, elliptic flow, triangular flow, etc) of the same type and magnitude as in heavy-ion collisions are present in small systems \cite{CMS:2012qk,Abelev:2012ola,Aad:2012gla,Chatrchyan:2013nka,Adare:2013piz,Adare:2015ctn,Aidala:2016vgl} (see Ref.~\cite{Nagle:2018nvi} for a recent review).

Based on the traditional view that hydrodynamics necessarily requires systems to come to near local equilibrium, a number of ``initial state'' mechanisms have been suggested in an attempt to explain the experimental data without hydrodynamics (see e.g. Ref.~\cite{Dusling:2015gta} for an overview). However, in light of the modern understanding of the applicability of fluid dynamics out of equilibrium as outlined in section \ref{sec:outoff}, a hydrodynamic origin of the observed flow signals in small systems seems natural. Indeed, hydrodynamic model results in small systems have proven just as successful as in more traditional large heavy-ion collisions systems \cite{Bozek:2011if,Nagle:2013lja,Schenke:2014zha,Kozlov:2014fqa,Werner:2014xoa,Romatschke:2015gxa,Bozek:2015qpa,Weller:2017tsr,Mantysaari:2017cni}.

\begin{figure}[t]
  \begin{center}    
    \includegraphics[width=.7\linewidth]{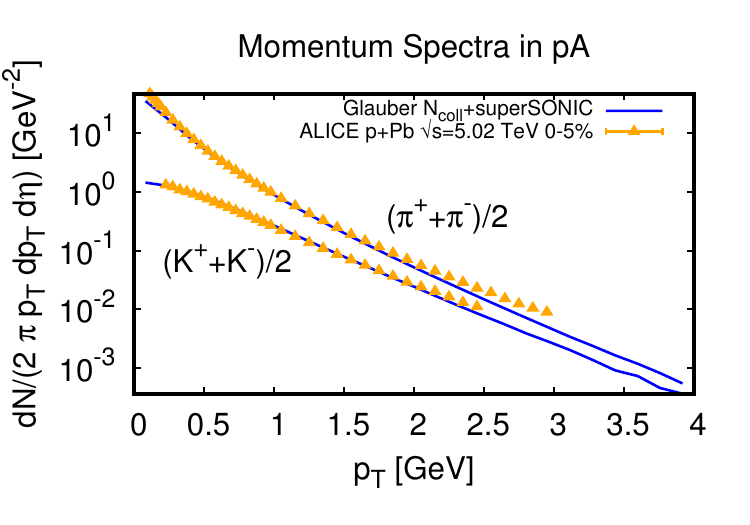}
    \includegraphics[width=.7\linewidth]{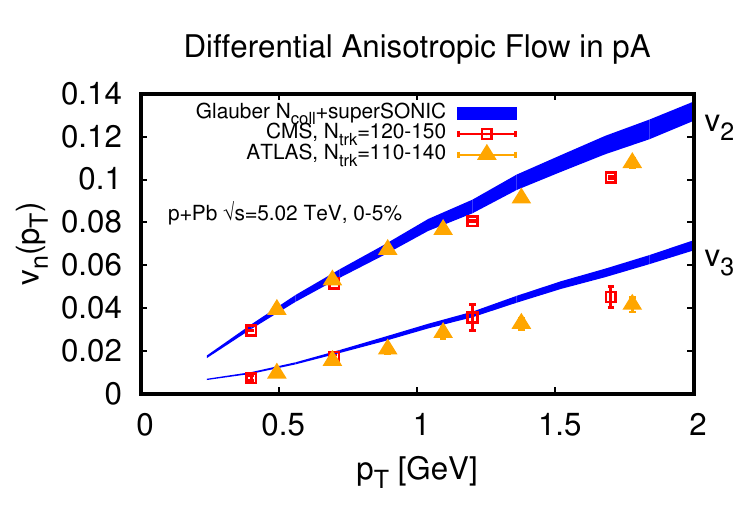}

     \end{center}
  \caption{\label{fig5:IdpA} Momentum spectra for pions and kaons (upper panel) and anisotropic flow coefficients (lower panel) as a function of transverse  momentum from theory simulations (``superSONIC'', using $\eta/s=0.12$, $\zeta/s=0.02$) for central p+Pb collisions compared to experimental data from Refs.~\cite{Abelev:2013haa,Chatrchyan:2013nka,Aad:2014lta}. Theory uncertainty bands indicate sensitivity to non-hydrodynamic modes.
  }
  \end{figure}

\begin{figure}[t]
  \begin{center}
    \includegraphics[width=.8\linewidth]{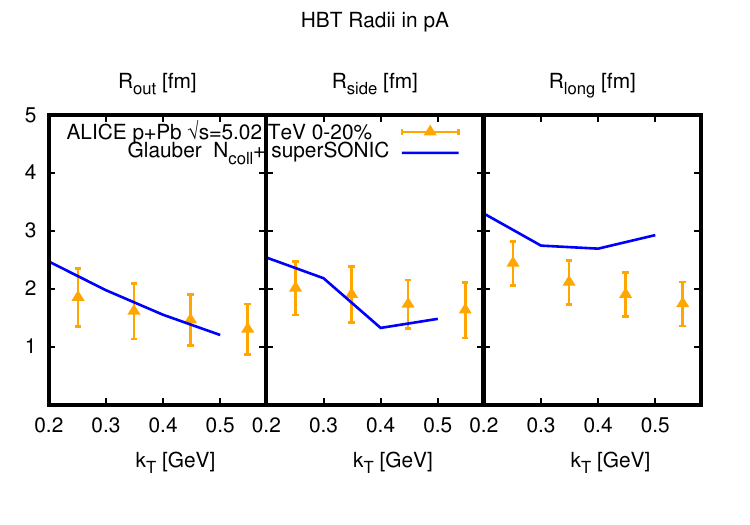}
     \end{center}
  \caption{\label{fig5:hbtpA} Pion HBT radii as a function of transverse pair momentum  from theory simulations (``superSONIC'', using $\eta/s=0.12$, $\zeta/s=0.02$) for central p+Pb collisions compared to experimental data from Ref.~\cite{Adam:2015pya}.
  }
  \end{figure}

Figs.~\ref{fig5:IdpA}, \ref{fig5:hbtpA} show data/theory comparisons for identified particle spectra\footnote{No comparison to protons/anti-protons are shown because baryon annihilation/regeneration is not currently implemented in superSONIC's hadron cascade code. Since annihilation reduces the proton yield by approximately 50 percent, no sensible data-theory comparison can be made without it.}, anisotropic flow coefficients and HBT radii in central p+Pb collisions at $\sqrt{s}=5.02$ TeV. These are the same type of observables used in the equivalent comparison in heavy-ion collisions, and Fig.~\ref{fig5:IdpA} indicates that theory simulations based on hydrodynamics offer reasonable agreement also for the description of p+Pb collisions.

The main difference with respect to simulations of heavy-ions is the sensitivity to the choice of $C_\pi$ seen in simulations of p+Pb collisions. 
%
Systems with stronger gradients should exhibit a stronger sensitivity to non-hydrodynamic modes, and indeed direct comparison of the uncertainty bands in Figs.~\ref{fig5:flowAA}, \ref{fig5:IdpA} (generated by varying $C_\pi\in [4,6]$) indicates that smaller systems are more sensitive than larger systems, as expected. Nevertheless, the uncertainty bands resulting from varying $C_\pi$ are still small compared to the overall magnitude of the signal, suggesting that according to the Central Lemma of fluid dynamics, simulations are still quantitatively reliable.
The overall agreement between theory simulations and experimental data is not limited to p+Pb collisions. Indeed, a wealth of experimental data at lower energies such as p+Au, d+Au and $^3{\rm He}$+Au collisions at $\sqrt{s}=200$ GeV and d+Au collisions at $\sqrt{s}=19.6,39,62.4$ GeV has become available in recent years \cite{Adare:2013piz,Adare:2015ctn,Aidala:2016vgl,Aidala:2017pup,Adare:2017wlc}, all either successfully matched \cite{Nagle:2013lja,Bozek:2014cya,Romatschke:2015gxa,Bozek:2015qpa} and in some cases even predicted by fluid dynamics simulations \cite{Nagle:2013lja,Romatschke:2015gxa,Koop:2015trj}. In particular, it has been realized  that the availability of experimental data for different collision systems such as p+Au, d+Au and $^3{\rm He}$+Au at the same energy is able to help disentangle initial geometry (such as initial eccentricities, cf. section \ref{sec:eccs}) from hydrodynamic flow (see e.g. section \ref{sec:simplified}) \cite{Nagle:2013lja}. Furthermore, since the lifetime of small systems created in heavy-on-light-ion collisions is typically considerably shorter than that of heavy-ion collisions \cite{Koop:2015trj}, it has been suggested that small systems offer an experimental handle on pre-hydrodynamic flow \cite{Romatschke:2015gxa}. This exciting line of research is ongoing at the time of writing.

\section{Proton-Proton Collisions}

Traditionally thought to be void of any type of medium-like effects, systems created in proton-proton collisions are perhaps
one of the more exotic systems to consider in the context of fluid dynamics. Nevertheless, recent experimental data unequivocally demonstrates the presence of flow patterns (radial flow, elliptic flow, triangular flow) in high-multiplicity proton-proton collisions \cite{Khachatryan:2010gv,Aad:2015gqa,Khachatryan:2015lva,Khachatryan:2016txc,Aaboud:2016yar}. Furthermore, some of these signals even extend to lower multiplicities, into a regime where experimental techniques designed to extract correlations in heavy-ion collisions with many particles are becoming challenging. Moreover, other experimental signatures such as strangeness enhancement -- traditionally expected to occur only in heavy-ion collisions -- have been observed in proton-proton collisions \cite{ALICE:2017jyt}. Possible explanations for the observed experimental signals have been suggested in approaches that do not involve fluid dynamics, see e.g. Refs.~\cite{Schenke:2016lrs,Iancu:2017fzn,Blok:2017pui}. However, just as in the case of heavy-on-light-ion collisions discussed in the previous section, theory simulations based on fluid dynamics are able to successfully match \cite{Weller:2017tsr} and in some case successfully predict the observed experimental flow signals \cite{Prasad:2009bx,Ortona:2009yc,Werner:2010ss,Bozek:2010pb}.

\begin{figure}[t]
  \begin{center}    
    \includegraphics[width=.7\linewidth]{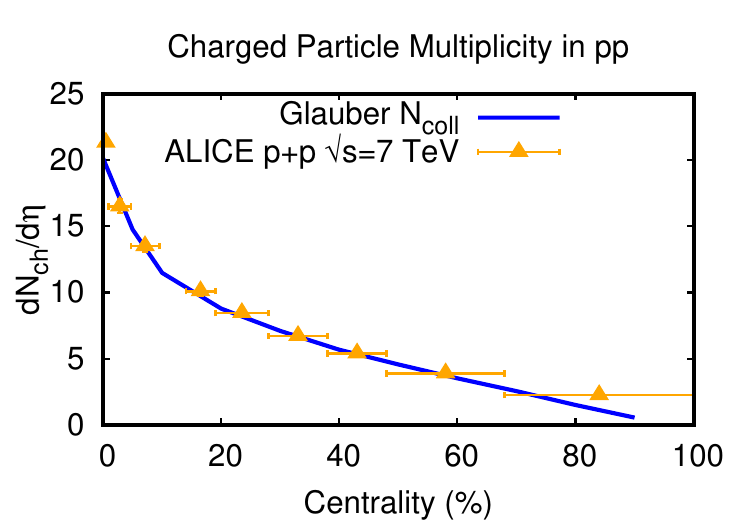}
    \includegraphics[width=.75\linewidth]{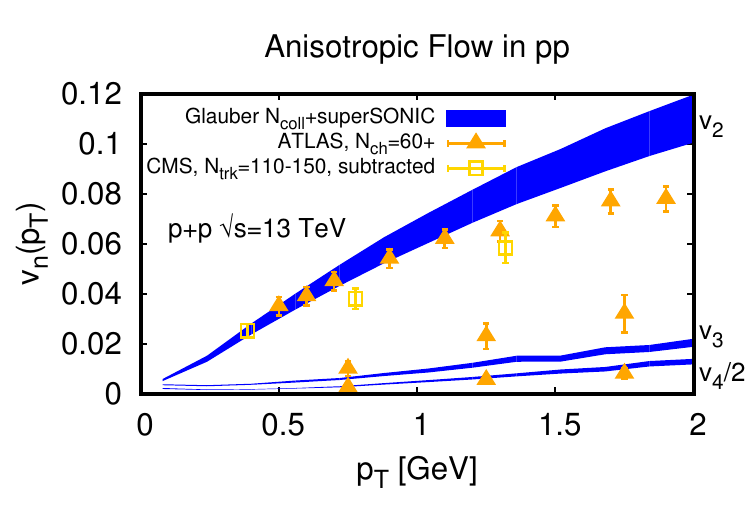}
     \end{center}
  \caption{\label{fig5:pp1} Charged particle multiplicity as a function of centrality (upper panel) and anisotropic flow coefficients as a function of transverse momentum  (lower panel) from theory simulations (``superSONIC'', using $\eta/s=0.12$, $\zeta/s=0.02$) for central proton-proton collisions compared to experimental data from Refs.~\cite{Bianchi:2016szl,Khachatryan:2016txc,Aaboud:2016yar}. Theory uncertainty bands indicate sensitivity to non-hydrodynamic modes.}
\end{figure}

The key challenge to successful fluid dynamics descriptions of experimental data in proton-proton collisions traditionally has been the question of applicability of hydrodynamics to small systems with sizable gradients that are not near local equilibrium. In view of the progress made in the understanding of fluid dynamics over the past few years (see section \ref{sec:modern}), this question can now be made precise by studying the effect of non-hydrodynamic modes on observables calculated from a fluid dynamic approach. Varying the damping rate of the non-hydrodynamic mode (\ref{eq:rbrsssnonhydro}) through variation of $C_\pi$ provides a practical means of testing the sensitivity of fluid dynamic results to non-hydrodynamic modes, and hence allows one to assign quantitative uncertainties to observables concerning the applicability of fluid dynamics itself. These tests have been performed for theoretical simulations of proton-proton collisions, finding fluid dynamics to be quantitatively applicable for momentum spectra and elliptic flow coefficients for charged-particle multiplicities exceeding $\frac{dN_{\rm ch}}{dY}>2$ \cite{Habich:2015rtj}, consistent with the arguments from Ref.~\cite{Spalinski:2016fnj}.

To give a flavor of the success of fluid dynamics simulations, data-theory comparisons for charged particle multiplicity and pion momentum spectra in proton-proton collisions at $\sqrt{s}=5.02,7,13$ TeV are shown in Fig.~\ref{fig5:pp1}, where the theoretical uncertainty bands shown indicate the sensitivity of hydrodynamic results to the non-hydrodynamic mode damping rate. As was the case in heavy-ion collisions and heavy-on-light-ion collisions, the results shown in Fig.~\ref{fig5:pp1} suggest that simulations based on hydrodynamics are quantitatively reliable, because the sensitivity to the non-hydrodynamic modes is under control. Furthermore, the hydrodynamic simulations offer a reasonable description of experimental data in proton-proton collisions.

\chapter{Conclusions}

\section{Relativistic Fluid Dynamics Theory in the 21st Century}

Research accomplishments made in the past decade have given rise to the theory of relativistic fluid dynamics as presented in this work: a first-principles framework based on effective field theory that is free from the traditional requirement of local
thermodynamic equilibrium. No longer relegated to the description of thermalized systems, this theory of fluid dynamics out of equilibrium holds the promise of becoming a powerful tool in describing out-of-equilibrium phenomena in nature. Besides the application to relativistic nuclear collisions mentioned in chapter \ref{chap:experiment}, other systems,  such as the strongly correlated electron fluid in graphene \cite{2016Sci...351.1055B} and non-relativistic applications such as cold quantum gases near unitarity \cite{Cao:2010wa}, and high-temperature superconductors \cite{PhysRevB.90.134509} are likely to profit from these recent theory developments.

In our opinion, the fundamental reason why fluid dynamics out of equilibrium is quantitatively reliable stems from the observation that the effective fluid degrees of freedom, while originally formulated for near-equilibrium systems, not only survive, but dominate the bulk dynamics of systems also significantly away from equilibrium. As long as these long-lived hydrodynamic modes are dominant, non-equilibrium transport will be governed by out of equilibrium fluid dynamics as described in this work, and hence follow ``universal'' laws of fluid dynamics. Conversely, other, non-hydrodynamic degrees of freedom are always present in physical systems, and it is the competition between these modes and the hydrodynamic modes that ultimately leads to the breakdown of fluid dynamics. Once the life-time of non-hydrodynamic modes exceeds that of the hydrodynamic modes in a system out of equilibrium, fluid dynamics has broken down, and the bulk dynamics of the system will follow (non-universal) dynamics governed by the non-hydrodynamic modes. In essence, this is the Central Lemma of fluid dynamics described in section \ref{sec:outoff}, which allows to lift the boundaries that fluid dynamics had been confined to in the past.

Given the major changes that our understanding of the theory of relativistic fluid dynamics has gone through in the past decade, it is likely that rapid development will continue in the coming years, in particular on the subject of hydrodynamic attractors and the characterization of non-hydrodynamic modes. Some of the open questions in this regard are outlined in section \ref{sec:problems}.

\section{Fluid Dynamics Simulations of Relativistic Nuclear Collisions}

Devised as a means to study QCD and the Standard Model of Physics, relativistic nuclear collision
experiments have been transformative in our understanding of strongly coupled system dynamics. The field has gone through major paradigm shifts. For instance, instead of exhibiting dynamics of weakly coupled quarks and gluons expected from QCD's property of asymptotic freedom, 
the bulk of experimental data from the first heavy-ion collisions at RHIC turned out to be much better described
by (ideal) fluid dynamics calculations \cite{Kolb:2000fha,Teaney:2000cw,Huovinen:2001cy,Hirano:2002ds}. The field
adjusted by formally separating results obtained in heavy-ion collision (``quark-gluon plasma'' or ``hot matter'',
described in terms of fluid dynamics) from results obtained in small
collision systems such as d+Au and proton-proton collisions
(``cold matter'', described in terms of weakly interacting degrees of freedom). The original
idea behind this separation was that large systems created in heavy-ion collisions would have time to
reach thermal equilibrium, thus leading to a near-equilibrium fluid description, whereas small systems, being more short-lived, would not have time to equilibrate and hence should be governed by the dynamics
of ``fundamental'' degrees of freedom in the absence of a medium.

In light of the treasure trove of experimental data in small collision systems collected in the past decade, the advances made in the understanding of fluid dynamics out of equilibrium, and last but not least, the seemingly infallible success of relativistic fluid dynamics simulations in describing and predicting
this very experimental data, the separation of large and small collision systems into different physics categories has become very hard to maintain.

In our opinion, the success of fluid dynamics in describing relativistic nuclear collisions can be potentially
understood in view of the theory developments in the past decade that were reviewed in this work.
First, the Central Lemma of fluid dynamics guarantees the quantitative applicability of fluid dynamics even
in systems away from equilibrium as long as hydrodynamic modes are dominant over non-hydrodynamic modes. Second, calculations in weak coupling kinetic theory as well as strong coupling gauge/gravity duality suggest that hydrodynamic modes remain dominant for QCD fluid droplets larger than $0.15$ fm,
considerably smaller than the size of a single proton \cite{Romatschke:2016hle}.
Moreover, fluid dynamics offers a simple explanation for the experimentally observed flow signals (initial anisotropies in geometry lead to anisotropies in fluid flow). In particular, the results presented in section \ref{chap:experiment} demonstrate that a single fluid dynamic
simulation model with a single set of initial conditions and a single set of transport parameters is able to
successfully describe the bulk of experimental results at low momenta in large (heavy-ion), medium (heavy-on-light-ion)
and small (proton-proton) collision systems. It certainly appears as if relativistic fluid dynamics could be the ruling and unifying mechanism behind the dynamics of relativistic nuclear collisions \cite{Weller:2017tsr}. 

Nevertheless, fluid dynamic descriptions of relativistic nuclear collision data are not perfect. The results presented in section \ref{chap:experiment} do not offer precision fits to experimental data (nor should they be expected to, given the minimal number of free parameters involved). Simultaneous descriptions of the experimentally measured photon yield and anisotropic flow pose challenges to the framework, as do simultaneous descriptions of ultra-central collisions \cite{Goldschmidt:2015kpa,Adare:2015lcd}. The framework presented here does not include a coupling to electromagnetic fields, nor does it include the presence of quantum anomalies \cite{Hongo:2013cqa,Roy:2015kma}.

Yet, it is remarkable that despite its simplicity, a single fluid description with a minimum number of parameters based on first-principles calculations of energy deposition  at weak and strong coupling is able to describe such a broad set of experimentally probed systems and observables without the need of additional components, explanations or fine-tuning. In our opinion, it is time to fully embrace fluid dynamics as the unified paradigm of all relativistic nuclear collision systems.

\section{Challenges and Open Problems}
\label{sec:problems}

In this section, we list several challenges and open problems that we recognize as important and unsolved at the time of writing, without any pretense of completeness, order of importance, or guarantee that a solution exists. It is our hope that some of these problems and challenges will be overcome in the near future.

  \begin{enumerate}[label*=\arabic*.]
\item
\textbf{Observational consequences for absence of LRF} The entirety of this work is based on the assumption that the energy-momentum tensor possesses a single and real time-like eigenvector. It is known that this assumption is violated in certain quantum systems out of equilibrium because an LRF does not exist, see Ref.~\cite{Arnold:2014jva}. It should be studied if the absence of an LRF has observable consequences.\index{Local rest frame (LRF)! Absence of}
  \item
    \textbf{Study of Non-hydrodynamic Modes in Nature.} Unlike hydrodynamic modes, which exhibit universal dispersion relations such as (\ref{eq:soundmode}), non-hydrodynamic mode signatures are system-dependent. Very little experimental input exists on the properties of non-hydrodynamic modes, despite the possibility of using non-hydrodynamic modes as a type of unique characterization of the system's relevant degrees of freedom \cite{Brewer:2015ipa,Bantilan:2016qos}. Experimental handles on non-hydrodynamic modes in different systems should be studied and analyzed.
  \item
    \textbf{Universality of Hydrodynamic Attractors.} Hydrodynamic attractors solutions have been found and studied in several theory systems (see section \ref{sec:howgen}). It is not known if these attractors are specific to the type of gradients applied to the model system, or exhibit a universal form independent from the forcing, dimensionality or symmetry. More studies on hydrodynamic attractors should be done to answer this question.
  \item
    \textbf{Attractors for Hydrodynamic Action.} In the context of the equations of motion of classical fluid dynamics, the program of gradient expansion and resummation has led to the discovery of hydrodynamic attractors which define the theory non-perturbatively out of equilibrium. Fluid dynamics in the presence of fluctuations has been set up as a gradient expansion of a hydrodynamic action (see section \ref{sec:hydrofluc}). This program has only been carried out to low orders in the gradient expansion. It should be studied if the gradient expansion of the hydrodynamic effective action is divergent (this would be expected), and if a Borel-type resummation of the divergent gradient series for the action is possible and will give rise to an hydrodynamic attractor for fluctuating hydrodynamics.
  \item \textbf{Poles vs. Branch Cuts}
    In section \ref{sec:ggnonhydro}, it was hypothesized that the analytic structure of finite temperature retarded correlators of the energy-momentum tensor for a large $N_c$ gauge theory exhibits poles at any non-vanishing value of the gauge coupling. To test this hypothesis the corresponding correlators should be calculated for a large $N_c$ gauge theory at any \textit{fixed, but non-vanishing} value of the gauge coupling, possibly along the lines of Ref.~\cite{Larkoski:2015lea}. 
    \item \textbf{Spatial Geometry of a Proton}
      One of the biggest uncertainties in hydrodynamic simulations of proton-proton collisions is the spatial geometry of a boosted proton. Lattice gauge theory simulations of the proton structure should be used to reduce this uncertainty.
    \item \textbf{Real Time Dynamics of Collisions}
      Sections \ref{sec:weak}, \ref{sec:strong} offer first-principles calculations of the energy-deposition following a collision in a gauge theory at infinitely weak and infinitely strong coupling. Lattice gauge theory simulations in real time, possibly along the lines of Refs.~\cite{Berges:2006xc,Alexandru:2017lqr} should be used to study energy deposition in gauge theories at finite coupling.
  \item
    \textbf{Stable Lattice-Boltzmann Solver} All currently available algorithms for simulating out of equilibrium relativistic nuclear collisions employ patches to deal with numerical instabilities at large gradients (see section \ref{sec:numalgo}). Algorithms based on the Lattice Boltzmann approach do not suffer from this problem, but show numerical instabilities for supersonic flows for different reasons. A stable Lattice Boltzmann algorithm should be developed to allow for simulations of nuclear collisions in the presence of shear and bulk viscosities without any additional patches.
  \item
    \textbf{Analytic Solutions for Holographic Shock Wave Collisions}
    In section \ref{sec:strong}, approximate analytic solutions for the collision of holographic shock waves were discussed. In the general relativity literature, full analytic solutions for shock wave collisions exist \cite{Khan:1971vh,Kajantie:2008rx}. It should be studied if full analytic solutions for shock wave collisions in asymptotic AdS space-times are possible if the system obeys boost-invariance (e.g. by considering the collision of $\delta^2$ shocks). 
  \item
    \textbf{Two-phase Simulations} In current hydrodynamic simulations of nuclear collisions, the ``hot'' phase is treated hydrodynamically with information on the isothermal switching hypersurface stored and later used to initialize the ``cold'' hadronic phase simulations. This treatment does not allow for hadrons to be re-absorbed into the ``hot'' phase. Two-phase simulations where both the fluid and hadron gas phase are solved simultaneously should be performed, possibly based on successful approaches in the non-relativistic literature \cite{INAMURO2004628}.        
\item
\textbf{Pre-hydrodynamic flow} It is known that both weakly and strongly coupled systems build up flow already in the pre-hydrodynamic stages following a nuclear collision. Small collision systems are sensitive to this pre-hydrodynamic flow because their hydrodynamic lifetime is considerably shorter than for large collision systems. Observational constraints on pre-hydrodynamic flow should be studied, possibly along the lines of Ref.~\cite{Romatschke:2015gxa}.
  \item
    \textbf{Flow in ${\rm e^+}-{\rm e^-}$ Collisions}
    The out-of-equilibrium fluid framework set up as part of this work suggests the applicability of fluid dynamics to system sizes at or below the femtoscale. As a consequence, it is possible that also systems created in ${\rm e^+}-{\rm e^-}$ collisions evolve according to the laws of out of equilibrium fluid dynamics, suggesting that low-momentum particle spectra should follow Eq.~(\ref{eq:KTmodel2}) \cite{Hoang1987,Becattini:2001fg,Ferroni:2011fh} and there should be anisotropic momentum flow. Experimental data for ${\rm e^+}-{\rm e^-}$ collisions should be re-analyzed using modern techniques to search for these effects, and theoretical studies aiming to quantify these effects should be performed \cite{Nagle:2017sjv}.
  \end{enumerate}

\chapter*{Acknowledgments}
 
This work was supported in part by the Department of Energy, DOE award No DE-SC0017905. The National Center for Atmospheric Research is sponsored by the National Science Foundation.
We would like to thank Nicolas Borghini, Jamie Nagle, Jean-Yves Ollitrault, Parisa Rezaie, Bj\"orn Schenke, Mike Strickland, Prithwish Tribedy, Wilke van der Schee and Bill Zajc for helpful discussions. Furthermore, we would like to thank
Prithwish Tribedy for providing tabulated IP-Glasma data from Ref.~\cite{Schenke:2012fw} shown in Fig.~\ref{fig4:six}.


\begin{appendices}
\addtocontents{toc}{\protect\setcounter{tocdepth}{1}}
\makeatletter
\addtocontents{toc}{%
  \begingroup
  \let\protect\l@chapter\protect\l@section
  \let\protect\l@section\protect\l@subsection
}
\makeatother
 
\chapter{Relativistic Velocities}
  \label{chap:vel1}

  For a relativistic system, the ordinary (space) velocity vector ${\bf v}\equiv \frac{d {\bf x}}{dt}$ is not a good degree of freedom, because it does not transform appropriately under the symmetries of relativity.
  The relativistic generalization of the space-velocity is
  \begin{equation}
    u^\mu \equiv \frac{d x^\mu}{ds}\,,
    \end{equation}
  where $ds$ is the proper time increment given by
  \begin{equation}
    (ds)^2=-g_{\mu\nu}dx^\mu dx^\nu\,.
  \end{equation}

  To elucidate its role, consider the case of Minkowski space-time where $g_{\mu\nu}={\rm diag}(-1,1,1,\ldots,1)$. In this case
  \begin{equation}
    (ds)^2=(dt)^2-(d{\bf x})^2=(dt)^2\left[1-\left(\frac{d{\bf x}}{dt}\right)\right]^2=(dt)^2\left[1-{\bf v}^2\right]\,.
  \end{equation}
  Thus
  \begin{equation}
    u^\mu=\frac{d t}{ds}\frac{d x^\mu}{dt}=\frac{1}{\sqrt{1-{\bf v}^2}}\left(\begin{array}{c}1\\ {\bf v}\end{array}\right)=\gamma({\bf v})\left(\begin{array}{c}1\\ {\bf v}\end{array}\right)\,,
  \end{equation}
  where $\gamma({\bf v})$ is the Lorentz ``gamma'' factor. With this explicit form for $u^\mu$, it is straightforward to take the non-relativistic limit (small velocities compared to the speed of light), $|{\bf v}|\ll 1$. One finds for the relativistic velocity 
  \begin{equation}
    \label{eq:uexp}
    u^\mu\simeq (1,{\bf v})+{\cal O}\left(|{\bf v}|^2\right)
    \end{equation}

  \chapter{Riemann, Ricci, Christoffels and all that}
\label{chap:aGR}

Since the dynamics of energy and momentum and the dynamics of space-time are intimately intertwined, some basic familarity with general relativity is useful. For this reason, this appendix is meant to provide a bare-bones ``crash-course'' in general relativity, which readers familiar with the subject can easily skip.

To get started, the central object for general relativity is the space-time metric tensor denoted by $g_{\mu\nu}$ where $\mu$ is an index running over space-time coordinates such as $x^\mu=\left(t, x, y, z\right)$ in four dimensions. For Minkowski space-time in the mostly plus sign convention adopted here, the metric can be expressed as $g_{\mu\nu}={\rm diag}(-1,+1,+1,+1)$ where ``${\rm diag}$'' denotes a square matrix with non-vanishing entries in the diagonal only.

In a Euclidean space, with coordinates $x^i=\left(x,y,z\right)$, the length of a vector $x^i$ may be calculated using the Euclidean metric $\delta^{ij}={\rm diag}(+1,+1,+1)$ as ${\bf x}^2=x^i x^i=x^i \delta^{ij} x^j=x^2+y^2+z^2$, using the Einstein sum convention $x^i x^i\equiv\sum_{i=1}^3 x^i x^i$. By analogy, one may calculate the space-time distance of $x^\mu$ from the origin in Minkowski space using the metric tensor as $x^\mu g_{\mu\nu} x^\nu=-t^2+x^2+y^2+z^2$. For an infinitesimal space-time distance $d x^\mu=\left(dt,dx,dy,dz\right)$ this leads to
\begin{equation}
  ds^2\equiv dx^\mu g_{\mu\nu} dx^\nu=-dt^2+dx^2+dy^2+dz^2\,.
  \end{equation}
Note that it is common practice to introduce the short-hand notation of lower indices as
\begin{equation}
  A_\mu\equiv g_{\mu\nu}A^\nu\,,
\end{equation}
for an arbitrary vector $A^\mu$ (such as $dx^\mu$), which leads to $ds^2=dx^\mu dx_\mu$. Equivalently, we may write $ds^2=dx_\mu dx_\nu g^{\mu\nu}$ with $g^{\mu\nu}$ the inverse metric that has to satisfy
\begin{equation}
  g^{\mu \lambda} g_{\lambda \nu}=\delta^\mu_\nu={\rm diag}(+1,+1,+1,+1)\,.
  \end{equation}

For a stationary observer at the origin of Minkowski space, $dx^\mu=\left(dt,{\bf 0}\right)$ so that $ds^2=-dt^2$ and hence the proper time for this observer is given by $\tau=\int \sqrt{-ds^2}$. Since proper time can be measured by the observer, it should be independent from the particular coordinates chosen to express $ds^2$, hence $ds^2$ should be invariant under (not necessarily linear) coordinate changes $x^\mu\rightarrow \tilde x^\mu(x^\nu)$. Using $dx^\mu=\frac{\partial x^\mu}{\partial \tilde x^\nu} d\tilde x^\nu$ and the definition of $ds^2$, this immediately leads to the transformation of the metric under coordinate transformations as
\begin{equation}
  \label{eq:gmntrafo}
  g_{\lambda \kappa}\rightarrow \tilde g_{\lambda \kappa}= \frac{\partial x^\mu}{\partial \tilde x^\lambda} \frac{\partial x^\nu}{\partial \tilde x^\kappa} g_{\mu\nu}\,,
\end{equation}
confirming that $g_{\mu\nu}$ indeed transforms as a tensor. A simple exercise shows that the inverse metric transforms as $
  g^{\lambda \kappa}\rightarrow \tilde g^{\lambda \kappa}= \frac{\partial \tilde x^\lambda}{\partial x^\mu} \frac{\partial \tilde x^\kappa}{\partial x^\nu} g^{\mu\nu}$.

Eq.~(\ref{eq:gmntrafo}) is the transformation rule for rank two tensors, and arbitrary rank $n$ tensors transform by straightforward generalization of (\ref{eq:gmntrafo}), cf. appendix \ref{chap:coor}. These tensors are said to transform covariantly (``general covariance'').

One of the main results of general relativity is the equation of motion of a particle in curved space-time (``geodesic equation''). To derive this equation, first consider a massive particle in Minkowski space that is moving in the absence of any forces such that
\begin{equation}
  \label{eq:geodesic1}
  \frac{d^2 \tilde x^\mu}{d\tau^2}=0\,,\quad d\tau^2=-\tilde g_{\mu\nu} d\tilde x^\mu d\tilde x^\nu\,.
\end{equation}
Let us now consider the same particle in the presence of the gravitational force. Einstein's weak principle of equivalence then suggests that locally, the action of the gravitational force cannot be distinguished from a ``pseudo-force'' experienced by an observer in a non-inertial (e.g. accelerated) rest-frame. Since non-inertial and inertial rest frames are related by a general coordinate transformation, this implies that -- upon choosing suitable coordinates $x^\mu$ --- the equations of motion for a massive particle in the presence of gravity are also given by (\ref{eq:geodesic1}), albeit with a metric tensor $\tilde g_{\mu\nu}$ that is not that of Minkowski space-time, but curved space-time. If desired, Eq.~(\ref{eq:geodesic1}) may be transformed into more suitable coordinates $x^\mu=x^\mu(\tilde x^\nu)$ such that
\begin{equation}
  \label{eq:geodesic2}
\frac{\partial \tilde x^\mu}{\partial x^\lambda} \left[ \frac{d^2 x^\lambda}{d\tau^2}+\frac{dx^\nu}{d\tau}\frac{d x^\sigma}{d\tau} \frac{\partial^2 \tilde x^\mu}{\partial x^\nu \partial x^\sigma} \frac{\partial x^\lambda}{\partial \tilde x^\mu}\right]=0\,,\quad d\tau^2=-g_{\mu\nu} dx^\mu dx^\nu\,.
\end{equation}
Introducing the ``connection coefficients'' $\Gamma^\lambda_{\nu \sigma}$ as
\begin{equation}
  \Gamma^\lambda_{\nu \sigma}\equiv\frac{\partial^2 \tilde x^\mu}{\partial x^\nu \partial x^\sigma} \frac{\partial x^\lambda}{\partial \tilde x^\mu}\,,
\end{equation}
this leads to the geodesic equation for a massive particle as
\begin{equation}
  \frac{d^2 x^\lambda}{d\tau^2}+\Gamma^\lambda_{\nu \sigma}\frac{dx^\nu}{d\tau}\frac{d x^\sigma}{d\tau}=0\,,
\end{equation}
for arbitrary coordinates $x^\mu$. Note that the action of gravity is encoded in the connection coefficients, also known as ``Christoffel symbols'', which may be calculated from the metric tensor as
\begin{equation}
    \label{eq:christoffel}
    \Gamma_{\mu\nu}^\lambda=\frac{1}{2}g^{\lambda \rho}\left(\partial_\mu g_{\rho \nu}+\partial_\nu g_{\rho \mu}-\partial_\rho g_{\mu\nu}\right)\,.
\end{equation}
(Note that this formula can be obtained by considering the partial derivative of the transformation rule (\ref{eq:gmntrafo})). From (\ref{eq:christoffel}), it is obvious that the Christoffel symbols are symmetric in the lower indices, e.g. $\Gamma^\lambda_{\mu\nu}=\Gamma^\lambda_{\nu \mu}$.

One may verify that the Christoffel symbols transform as 
\begin{equation}
  \label{eq:gammarule}
\Gamma^\mu_{\lambda \kappa}\rightarrow  \tilde \Gamma^\mu_{\lambda \kappa} =\frac{\partial \tilde x^\mu}{\partial x^\nu} \frac{\partial x^\rho}{\partial \tilde x^\lambda} \frac{\partial x^\sigma}{\partial \tilde x^\kappa} \Gamma^\nu_{\rho\sigma} -  \frac{\partial \tilde x^\mu}{\partial x^\rho \partial x^\sigma} \frac{\partial x^\sigma}{\partial \tilde x^\lambda} \frac{\partial x^\rho}{\partial \tilde x^\kappa}\,,
\end{equation}
under coordinate transformations $x^\mu\rightarrow \tilde x^\mu$. This means that the Christoffel symbols are \textit{not} tensors under coordinate transformations.

To derive Einstein's equations, first note that for a given tensor $X^{\mu\nu\lambda\ldots}$, its space-time derivative $\partial_\kappa X^{\mu\nu\lambda\ldots} $ will in general not be a tensor under coordinate transformations, because the transformation matrix is itself space-time dependent. However, it is possible to construct a tensor out of the derivative of any tensor by adding suitable terms. This is most easily elucidated by considering the derivative of a rank one tensor (a vector) $A^\mu$ which transforms as
\begin{equation}
\partial_\lambda A^\mu\rightarrow \tilde \partial_\lambda \tilde A^\mu=\frac{\partial}{\partial \tilde x^\lambda} \left[\frac{\partial \tilde x^\mu}{\partial x^\nu} A^\nu\right]=\frac{\partial \tilde x^\mu}{\partial x^\nu} \frac{\partial x^\rho}{\partial \tilde x^\lambda} \partial_\rho A^\nu + A^\nu \frac{\partial^2 \tilde x^\mu}{\partial x^\nu \partial x^\rho}\frac{\partial x^\rho}{\partial \tilde x^\lambda}\,,
\end{equation}
where the last term spoils general covariance. However, using the transformation rule (\ref{eq:gammarule}), it is straightforward to verify that the combination
\begin{equation}
\nabla_\lambda A^\mu\equiv  \partial_\lambda A^\mu+\Gamma^\mu_{\lambda \rho} A^\rho\,,
\end{equation}
transforms covariantly under coordinate transformations. The combination $\nabla_\mu$ of partial derivative and Christoffel symbols is called ``geometric covariant derivative''. The covariant derivative may be generalized for a tensor $X^{\mu_1 \mu_2\ldots \mu_n}$ of rank n as
  \begin{equation}
    \nabla_\nu X^{\mu_1 \mu_2\ldots \mu_n}=\partial_\nu X^{\mu_1 \mu_2\ldots \mu_n}+\Gamma_{\nu \nu_1 }^{\mu_1}X^{\nu_1 \mu_2\ldots \mu_n}+\Gamma_{\nu \nu_2 }^{\mu_2}X^{\mu_1 \nu_2\ldots \mu_n}+\ldots +\Gamma_{\nu \nu_n }^{\mu_n}X^{\mu_1 \mu_2\ldots \nu_n}\,,
  \end{equation}
  and a similar formula holds for tensors with downstairs indices when replacing $\Gamma\rightarrow - \Gamma$. As particular examples, note that the geometric covariant derivative of a scalar (rank zero tensor) $X$, and that of a vector (rank one tensor) $X^\mu$ simply are given by
  \begin{equation}
    \nabla_\nu X=\partial_\nu X\,,\quad
    \nabla_\nu X^\mu=\partial_\nu X^\mu+\Gamma_{\nu\rho}^\mu X^\rho\,.
  \end{equation}

  Using the Christoffel symbols, one may verify that the covariant derivative of the metric tensor exactly vanishes,
  \begin{equation}
    \label{eq:vanishingmetric}
    \nabla_\lambda g_{\mu\nu}=\nabla_\lambda g^{\mu\nu}=0\,.
  \end{equation}
  Similarly, general covariance dictates that the conservation of the energy momentum-tensor $T^{\mu\nu}$ in curved space takes the form
  \begin{equation}
    \label{eq:tmunucons}
    \nabla_\mu T^{\mu\nu}=0\,,
  \end{equation}
  because combinations such as $\partial_\mu T^{\mu\nu}$ would not transform properly under general coordinate transformations. One may then ask if there are rank two tensors other than the metric and the energy-momentum tensor which have vanishing covariant derivatives. The answer is affirmative, since one can construct a symmetric rank two tensor out of the metric tensor and derivatives as follows: First note that
  \begin{equation}
    \left[\nabla_\mu, \left[\nabla_\nu,\nabla_\lambda\right]\right] V^\sigma+
    \left[\nabla_\nu, \left[\nabla_\lambda,\nabla_\mu\right]\right] V^\sigma+
    \left[\nabla_\lambda, \left[\nabla_\mu,\nabla_\nu\right]\right] V^\sigma=0\,,
  \end{equation}
  where $\left[A_\mu,B_\nu\right]\equiv A_\mu B_\nu-B_\nu A_\mu$, which is known as Bianchi identity. Defining the Riemann curvature tensor $R_{\sigma \rho \mu \nu}=-R_{\rho \sigma \mu \nu}=R_{\rho \sigma \nu\mu}$ through
  \begin{equation}
    \left[\nabla_{\mu},\nabla_\nu\right]V_\rho=R_{\sigma \rho \mu \nu}V^\sigma\,,
  \end{equation}
  the above Bianchi identity can be written as $\nabla_\mu R^\sigma_{\ \rho \nu\lambda}+\nabla_\nu R^\sigma_{\ \rho \lambda\mu}+\nabla_\lambda R^\sigma_{\ \rho \mu\nu}=0$ because the identity must be satisfied for any choice of $V^\rho$. Contracting this result with $g_\sigma^\mu g^{\rho \nu}$ leads to
  \begin{equation}
    \label{eq:biaid}
    \nabla_\mu \left(R^{\mu \lambda}-\frac{1}{2}g^{\mu \lambda} R \right)=0\,,
  \end{equation}
  where the contractions $R_{\mu\nu}\equiv R^{\rho}_{\ \mu \rho \nu}$ known as ``Ricci tensor'' and $R\equiv R^\mu_\mu$ known as ``Ricci scalar'' have been introduced.
  The value of the Ricci scalar indicates the curvature of the space-time described by the metric tensor. $R=0$ implies flat space-time, while positive or negative $R$ imply positive curvature (de-Sitter or ``dS'' space-time) or negative curvature (anti-de-Sitter or ``AdS'' space-time).

  In modern times, calculating the entries of the above tensors does not need to be done by hand. Powerful computer algebra packages (such as the RG\&TC package \cite{RGTC}) exist that let one input the coordinates and metric tensor to automatically calculate these (and other) general relativity objects.

  The Einstein field equations may be constructed out of the above symmetric rank two tensors that have vanishing covariant derivatives, namely (\ref{eq:vanishingmetric}), (\ref{eq:tmunucons}) and (\ref{eq:biaid}), as
  \begin{equation}
    \label{eq:einstein}
    R^{\mu\nu}-\frac{1}{2} R g^{\mu\nu}+\Lambda g^{\mu\nu}=8 \pi G T^{\mu\nu}\,,
    \end{equation}
  where $G$ is Newton's constant, $T_{\mu\nu}$ is the energy-momentum tensor of the matter under consideration and $\Lambda$ is the cosmological constant.
  
  The Einstein equations are the classical equations of motion for the metric tensor $g_{\mu\nu}$. As such, they can be derived from a classical Lagrangian via the standard Euler-Lagrange formalism. In particular, when considering the action
  \begin{equation}
    \label{eq:gravityaction}
    S=\frac{1}{16 \pi G}\int dt d^3x \sqrt{-g}\left[R-2 \Lambda\right]+S_{\rm eff}\,,\quad g={\rm det} g_{\mu\nu}\,,
  \end{equation}
  with $S_{\rm eff}$ containing all matter degrees of freedom,   the equations of motion for the metric follow from the variation
  \begin{equation}
    0=    \frac{\delta S}{\delta g_{\mu\nu}}=-\frac{1}{16 \pi G} \int dt d^3x
    \sqrt{-g}\left[R^{\mu\nu}-\frac{1}{2}g^{\mu\nu} R-g^{\mu\nu}\Lambda - 8 \pi G T_{\mu\nu}\right]\,,
  \end{equation}
  when using
  \begin{equation}
    \label{eq:traforules}
    \frac{\delta \sqrt{-g}}{\delta g_{\mu\nu}}=\frac{1}{2}\sqrt{-g}\, g^{\mu\nu} \,,\quad\quad \frac{\delta R}{\delta g_{\mu\nu}}=-R^{\mu\nu}\,,\quad
    \frac{\delta S_{\rm eff}}{\delta g_{\mu\nu}}=\frac{1}{2} \sqrt{-g}\, T^{\mu\nu}\,.
    \end{equation}

 \chapter{Coordinate systems used}
 \label{chap:coor}

 In this appendix, we collect information about some of the coordinate systems used in this work. For simplicity, the treatment is limited to d=4 space-time dimensions, but generalization to other space-time dimensions should be straightforward.

 \section{Coordinate Transformations}

 Given a reference set of coordinates $x^\mu$ equipped with a metric tensor $g_{\mu\nu}$, let us consider a (not necessarily linear) change of coordinates to new coordinates $\tilde x^\mu$ with $x^\mu=x^\mu\left(\tilde x^\mu\right)$. This coordinate transformation gives rise to a transformation matrix
 \begin{equation}
   \label{eq:Rtrafo}
   R^\mu_{\ \nu}\equiv \frac{\partial x^\mu}{\partial \tilde x^\nu}\,.
 \end{equation}
 Using this transformation matrix, the metric in the new coordinate system is given as
 \begin{equation}
   \label{eq:appmetrictrafo}
   \tilde g_{\mu\nu}=\left(R^T\right)_\mu^{\ \lambda} g_{\lambda \kappa} R^{\kappa}_{\ \nu}\,,
 \end{equation}
 and indeed any vector $v^\mu$ or tensor $t^{\mu\nu}$ can be transformed into the new coordinate system using
 \begin{equation}
   \label{eq:vectrafors}
   \tilde v_\mu=\left(R^T\right)_\mu^{\ \lambda} v_\lambda\,,\quad
   \tilde t_{\mu\nu}=\left(R^T\right)_\mu^{\ \lambda} t_{\lambda \kappa} R^{\kappa}_{\ \nu}\,,
   \end{equation}
 Transformation of quantities with upstairs indices can either be obtained by raising indices using the new metric $\tilde g^{\mu\nu}$ or alternatively through
 \begin{equation}
   \tilde t^{\mu\nu}=\left(R^{-1}\right)^\mu_{\ \lambda} \left(R^{-1}\right)^\nu_{\ \kappa} t^{\kappa\lambda}\,,
 \end{equation}
 where $R^{-1}$ is the inverse of the matrix $R$, also given by
 \begin{equation}
   \left(R^{-1}\right)^{\mu}_{\ \nu}=\frac{\partial \tilde x^\mu}{\partial x^\nu}\,.
   \end{equation}

 As part of the coordinate transformation, also the volume element $dV\equiv \prod_{\mu}d x^\mu$ gets transformed by the Jacobian ${\rm det} R$ of the transformation. Using the result (\ref{eq:appmetrictrafo}), the volume element in any coordinates $\tilde x^\mu$ may be written as
 \begin{equation}
   d\tilde V=\sqrt{|{\rm det}\tilde g|} \, \prod_{\mu}d\tilde x^\mu\,,
 \end{equation}
 where ${\rm det}\tilde g$ is the determinant of the metric tensor $\tilde g_{\mu\nu}$.
 

 \section{Minkowski Space-time}

 Minkowski space-time in $d=4$ is given by the coordinates $x^\mu=(t,x,y,z)$ equipped with the metric tensor $g_{\mu\nu}={\rm diag}\left(-1,1,1,1\right)$. Because this space-time is flat, the Ricci and Riemann tensors vanish identically, as do all the Christoffel symbols.

 \section{Light-cone Coordinates $x^\pm$}

 Light-cone coordinates $x^\pm$ are defined by the coordinate transformation from Minkowski space-time as
 \begin{equation}
   x^\pm\equiv \frac{t\pm z}{\sqrt{2}}\,,
 \end{equation}
 and the inverse transformations are
 \begin{equation}
   t=\frac{x^++x^-}{\sqrt{2}}\,,\quad z=\frac{x^+-x^-}{\sqrt{2}}\,.
   \end{equation}
The full set of light-cone coordinates is thus $\tilde x^\mu=\left(x^+,x,y,x^-\right)$ and the transformation matrix (\ref{eq:Rtrafo}) can be calculated as
 \begin{equation}
   R^{\mu}_{\ \nu}=\left(\begin{array}{cccc}
     \frac{1}{\sqrt{2}} & 0 & 0 & \frac{1}{\sqrt{2}}\\
     0 & 1 & 0 & 0\\
     0 & 0 & 1 & 0\\
     \frac{1}{\sqrt{2}} & 0 & 0 &-\frac{1}{\sqrt{2}}
   \end{array}\right)\,,
 \end{equation}
 such that the light-cone coordinate metric becomes
 \begin{equation}
   \tilde g_{\mu\nu}=\left(\begin{array}{cccc}
     0 & 0 & 0 & -1\\
     0 & 1 & 0 & 0\\
     0 & 0 & 1 & 0\\
     -1 & 0 & 0 & 0
   \end{array}\right)\,.
   \end{equation}
Hence the line-element for the light-cone metric is
 \begin{equation}
   ds^2=\tilde g_{\mu\nu}d\tilde x^\mu d\tilde x^\nu=-2 dx^+ dx^-+dx^2+dy^2\,.
 \end{equation}
 For this metric, all Christoffel symbols vanish and the space-time is flat.

 \section{Milne coordinates}
\label{sec:milnecoo}
 
 \subsection{Minkowski to Milne}

 Milne coordinates $\tau,\xi$ are defined by the coordinate transformation from Minkowski space-time as
 \begin{equation}
   \label{eq:milne}
   \tau\equiv \sqrt{t^2-z^2}\,,\quad
   \xi\equiv {\rm arctanh}(z/t)\,,
 \end{equation}
 and the inverse transformations are
 \begin{equation}
  t=\tau \cosh\xi\,,\quad z=\tau \sinh\xi\,.
 \end{equation}
 The full set of Milne coordinates is thus $\tilde x^\mu=\left(\tau,x,y,\xi\right)$ and the transformation matrix (\ref{eq:Rtrafo}) can be calculated as
 \begin{equation}
   R^{\mu}_{\ \nu}=\left(\begin{array}{cccc}
     \cosh\xi & 0 & 0 & \tau \sinh\xi\\
     0 & 1 & 0 & 0\\
     0 & 0 & 1 & 0\\
     \sinh\xi & 0 & 0 &\tau \cosh\xi
   \end{array}\right)\,,
   \end{equation}
 such that the Milne metric becomes
 \begin{equation}
   \tilde g_{\mu\nu}={\rm diag}\left(-1,1,1,\tau^2\right)\,.
 \end{equation}
 Hence the line-element for the Milne metric is
 \begin{equation}
   ds^2=\tilde g_{\mu\nu}d\tilde x^\mu d\tilde x^\nu=-d\tau^2+dx^2+dy^2+\tau^2 d\xi^2\,.
   \end{equation}
 For this metric, the non-vanishing Christoffel symbols are
 \begin{equation}
   \tilde \Gamma^\xi_{\xi \tau}=\tilde \Gamma^\xi_{\tau \xi}=\frac{1}{\tau}\,,\quad \tilde \Gamma^\tau_{\xi\xi}=\tau\,.
  \end{equation}
Despite the fact that there are non-vanishing Christoffel symbols, it can be shown that Ricci and Riemann tensors are still identically zero, implying that Milne space-time is flat.

\subsection{Light-cone to Milne}

Sometimes it is convenient to consider the transformation from light-cone coordinates $x^\mu=\left(x^+,x^-\right)$ to Milne coordinates $\tilde x^\mu=\left(\tau,\xi\right)$ which is defined by the transformations
\begin{equation}
  \tau=\sqrt{2 x^+ x^-}\,,\quad \xi=\frac{1}{2}\ln \left(\frac{x^+}{x^-}\right)\,,
\end{equation}
and the inverse transformations
\begin{equation}
  x^+=\frac{\tau}{\sqrt{2}}e^{\xi}\,,\quad
  x^-=\frac{\tau}{\sqrt{2}}e^{-\xi}\,.
\end{equation}
The transformation matrix is given by
\begin{equation}
  R^\mu_{\ \nu}=\frac{1}{\sqrt{2}}\left(\begin{array}{cc}
    e^\xi & \tau e^{\xi}\\
    e^{-\xi}& -\tau e^{-\xi}
  \end{array}\right)\,.
\end{equation}
Of course, the Milne metric and Christoffel symbols are the same as those in the previous subsection.
The transformation matrix implies that the energy-momentum tensor in Milne and light-cone coordinates are related as in (\ref{eq:vectrafors}), leading to 
\begin{equation}
  \label{eq:tmntrafo}
 \tilde T_{\tau\tau}=\frac{1}{2} e^{-2 \xi}T_{--}+T_{+-}+\frac{1}{2} e^{2 \xi}T_{++}\,.
  \end{equation}

\addtocontents{toc}{\endgroup}
\end{appendices}



\bibliographystyle{hunsrt}
\bibliography{pp-hydro}

\printindex

\end{document}